UNIVERSIDAD TECNOLÓGICA DE PANAMÁ

FACULTAD DE CIENCIAS Y TECNOLOGÍA

COORDINACIÓN POSTGRADO Y EXTENSIÓN

DOCTORADO REGIONAL EN CIENCIAS FÍSICAS

"EVOLUCIÓN DE LA FORMACIÓN DE BARRAS EN GALAXIAS EN LOS

ÚLTIMOS 6 GIGA-AÑOS"

TRABAJO PRESENTADO POR:

MANUEL ALEJANDRO CHACÓN

ASESOR:

RODNEY DELGADO SERRANO

CO-ASESOR:

BERNARDO CERVANTES SODI

TRABAJO DE GRADUACIÓN PRESENTADO EN LA UNIVERSIDAD

TECNOLÓGICA DE PANAMÁ COMO REQUISITO PARA OPTAR POR EL

TÍTULO DE DOCTOR EN CIENCIAS FÍSICAS

PANAMÁ, REPÚBLICA D EPANAMÁ

2025

# Agradecimientos

Son muchas las personas que me han ayudado a lo largo de estos años en los que desarrollé mi trabajo de doctorado y que han contribuido, en mayor o menor medida, a la realización de esta tesis.

En primer lugar, quiero agradecer a mi director de tesis, Dr. Rodney Delgado Serrano, por brindarme la oportunidad de realizar este trabajo, por su cordialidad y apoyo constante. A mi co-asesor, Dr. Bernardo Cervantes Sodi, por confiar en mí desde el principio y permitirme desarrollar un proyecto de tesis que fue tomando forma a medida que los resultados emergían.

Agradezco a la Secretaría Nacional de Ciencia, Tecnología e Innovación (SENACYT) de la República de Panamá, por el apoyo a través del programa de Movilidad e Investigación en el marco del proyecto "Formación de Capacidades Investigativas en la Línea de Astronomía Extragaláctica". También agradezco el apoyo del Programa de Apoyo a Proyectos de Investigación e Innovación Tecnológica (PAPIIT) IN108323 (Galaxias de bajo brillo superficial, medio ambiente, evolución y el caso de galaxias barradas), de la Dirección General de Asuntos del Personal Académico (DGAPA) de la Universidad Nacional Autónoma de México (UNAM). Agradezco al Departamento de Física de la Universidad Técnica Federico Santa María (USM) en Valparaíso, donde trabajé bajo la supervisión del Dr. Hugo Alarcón, a la Dra. Yara Jaffé y al Dr. Diego Pallero.

A mis compañeros de la Dirección Nacional de Ciencias Espaciales de la Universidad Tecnológica de Panamá (DINACE-UTP), quienes, de una u otra manera, me han apoyado como parte de un equipo de trabajo, les extiendo mi gratitud.

Quiero expresar mi más profundo agradecimiento a mi familia por el apoyo y cariño que me han brindado a lo largo de estos años. Con todo mi amor, a mi hijo Thiago, quien ha sido mi mayor fuente de felicidad y energía. Especialmente agradezco a Ariadna, por compartir su vida conmigo y ser mi compañera incondicional. Tu apoyo constante y tu ánimo para seguir adelante, con la única recompensa de verme feliz y realizado, han sido fundamentales. Esta tesis no hubiese sido posible sin ustedes.  En especial, agradezco a mi madre Adelina, quien, con su sacrificio y fortaleza, ha logrado que sus dos hijos pudieran llevar a cabo sus sueños. A mi hermano Ulises, por su capacidad de hacer feliz a los demás. A mi tía Olimpia y mi prima Julia, y en especial a mi tío Javier, quien me ha apoyado desde que inicié mis estudios. Quiero que sepan que mi mayor alegría es poder contar con su cariño en todo momento, razón por la cual les dedico este trabajo

Deseo honrar la memoria de mis abuelos, en cuyo amor y cuidado me crié. A mi suegra Petra y, en especial, a mi tío Domiciano, que, aunque ya no estén con nosotros, dejaron profundas enseñanzas que llevo marcadas en mi vida.

Finalmente, agradezco a todas aquellas personas que, en algún momento de mi vida, se cruzaron en mi camino y contribuyeron a que yo sea quien soy.

# Dedicatoria

*A mi familia,*

*a mi hijo Thiago,*

*el sol de mi vida.*

# Resumen

En este trabajo se investiga la evolución de la formación de barras en galaxias de disco a lo largo de los últimos 6 giga-años, analizando la fracción de galaxias barradas en función de parámetros estructurales. Se utiliza una muestra representativa de galaxias de campo en diversas épocas cósmicas. Se analizaron galaxias espirales y lenticulares, tanto locales como distantes, empleando métodos para la caracterización de las barras, como isofotas elípticas y descomposición de Fourier. Los resultados ofrecen una visión detallada de la evolución de las barras en galaxias de disco, con énfasis en su relación con la masa estelar, el índice de color y el cociente bulbo-disco ($B/T$).

En las galaxias locales, la fracción de barras aumenta notablemente con la masa estelar, alcanzando un máximo en las galaxias más masivas. Esto sugiere que la estabilidad dinámica de los discos es crucial para la formación de barras. Las galaxias más evolucionadas y de colores más rojos presentan una mayor prevalencia de barras, lo que indica que estas estructuras son más comunes en sistemas que han tenido tiempo suficiente para estabilizar sus discos. En galaxias distantes, la fracción de barras es significativamente menor, lo que sugiere que, en el universo temprano, las condiciones dinámicas eran menos favorables para la formación de barras debido a la mayor inestabilidad de los discos, lo que dificultaba tanto la formación como el mantenimiento de barras.

El análisis en función del índice de color revela que, en las galaxias locales, las barras son más comunes en sistemas de color rojo, asociados a una evolución estelar pasiva y estabilidad en los discos. En las galaxias distantes, las barras son escasas en todo el espectro de colores, aunque se observa una ligera alza en las galaxias más azules, lo que indica que en el universo temprano la formación de barras estaba menos condicionada por la madurez estelar.

Al considerar la fracción de barras en función de $B/T$, se observa que las galaxias con bulbos menos prominentes y discos dominantes son más propensas a desarrollar barras, tanto en galaxias locales como distantes. Sin embargo, en las galaxias distantes, incluso aquellas con bajo $B/T$, a fracción de barras es considerablemente menor, lo que refuerza la idea de una menor estabilidad en los discos galácticos en épocas tempranas.

Estos resultados aportan evidencia crucial sobre los mecanismos que dirigen la formación de barras en diferentes etapas cósmicas, sugiriendo que la estabilización de los discos galácticos juega un papel central en la evolución morfológica.

# Abstract

This work investigates the evolution of bar formation in disk galaxies over the past 6 giga-years, analyzing the fraction of barred galaxies as a function of structural parameters. A representative sample of field galaxies from various cosmic epochs is used. Spiral and lenticular galaxies, both local and distant, were analyzed using methods for bar characterization, such as elliptical isophotes and Fourier decomposition. The results provide a detailed view of the evolution of bars in disk galaxies, with emphasis on their relationship with stellar mass, color index, and the bulge-to-total ratio ($B/T$).

In local galaxies, the fraction of bars increases significantly with stellar mass, reaching a maximum in the most massive galaxies. This suggests that the dynamic stability of disks is crucial for bar formation. More evolved, redder galaxies show a higher prevalence of bars, indicating that these structures are more common in systems that have had enough time to stabilize their disks. In distant galaxies, the fraction of bars is significantly lower, suggesting that in the early universe, dynamic conditions were less favorable for bar formation due to greater disk instability, which hindered both the formation and maintenance of bars.

The color index analysis reveals that in local galaxies, bars are more common in red systems, associated with passive stellar evolution and disk stability. In distant galaxies, bars are scarce across the color spectrum, although a slight increase is observed in bluer galaxies, indicating that in the early universe, bar formation was less conditioned by stellar maturity.

When considering the bar fraction as a function of $B/T$, it is observed that galaxies with less prominent bulges and dominant disks are more likely to develop bars, both in local and distant galaxies. However, in distant galaxies, even those with low $B/T$, the bar fraction is considerably lower, reinforcing the idea of less stable galactic disks in early epochs.

These results provide crucial evidence about the mechanisms driving bar formation at different cosmic stages, suggesting that the stabilization of galactic disks plays a central role in morphological evolution.

# Contenido

## Índice General

# Índice de Tablas

# Índice de Figuras

# Lista de Abreviaturas

**ACS** – Advanced Camera for Surveys

**AGN** – Active Galactic Nuclei

**ASTROPY** – Astronomy Python Library

**BASS** – Beijing-Arizona Sky Survey

**CANDELS** – Cosmic Assembly Near-infrared Deep Extragalactic Legacy Survey

**CEERS** – Cosmic Evolution Early Release Science Survey

**COSMOS** – Cosmic Evolution Survey

**DECam** – Dark Energy Camera

**EROs** – Early Release Observations

**ESO** – European Southern Observatory (Observatorio Europeo Austral)

**EW[Hα]** – Ancho Equivalente de la línea Hα (Equivalent Width of Hα)

**F160W** – Filtro óptico del HST

**F444W** – Filtro en el infrarrojo cercano del JWST

**FWHM** – Full Width at Half Maximum

**GOODS** – Great Observatories Origins Deep Survey

**HST** – Hubble Space Telescope

**ISM** – interstellar medium

**JWST** – James Webb Space Telescope

**LSS** – Large-scale structure of the Universe

**MPIA** – Max Planck Institute for Astronomy

**NASA** – National Aeronautics and Space Administration

**NIR** – Near-Infrared

**NIRCam** – Near Infrared Camera

**P2DFFT** – Parallel 2D Fast Fourier Transform

**PRIMER** – Public Release Imaging for Extragalactic Research

**SDSS** – Sloan Digital Sky Survey

**SEXTRACTOR** – Source Extraction

**SFR** – Tasa de Formación Estelar

**SMACS** – Southern Massive Cluster Survey

**TNG50** – The Next Generation simulations at 50 Mpc

# Capítulo 1

## Introducción

El estudio de los cielos ha sido una constante a lo largo de la historia de la humanidad. Desde las primeras civilizaciones, que intentaron interpretar los fenómenos celestes a través de mitos y creencias religiosas, hasta el desarrollo de teorías científicas que han transformado nuestra comprensión del cosmos, el interés por el universo ha sido inquebrantable. La transición de una visión geocéntrica a un modelo heliocéntrico marcó un punto de inflexión en la historia del pensamiento humano. A través del trabajo de figuras como Copérnico [1], Galileo y Newton, se consolidó la idea de que las leyes físicas que rigen la Tierra son las mismas que gobiernan el universo. Este cambio de paradigma fue esencial para el desarrollo de la astrofísica moderna, proporcionando una base científica para el estudio de los cuerpos celestes y permitiendo una visión más profunda del lugar de la humanidad en el cosmos. Aunque el avance científico permitió explicar el movimiento de los cuerpos celestes mediante la gravedad, no eliminó completamente la idea de una causa trascendente, como reconoció el propio Newton en sus reflexiones [2]. Hoy en día, la astrofísica sigue desafiando y expandiendo los límites del conocimiento, buscando respuestas a preguntas fundamentales sobre el origen, la estructura y el destino del universo.

Con el progreso de las observaciones nocturnas, se descubrieron nuevas y distantes estructuras que inicialmente se pensaban parte de la Vía Láctea. A principios del siglo XX, Edwin Hubble reveló un reino completamente nuevo de estos objetos, denominados

"galaxias". Este descubrimiento marcó un hito en nuestra comprensión del universo y dio origen a una nueva rama de la astronomía llamada Astronomía Extragaláctica. Este avance despertó un gran interés, impulsando a los astrónomos a investigar la formación y evolución de objetos a gran escala en el universo.

En este capítulo, se presenta y justifica la necesidad de la investigación realizada en esta tesis. En la Sección 1.1, se discute la motivación detrás del estudio y su relevancia. La Sección 1.2 ofrece una visión general sobre la formación y evolución de las galaxias, proporcionando el contexto necesario para entender el tema de investigación. La Sección 1.3 detalla los sistemas de clasificación de galaxias, fundamentales para nuestro análisis. En la Sección 1.4, se examinan las galaxias en el universo visible, con un enfoque en las observaciones más relevantes. Finalmente, en la Sección 1.5, se explora la importancia de las barras en la evolución de las galaxias, destacando su impacto en la dinámica galáctica.

## 1.1 Motivación

La exploración de la estructura y evolución de las galaxias ha sido un tema central en la Astronomía y la cosmología. En particular, el estudio de las barras galácticas ha demostrado ser crucial para comprender la formación y evolución de las galaxias espirales. Las barras son estructuras alargadas de estrellas, gas y polvo que se extienden a través del centro de muchas galaxias espirales y su presencia puede influir significativamente en la dinámica y la evolución de la galaxia en su conjunto. A pesar de su importancia, aún no se conoce completamente la fracción de galaxias que albergan barras y cómo esta fracción evoluciona a lo largo del tiempo cósmico, lo que implica la necesidad de estudiarla más a fondo. Por lo tanto, la motivación principal de este proyecto de tesis doctoral es abordar estas cuestiones mediante un enfoque observacional. Se busca determinar la fracción de galaxias barradas en el universo local y comparar estos resultados con datos de la fracción de galaxias barradas en el universo distante. Esto permitirá una comprensión más completa de cómo la presencia y la frecuencia de las barras galácticas varían a lo largo del tiempo y cómo esto se relaciona con la formación y evolución de las galaxias de campo de masa intermedia.

## 1.2 Sistemas de Clasificación de Galaxias

Es posible clasificar a las galaxias en dos grandes grupos: las galaxias regulares y las galaxias peculiares. Las galaxias regulares son aquellas cuyos subsistemas son concéntricos y coplanares, y cuya emisión energética proviene principalmente de procesos térmicos. Por otro lado, las galaxias peculiares son sistemas que han sido afectados por interacciones gravitacionales o hidrodinámicas intensas, o que contienen un núcleo activo (AGN), emitiendo la mayor parte de su energía a través de procesos no térmicos. Se estima que estas últimas representan entre el 5% y el 10% de la población total de galaxias conocidas en el universo local [3]. En esta tesis, solo se incluyen las galaxias regulares en el análisis.

A finales del siglo XVIII, el astrónomo William Herschel llevó a cabo uno de los primeros intentos sistemáticos de clasificar las nebulosas, que eran como se conocían entonces las galaxias, en términos de su brillo y tamaño. Utilizó un sistema de letras mayúsculas y minúsculas para etiquetarlas, como se muestra en la Tabla 1.1 [4]. Esta clasificación fue un paso importante en la comprensión temprana de la diversidad y la naturaleza de estas misteriosas nubes cósmicas. Posteriormente, el hijo de Herschel, John

Herschel, amplió y refinó este sistema, contribuyendo así al desarrollo de la taxonomía de las nebulosas y sentando las bases para futuras investigaciones en el campo de la galaxia y la cosmología.

| B = Bright | v = very |
|:---:|:---:|
| F = Faint | c = considerable |
| L = Large | p = pretty |
| S = Small | e = extremely |

Tabla 1.1: Sistema de clasificación de "nebulosas" de Herschel.

Sin embargo, el primer sistema de clasificación que ganó aceptación a nivel mundial fue publicado por Hubble. Apodado como "Turning Fork" o la "Secuencia de Hubble", este sistema de clasificación organiza las galaxias en algunas categorías amplias: elípticas, espirales e irregulares. En aquel entonces, la astronomía se limitaba al espectro óptico, por lo tanto, las galaxias solo se estudiaban visualmente o mediante sus espectros. La clasificación visual de la morfología de las galaxias consistía en observarlas utilizando imágenes de banda óptica (que eran sensibles a la región azul del espectro de luz) y categorizarlas según el juicio del observador sobre su apariencia. El sistema de Hubble tiene sus desventajas, y se han propuesto numerosas modificaciones al esquema original [5][6][7]. Sin embargo, sigue siendo el método de clasificación más popular en la actualidad. Es importante destacar que, hasta el momento, no se ha encontrado un método de clasificación ideal.

### 1.2.1 Secuencia de Hubble

La primera y más famosa clasificación fue hecha por Hubble en 1926, conocida como la secuencia de Hubble. En la primera mitad del siglo XX, Edwin Hubble clasificó las morfologías de galaxias observando la apariencia óptica de imágenes en placas fotográficas [8]. En la primera versión, Hubble propuso tres clases generales de galaxias: Elípticas (E), Espirales (S o SB) e irregulares (Ir). Más tarde en 1936, Hubble mejora el esquema que se muestra en la Figura 1.1, pensando en principio como una secuencia evolutiva de galaxias. Hubble agregó el tipo de galaxia llamada lenticular (S0), y lo presenta como un estado intermedio necesario entre las galaxias elípticas y galaxias espirales [9]. Así, el esquema final de la Secuencia de Hubble consiste en cuatro tipos principales de galaxias: las galaxias elípticas (E); las galaxias lenticulares (S0), las galaxias espirales barradas (SB) y espirales no barradas (S) y las galaxias irregulares (Ir).

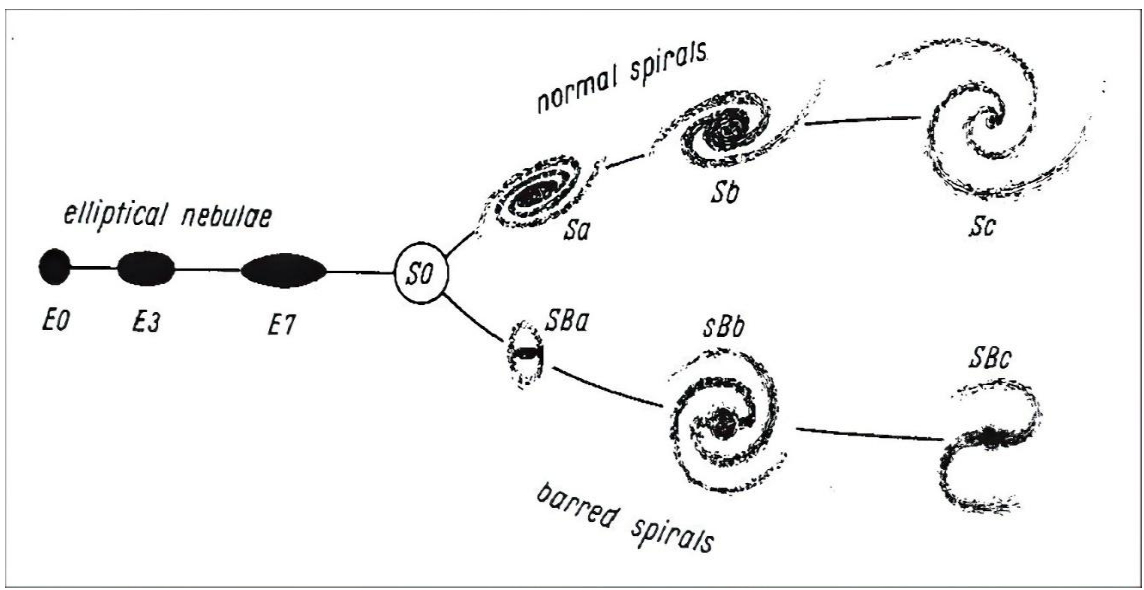

Figura. 1.1. La versión de 1936 del diagrama de Tuning Fork de Hubble, revisada para incluir galaxias lenticulares (S0), lo que facilita la clasificación de las galaxias elípticas y espirales. Imagen tomada de [6].

En el extremo izquierdo de la Secuencia de Hubble (Figura 1.1) se ubican las galaxias elípticas (*E*) que son los sistemas más suaves y sin estructura, con una componente esférica dominante y sin disco estelar. Estas galaxias varían en forma de proyección, desde redondas ($E_0$) hasta bastante alargadas ($E_7$) (Figura 1.2), y se designan por $En$ donde $n$ es un número entero que indica el grado de elipticidad $e = 1 - (b/a)$, donde $a$ es el semieje mayor y $b$ el semieje menor. Por convenio, $n$ es 10 veces la elipticidad de la galaxia, redondeado al número entero más cercano, a través de la siguiente ecuación $n = 10\,[1 - (b/a)]$.

Luego, encontramos en el diagrama las galaxias lenticulares ($SO$), las cuales son una transición entre las galaxias elípticas y las espirales. Pueden distinguirse dos componentes, un bulbo y un disco poco aplanado y sin estructura espiral. También, se introduce la definición de galaxias lenticulares no barradas ($SO$) y galaxias lenticulares barradas ($SBO$).

La clasificación morfológica de las galaxias espirales ($S$) se caracteriza por la presencia de un patrón espiral superpuesto a una estructura de disco, generalmente alrededor de un núcleo brillante. El tamaño del núcleo, la estrechez del enrollamiento de los brazos alrededor del núcleo y la densidad de los brazos espirales constituyen los criterios utilizados para clasificar estas galaxias. Así, las galaxias clasificadas como $Sa$ tienen bulbos conspicuos y brazos suaves y altamente enrollados, mientras que aquellas clasificadas como $Sc$ tienen condensaciones centrales de brillo pequeñas y brazos menos enrollados y muy resueltos.

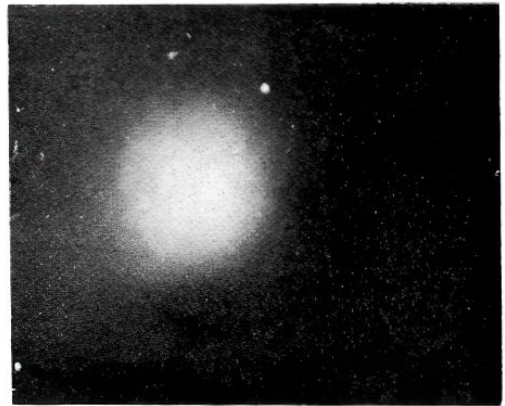
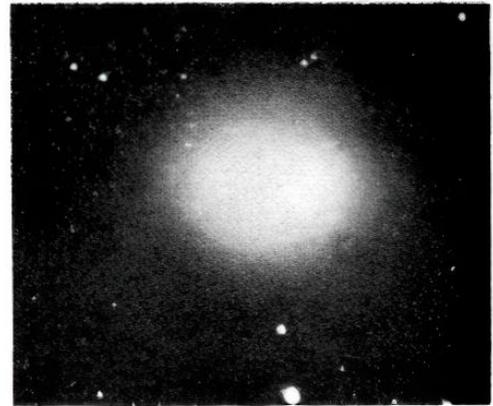

EO NGC 3379    E2 NGC 221 (M32)

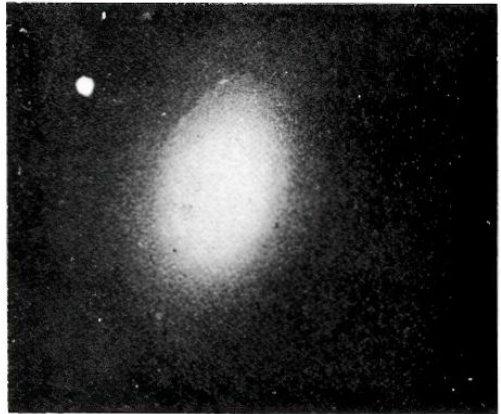
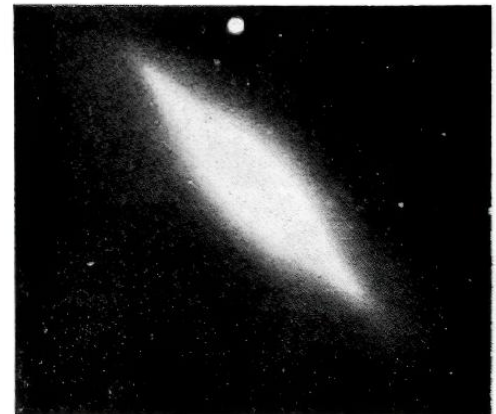

E5 NGC 4621 (M59)    E7 NGC 3115

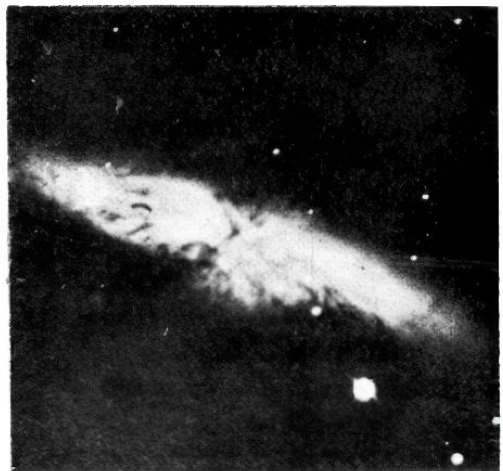
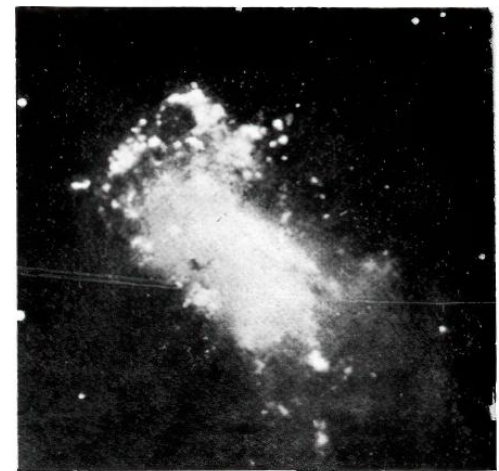

NGC 3034 (M82)    NGC 4449

Figura 1.2. Galaxias Elípticas e Irregulares publicadas en "The Realm of the Nebula" publicado en 1936. Imagen tomada de [9].

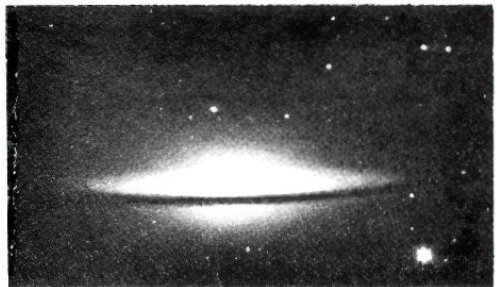
Sa NGC 4594

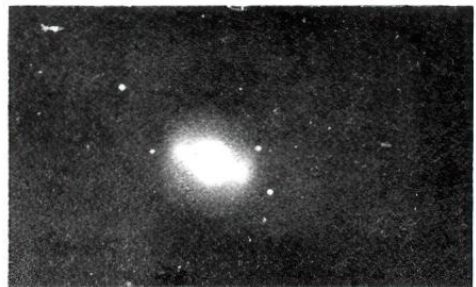
SBa NGC 2859

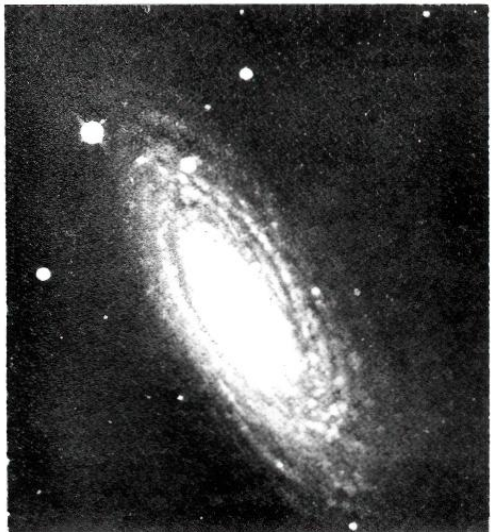
Sb NGC 2841

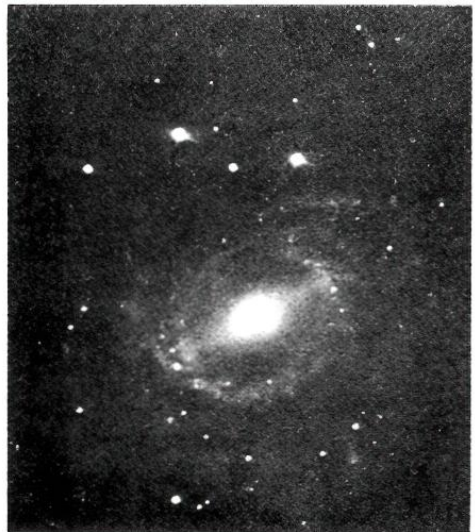
SBb NGC 5850

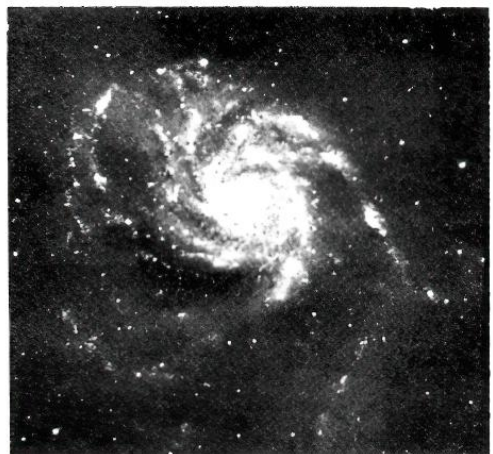
Sc NGC 5457 (MIOI)

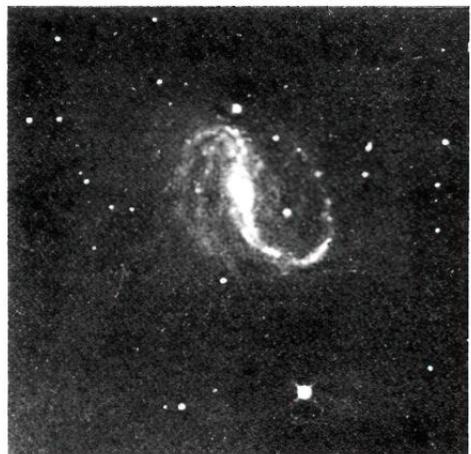
SBc NGC 7479

Figura 1.3. Galaxias espirales normales y barradas publicadas en el trabajo de Hubble "The Realm of the Nebula" publicado en 1936. Imagen tomada de [9].

Como puede verse en la Figura 1.3, esta categoría tiene dos grupos: espirales normales (*S*) y espirales barradas (*SB*). La barra es un sistema estelar suave y alargado que se asemeja a un listón rígido que rota en el centro del disco y de sus extremos emanan los brazos espirales.

Las galaxias irregulares están separadas del Tuning Fork. Las galaxias (*Ir*) típicamente muestran una estructura irregular y fragmentada, sin evidencia de simetría rotacional ni presencia de brazos espirales definidos (Figura 5). Esta tendencia continúa hacia objetos muy débiles en los cuales las estrellas jóvenes se encuentran dispersadas caóticamente.

### 1.2.2 Sistema Clasificación de de Vaucouleurs

Una modificación al sistema de clasificación de Hubble fue propuesta por de Vaucouleurs [10], quien introdujo divisiones más finas a las categorías de clasificación de Hubble, especialmente a la clase amplia de galaxias espirales de Hubble. El sistema mantiene la mayoría de las mismas clases de clasificación, pero además de agregar más detalle, puede ser visualizado tridimensionalmente (Figura 1.4). Este nuevo "volumen de clasificación" se explica a continuación.

La presencia de una barra se denota con una "*B*", mientras que su ausencia con una "*A*". Luego, una galaxia espiral se representa por "*SB*" o "*SA*" cuando tiene o no tiene una barra en su centro, respectivamente. Se adopta una representación "SBA" como clase intermedia, cuando la galaxia tiene características mixtas (débilmente barrada). De manera similar, las galaxias lenticulares podrían ser "*SA0*" (sin barra), "*SB0*" (barrada) o "*SAB0*" (clase intermedia). Sin embargo, cuando es imposible determinar si la galaxia tiene una barra o no, se nota solo con una "*S*" en el caso de una galaxia espiral o con "*S0*" si la galaxia es lenticular.

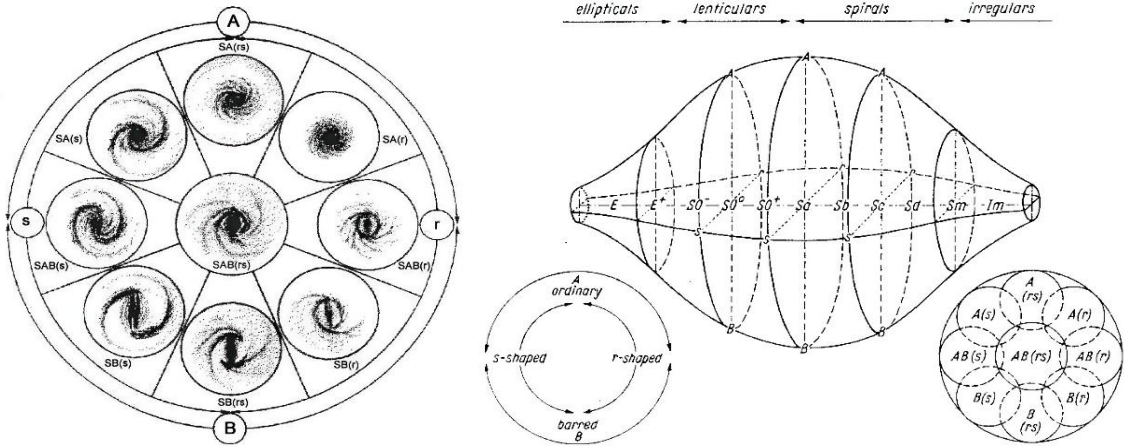

Figura 1.4. La representación tridimensional de la revisión de de Vaucouleurs en 1959 al diagrama de Hubble. Imagen tomada de [31].

La presencia de un anillo se identifica con "(*r*)", mientras que su ausencia se señala con "(*s*)". Un tipo intermedio se denota como "(*rs*)". Sin embargo, de Vaucouleurs precisó que el símbolo "(*s*)" advierte, más específicamente, sobre la presencia de espirales (un tipo "en forma de *S*"). Luego, "la distinción entre las dos familias *A* y *B* y entre las dos variedades (*r*) y (*s*) está más claramente marcada en la etapa de transición *S0/a* entre las clases *S0* y *S*. Desaparece en la etapa de transición entre *E* y *S0*, por un lado, y en la etapa de transición entre *S* e *I*, por otro".

Posteriormente, de Vaucouleurs introdujo un nuevo parámetro numérico llamado el tipo T, que está correlacionado con el eje principal de su sistema de clasificación tridimensional (ver Tabla 1.2). Las definiciones para los valores de T se derivan directamente de las definiciones de los ejes principales de de Vaucouleurs establecidas en [11]. El parámetro T varía desde T = -6, para las elípticas compactas (*cE*), hasta T = 10, para las irregulares magallánicas (*Im*). Otra característica de este esquema de clasificación fue asignar a cada etapa a lo largo del eje principal un valor entero numérico entre -6 y 11. Las galaxias E se representan con los valores -6 a -4, las lenticulares de -3 a -1, las espirales de 0 a 9 y las irregulares de 10 a 11 para un enfoque más cuantitativo de la clasificación. Además del enfoque cuantitativo para la clasificación de galaxias, de Vaucouleurs [12] también introdujo parámetros medibles que muestran un aumento o disminución medio consistentes a lo largo de la secuencia de clasificación actual. Las características incluyen las relaciones bulbo-disco, la luminosidad integrada en la banda B, la relación de diámetros de apertura, magnitudes totales o efectivas, brillo superficial medio e índice de hidrógeno. Este último se refiere a la cantidad de hidrógeno neutro presente en una galaxia, típicamente medido a través de la emisión en la línea de 21 cm del hidrógeno HI, lo que proporciona información sobre el contenido de gas y su relación con la formación estelar y la evolución galáctica [13].

| Clase Tipo *T* | −6 | −5 | −4 | −3 | −2 | −1 | 0 | 1 | 2 | 3 | 4 | 5 | 6 | 7 | 8 | 9 | 10 |
|---|---|---|---|---|---|---|---|---|---|---|---|---|---|---|---|---|---|
| Clase de de Vaucouleurs | cE | E | E⁺ | S0⁻ | S0⁰ | S0⁺ | S0/a | Sa | Sab | Sb | Sbc | Sc | Scd | Sd | Sdm | Sm | Im |
| Clase aproximada Hubble | | E | | | S0 | | S0/a | Sa | Sab | Sb | Sbc | | Sc | | Sdlrr | Irr I | |

Tabla 1.2. Clasificación de galaxias de de Vaucouleurs con el tipo T correspondiente de de Vaucouleurs y la aproximación con la Clasificación de Hubble.

En esta investigación utilizamos el esquema presentado por Hubble, y es nuestro interés es el poder establecer una estadística de la fracción de barras en galaxias de disco. En los siguientes subtítulos hablaremos de la distribución de las galaxias, distancia y la implicación de las barras en las galaxias.

## 1.3 Galaxias en el Universo Visible

El estudio de las galaxias en el universo visible, a través de diferentes corrimientos al rojo, permite comparar su evolución a lo largo del tiempo cósmico. Las galaxias cercanas (universo local) muestran estructuras maduras, mientras que las distantes (universo distante) revelan etapas formativas y procesos iniciales de formación estelar. Las galaxias cercanas revelan la configuración actual de aisladas de campo, cúmulos, supercúmulos y vacíos, mientras que las galaxias distantes nos muestran cómo estas estructuras se formaron y evolucionaron en el pasado. Esta comparación es esencial para comprender cómo las propiedades y morfologías galácticas cambian con el tiempo, proporcionando una visión integral de la evolución del universo.

### 1.3.1 Distribución de las Galaxias en el Universo Observable

Las galaxias son sistemas fundamentales para proyectar la distribución de materia del universo a gran escala. Por medio de observaciones astronómicas se han identificado galaxias en una gran variedad de formas, desde estructuras simples hasta complejas, evidenciando su formación y evolución.

La acreción de materia es facilitada por la red de materia oscura [14], como se ilustra en la Figura 1.5. En esta red, la materia oscura forma filamentos que actúan como conductos para la acumulación de materia bariónica, la cual eventualmente forma galaxias. Aunque en la Figura 1.5 solo se muestra la distribución de materia oscura, esta estructura subyacente es clave en la evolución cósmica, ya que las galaxias tienden a formarse a lo largo de los filamentos de materia oscura. Las más masivas suelen ubicarse en los nodos donde se interceptan varios filamentos [15][16].

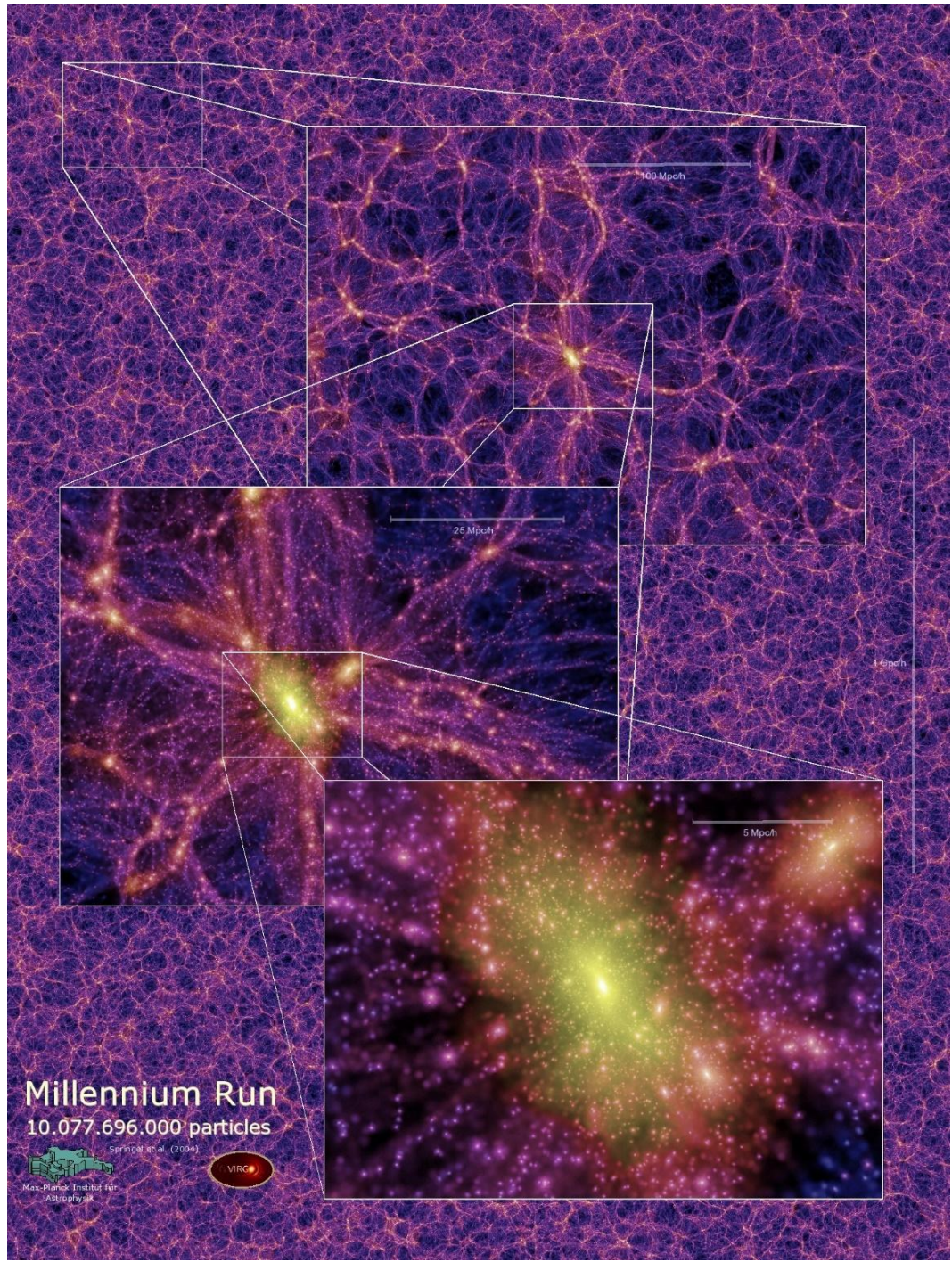

Figura 1.5. Simulación del Millennium Run, que muestra la estructura a gran escala del universo. La imagen destaca el patrón de red cósmica formada por filamentos de materia oscura, con cúmulos de galaxias concentrándose en los nodos de la estructura. Esta simulación es una de las más grandes y detalladas realizadas para comprender la formación y evolución de las estructuras en el universo, desde la escala de 125 Mpc/h hasta 5 Mpc/h. Imagen de [15] (©Simulación Millenium).

La manera en que las galaxias se distribuyen en el universo está directamente relacionada con esta red de materia oscura. Los filamentos de materia oscura actúan como conductos a través de los cuales el gas fluye hacia las galaxias, facilitando su crecimiento y distribución. Las estructuras a gran escala del universo, como cúmulos y supercúmulos de galaxias, se forman en las intersecciones de estos filamentos, evidenciando la influencia crucial de la red de materia oscura en la organización del cosmos. Entre los filamentos y nodos se encuentran los vacíos (voids), que son regiones de baja densidad donde hay muy pocas galaxias [17].

El gas que fluye por la red es capturado por las galaxias, aumentando así su masa. Existen dos modos principales de acreción: el modo caliente y el modo frío [18]. Uno de los principales desafíos de la acreción es que el gas debe estar a una temperatura suficientemente baja para ser capturado por la galaxia. En el modo caliente, el gas entrante tiene una temperatura ($T_{gas}$) mayor que la temperatura virial de la galaxia ($T_{virial}$), es decir, $T_{gas} > T_{virial}$. La temperatura virial es la temperatura a la cual el gas, bajo la influencia de la gravedad de la galaxia, puede alcanzar el equilibrio térmico. El gas en el modo caliente se calienta por choques a medida que cae en los pozos de potencial gravitacional de la galaxia y debe enfriarse, principalmente a través de la radiación, antes de integrarse en la galaxia. Este modo de acreción es casi esférico y prevalece en entornos de alta densidad a bajos desplazamientos al rojo ($z < 1$). En el modo frío, el gas acretado se canaliza a lo largo de los filamentos de la red de materia oscura, permitiendo que las galaxias atraigan gas desde grandes distancias. Aunque el gas se calienta debido a los choques y la compresión adiabática, irradia su calor rápidamente y su temperatura nunca alcanza ($T_{virial}$). Este modo de acreción es dominante a altos desplazamientos al rojo y en entornos de baja densidad a ($z \sim 0$) [19]. Es importante notar que estos conceptos se basan en investigaciones y modelos reconocidos en la literatura científica. Sin embargo, la ciencia está en constante evolución, y estos modelos pueden ser ajustados a medida que se dispone de más datos y mejores herramientas de análisis.

De lo descrito en esta sección, podemos dividir la distribución de galaxias en el universo observable en tres grupos:

Las **galaxias en cúmulos** se agrupan en grandes conjuntos llamados cúmulos galácticos, que pueden contener desde unas pocas docenas hasta miles de galaxias, todas unidas por la gravedad. En el corazón de estos cúmulos, a menudo se encuentran galaxias elípticas gigantes, rodeadas por galaxias espirales y galaxias irregulares más pequeñas. Las **galaxias de campo** se distribuyen de manera dispersa en el espacio, sin formar parte de cúmulos o supercúmulos galácticos. Estas galaxias se encuentran en regiones de baja densidad, a diferencia de las agrupadas en cúmulos o supercúmulos. Su distribución tiende a ser relativamente uniforme a gran escala. En cuanto a las **galaxias en vacíos**, residen en regiones extremadamente grandes del espacio con una densidad de galaxias mucho menor que el promedio del universo. Estas áreas, conocidas como vacíos, pueden tener dimensiones de cientos de millones de años luz y las galaxias en ellos suelen estar más separadas entre sí que en otras partes del universo. A pesar de la baja densidad de galaxias, los vacíos aún pueden contener materia oscura y gas interestelar.

De los grupos mencionados, es nuestro interés en el desarrollo de este trabajo ver el comportamiento de las galaxias de campo tanto en universo cercano como el lejano.

### 1.3.2. Corrimiento al Rojo y Ley de Hubble

Cuando se observa una galaxia distante, es común que conjuntos bien conocidos de líneas de emisión o absorción se desplacen hacia el extremo rojo del espectro. Por ejemplo, la Serie de Lyman del hidrógeno, que consiste en las longitudes de onda emitidas por un átomo de hidrógeno cuando los electrones caen al estado fundamental, experimenta este fenómeno. A pesar de que estas longitudes de onda están bien medidas para cada transición, en objetos distantes, incluyendo la Serie de Lyman y otras series, tienden a desplazarse hacia el extremo rojo del espectro desde donde se miden en la Tierra. Este desplazamiento es resultado de la expansión del universo (Figura 1.6), que predice que la radiación de fuentes que se acercan a nosotros se contrae hacia longitudes de onda más cortas, mientras que las fuentes que se alejan se alargan hacia longitudes de onda más largas. Este fenómeno se describe mediante el corrimiento al rojo ($z_{obs}$), definido como el cambio en la longitud de onda observada ($\lambda_{obs}$) dividido por la longitud de onda en reposo ($\lambda_{rest}$):

$$z_{obs} = \frac{\lambda_{obs} - \lambda_{rest}}{\lambda_{rest}} \qquad 1.3$$

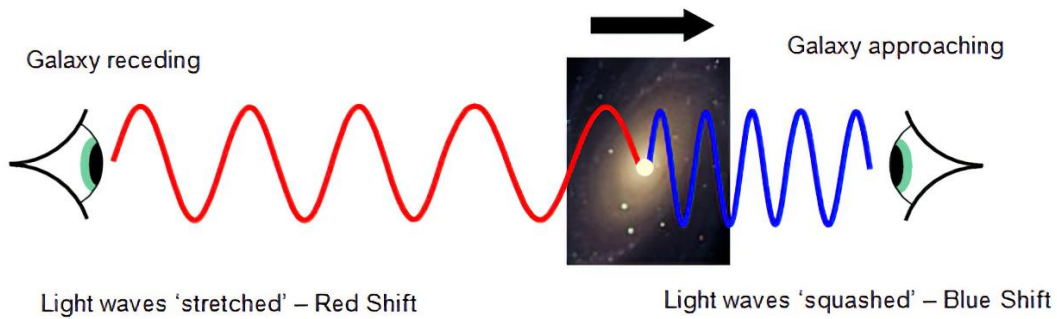

Figura 1.6.  Cuando observamos galaxias lejanas, las líneas espectrales se desplazan hacia el rojo debido a la expansión del Universo, que predice que la radiación se desplazará hacia longitudes de onda más cortas para las fuentes que se acercan y hacia longitudes de onda más largas para las que se alejan.

Actualmente, existen dos métodos para medir el corrimiento al rojo (Figura 1.7):

- **Espectroscopía**: Se obtiene el espectro de una galaxia y se comparan las líneas de emisión o absorción con sus valores esperados. En lugar de considerar un solo par $\lambda_{0bs}$ y $\lambda_{rest}$, se utilizan varios, lo que proporciona resultados bastante precisos. Sin embargo, obtener el espectro es un proceso laborioso, por lo que este método solo se aplica a un subconjunto de las galaxias observadas.

- **Fotometría**: Utilizando dos imágenes de una galaxia con diferentes filtros de colores en el telescopio (generalmente rojo y azul), es posible estimar su corrimiento al rojo. Aunque esta técnica permite observar múltiples galaxias simultáneamente, es menos precisa y puede tener errores de hasta aproximadamente $\sigma_{cz}/cz \approx 50$, lo que se traduce en 15 000 km/s para el rango de corrimientos al rojo de la materia [20].

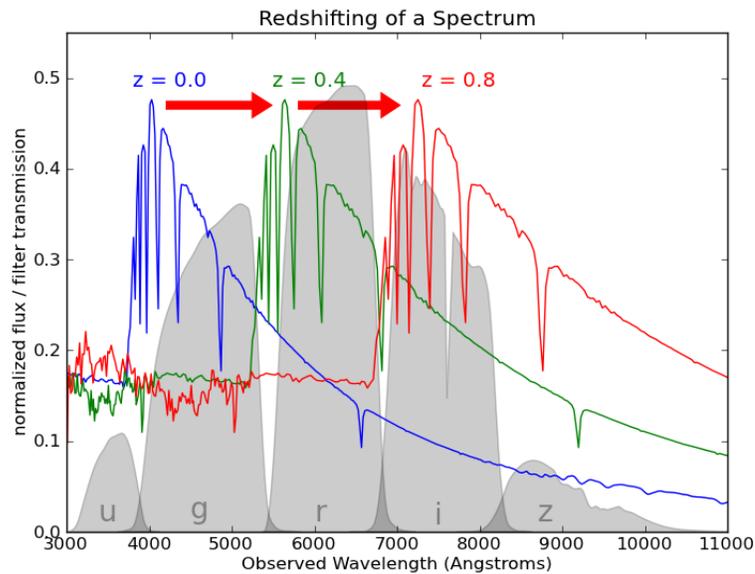

Figura 1.7. El espectro de la estrella Vega a tres diferentes corrimientos al rojo. Los filtros ugriz del SDSS se muestran en gris para referencia. Imagen tomada de [21]

El descubrimiento de la expansión cósmica fue posible gracias al desarrollo de un método para medir distancias extragalácticas. Se remonta a Hubble [22], quien utilizó observaciones de estrellas variables Cefeidas con el propósito de servir como "velas estándar". Descubrió que en escalas lo suficientemente grandes, todas las galaxias se están alejando de nosotros, y que la velocidad radial observada $v_r^{obs}$ aumenta con la distancia al observador $r$ de manera proporcional a una constante $H_o$. Esta constante de Hubble se interpreta como el valor actual de la tasa de expansión del Universo. Se puede estimar a partir de la relación de las velocidades radiales observadas $v_r^{obs}$ con las distancias medidas $r$ hasta escalas lo suficientemente grandes, lo cual será una relación lineal con la pendiente

$$H_o = \frac{v_r^{obs}}{r} \qquad\qquad 1.4$$

Esta ley ha sido confirmada en el universo local mediante mediciones posteriores [23]. Sin embargo, para galaxias más distantes, es necesario resolver las ecuaciones de Friedmann, que son fundamentales en cosmología para describir la expansión del universo. Estas ecuaciones relacionan la tasa de expansión del universo con la densidad de materia y energía en él. La constante de Hubble, que define la tasa de expansión actual del universo, es un parámetro crucial en estas ecuaciones. No obstante, el valor preciso de esta constante sigue siendo motivo de controversia, debido a las discrepancias entre diferentes métodos de medición y modelos teóricos. Las estimaciones de $H_o$ a partir del fondo cósmico de microondas (CMB) se encuentran alrededor de 70 km s⁻¹ Mpc⁻¹. Según las observaciones de WMAP, $H_o = 69.32 \pm 0.80$ km s⁻¹ Mpc⁻¹ [24] y la misión Planck, $H_o = 67.80 \pm 0.77$ km s⁻¹ Mpc⁻¹ [25]; lo cual es mucho más bajo que las estimaciones del universo local de $H_o = 74.6 \pm 0.8$ km s⁻¹ Mpc⁻¹ [26].

En cosmología teórica, a menudo se utiliza un sistema de unidades que parametriza esta incertidumbre definiendo la constante de Hubble como:

$$H_o = 100 \, h \text{ km s}^{-1} \text{ Mpc}^{-1} \qquad 1.8$$

donde $h$ es un parámetro adimensional que representa la constante de Hubble normalizada. Este enfoque simplifica los cálculos y permite una comparación más fácil entre diferentes estudios y modelos cosmológicos, ya que la constante de Hubble $H_o$ puede variar en función de las mediciones y métodos utilizados para determinarla.

**Relación entre la constante de Hubble y el corrimiento al rojo**

La dispersión alrededor de esta relación se debe a los movimientos peculiares $v_{pec}$ de las galaxias. La velocidad radial total observada $v_r^{obs}$ de una galaxia es, por lo tanto, la suma de su movimiento peculiar y el "arrastre de Hubble" debido a la expansión,

$$v_r^{obs} = v_{pec} + H_o r \qquad 1.5$$

El corrimiento al rojo se relaciona con el factor de escala del Universo $a(t)$, donde $a(t_{obs})$ y $a(t_{rest})$ son los factores de escala ahora y en el momento de emisión, respectivamente. Así:

$$\frac{\lambda_{obs}}{\lambda_{rest}} = \frac{a(t_{obs})}{a(t_{rest})} \qquad 1.6$$

En un Universo en expansión, esto implica que $\lambda_{obs} > \lambda_{rest}$, lo que lleva a longitudes de onda más largas y enrojecidas. La velocidad radial total se observa directamente a través del corrimiento al rojo $z$ del espectro electromagnético de un objeto que se aleja, desde longitudes de onda $\lambda_{rest}$ a longitudes de onda observadas $\lambda_{obs}$, lo cual puede interpretarse como ya mencionamos en el desplazamiento Doppler:

$$v_r^{obs} = c z_{obs} \qquad 1.7$$

donde la c es la velocidad de luz en el vacío. Esta velocidad no necesariamente tiene una interpretación física, ya que puede exceder la velocidad de la luz. Por una cuestión de notación, los corrimientos al rojo suelen expresarse en km/s. Los movimientos de los objetos en relación entre sí también crean velocidades "peculiares" ($v_{pec}$), debido a las cuales existen diferencias entre el corrimiento al rojo medido de un objeto, $z_{obs}$, y su corrimiento al rojo cosmológico, $z_{cos}$. Los corrimientos al rojo cosmológicos son aquellos corrimientos al rojo que se deben únicamente a la expansión del Universo. La relación entre el corrimiento al rojo observado $z_{obs}$, el corrimiento al rojo cosmológico $z_{cos}$ , el corrimiento al rojo peculiar ($z_{pec}$, corrimiento al rojo causado por la velocidad peculiar) se expresa mediante:

$$1 + z_{obs} = (1 + z_{obs})(1 + z_{pec}) \qquad\qquad 1.8$$

La razón por la que el corrimiento al rojo es uno de los observables más importantes es porque la mayoría de las distancias a fuentes extragalácticas se miden a través del corrimiento al rojo, y la distancia, como sabemos, es una información primaria necesaria para estudiar cualquier fuente astronómica. En resumen, la relación entre el corrimiento al rojo y la distancia de las galaxias es fundamental para comprender la evolución cósmica y las propiedades de las galaxias a lo largo del tiempo.

La siguiente sección, "La Importancia de las Barras en la Evolución de las Galaxias", abordará en detalle la definición y las principales características de las galaxias con barras estelares, así como sus efectos en la estructura galáctica.

## 1.4 La Importancia de las Barras en la Evolución de las Galaxias

En este estudio, nos enfocaremos principalmente en las galaxias de disco, que incluyen tanto a las galaxias espirales como a las lenticulares. Las galaxias de disco se forman y desarrollan dentro de halos de materia oscura, al igual que otras galaxias. Estos halos son cruciales para la evolución de las diferentes componentes de las galaxias.

### 1.4.1. Definición y principales características

Las barras estelares son estructuras características comunes e identificables en las galaxias de disco (Figura 1.8). Exhiben una estructura sólida compuesta principalmente por estrellas que giran rápidamente alrededor de un eje perpendicular al plano del disco galáctico [27]. Estas estructuras juegan un papel crucial en la evolución de las galaxias y están asociadas con la formación de brazos espirales [28] y/o pseudobulbos [29]. La relevancia de estos componentes impulsa la necesidad de nuevos estudios dirigidos a analizar y comprender mejor su naturaleza y cómo evolucionan a lo largo del tiempo cósmico.

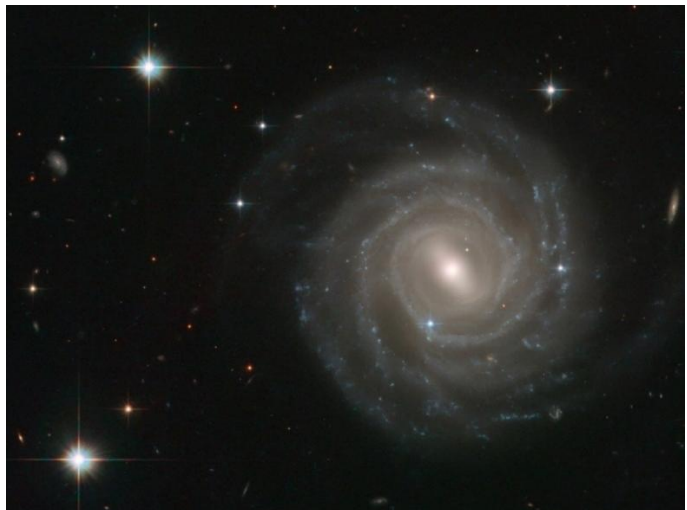

Figura 1.8. Galaxia espiral barrada UGC 12158. Créditos de imagen: ESA/Hubble & NASA. Imagen Tomada de [50].

Ópticamente, las galaxias barradas representan aproximadamente el 50% de las galaxias de disco en el universo local [30] [31] [32]. Estudios basados en imágenes en el infrarrojo cercano sugieren que la proporción de galaxias barradas aumenta a aproximadamente el 70% [33] [34]. También, algunos estudios muestran que esta proporción se mantiene constante hasta un corrimiento al rojo de $z \approx 1$ [35]. No obstante, también hay evidencia significativa que sugiere lo contrario, con la proporción de barras disminuyendo rápidamente a medida que aumenta el corrimiento al rojo (Figura 1.9) [36] [37]. Esta es la razón, por la cual nos enfocaremos en esta tesis en ver cómo cambia la fracción de barras desde el universo local hasta un corrimiento al rojo de $z \sim 0.7$ en galaxias de campo. También existe evidencia de que la fracción de barras está relacionada con el tipo Hubble de las galaxias, aunque la naturaleza exacta de esta relación es ampliamente debatida. En algunos estudios, se ha observado que la fracción de barras aumenta en las galaxias espirales de tipo temprano, que son más masivas, rojas, pobres en gas y dominadas por el bulbo [38] [39] [40]. Sin embargo, otros estudios sugieren lo contrario, mostrando que las barras son más comunes en las espirales de tipo tardío, que son menos masivas, azules, ricas en gas y dominadas por el disco [41] [42]. Alternativamente, es posible que ambas observaciones sean correctas. De hecho, hay evidencia de un pico bimodal en la fracción de barras en correlación con la secuencia Hubble, con un aumento tanto en las espirales de tipo temprano como en las de tipo tardío [43] [44] [45].

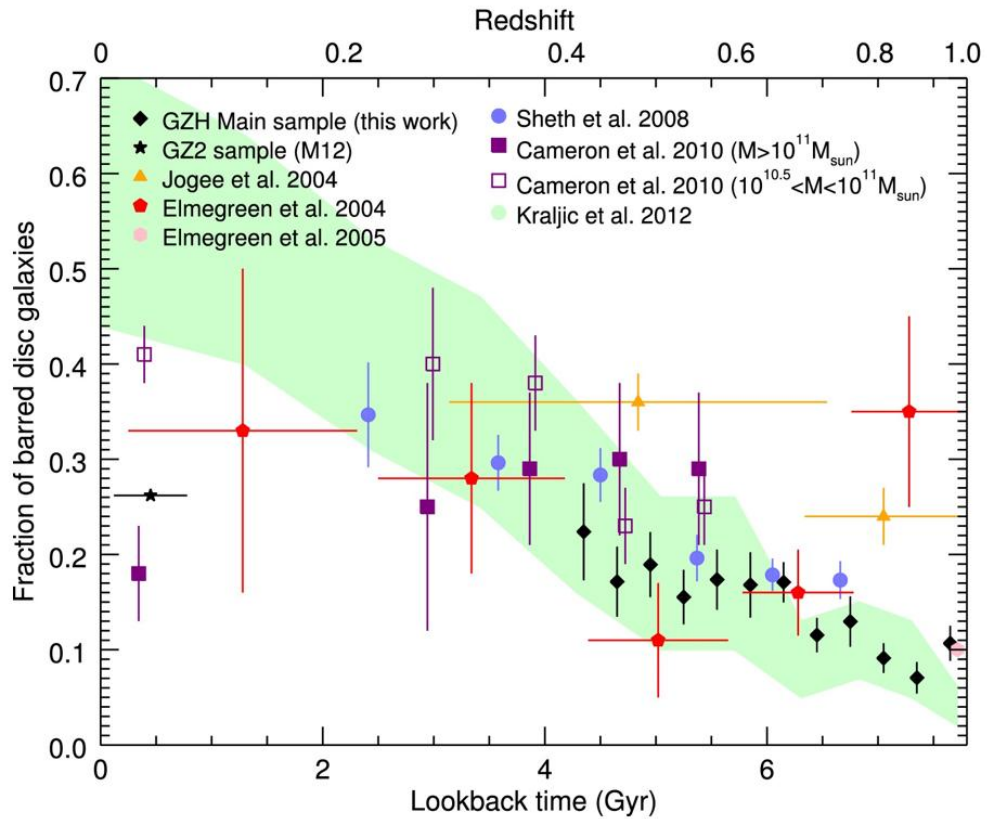

Figura 1.9. Evolución de la fracción de galaxias barradas a lo largo del tiempo, con varios estudios mostrando una tendencia decreciente a medida que aumenta el corrimiento al rojo (disminuyendo el tiempo de retraso). Los datos sugieren que las barras eran menos frecuentes en el universo temprano. Imagen tomada de [47]

### 1.4.2. Efectos de las Barras en la Estructura Galáctica

Las galaxias barradas evolucionan mediante el transporte de momento angular desde el interior hacia las regiones externas. Esta redistribución eficiente del momento angular hace que las barras jueguen un papel crucial en la evolución de las galaxias de disco. A continuación, mostramos los procesos más importantes que influyen las barras en la estructura interna de las galaxias de disco:

- **Intercambio de momento angular**: Las características no axisimétricas en las galaxias de disco facilitan la transferencia de momento angular, con las barras desempeñando un papel clave en este proceso [48]. Las barras generan fricción dinámica contra el halo de materia oscura, transfiriendo energía y momento, lo que las hace ralentizarse y crecer (Figura 1.10). Al crecer, las barras atrapan material en órbitas alargadas y requieren la transferencia de momento angular a otras partes de la galaxia para conservar el momento angular del disco.

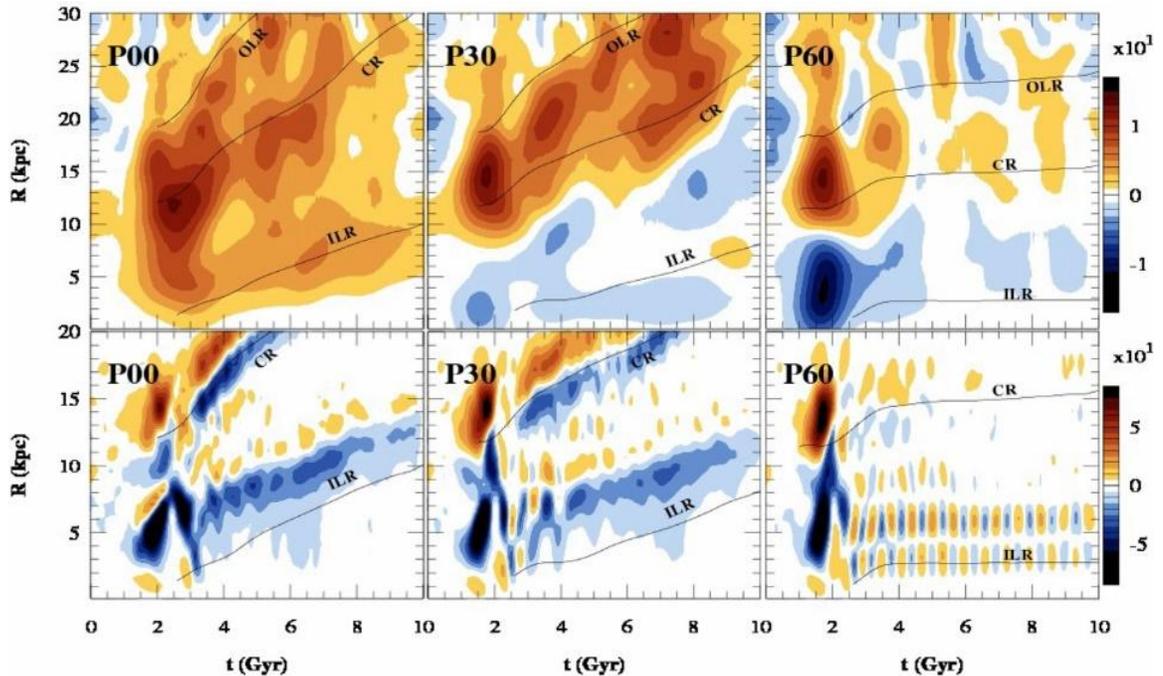

Figura 1.10. Los paneles destacan la interacción de las barras galácticas con los halos de materia oscura en los modelos P00, P30 y P60, mostrando la tasa de flujo de momento angular a lo largo del tiempo y el radio. Esta transferencia de momento angular, impulsada por la fricción dinámica entre las barras y el halo, redistribuye la energía, lo que ralentiza y hace crecer las barras con el tiempo. Las resonancias principales del disco (ILR, CR y OLR) están marcadas para identificar los puntos donde este intercambio de momento angular es más significativo. Imagen Tomada de [49]

- **Formación de pseudobulbos**: El material que las barras transportan hacia las regiones centrales contribuye a la formación de pseudobulbos, que se diferencian de los bulbos clásicos en varios aspectos. Los pseudobulbos tienen un perfil de brillo superficial casi exponencial, descrito por el perfil de Sersic [50], donde $n < 2$ [51], mientras que los bulbos clásicos tienen $n > 2$ [52]. Además, los pseudobulbos muestran formación estelar continua y poblaciones jóvenes de estrellas [53], a diferencia de los bulbos clásicos, que son generalmente antiguos y rojos.

- **Estimulando la formación de estrellas en la región central**: Cuando hay una barra en una galaxia, el gas dentro de una región llamada corrotación tiende a seguir órbitas específicas llamadas x1, alineadas con el eje mayor de la barra [54]. Mientras tanto, el gas fuera de esta región permanece mayormente en el disco galáctico [55]. Debido a que el gas puede chocar entre sí (a diferencia de las estrellas), el gas en las órbitas x1 experimenta choques, lo que hace que pierda su momento angular. Esto causa que el gas caiga hacia las regiones centrales de la galaxia y se establezca en órbitas llamadas x2. Este proceso de choque se repite varias veces hasta que el gas se asienta cerca de lo que se llama la resonancia de Lindblad interna [56]. La acumulación de gas frío en el centro de la galaxia desencadena episodios intensos de formación estelar. Observaciones muestran que las galaxias barradas típicamente tienen concentraciones de gas más altas en sus centros y tasas de formación estelar más elevadas en comparación con galaxias de masa similar, pero sin barra [57].

En el desarrollo de este trabajo, nos centramos en el análisis de grupos morfológicos de galaxias de discos en el universo cercano y lejano, con un enfoque particular en la detección y caracterización de barras estelares. Las galaxias de campo, conocidas por su distribución dispersa y baja densidad de vecinos, presentan una oportunidad única para estudiar las barras estelares en diferentes contextos cósmicos. Investigaciones recientes indican que aproximadamente el $65 - 70\%$ de las galaxias clasificadas como aisladas a $z = 0$ han mantenido su aislamiento desde $z = 1$ [58], evitando interacciones significativas durante largos períodos. Esta característica de aislamiento permite observar con mayor claridad las estructuras internas, como las barras estelares, sin las complicaciones de interacciones cercanas típicas de los cúmulos galácticos. En este estudio, buscamos identificar y analizar las barras estelares en galaxias de disco de diferentes tipos y épocas, para comprender cómo estas estructuras evolucionan a lo largo del tiempo en el universo cercano y lejano. En particular, se observa la presencia de barras en galaxias de disco de tipo temprano y tardío, lo que aporta información clave sobre su distribución y evolución. Para el grupo local, utilizamos imágenes obtenidas del Sloan Digital Sky Survey (SDSS) [59], con corrimientos al rojo en el rango de $0.0207 \leq z \leq 0.030$. Para el grupo del universo intermedio, empleamos datos del Great Observatories Origins Deep Survey (GOODS) [60], con corrimientos al rojo en el rango de $0.4 \leq z \leq 0.8$.

En esta tesis, examinamos la fracción de barras estelares y su evolución en los últimos 6 giga-años, centrándonos en galaxias de campo. La tesis está organizada de la siguiente manera: en el capítulo 2 se describen los métodos para la detección de barras estelares; en el capítulo 3, la muestra observacional de galaxias; en el capítulo 4, se presentan los resultados y análisis de datos; en el capítulo 5, se exponen las conclusiones; y finalmente, el capítulo 6 ofrece perspectivas para futuros trabajos. Además, se incluye una sección de Apéndices que ofrece información complementaria, incluyendo una visión general de la formación y evolución de las galaxias, detalles sobre los métodos estadísticos y resultados adicionales de las técnicas utilizadas en esta investigación.

# Capítulo 2

## Métodos para la Detección de Barras Estelares

Para investigar la relación entre la formación de discos y barras en las galaxias, no solo es crucial determinar la fracción de galaxias de disco que presentan barras, sino también vincular las propiedades de estas barras con las del disco. En esta sección, exploraremos los diversos métodos disponibles para identificar y caracterizar las barras estelares. La fotometría superficial de galaxias es una herramienta esencial utilizada en todos estos métodos, ya que permite analizar detalladamente la estructura de las barras y su interacción con el disco galáctico (Figura 2.1).

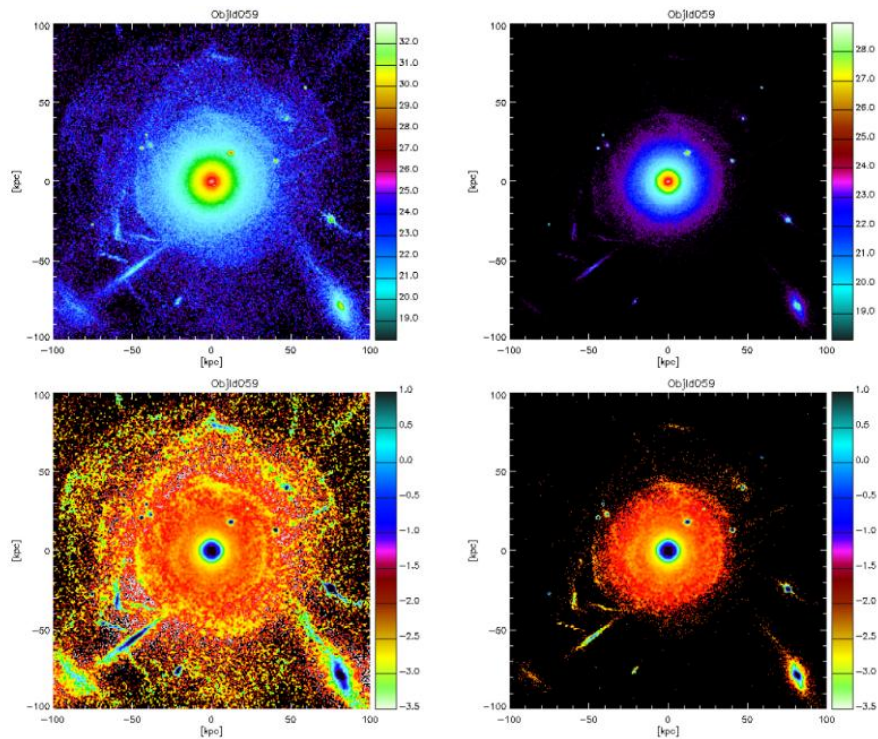

Figura 2.1. Mapa de Brillo Superficial: La figura muestra el brillo superficial de la galaxia central en la banda g con dos escalas de colores (paneles superior e inferior) y cortes a 33 mag arcsec⁻² (izquierda) y 29 mag arcsec⁻² (derecha). Estos mapas permiten evaluar cómo se distribuye la luz y analizar características morfológicas de las galaxias. Imagen tomada de [61]

Este enfoque es fundamental para extraer la distribución de brillo superficial, lo que facilita tanto las clasificaciones visuales como los análisis cuantitativos. En cuanto a la detección visual, las barras pueden ser identificadas y clasificadas basándose en características como su longitud, prominencia y la configuración de la región central. Sin embargo, para estudios más cuantitativos y precisos, utilizamos métodos como el análisis de isofotas elípticas y el análisis de Fourier. Estos enfoques permiten obtener descripciones más detalladas de las propiedades de las barras, como su elipticidad, orientación y la distribución de la densidad superficial, proporcionando una visión más robusta de su dinámica y evolución.

## 2.1 Principales Propiedades Físicas para la Detección de Barras

La detección de barras estelares en galaxias de disco es clave para comprender la estructura y evolución de las galaxias. Estas barras, que son estructuras alargadas formadas por estrellas y material interestelar, juegan un papel crucial en la dinámica galáctica y en la redistribución de materia en los discos.

Las barras se pueden describir generalmente mediante tres propiedades principales: la longitud de la barra, la fuerza de la barra, y la velocidad de patrón de la barra. Estas características determinan cómo influyen las barras en el movimiento del gas y las estrellas dentro de la galaxia, así como su capacidad para desencadenar procesos de formación estelar. A continuación, se detallan estas propiedades:

### 2.1.1 Longitud de Barra $(R_{bar})$

La longitud de la barra representa la extensión radial máxima de las órbitas estelares que sostienen la estructura de la barra y corresponde al eje semi-mayor de la misma [62]. Determinar $(R_{bar})$ no es una tarea sencilla, ya que los bordes de la barra no siempre están bien definidos. Además, la barra suele estar asociada con otras estructuras, como anillos o brazos espirales, lo que puede complicar la medición de su longitud. En la literatura, se han desarrollado diversos métodos para estimar $R_{bar}$, y aunque cada uno tiene sus limitaciones, la combinación de varios enfoques suele proporcionar una estimación más robusta. Entre los métodos más comunes para estimar $R_{bar}$ se incluyen la inspección visual directa de imágenes de galaxias con barras [63], aunque esta técnica es menos útil en imágenes con baja resolución espacial; el análisis de isofotas, donde la elipticidad $\varepsilon$ y el ángulo de posición $PA$ de las elipses ajustadas permiten medir $R_{bar}$ [64] ; y el análisis de Fourier, que identifica picos en la amplitud para $m = 2$, los cuales corresponden a las características de la barra cerca del centro y proporcionan información sobre su distribución de luz en la galaxia [65].

### 2.1.2 Fuerza de la Barra $(S_{bar})$

La fuerza de la barra se refiere a la contribución de la barra al potencial gravitacional total de la galaxia [66]. Esta fuerza puede interpretarse como la fracción de estrellas en la región de la barra que necesitarían ser reorganizadas para transformar dicha estructura en una configuración aximétrica. Al igual que con $R_{bar}$ se han propuesto varios métodos para medir $S_{bar}$. Uno de estos métodos utiliza la forma de la barra, basada en la relación axial de la misma, $Q_{bar}$, que puede calcularse mediante un análisis de isofotas, midiendo la elipticidad de la barra $\varepsilon_{bar} = 1 - Q_{bar}$ en el radio de la barra [67]. Otro enfoque, basado en componentes de Fourier, define $S_{bar}$ como el valor medio de la amplitud relativa de Fourier para $m = 2$ $(I_2/I_0)$, calculado en la región de la barra [68]. Por último, se utiliza el mapa de la relación entre la fuerza tangencial y la radial, $Q_T = (R, \theta)$, que se calcula como el promedio de los valores máximos de los picos en el perfil azimutal de $Q_{T/R}$ para $R_{bar}$ [69].

### 2.1.3 Velocidad patrón de la Barra ($\Omega_{bar}$)

La velocidad del patrón de la barra es la frecuencia angular de rotación de la barra alrededor del centro galáctico y requiere mediciones de la cinemática estelar. Se parametriza mediante la tasa de rotación$(R)$, que es la relación entre el radio de corotación ($R_{cor}$) y el radio de la barra ($R_{bar}$) [70]. Esta velocidad se puede determinar a través de las bandas de polvo, que indican la ubicación de las resonancias de Lindblad y muestran áreas de acumulación de gas que se extienden desde el núcleo hasta los brazos espirales [71]. También se estima comparando las propiedades fotométricas y cinemáticas de las galaxias barradas con simulaciones, utilizando modelos hidrodinámicos del flujo de gas para estudiar la cinemática del gas en la barra [72].

Para el desarrollo de esta tesis, hemos elegido los análisis de isofotas elípticas y el análisis de Fourier. El análisis de isofotas elípticas permite evaluar la forma de la barra a partir de la elipticidad y el ángulo de posición, proporcionando una medida de la longitud de la barra. Este método es particularmente útil para caracterizar la geometría global de la barra. Por otro lado, el análisis de Fourier ofrece una medición precisa de las amplitudes y fases, lo cual es esencial para determinar la fuerza de la barra y corroborar la longitud estimada. Mientras que ambos métodos miden la longitud de la barra, el análisis de Fourier permite una evaluación más detallada de las componentes estructurales de la barra galáctica. En la siguiente sección, abordaremos la fotometría superficial y presentaremos en detalle los métodos y programas utilizados para realizar estos análisis.

## 2.2 Fotometría Superficial en galaxias

La fotometría superficial proporciona información crucial sobre parámetros como el tamaño y la forma del potencial gravitacional en galaxias barradas. Estos aspectos son esenciales para estudiar la dinámica y la estructura de las barras galácticas, así como para entender su influencia en la evolución de la galaxia [73]. La fotometría superficial permite caracterizar la geometría de las barras y su impacto en la distribución de estrellas y gas, proporcionando así una visión integral de cómo las barras afectan la dinámica galáctica. La simplificación de la reducción de imágenes galácticas a fotometría superficial se debe a que, en la mayoría de los tipos morfológicos, las isofotas de las galaxias se pueden aproximar de manera precisa con elipses. Esta aproximación es vital para un análisis fotométrico detallado y se aplica incluso en galaxias irregulares, donde una elipse puede describir adecuadamente las isofotas que no están distorsionadas por la formación estelar activa o la extinción del polvo [74]. Esta metodología es crucial para obtener información detallada sobre los colores galácticos, sus implicaciones en las edades y gradientes de metalicidad [75], las poblaciones estelares [76], el contenido de polvo y su extinción [77], así como para comprender mejor la estructura, formación y evolución de las galaxias [78]. Los avances en tecnología de detectores CCD y sistemas informáticos han mejorado significativamente nuestra capacidad para mapear grandes áreas del cielo y medir con precisión las propiedades fotométricas y estructurales de numerosas fuentes extendidas, incluso con exposiciones cortas. Estos desarrollos han revolucionado el estudio de la estructura y evolución de objetos astronómicos, permitiendo un análisis detallado a escalas previamente inalcanzables. Entre las herramientas más ampliamente utilizadas para el análisis de fotometría superficial y de apertura de galaxias barradas se encuentran Source Extractor [79], GIM2D [80], ISOPHOTE [81], GALPHOT [82], GASPHOT [83], GALFIT [84], P2DFFT [85] entre otros. Estas herramientas permiten una evaluación precisa de las propiedades estructurales de las galaxias, facilitando el estudio de sus barras galácticas y proporcionando datos valiosos sobre su dinámica y evolución.

Para la detección y caracterización de barras estelares, utilizaremos dos métodos principales: el análisis de isofotas elípticas (ISOPHOTE) y el análisis de Fourier (P2DFFT). Además, como apoyo se utiliza Source Extractor para parámetros iniciales.

### 2.2.1 Análisis de Isofotas Elípticas

Entre los primeros esfuerzos para expresar desviaciones de formas puramente elípticas en forma matemática, se encontraba el trabajo que se volvió fundamental en el modelado de isofotas más refinadas y realistas de galaxias [86]. Este enfoque propuso agregar perturbaciones a una elipse como función del ángulo azimutal φ, de manera análoga a una descomposición en series de Fourier.

$$I(\varphi) = \langle I_{ell} \rangle + \sum_n [A_n sen(n\varphi) + B_n cos(n\varphi)] \qquad 2.1$$

El perfil de intensidad a lo largo de la isofota, $I(\varphi)$, se expresa como una función del ángulo azimutal (central), donde $\langle I_{ell} \rangle$ representa la intensidad promedio a lo largo de la trayectoria puramente elíptica. La suma representa las perturbaciones armónicas de Fourier a $\langle I_{ell} \rangle$, siendo $n$ el orden del armónico (entero). Este formalismo resulta particularmente elegante debido a que los coeficientes de los armónicos de Fourier ($A_n$ y $B_n$) tienen un significado físico sobre las galaxias. Más tarde, se desarrolló un algoritmo para ajustar las isofotas en las imágenes de galaxias [81].

En este método, las isofotas, que no necesariamente son concéntricas, se ajustan en varios puntos a lo largo del semieje mayor (denotado como $a$) en la imagen de una galaxia. En cada uno de estos puntos, la isofota inicialmente se modela como una elipse pura, con parámetros geométricos estimados para el centro ($x_0$, $y_0$), el ángulo de posición ($PA$., por sus siglas en inglés), y la elipticidad ($e = 1 - b/a$, donde $b$ es el semieje menor). Esto se muestra en la Figura 2.2, para los primeros dos armónicos. Luego, se toma una muestra de la imagen a lo largo de esta trayectoria elíptica, proporcionando una distribución unidimensional de intensidad en función del ángulo azimutal $I_{imagen}(\varphi)$, inicialmente modelada según la ecuación mencionada, pero limitada a $n \in \{1,2\}$. Así la ecuación (2.1) queda como:

$$I(\varphi) = \langle I_{ell} \rangle + \sum_{n=1}^{2}[A_n sen(n\varphi) + B_n cos(n\varphi)] \qquad 2.2$$

De manera general, la estructura de una barra se identifica al aumentar el valor de la elipticidad ($e$) hasta alcanzar su máximo en la longitud de la barra, preservando el ángulo de posición ($PA$) de la misma. Esto se determina a partir de un análisis de isofotas en la imagen de la galaxia, para más detalle ver [87]. Para determinar estos parámetros, se emplea un algoritmo de mínimos cuadrados iterativo que ajusta los parámetros del modelo de isofota, como el centro, la orientación y la elipticidad de la elipse, además de los coeficientes de Fourier $A_n$ y $B_n$. En cada iteración del algoritmo, se evalúa la función objetivo $S$, definida como:

$$S = \sum_i \left[ I_{imagen}(\varphi_i) - I(\varphi_i) \right]^2 \qquad 2.3$$

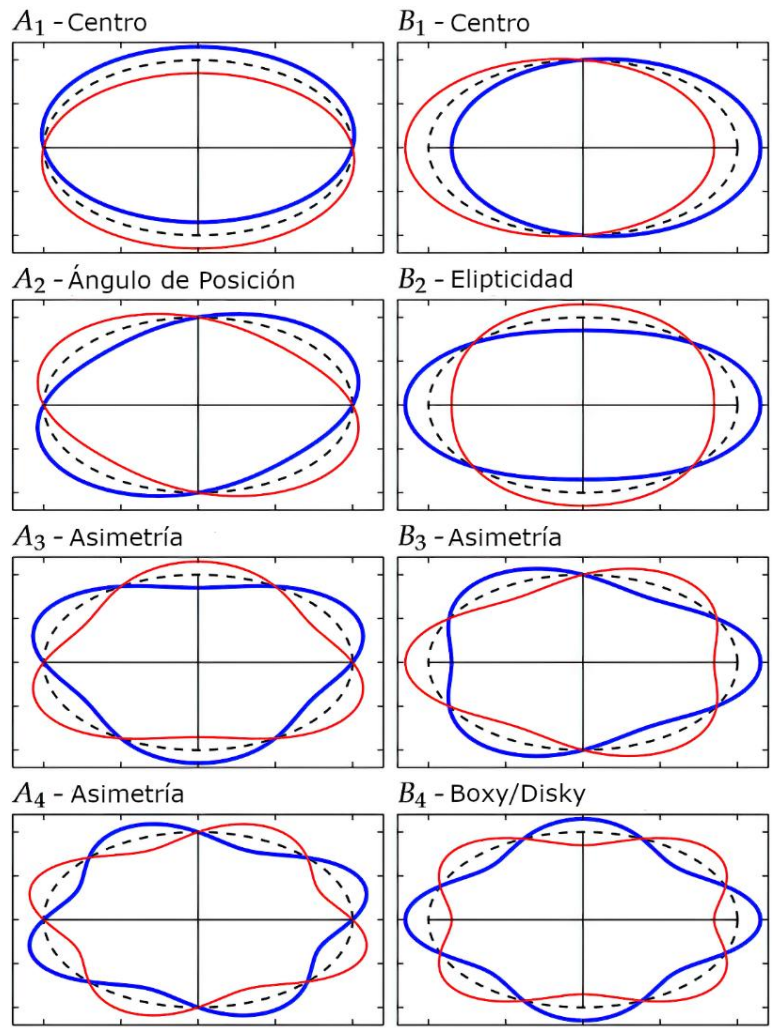

Figura 2.2. Se ilustra la significancia física de las correcciones armónicas de los primeros cuatro armónicos a una isofota elíptica. Los coeficientes positivos se muestran en la línea azul, mientras que los coeficientes negativos se representan en la roja. La isofota de referencia se muestra en la línea negro (punteado) en cada panel. Imagen tomada de [88].

Los parámetros del modelo se ajustan iterativamente hasta alcanzar un criterio de convergencia, típicamente basado en una reducción adecuada del error cuadrático medio. En la tabla 2.1 se muestra diferentes estudios utilizando el método de isofotas elípticas para la caracterización de barras, así como los valores adoptados para la identificación de galaxias barradas.

| Estudio | $\Delta PA_{\mathrm{bar}}$ | $e_{\mathrm{bar}}$ |
|---|---|---|
| Wozniak et al. (1995) [89] | const | - |
| Laine et al. (2002) [90] | $\pm 10^{o}$ | $\geq 0.45$ |
| Jogee et al. (2004) [91] | $\pm 20^{o}$ | $\geq 0.4$ |
| Marinova et al. (2012) [92] | $\pm 10^{o}$ | $\geq 0.25$ |
| Lee et al. (2019) [93] | $\pm 5^{o}$ | $\geq 0.25$ |

Tabla 2.1. Criterios diversos para la detección de barras.

Los criterios de selección para una galaxia barrada que nosotros utilizaremos en el análisis de isofotas elípticas son los siguientes: (a) La elipticidad ($e$) aumenta hasta un máximo global ($e_{bar} \geq 0.25$) mientras el ángulo de posición ($PA$) se mantiene constante dentro de $\pm 10°$. (b) Al pasar de la barra al disco, la elipticidad disminuye al menos en $\Delta e \geq 0.1$ y el ángulo de posición cambia más de ($\Delta PA \geq 10°$). En la Figura 2.3 se muestra un ejemplo de una galaxia barrada en la que el perfil de elipticidad muestra un aumento superior a $e \geq 0.25$ y el PA en la región de la barra permanece casi constante, seguido de una disminución de $e$ y cambio en $PA$ al pasar al disco.

Es importante mencionar que, en este trabajo, el análisis de isofotas elípticas se inició utilizando la tarea ISOPHOTE de IRAF [94], una herramienta consolidada y ampliamente utilizada en astrofísica para el ajuste preciso de isofotas. Sin embargo, con el objetivo de optimizar y modernizar el proceso, se decidió migrar a la librería PHOTUTILS de Python, utilizando su función ISOPHOTE [95]. Esta transición no fue motivada por limitaciones de IRAF, sino por las ventajas que ofrece Python en términos de flexibilidad, integración con bibliotecas científicas y automatización del análisis. La utilización de PHOTUTILS permitió mantener la precisión y rigor científico en el ajuste de las isofotas, mientras se aprovechaba un entorno más versátil y adecuado para flujos de trabajo contemporáneos en la astrofísica.

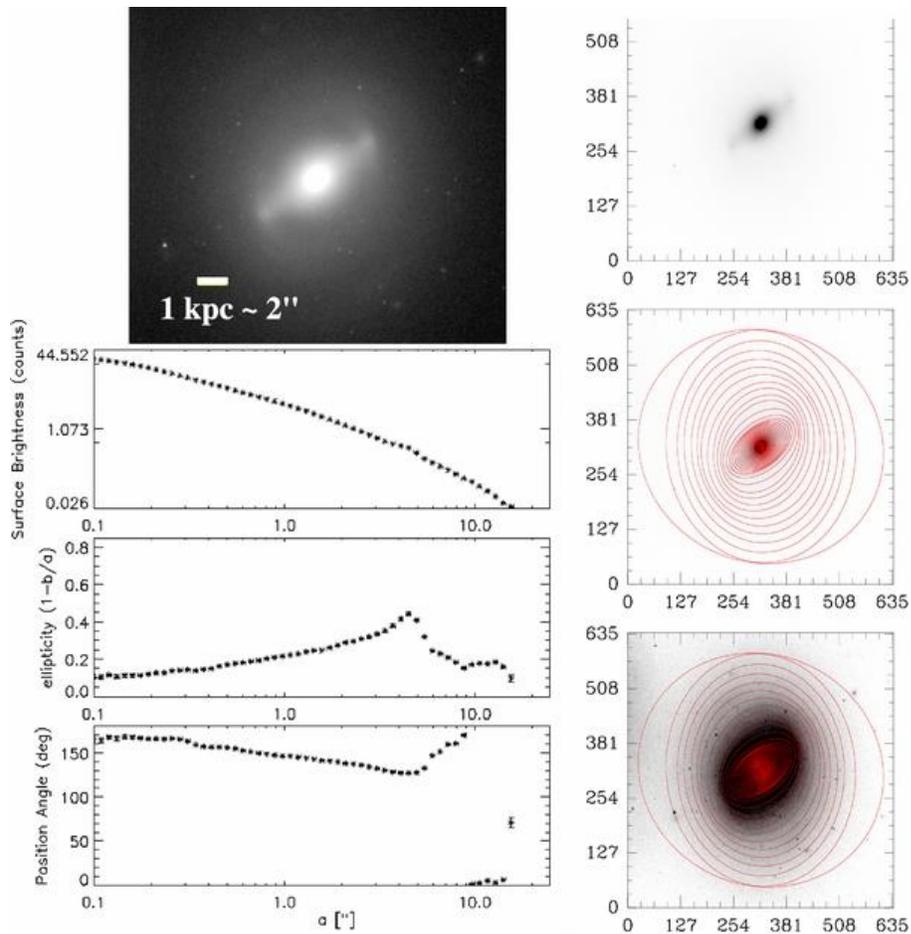

Figura 2.3. Izquierda: Imagen de una galaxia con barra y sus perfiles radiales de brillo superficial, $e$ y $PA$. La firma de la barra se muestra en el aumento superior a $e \geq 0.25$ y la constancia de PA en la región de la barra, seguido de una disminución de $e$ y cambio en $PA$ al pasar al disco. Derecha: Ajustes de elipses sobre la imagen de la galaxia, con estiramiento para resaltar las regiones internas del disco y la barra, y el panel inferior muestra el disco exterior. Imagen tomada de [92].

### 2.2.2 Análisis de Fourier

Un método alternativo para caracterizar la barra es mediante un análisis de Fourier [96][97], utilizando la aplicación P2DFFT [85]. Esto implica descomponer nuestros perfiles de luz azimutal mediante una transformada de Fourier, dada como:

$$F(r) = \int_{-\pi}^{\pi} I_r(\theta) e^{(-2i\theta)} d\theta \qquad 2.3$$

Donde $I_r(\theta)$ son los perfiles de luz azimutal y $\theta$ es el ángulo azimutal. Los coeficientes de Fourier se calculan como:

$$A_m(r) = \frac{1}{\pi} \int_0^{2\pi} I_r(\theta) \cos(m\theta) \, d\theta \qquad 2.4$$

$$B_m(r) = \frac{1}{\pi} \int_0^{2\pi} I_r(\theta) \sin(m\theta) \, d\theta \qquad 2.5$$

y las amplitudes se determinan como:

$$I_0(r) = \left( A_0(r^2) \right) \qquad 2.6$$

$$I_m(r) = \sqrt{A_m^2(r) + B_m^2(r)} \qquad 2.7$$

La relación $I_m/I_0$ de estas amplitudes se pueden usar tanto para confirmar la presencia de una barra como para medir la longitud de la barra. En la Figura 2.4, se muestran las amplitudes relativas de los componentes de Fourier para varias galaxias. El radio en el eje x refleja la distancia radial en segundos de arco, mientras que el eje y muestra la amplitud relativa de Fourier, destacando los componentes de mayor orden (como $m = 2$, $m = 4$, $m = 6$ y $m = 8$), que indican la estructura de las barras galácticas.

En este trabajo, nos enfocamos en los componentes pares de Fourier, específicamente en $m = 2$, $m = 4$ y $m = 6$, ya que estos describen de manera efectiva la simetría y la intensidad de las barras estelares en las galaxias. El componente $m = 2$ es el más relevante para identificar la estructura de una barra, al representar la forma elíptica o bipolar de las galaxias barradas. Los componentes $m = 4$ y $m = 6$ proporcionan información adicional sobre variaciones más sutiles en la forma de la barra, como su grosor o estructuras internas complejas. Estos modos pares son fundamentales para detectar y caracterizar la barra, permitiendo una comprensión detallada de su morfología y su influencia en la dinámica galáctica. Por otro lado, los componentes impares, aunque menos prominentes, pueden ser indicativos de asimetrías en la galaxia, pero su análisis queda fuera del enfoque principal de este trabajo, ya que nos centramos en los modos que mejor describen la estructura de barras simétricas.

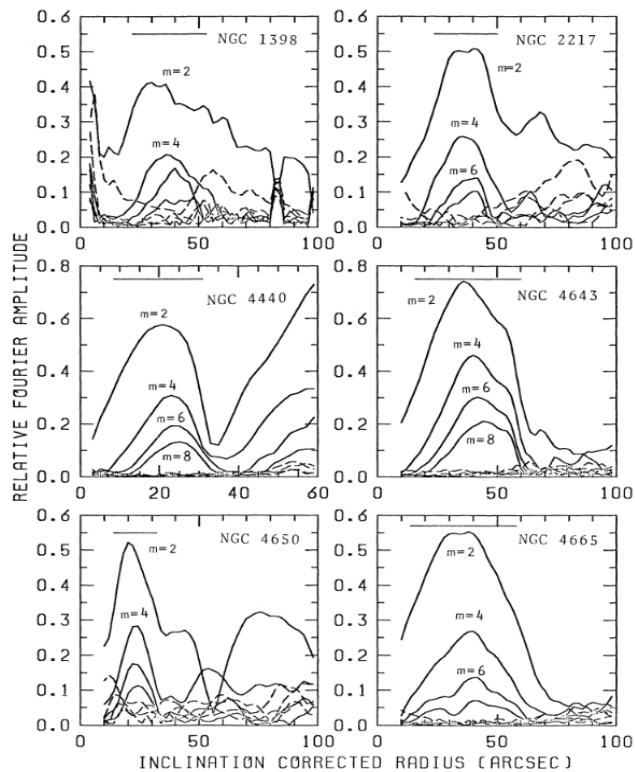

Figura 2.4. Las amplitudes relativas de los componentes de Fourier ($I_m / I_0$; $m = 1, ..., 8$) se presentan en función del radio. Las líneas sólidas y las líneas discontinuas representan los componentes pares e impares, respectivamente. Las líneas horizontales indican las regiones de la barra, definidas utilizando las ecuaciones (2.6) y (2.7). Imagen tomada de [96].

Para medir la longitud de la barra se utilizan las intensidades de Fourier de la barra e inter-barra ($I_b$ e $I_{ib}$, respectivamente), definidas como:

$$I_b = I_0 + I_2 + I_4 + I_6 \qquad\qquad 2.8$$

$$I_{ib} = I_0 - I_2 + I_4 - I_6 \qquad\qquad 2.9$$

La región de la barra se define como [115]:

$$\frac{I_b}{I_{ib}} = \frac{1}{2}\left[\left(\frac{I_b}{I_{ib}}\right)_{max} - \left(\frac{I_b}{I_{ib}}\right)_{min}\right] + \left(\frac{I_b}{I_{ib}}\right)_{min} \qquad\qquad 2.10$$

y el último radio en el que se cumple esta condición (dentro de la región de la barra) se toma como el radio de la barra. En la Figura 2.5 se muestra esta región la barra/inter-barra para una galaxia barrada.

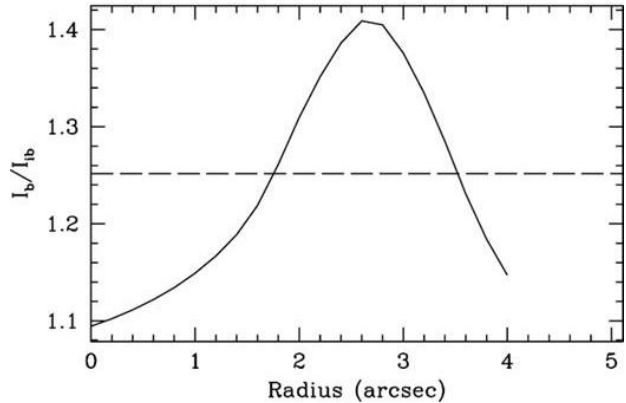

Figura 2.5. Intensidades de Fourier de la barra/inter-barra para una galaxia barrada. La línea horizontal discontinua indica el valor límite para identificar la presencia de una barra, dada por la ecuación (2.10). Imagen tomada de [116].

Una mejor estimación de la influencia de una barra es cuando se normaliza el componente $m = 0$, teniendo en cuenta tanto el disco como el bulbo subyacente. Utilizamos los componentes $m = 2$ ($m_2$) y $m = 4$ ($m_4$), que son los modos dominantes en las barras y tienen la ventaja de aparecer por encima del nivel de ruido en todas las galaxias barradas [98]. Además, la fuerza de una barra se puede considerar un indicador del potencial gravitacional subyacente de la misma. Aunque originalmente solo era una clasificación visual basada en cuán brillante o grande parecía la barra [99], con barras fuertes simplemente clasificadas como SB y barras más débiles como SAB, la fuerza de la barra ahora es un parámetro cuantificable. En este trabajo, tomamos como referencia [100] que define que la fuerza de la barra de la siguiente manera:

$$S_b = \frac{1}{r_{bar}} \sum_{m=2,4,6} \int_0^{r_{bar}} \frac{I_m}{I_0} dr \qquad 2.11$$

donde $m = 2,4,6$ son los modos pares, y $r_{bar}$ es el radio de la barra según el análisis de Fourier. Para realizar el ajuste con el método de Fourier primero hay que hacer una preparación previa. Consiste en desproyectar la imagen a una orientación de frente, asumiendo que una galaxia con su disco paralelo al plano del cielo es circular [101]. Desproyectamos cada una de las imágenes de las galaxias espirales, como se muestra en el ejemplo de la Figura 2.6, a través de un programa que realizamos con Python utilizando la función `resize` de `skimage.transform`, que por defecto emplea interpolación bilineal [102].

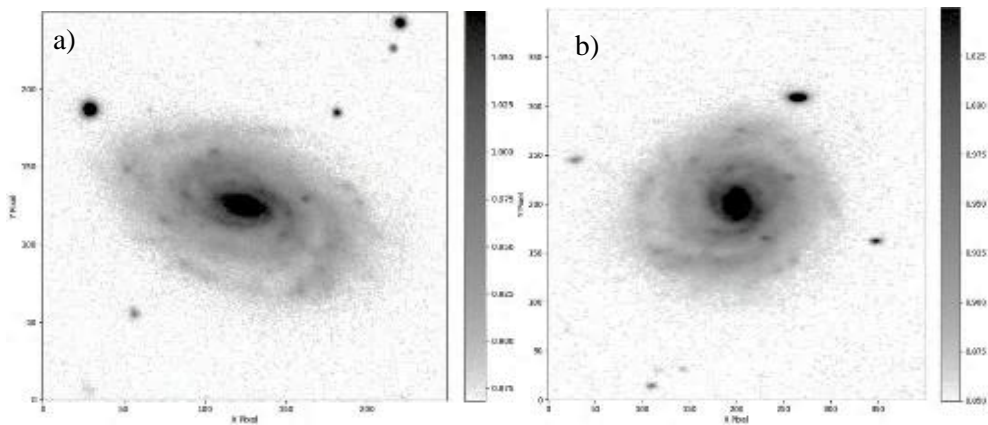

Figura 2.6. Deproyección de imagen de una galaxia espiral. De izquierda a derecha, la figura a) muestra la imagen original de una galaxia espiral, mientras que b) presenta el resultado de la deproyección de la imagen, lo cual circulariza la forma elíptica original de la galaxia. Posteriormente, la imagen se recorta para hacerla cuadrada, con el centro de la galaxia posicionado en el centro de la imagen.

Este proceso alinea el semieje mayor con el eje *y* de la imagen, y estira el semieje menor hasta igualarlo con el semieje mayor, creando un círculo. Es importante recalcar que realizamos un programa completo en Python donde se ejecutan ambos métodos con el fin de complementar los resultados que nos brinda cada uno. En el caso del método de ajustes de elipse nos da la longitud de la barra y el caso de análisis de Fourier nos da la longitud y la fuerza de la barra. Los criterios de selección para una galaxia barrada en el análisis de Fourier son los siguientes: (a) el primer factor que se debe observar es que la intensidad del modo $m_2 > 0.12$ [98] y b) se verifica si en la gráfica de $I_b/I_{ib}$ vs radio se observa que la región inter-barra supere el umbral calculado con la ecuación (2.10). La intensidad de la barra $I_b$ y la inter-barra $I_{ib}$ se calculan sumando y alternando las intensidades de Fourier, respectivamente. La longitud de la barra se determina donde la relación $I_b/I_{ib}$ alcanza su máximo ($r_{bar}$), siguiendo la ecuación (2.10). (c) Si se desea, se puede determinar la fuerza de la barra ($S_b$), siguiendo la ecuación (2.11), integrando las intensidades de los modos pares ($m = 2,4,6$) a lo largo del radio de la barra ($r_{bar}$).

## 2.3 Complementación de los Métodos de Detección

En la detección y caracterización de barras estelares, la combinación de diferentes técnicas analíticas proporciona una visión más completa y detallada de sus propiedades. En esta sección, discutimos cómo el análisis isofotas elípticas y el método de Fourier se complementan y se utilizan de manera conjunta para una evaluación más precisa.

El método de isofotas elípticas, abordado en la sección 2.2.1, se centra en la evaluación de la distribución de brillo superficial y permite una detallada identificación de la forma y la longitud de la barra a través de la geometría local de las isofotas. Este método es esencial para entender la estructura y orientación de la barra dentro de la galaxia. Por otro lado, el método de Fourier, tratado en la sección 2.2.2, ofrece una perspectiva complementaria al descomponer la imagen en componentes espaciales. Este análisis revela patrones de simetría y características estructurales de la barra que no siempre son evidentes en el análisis de isofotas elípticas.

Al integrar ambos enfoques, obtenemos una caracterización más completa de las barras estelares. El método de isofotas elípticas proporciona detalles sobre la geometría específica de la barra, mientras que el análisis de Fourier aporta información sobre su simetría y estructura global, permitiendo una evaluación más robusta de las características y propiedades de las barras en galaxias.

### 2.3.1 Resultados en Estudios Previos en la Comparación de los Métodos

En una investigación sobre barras en galaxias realizada en [93], se emplean los métodos de isofotas elípticas y el análisis de Fourier para caracterizar las estructuras de las barras en galaxias. Los autores destacan que, aunque ambos métodos aportan enfoques distintos, presentan fortalezas y limitaciones particulares que afectan su eficacia en diferentes contextos.

**Detección de barras:** Los resultados muestran que el ajuste de isofotas elípticas tiende a identificar una mayor fracción de galaxias con barras, alcanzando el 48%. Este método capta prácticamente todas las barras fuertes (SB) y alrededor de la mitad de las barras intermedias (SAB) identificadas visualmente. En contraste, el análisis de Fourier, utilizando los criterios de amplitud relativa y fase constante propuestos por [98], detecta principalmente barras fuertes, resultando en una fracción menor del 36%.

**Fortalezas y limitaciones de los métodos:**

- Isofotas elípticas: Es eficaz para identificar barras en galaxias con estructuras claras, pero tiende a fallar en aquellas con bulbos prominentes, ya que el bulbo puede enmascarar la transición entre barra y disco.

- Análisis de Fourier: Facilita la identificación de barras robustas y cuantifica su fuerza, pero también tiene limitaciones en galaxias con bulbos prominentes, donde puede confundir el bulbo con la barra.

**Dependencia de las propiedades de las galaxias:** El método de isofotas elípticas identifica más barras en galaxias de tipo tardío, con bulbos menos prominentes, mientras que el análisis de Fourier detecta barras con mayor frecuencia en galaxias de tipo temprano, donde las barras suelen ser más fuertes.

**Comparación de precisión:** Las isofotas elípticas son más eficaces en la detección de barras débiles, mientras que Fourier mide con precisión la fuerza de las barras prominentes. Así, ambos métodos se complementan, proporcionando información valiosa en distintos contextos.

### 2.3.2 Eficacia, Precisión y Complementariedad de los Métodos

En esta sección, se analiza la eficacia, precisión y complementariedad de los métodos de isofotas elípticas y Fourier en el contexto de la detección y caracterización de barras estelares en galaxias. Ambos ofrecen herramientas valiosas para medir tanto la longitud como la fuerza de las barras, pero cada uno presenta enfoques distintos que, al combinarse, permiten obtener un análisis más robusto y completo.

- **Eficacia:** El método de isofotas elípticas permite identificar la longitud de las barras mediante los perfiles radiales de elipticidad. Es especialmente útil en la detección de barras débiles que otros métodos podrían no identificar. Por otro lado, el análisis de Fourier, al descomponer la imagen en componentes de frecuencia, facilita la evaluación detallada de la estructura de las barras, destacando por su capacidad para cuantificar la fuerza de las barras.

- **Precisión:** La combinación de ambos métodos mejora la precisión en la caracterización de las barras estelares. El análisis de isofotas elípticas es eficaz en la detección de barras, pero su precisión depende de la calidad de la imagen y el ajuste de los perfiles. El análisis de Fourier, por su parte, proporciona una visión más robusta y cuantitativa al descomponer la imagen en componentes de frecuencia. Al combinar ambos enfoques, se asegura una caracterización más completa y precisa.

- **Complementariedad y Flexibilidad:** El uso conjunto de ambos métodos proporciona una mayor flexibilidad en el análisis de diferentes tipos de galaxias. Las isofotas elípticas son más efectivas en galaxias con barras débiles o difusas, mientras que Fourier sobresale en galaxias con barras prominentes. Esta combinación permite adaptarse a una variedad de morfologías galácticas y diferentes calidades de imagen, asegurando un análisis más confiable.

En resumen, el análisis comparativo entre los métodos de isofotas elípticas y el análisis de Fourier demuestra que cada uno aporta ventajas complementarias en la caracterización de las barras estelares. Las isofotas elípticas se destacan por su capacidad para describir de manera precisa la longitud y geometría de las barras, particularmente en galaxias con estructuras bien definidas y sin bulbos dominantes. Este enfoque es ideal para detectar barras más sutiles y de menor intensidad. Por otro lado, el análisis de Fourier se posiciona como una herramienta más robusta para cuantificar la fuerza de las barras estelares, siendo especialmente efectivo en la identificación de barras prominentes en galaxias con bulbos más masivos. La combinación de ambos métodos no solo mejora la precisión en la detección y caracterización de las barras estelares, sino que también permite una interpretación más rica y matizada de las propiedades morfológicas de las galaxias. Al utilizar ambos enfoques, logramos abarcar tanto las barras débiles y difusas como las estructuras más prominentes y dominantes, lo que proporciona una visión integral del papel de las barras en la evolución galáctica.

En la siguiente sección, se describen los datos utilizados en nuestra investigación, los cuales son fundamentales para validar y expandir los resultados obtenidos con estos métodos complementarios.

# Capítulo 3

## Muestra Observacional de Galaxias

En este capítulo, presentamos la muestra observacional de galaxias utilizadas en esta tesis, enfocándonos en aquellas con corrimientos al rojo en el rango de 0.027 a 0.7. Esta muestra ha sido seleccionada de dos importantes sondeos astronómicos: el Sloan Digital Sky Survey (SDSS) [59] (Figura 3.1a) y el Great Observatories Origins Deep Survey (GOODS) [60] (Figura 3.1b).

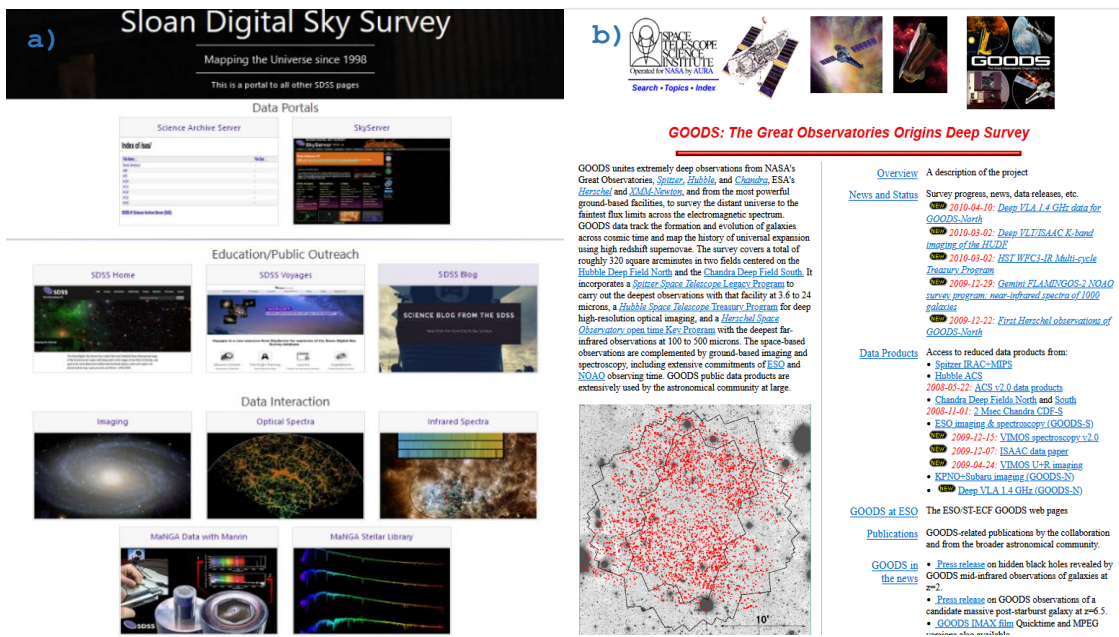

Figura 3.1. Sitios en línea para el acceso a las bases de datos de a) Sloan Digital Sky Survey (SDSS) [103], b) Great Observatories Origins Deep Survey (GOODS) [104].

El objetivo principal de este trabajo es investigar la evolución de las barras estelares a lo largo del tiempo cósmico, un aspecto clave en la comprensión de la evolución morfológica y dinámica de las galaxias.

Dado que el SDSS cubre un rango de corrimientos al rojo más cercano, proporciona una visión detallada de las galaxias en el universo local ($z \leq 0.3$), mientras que el GOODS, con su enfoque en corrimientos al rojo más altos ($z \geq 0.3$), nos permite extender nuestro análisis hacia épocas más tempranas del universo. Este enfoque combinado nos brinda la oportunidad de estudiar la evolución de las barras estelares a lo largo de un intervalo de tiempo significativo, permitiendo un análisis comparativo entre el universo cercano y el universo distante.

## 3.1 Clasificación Morfológica

Para definir las muestras en estudio, es esencial primero considerar el contexto del programa Estudio de Evolución de Galaxias de Masa Intermedia (IMAGES por sus siglas en inglés) (Figura 3.2). Esta encuesta, establecida por la Agencia Espacial Europea (ESO), tiene como objetivo obtener una visión integral de una muestra representativa de galaxias distantes. El estudio IMAGES busca establecer la evolución cinemática y morfológica de las galaxias, evaluar los procesos físicos que han dado forma a la secuencia de Hubble actual, y comprender la historia de formación estelar de galaxias individuales. Además, examina la evolución de la relación masa-metalicidad, el momento angular, el tamaño y la masa de las galaxias [105] [106] [107] [108] [109].

El estudio IMAGES se ha aplicado inicialmente a galaxias distantes utilizando datos de GOODS, proporcionando una visión detallada de galaxias en etapas tempranas de evolución. Para complementar la muestra local, se ha aplicado el mismo enfoque del estudio IMAGES a una muestra de galaxias cercanas obtenidas del SDSS. Esto permite una comparación directa entre galaxias locales y galaxias distantes, cubriendo así dos periodos clave en la evolución de las galaxias.

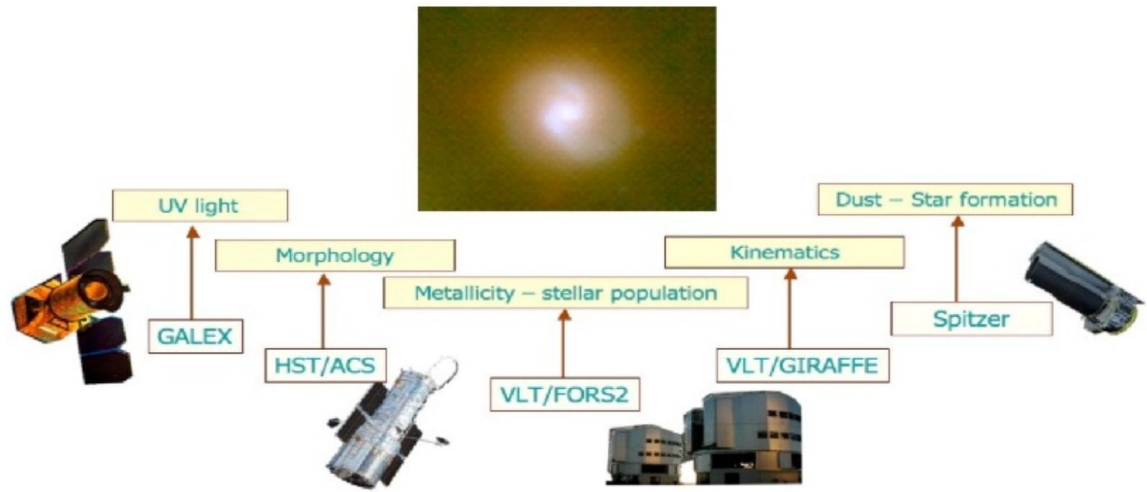

Figura 3.2. Muestra lo instrumentos utilizados por IMAGES y las cantidades observadas relacionadas permiten estudiar la cinemática, la morfología, las propiedades de las líneas de emisión/absorción y las propiedades químicas integrales de las galaxias. Imagen tomada de [110].

Uno de los componentes clave para estos estudios es la clasificación morfológica precisa de las galaxias. Utilizando un esquema sistemático, el cual se describirá en esta misma sección más adelante, se ha logrado reducir la subjetividad en la clasificación de galaxias en elípticas, lenticulares, espirales y peculiares. Este enfoque es fundamental para identificar de manera confiable las características morfológicas clave, lo que es esencial para estudios como la detección de barras tanto en galaxias cercanas como en lejanas.

Nuestro trabajo se basa en esta clasificación morfológica para detectar la presencia de barras en galaxias de ambos grupos, cercanas y lejanas, agrupadas según su morfología. Esta distinción es crucial para analizar cómo la presencia de barras está relacionada con el proceso evolutivo de cada tipo de galaxia y cómo estos elementos estructurales varían con el tiempo. El uso combinado de datos del SDSS para galaxias cercanas y GOODS para galaxias lejanas proporciona una base sólida para examinar la evolución de las propiedades físicas de las galaxias a lo largo del tiempo cósmico. La aplicación del enfoque del estudio IMAGES en ambas muestras permite trazar conexiones entre las galaxias locales actuales y sus contrapartes más distantes, ofreciendo una comprensión más completa de cómo las galaxias han llegado a su configuración actual mediante procesos evolutivos y cómo la formación de barras ha influido en esa evolución.

Para asegurar que el proceso de clasificación morfológica sea reproducible y reducir al mínimo la subjetividad inherente, se utilizó un árbol de decisiones basado en [108], que fue mejorado posteriormente en [111]. Este último estudio es el responsable de la clasificación morfológica de la muestra empleada en este trabajo. El árbol de decisiones está diseñado específicamente para clasificar galaxias en tres tipos principales: elípticas (*E*), lenticulares (*S0*) y espirales (*Sp*). Integra diversos parámetros como la relación de luminosidad entre el Bulbo y la galaxia total (B/T), el radio de media luz, los parámetros del modelo GALFIT, las imágenes residuales y de error, los perfiles bulbo-disco-galaxia, los mapas a dos colores y las imágenes en tres colores. El objetivo de este enfoque sistemático es aplicar de manera coherente estos criterios a lo largo de las muestras, lo que asegura la solidez y consistencia del proceso de clasificación.

El método clasifica las galaxias en cuatro categorías principales:

- **Galaxias elípticas (*E*):** Caracterizadas por una relación B/T entre 0.8 y 1.0, estas galaxias presentan perfiles de brillo suaves y homogéneos, sin rasgos distintivos ni estructuras visibles de disco.

- **Galaxias lenticulares (*S0*):** Con una relación B/T entre 0.5 y 0.8, estas galaxias no muestran brazos espirales y se distinguen por un bulbo rojizo en contraste con el disco. Los centros del bulbo y el disco coinciden, y el disco es altamente simétrico y regular.

- **Galaxias espirales (*Sp*):** Definidas por una estructura de disco con brazos espirales visibles, estas galaxias presentan una relación B/T inferior a 0.5 y un bulbo rojizo en contraste con el disco. Los centros del bulbo y el disco están alineados, lo que contribuye a su apariencia simétrica.

- **Galaxias peculiares (*Pec*):** Esta categoría agrupa galaxias con características asimétricas o irregulares, subdividiéndose en posibles fusiones (Pec/M), estructuras tipo renacuajo (Pec/T), galaxias irregulares (Pec/Irr), y galaxias compactas (Pec/C), estas últimas con un radio de media luz inferior a 1.0 kpc.

Para nuestro estudio, nos enfocaremos únicamente en las galaxias de discos, seleccionando específicamente galaxias lenticulares (*S0*) y espirales (*Sp*). Dado que nuestro principal objetivo es la detección y caracterización de barras estelares, estas galaxias, con estructuras de disco bien definidas, proporcionan las condiciones óptimas para identificar y analizar con precisión las barras. En las galaxias elípticas es difícil que se formen barras, ya que requieren un disco dinámicamente frío para su formación. Por lo que, las galaxias elípticas y peculiares quedan fuera del alcance de este análisis debido a la ausencia de características morfológicas relevantes para la detección de barras.

## 3.2 Sloan Digital Sky Survey (SDSS)

Para realizar un estudio astronómico es esencial contar con la instrumentación adecuada, ya que la recolección de datos depende de tener los mecanismos y procedimientos correctos. Los telescopios, fundamentales para la detección y registro de fuentes, requieren un uso y una sincronización precisos para mapear correctamente las áreas del cielo asignadas. Antes de explorar el SDSS, es crucial entender cómo se recopilan y se ponen a disposición los datos. El SDSS ha sido fundamental en la astronomía moderna, con observatorios en dos ubicaciones clave: el Observatorio Apache Point en Nuevo México y el Observatorio Las Campanas en Chile, lo que permite que el estudio cubra ambos hemisferios (Figura 3.3). Estos observatorios albergan tres telescopios: el telescopio Sloan Foundation de 2.5 m y el telescopio NMSU de 1 m en Apache Point, y el telescopio Irénée du Pont en Las Campanas. El telescopio Sloan Foundation fue el primero en ser utilizado [112] y desempeñó un papel crucial en la toma de imágenes fotométricas y espectroscópicas, como se describe en [113].

El SDSS, dividido en cinco fases principales (SDSS I, II, III, IV, y V) [114] [115] [116] [117], ha evolucionado significativamente desde sus inicios. Las fases SDSS I y II se centraron en la recolección de datos ópticos, con lanzamientos de datos hasta el DR7 que abarcan estas primeras colecciones. Estos datos ópticos han sido esenciales para mapear el Universo en múltiples bandas de luz, proporcionando un valioso recurso para la astronomía observacional. Durante el SDSS III, que abarca los lanzamientos de datos del 8 al 12, se introdujeron mejoras que permitieron extender y profundizar los estudios, incluyendo nuevos instrumentos y técnicas espectroscópicas. SDSS IV, a su vez, es una versión actualizada y mejorada del SDSS III [138], y ha continuado con la tradición de lanzar datos acumulativos, que hasta la fecha incluyen desde el DR13 al DR17.

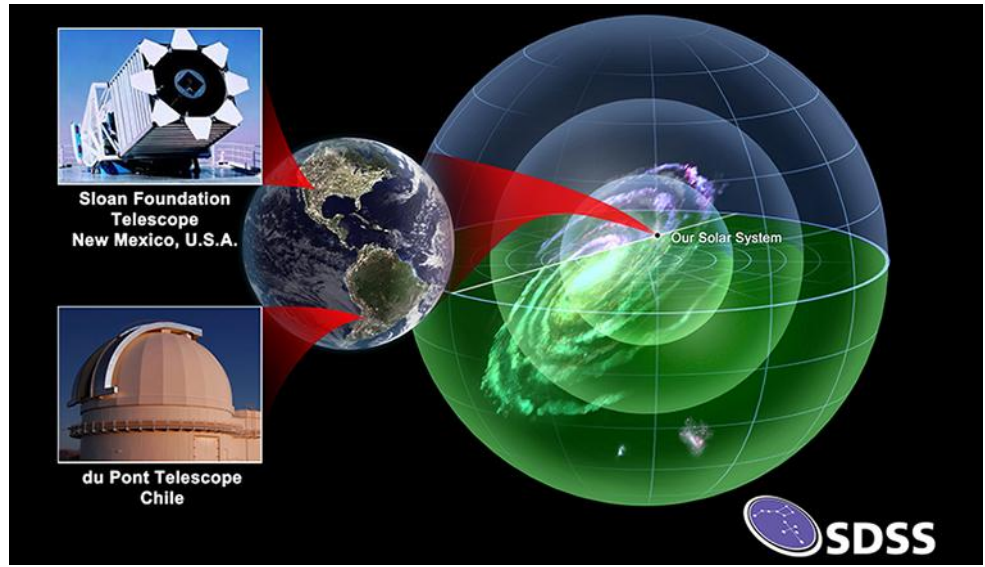

Figura 3.3. El SDSS utiliza tanto el Telescopio de la Fundación Sloan en el Observatorio Apache Point, como el Telescopio Irénée du Pont en el Observatorio de Las Campanas, en Chile, para sus observaciones. Con un telescopio en cada hemisferio, el SDSS cubre la observación de todo el cielo visible desde la Tierra. Créditos: Dana Berry / SkyWorks Digital Inc. and the SDSS collaboration. Imagen tomada de [119]

En particular, los datos obtenidos en SDSS I y II han sido cruciales para el estudio de la

estructura desde el Universo local hasta el Universo lejano y la identificación de galaxias

y cuásares. Aunque estos datos fueron recopilados en fases anteriores, siguen siendo de

gran relevancia para la comunidad científica. En esta tesis, se utilizan datos ópticos del

SDSS I, subrayando su importancia continua en la investigación astronómica.

### 3.2.1 Muestra Local de Galaxias

La muestra de galaxias locales utilizada en esta tesis fue previamente seleccionada

por [111], quienes emplearon el catálogo de galaxias [120] como base para sus criterios

de selección. La muestra inicial de galaxias fue elegida de un conjunto de 2253 galaxias

con magnitudes de Petrosian menores a $r_p = 16$, ubicadas en un área rectangular de 230

grados cuadrados en la región ecuatorial del hemisferio norte. El procesamiento de los datos se realizó utilizando el tercer lanzamiento de datos (DR3) del Sloan Digital Sky Survey (SDSS), excluyendo las galaxias que no contaban con espectros disponibles, dado que el estudio incluía un análisis morfo-cinemático. Las magnitudes relativas se obtuvieron del estudio 2MASS [121], lo que resultó en una selección inicial de 2113 galaxias. Posteriormente, la muestra se refinó a 1665 galaxias que cumplían con criterios específicos, como una magnitud absoluta en la banda $J$ de $M_J \leq -20.3$, con el objetivo de estudiar galaxias de masa intermedia.

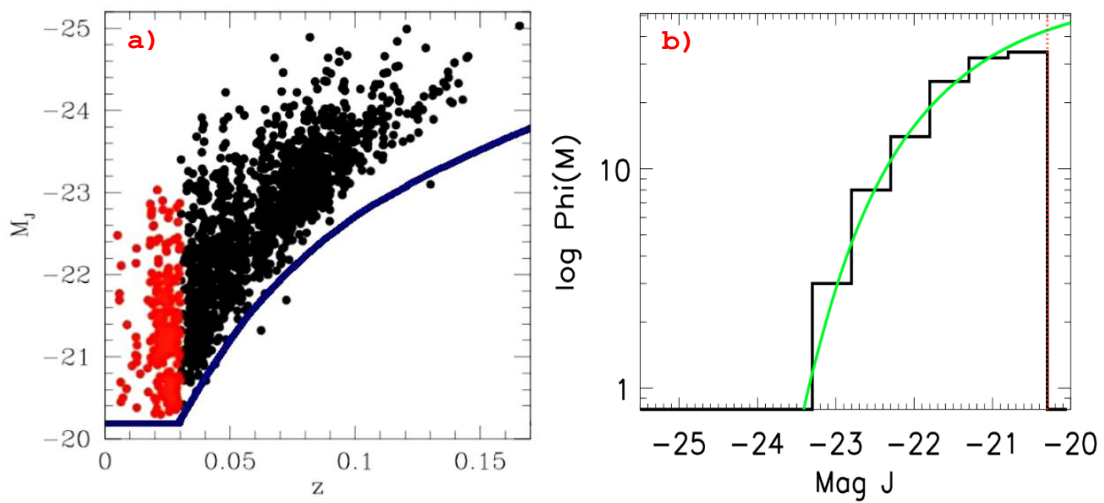

Figura 3.4. a) La muestra de galaxias se presenta en el plano $M_J$ versus $z$, donde los puntos rojos representan las galaxias incluidas en la submuestra representativa. b) La línea negra muestra el histograma de $M_J$ para las 116 galaxias. Como referencia, también se ha trazado la función de luminosidad observada en el rango de desplazamiento al rojo $0.0025 \leq z \leq 0.2$ (línea verde). Imagen tomada de [111].

La Figura 3.4a muestra la distribución de $M_J$ frente al corrimiento al rojo en el catálogo local del SDSS, como se reporta en [120]. Se observa una ausencia significativa de galaxias tenues en $z > 0.03$, en particular aquellas galaxias azules con magnitudes cercanas al límite de $r_p = 16$. Esta observación se refuerza con la curva azul, que muestra una planitud espectral en las bandas $r$ y $J$, similar a la de una galaxia con un brote estelar cerca del umbral de magnitud de la muestra. Además, todas las galaxias con $M_J =< -20.3$ están incluidas en esta muestra para $z < 0.03$, sumando un total de 218 galaxias (representadas por puntos rojos). Finalmente, de estas 218 galaxias, solo se seleccionaron aquellas con espectros que contenían la línea de emisión $[OII]\lambda 3727$, resultando en una muestra final de 116 galaxias con un corrimiento al rojo entre $0.0207 \leq z \leq 0.030$.

La validación de esta muestra de 116 galaxias incluyó comparaciones con la función de luminosidad local documentada por [122]. Se realizó una prueba de Kolmogorov-Smirnov para evaluar la similitud entre esta muestra y la función de luminosidad local (Figura 3.4b), mostrando una alta probabilidad del 98% de que ambas compartan una distribución similar. Por lo tanto, esta muestra de galaxias es representativa de las galaxias con $M_J =< -20.3$ en el universo local.

## 3.3 Great Observatories Origins Deep Survey (GOODS)

Ya habiendo definido la muestra local con datos del SDSS, presentamos ahora la procedencia de la muestra de galaxias lejanas. GOODS es uno de los estudios más importantes realizados con el Telescopio Espacial Hubble, en colaboración con otros grandes observatorios espaciales como Spitzer y Chandra. El objetivo principal de GOODS es investigar la formación y evolución de galaxias a lo largo de la historia cósmica, abarcando dos regiones del cielo conocidas como GOODS-North y GOODS-South, que en conjunto cubren un área total de aproximadamente 320 arcmin² (Figura 3.5).

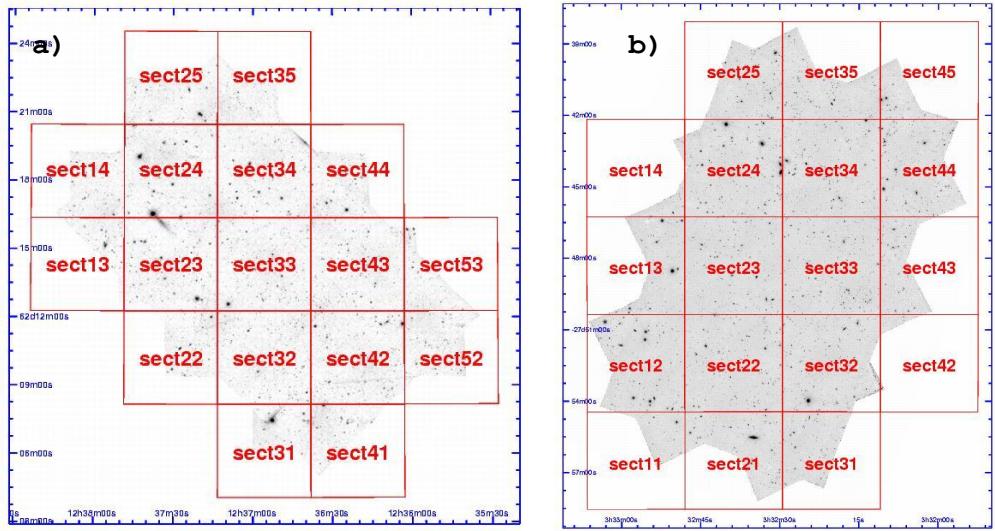

Figura 3.5. Las imágenes del programa GOODS del HST/ACS, calibradas y apiladas en los filtros F435W, F606W, F775W y F850LP, cubren de manera uniforme los campos GOODS: a) HDF-N y b) CDF-S, permitiendo una búsqueda profunda de galaxias en el universo [123].

El Hubble, a través de su cámara ACS (Advanced Camera for Surveys), realizó observaciones en múltiples filtros ópticos, proporcionando imágenes de alta resolución que han permitido estudiar galaxias desde el universo cercano hasta galaxias en formación en épocas muy tempranas, con corrimiento al rojo de hasta 6 y superiores [60]. Las observaciones de GOODS con el Hubble han sido fundamentales para identificar y caracterizar galaxias en diferentes etapas de su evolución, incluyendo la identificación de galaxias masivas en fases muy tempranas del universo y el estudio detallado de su estructura interna [124]. Además, las imágenes profundas obtenidas por el Hubble en el campo GOODS han permitido un análisis detallado de la morfología de las galaxias [125], ayudando a entender cómo las fusiones y otros procesos galácticos han influido en la evolución de las galaxias a lo largo del tiempo cósmico. GOODS también ha sido esencial para estudiar la formación estelar en galaxias distantes [126] y ha proporcionado un censo detallado de las poblaciones galácticas en diferentes épocas del universo [127].

### 3.3.1 Muestra de Galaxias Distantes

Basado en el estudio IMAGES, [111] se construyó una muestra de galaxias distantes combinando dos submuestras definidas por el índice de emisión equivalente $EW([OII]\lambda3727)$. Estas submuestras incluyen galaxias en formación estelar $EW([OII]\lambda3727) \geq 15\,\text{Å}$ y galaxias sin formación estelar $EW([OII]\lambda3727) < 15\,\text{Å}$ [128]. Las galaxias seleccionadas tenían una magnitud absoluta $M_J(AB) \leq -20.3$ y un desplazamiento al rojo $0.4 < z < 0.8$, en concordancia con los criterios del estudio IMAGES. Solo se incluyeron galaxias con imágenes de GOODS (ACS) v2.0 en al menos tres bandas $(V \rightarrow 5915\,\text{Å}, \quad i \rightarrow 7697\,\text{Å}, \quad z \rightarrow 9103\,\text{Å})$. Se observó una buena correspondencia entre las galaxias en formación estelar, que tienden a ser azules, y las galaxias sin formación estelar, que suelen ser rojas. La Figura 3.6 ilustra la comparación de los colores $(U - B)$ con el límite propuesto por [129] para separar galaxias azules y rojas.

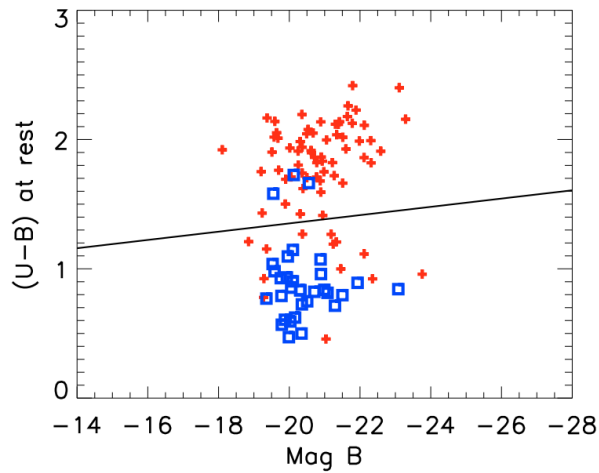

Figura 3.6. Distribución del color $U - B$ de la muestra distante (cuadrados azules: galaxias con formación estelar, cruces rojas: galaxias sin formación estelar). La línea negra representa la relación color-magnitud utilizada para separar las poblaciones azules y rojas. Imagen tomada [111].

Para las galaxias con formación estelar, se recopilaron 49 galaxias con $EW([OII]\lambda3727) \geq 15$ Å [111] en el estudio IMAGES. La Figura 3.7a muestra la distribución de magnitudes absolutas en la banda J de esta submuestra, comparada con la función de luminosidad en la banda $K$ de [129] para galaxias azules en el rango $0.25 < z < 0.75$. La corrección de magnitud $J - K$ es mínima, con un promedio de $-0.08 \pm 0.03$, y nuestra muestra sigue la misma distribución de la función de luminosidad (prueba Kolmogorov-Smirnov, probabilidad del 97%), confirmando que IMAGES es representativa de galaxias distantes en formación estelar. Para las galaxias sin formación estelar, se incluyeron 94 galaxias con $EW([OII]\lambda3727) < 15$ Å [129].

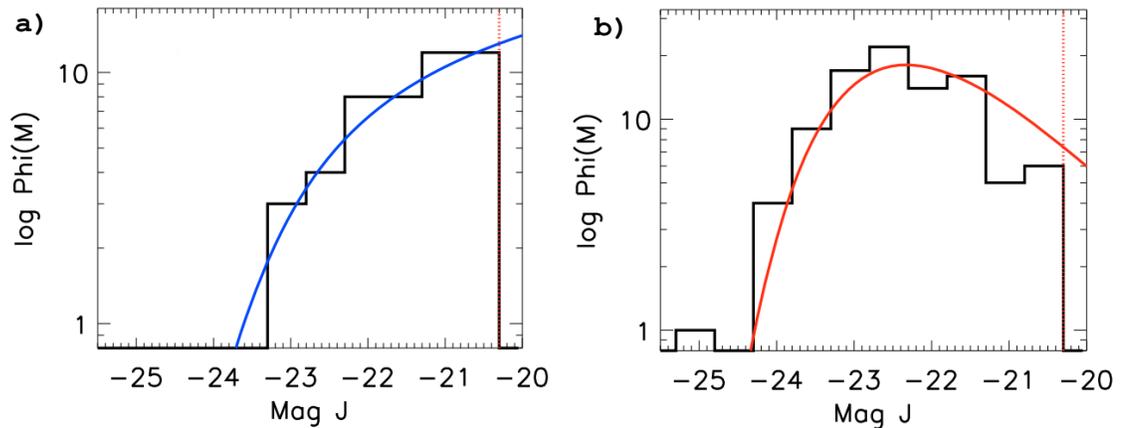

Figura 3.7. Muestra las distribuciones de magnitudes absolutas $M_J$ para dos submuestras de galaxias distantes: en formación estelar (panel a) y galaxias sin formación estelar (panel b). Para las galaxias en formación estelar, el histograma negro representa las 49 galaxias estudiadas, con una función de luminosidad observada (línea azul) para el rango de desplazamiento al rojo $0.025 < z < 0.75$. Para las galaxias quiescentes, el histograma negro muestra las 94 galaxias sin formación estelar, con una función de luminosidad observada (línea roja) para el mismo rango de z. En ambos paneles, la línea vertical punteada roja indica el límite de magnitud $M_J$ de $-20.3$ aplicado en el estudio IMAGES. Imagen tomada de [111].

La Figura 3.7b muestra su distribución en la banda J, comparada con la función de luminosidad en [129] para galaxias rojas en el mismo rango. La corrección $J - K$ se realizó de manera similar, y la prueba Kolmogorov-Smirnov indica una probabilidad del 94% de que ambas muestras provengan de la misma distribución, mostrando un claro pico por galaxias masivas de tipo temprano.

## 3.4 Evaluación Comparativa de las Muestras y su Entorno

### 3.4.1 Revisión de los Datos para la Comparación de Muestras

En [111], se llevó a cabo una comparación entre las muestras para evitar que estuvieran afectadas por sesgos relacionados con el muestreo, la profundidad y las correcciones k. Este análisis se basó en el estudio de [130], donde se creó un conjunto de galaxias del SDSS desplazadas artificialmente a corrimientos al rojo de $0.1 < z < 1.1$, con el fin de compararlas con galaxias de GEMS y COSMOS. En dicho estudio, no se encontraron diferencias significativas en la estimación de la magnitud absoluta, el radio de medio brillo y el índice de Sérsic. Las simulaciones realizadas indicaron que no existen efectos sistemáticos en los parámetros morfológicos y fotométricos. Al realizar una función gaussiana de la distribución de las galaxias de disco locales en fusión de corrimiento al rojo nos da un valor medio de $z = 0.027$ (Figura 3.8a); y para la muestra lejana $z = 0.7$ (Figura 3.8b). A $z = 0.7$ las condiciones de muestreo son óptimas, con el HST/ACS de GOODS proporcionando un FWHM de 0.108 arcsec (0.81 kpc), en comparación con un FWHM promedio de 1.4 arcsec (0.74 kpc) para el SDSS a $z = 0.027$. Además, es fundamental considerar tanto la profundidad óptica de las bases de datos como los efectos de la corrección k. En este contexto, la Tabla 3.1 compara las longitudes de onda observadas en las bandas $u$, $g$, $r$, $i$, $z$ del SDSS y las bandas del $V$, $i$, $z$ del HST/ACS de GOODS con las correspondientes bandas en el marco de reposo de las galaxias distantes, asumiendo un valor mediano de $z = 0.7$ para estas últimas. Esta comparación asegura una correspondencia adecuada entre las observaciones de las galaxias locales y distantes, lo que permite una evaluación fiable de sus propiedades.

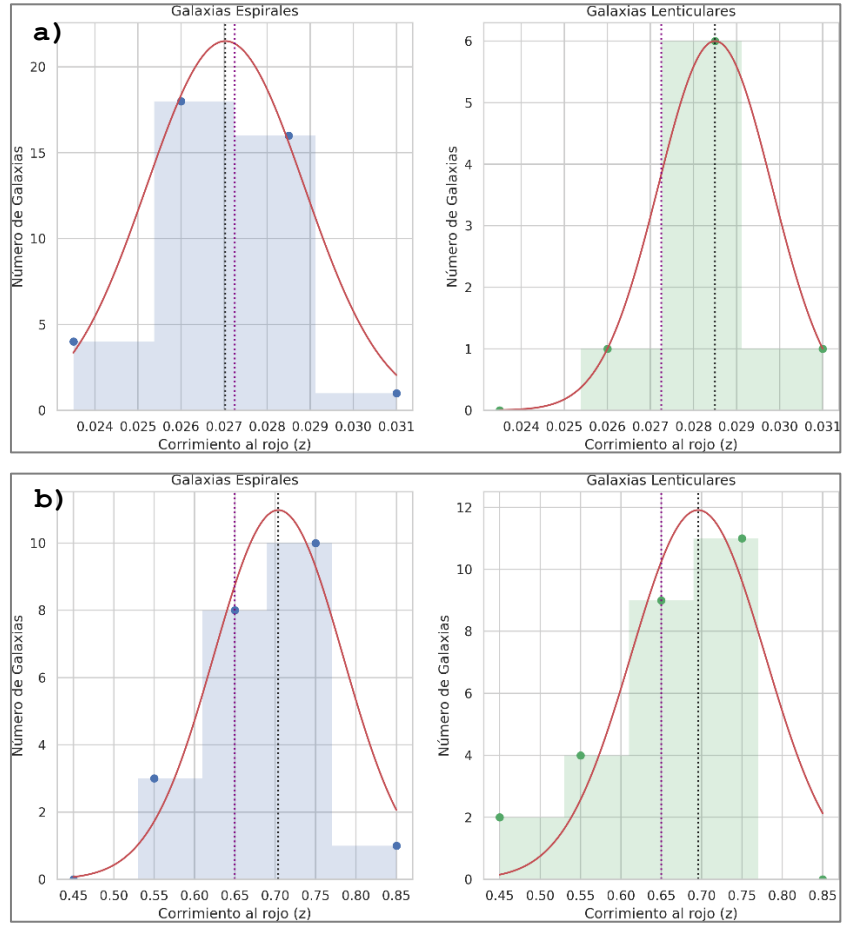

Figura 3.8. a) Distribución de las galaxias de disco locales en fusión de corrimiento al rojo, b) Distribución de las galaxias de disco lejanas en fusión de corrimiento al rojo

| Survey | — | u band | g band | r band | i band | z band |
|---|---|---|---|---|---|---|
| SDSS | — | 3551 Å | 4686 Å | 6165 Å | 7481 Å | 8931 Å |
| — | B band | V band | i band | z band | — | — |
| GOODS | 4312 Å | 5915 Å | 7697 Å | 9103 Å | — | — |
| Rest-frame | 2582 Å | 3542 Å | 4609 Å | 5451 Å | — | — |

Tabla 3.1. La tabla se presenta cómo las bandas de los surveys SDSS (primera fila) y GOODS (tercera fila) cubren longitudes de onda similares en el marco de reposo (cuarta fila) de las galaxias. En base a [111].

Dadas las condiciones óptimas de muestreo y la correspondencia en las bandas de observación entre las galaxias locales y distantes, es posible realizar una comparación robusta de las propiedades fotométricas de ambas muestras, particularmente en lo que respecta a la detección de estructuras como las barras estelares. Para asegurar una comparación coherente, hemos utilizado la *banda r* del SDSS para las galaxias locales y la *banda z* del GOODS para las galaxias distantes, lo que garantiza que las características morfológicas observadas en ambas muestras sean consistentes y comparables.

### 3.4.2 Muestras Representativas en Galaxias de Campo

Estudiar las barras en galaxias de campo resulta esencial para comprender de manera profunda su formación y evolución en diferentes entornos galácticos. Las galaxias de campo, a diferencia de las que se encuentran en cúmulos o en entornos más densos, permiten analizar cómo las barras se desarrollan en condiciones de aislamiento relativo, sin la influencia gravitacional significativa de grandes estructuras galácticas cercanas. Aunque la mayoría de los estudios sobre barras se han centrado en galaxias dentro de cúmulos o en combinaciones de entornos, la exploración de las características de las barras en galaxias de campo ofrece una perspectiva crítica, ya que estas presentan escenarios únicos en cuanto a su evolución dinámica y morfológica.

Para garantizar que la muestra seleccionada contenga solo galaxias de campo, se ha realizado una comparación exhaustiva con el catálogo de grupos de galaxias [131], el cual utiliza un método de búsqueda de grupos basado en halos aplicado a estudios de gran escala como el SDSS y el Two Micron All Sky Survey (2MRS). Este catálogo identifica grupos de galaxias y sus propiedades fundamentales, lo que nos permite determinar si las galaxias seleccionadas son galaxias centrales dentro de estos grupos. En este análisis, se constató que el 60% de las galaxias espirales y el 40% de las galaxias lenticulares en nuestra muestra son completamente aisladas, sin indicios de pertenecer a grandes estructuras. El resto de las galaxias presentan un número reducido de miembros en sus grupos, lo que refuerza la idea de que estas galaxias actúan como las centrales de sus respectivos sistemas. Además, como parte del análisis de entorno, se investigó la distancia a la galaxia vecina más cercana para cada objeto en la muestra, considerando aquellas galaxias que no fueran más débiles que la galaxia objetivo por más de 0.5 magnitudes en la $banda\ r$ y con una diferencia de velocidad radial de $\Delta v \leq 800\ km\ s^{-1}$. Este análisis reveló que las galaxias de la muestra no presentaban interacciones claras con otras galaxias cercanas, confirmando así que la muestra está compuesta por galaxias de campo relativamente aisladas. Esto es especialmente relevante para nuestro estudio, ya que permite asegurar que las propiedades morfológicas y fotométricas observadas, como la presencia de barras, no están siendo influenciadas por interacciones externas recientes o efectos de entorno denso.

En el caso de la muestra de galaxias más distantes, correspondientes a los datos del GOODS, hemos utilizado el trabajo [132] para garantizar que la muestra esté compuesta por galaxias de campo relativamente aisladas. Este catálogo, además de tener las propiedades como color, masa estelar, etc., permite identificar el vecino más cercano. Los resultados indicaron que las galaxias en la muestra del GOODS no están en interacción con otras galaxias cercanas, lo que confirma que también se trata de galaxias de campo. Este criterio, aplicado tanto en la muestra local del SDSS como en la muestra distante del GOODS, asegura que estamos comparando galaxias en entornos similares, libres de influencias externas que pudieran alterar sus estructuras.

Finalmente, la comparación entre ambas muestras, utilizando criterios homogéneos de selección en las *bandas r* del SDSS y *banda z* del GOODS, nos proporciona una visión completa de la evolución de las barras galácticas en galaxias de campo. La selección cuidadosa de estas galaxias, tanto locales como lejanas, libres de interacciones significativas con otras galaxias, nos permite evaluar de manera fiable cómo las barras evolucionan en entornos aislados, ofreciendo una oportunidad para investigar los procesos de formación y evolución de las barras sin la interferencia de factores externos.

### 3.4.3. Reducción del Sesgo por Inclinación en Galaxias con Barras

Para asegurar la precisión en la detección de barras galácticas y minimizar el sesgo asociado con la inclinación de las galaxias, aplicamos un filtro basado en la razón de ejes isofotales $b/a > 0.6$ en las bandas r SDSS y z GOODS. Esta restricción se basa en la recomendación de [133], quien sugirió que una razón de ejes mayor a 0.6 reduce significativamente el impacto de la inclinación en la visibilidad de las barras. (Figura 3.9)

La razón de ejes $b/a$ se calculó utilizando imágenes procesadas con SExtractor. Los valores de $a$ y $b$ fueron tomados del catálogo de resultados de SExtractor explicados en el capítulo 2. Esta restricción asegura que estamos considerando principalmente galaxias vistas de frente, minimizando así el sesgo de selección derivado de la inclinación de la galaxia. Este enfoque proporciona una base sólida para la estimación de la densidad de barras galácticas y permite una evaluación más precisa de la fracción de galaxias barradas en diferentes distancias cosmológicas.

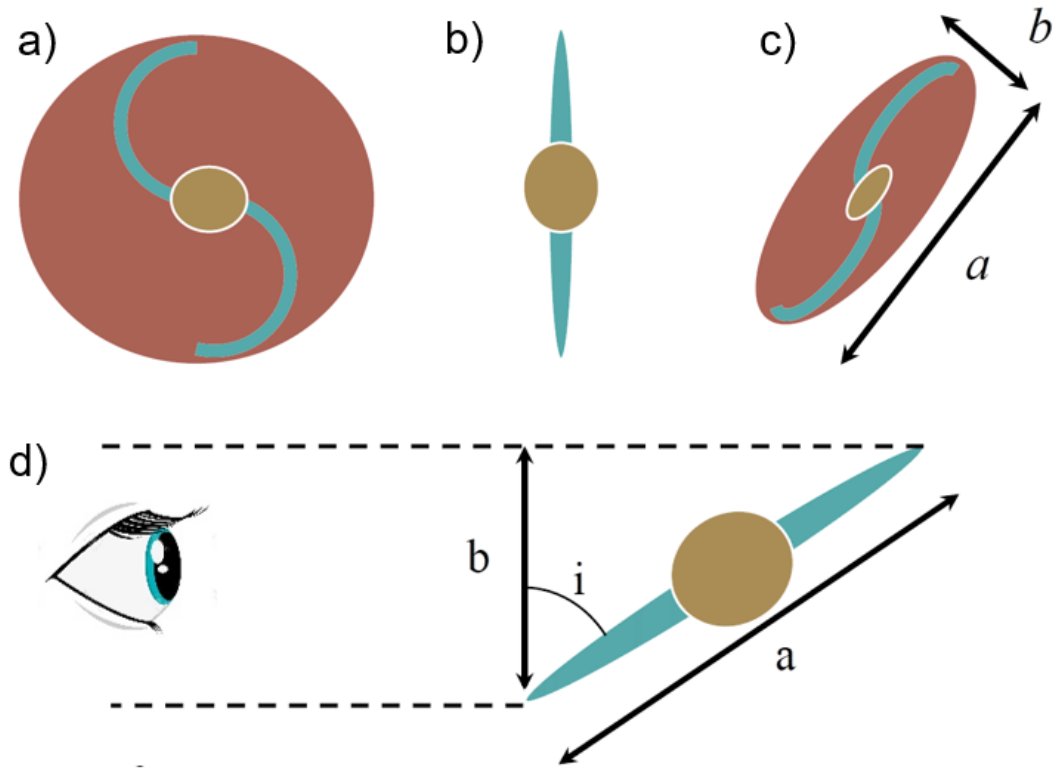

Figura 3.9. Razón de ejes $b/a$ de las galaxias. La figura muestra: (a) una galaxia vista de frente con $b/a \approx 1$; (b) una galaxia vista de perfil con $b/a \approx 0$; (c) definición de los ejes $a$ y $b$; y (d) el filtro aplicado que selecciona galaxias con $b/a > 0.6$ para minimizar el sesgo observacional

Al aplicar de esta reducción no solo minimiza el sesgo por inclinación y absorción interna, sino que también garantiza que las galaxias seleccionadas sean aquellas en las que las barras galácticas pueden ser identificadas de manera más precisa por ambos métodos. De esta manera, hemos establecido una metodología robusta para detectar y estudiar barras galácticas en nuestra muestra. Este enfoque riguroso sienta las bases para un análisis cuantitativo más detallado en las secciones posteriores, donde evaluaremos los resultados de la fracción y evolución de las barras a lo largo del tiempo cosmológico.

# Capítulo 4

## Resultados y Análisis de Datos

En esta sección, se ofrecerá una descripción concisa de los datos utilizados para el análisis de los resultados, centrándonos en los aspectos clave que influencian la interpretación de las fracciones de galaxias con barra. Se presentan los resultados y los análisis de la fracción de galaxias de disco con barra en dos muestras seleccionadas: una muestra local de la banda $r$ de SDSS y una muestra distante de la banda $z$ de GOODS, que ya fueron discutidas en el Capítulo 3.

La fracción de galaxias con barra es un parámetro crucial para entender la evolución morfológica de las galaxias, y en este trabajo se ha definido como:

$$f_{bar} = \frac{g_{bar}}{N},$$

donde $g_{bar}$ representa el número de galaxias barradas y $N$ el número total de galaxias.

## 4.1 Fracción de Barras en Función del Desplazamiento al Rojo

### 4.1.1 Resultados de Fracción de Barras por Isofotas Elípticas y Fourier

En esta subsección, se exponen las medidas de radio de la barra ($R_{bar}$), obtenidas con ambos métodos (análisis de los perfiles de isofotas elípticas y descomposición de Fourier), así como la fuerza de la barra ($S_b$) derivada del análisis de Fourier. Se han analizado tanto galaxias espirales como lenticulares en dos categorías: locales y distantes. En las Figuras 4.1 y 4.2, se muestran los resultados obtenidos mediante ambos métodos, el análisis de perfiles de isofotas elípticas y la descomposición de Fourier, aplicados a una galaxia espiral barrada de nuestra muestra representativa.

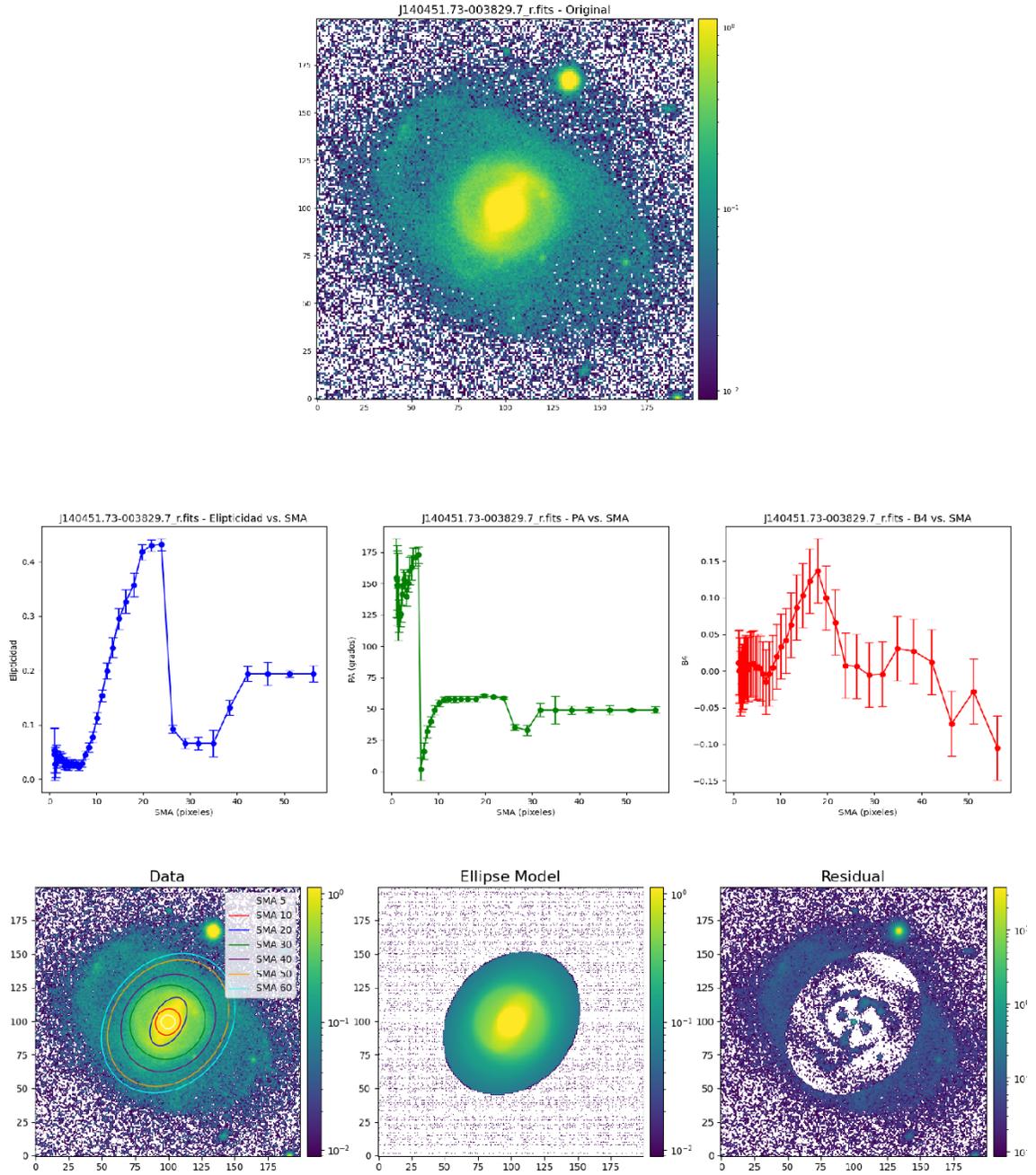

Figura 4.1. Resultado del método de isofotas elípticas a la galaxia espiral J140451.73-003829.7 de la muestra local $z \sim 0.027$. De arriba hacia abajo: se observa la galaxia a estudiar en el primer panel. En la segunda fila de imágenes, de izquierda a derecha, están las gráficas $e$ $vs$ $SMA$, $PA$ $vs$ $SMA$, $B_4 vs$ $SMA$, en donde se observa los cambios característicos para validar una galaxia barrada, explicados en el Capítulo 2. En la última fila de imágenes, de izquierda a derecha, se observa la aplicación de las isofotas a la galaxia, la galaxia modelo generada por la iteración y el residuo obtenido a partir de las dos anteriores.

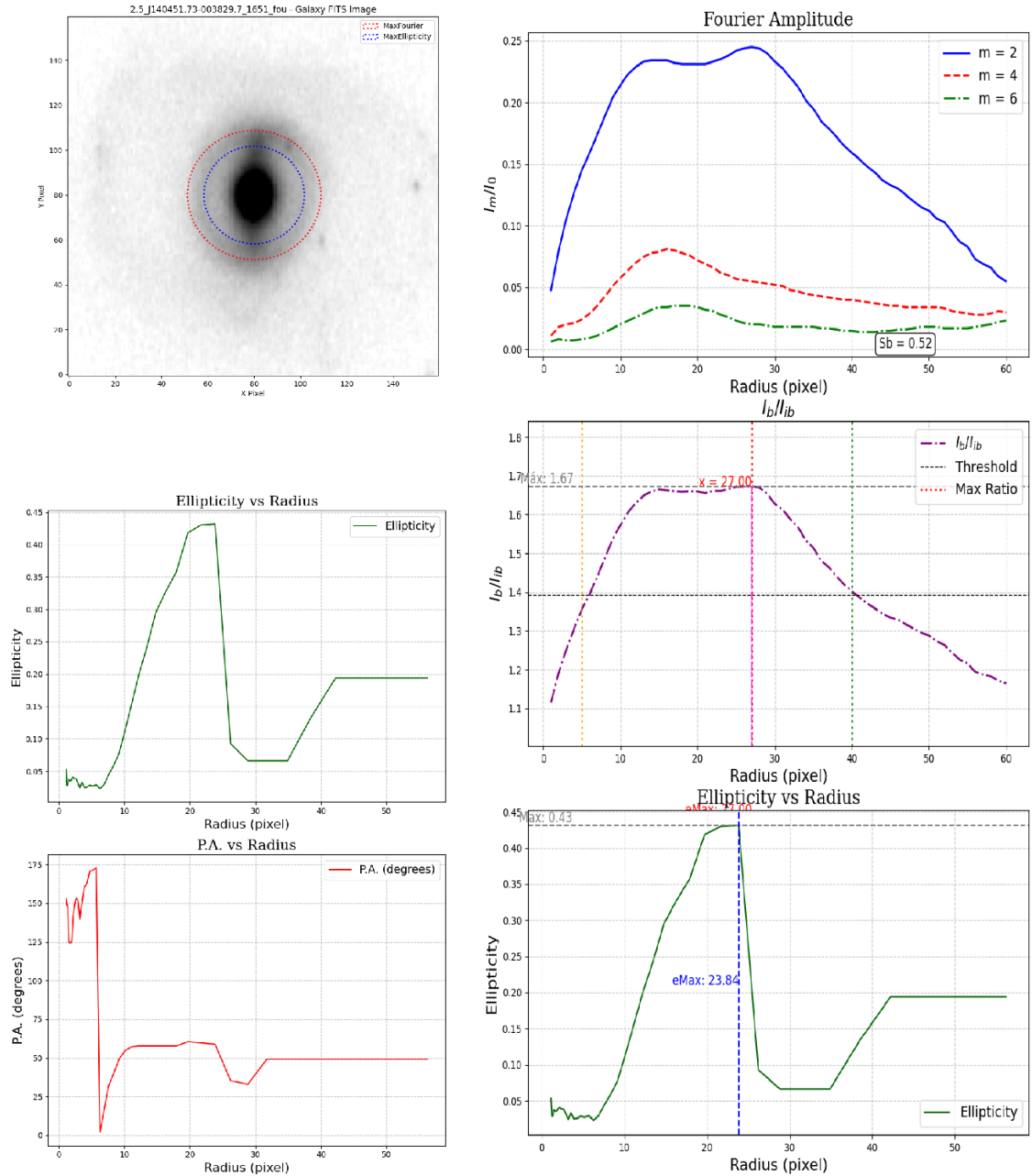

Figura 4.2. Resultado del método de Fourier para misma galaxia espiral J140451.73-003829.7 de la muestra local $z \sim 0.027$. De arriba hacia abajo: se observa, de izquierda a derecha, la galaxia con los círculos concéntricos que denotan la aproximación de ($R_{bar}$) del método de isofotas elípticas (línea punteada azul) y del método de Fourier (línea punteada roja); los modos de Fourier donde claramente se observa la presencia de la barra y el cuadro inferior derecho muestra la magnitud de la fuerza de la barra. En la segunda fila de imágenes, de izquierda a derecha, están las gráficas $e$ $vs$ $radio$ y $I_b/I_{ib}$ $vs$ $radio$, esta última marcando la línea puntea roja del $R_{bar}$. En la tercera fila están las gráficas de $PA$ $vs$ $SMA$ y $e$ $vs$ $radio$, la última marcando la línea puntea azul del $R_{bar}$. La validación de la barra esta explicado en Capitulo 2.

En las secciones de Apéndice C y D se incluyen las gráficas de las galaxias barradas y no barradas más relevantes para ambos métodos, proporcionando una comparación de los ajustes y resultados. Estas gráficas permiten observar con mayor claridad las diferencias entre los métodos, aportando un contexto visual que complementa los valores numéricos tabulados, facilitando una comprensión más profunda de las discrepancias observadas. A continuación, se muestras las tablas de:

- **Galaxias locales ($z = 0.027$)**

**Galaxias espirales locales barradas**

La Tabla 4.1 muestra los valores de $R_{bar}$ y $S_b$ para las galaxias espirales barradas locales. Las mediciones del radio de la barra obtenidas mediante el análisis de isofotas elípticas se presentan en la segunda columna, mientras que los valores calculados a través del método de Fourier se encuentran en la tercera columna. La fuerza de la barra, que refleja la intensidad estructural de la barra, se detalla en la última columna.

**Galaxias lenticulares locales barradas**

En la Tabla 4.2 se presentan los resultados correspondientes a las galaxias lenticulares locales barradas. Al igual que en el caso anterior, se incluyen las mediciones del radio de la barra obtenidas mediante ambos métodos (isofotas elípticas y análisis de Fourier), así como la fortaleza de la barra, calculada a través de la descomposición de Fourier.

| Galaxias espirales locales barradas | | | |
|---|---|---|---|
| *Galaxias* | $R_{bar}$ *(pixel)* *ISOFOTAS* | $R_{bar}$ *(pixel)* *FOURIER* | $S_b$ |
| J111339.88+004915.5 | 4 | 6 | 1.30 |
| J111202.58-001029.3 | 8 | - | - |
| J140451.73-003829.7 | 27 | 23 | 0.52 |
| J143411.25+003656.6 | 11 | 12 | 0.55 |
| J120127.92-004306.1 | 13 | 23 | 0.45 |
| J134822.64-004601.1 | 9 | 10 | 0.74 |
| J135102.22-000915.1 | 14 | 12 | 0.31 |
| J152254.81-005343.2 | 13 | - | - |
| J103657.37+001347.0 | 14 | - | - |
| J122411.84+011246.8 | 10 | - | - |
| J140320.74-003259.7 | 12 | 14 | 0.49 |
| J140452.62-003640.5 | 27 | 28 | 0.30 |
| J140601.42-001837.9 | 6 | - | - |
| J140831.60-000737.3 | 8 | - | - |
| J112418.64+003837.4 | - | 8 | 0.71 |
| J143519.72+002832.6 | - | 5 | 0.69 |
| J142720.36+010133.0 | - | 26 | 0.41 |
| J112535.07-004605.6 | - | - | - |
| J112738.15+003950.9 | - | - | - |
| J113116.65+001323.3 | - | - | - |
| J113209.23-005633.3 | - | - | - |
| J113438.36+001119.1 | - | - | - |
| J114453.04+005622.6 | - | - | - |
| J140013.27-005745.8 | - | - | - |
| J145052.33+010956.4 | - | - | - |
| J151926.88-005526.0 | - | - | - |
| J152024.55-001330.8 | - | - | - |
| J150648.62+005124.6 | - | - | - |
| J115912.88-003125.7 | - | - | - |
| J094847.11+010311.5 | - | - | - |
| J113439.15+000729.1 | - | - | - |
| J094913.51+011039.8 | - | - | - |
| J110840.90+002330.2 | - | - | - |
| J113833.27-011104.1 | - | - | - |
| J114006.20-005405.0 | - | - | - |
| J144613.34+005157.7 | - | - | - |
| J111849.55+003709.3 | - | - | - |
| J122337.48-002821.3 | - | - | - |
| J135807.05-002332.9 | - | - | - |
| J141814.91+005327.9 | - | - | - |
| J142804.53+010022.6 | - | - | - |
| J143101.75+011434.2 | - | - | - |
| J112535.07-004605.6 | - | - | - |
| J112738.15+003950.9 | - | - | - |
| J113116.65+001323.3 | - | - | - |

Tabla 4.1. Resultados de las galaxias espirales locales barradas.

| Galaxias lenticulares locales barradas | | | |
|---|---|---|---|
| *Galaxias* | $R_{bar}$ *(pixel)* *ISOFOTAS* | $R_{bar}$ *(pixel)* *FOURIER* | $S_b$ |
| J143124.59+011403.7 | 28 | 18 | 0.55 |
| J143540.07+001217.7 | 11 | 12 | 0.61 |
| J112437.05-005930.7 | - | - | - |
| J113523.27+000525.9 | - | - | - |
| J103534.47-002116.2 | - | - | - |
| J113420.50+001856.4 | - | - | - |
| J152327.26-010956.0 | - | - | - |
| J144258.24+001608.3 | - | - | - |

Tabla 4.2. Resultados de las galaxias lenticulares locales barradas.

Vemos ahora las tablas de:

- **Galaxias distantes ($z = 0.7$)**

**Galaxias espirales distantes barradas**

La Tabla 4.3 presenta los resultados obtenidos para las galaxias espirales distantes barradas. Siguiendo la misma estructura que en las tablas anteriores, se incluyen las mediciones del radio de la barra obtenidas mediante ambos métodos (isofotas elípticas y análisis de Fourier), así como la fortaleza de la barra derivada del análisis de Fourier.

**Galaxias lenticulares distante barradas**

Finalmente, la Tabla 4.4 presenta los valores obtenidos para las galaxias lenticulares distantes barradas, incluyendo las mediciones del radio de la barra obtenidas mediante ambos métodos (isofotas elípticas y análisis de Fourier), así como la fortaleza de la barra derivada del análisis de Fourier.

| Galaxias espirales distantes barradas | | | |
|---|---|---|---|
| *Galaxias* | $R_{bar}$ *(pixel)* *ISOFOTAS* | $R_{bar}$ *(pixel)* *FOURIER* | $S_b$ |
| J033221.99-274655.9 | 16 | 12 | 0.38 |
| J033243.21-274457.0 | 14 | - | 0.08 |
| J033222.58-274425.8 | 26 | 19 | 0.27 |
| J033223.40-274316.6 | 13 | 13 | 0.47 |
| J033233.08-275123.9 | 22 | 14 | 0.18 |
| J033219.68-275023.6 | - | - | - |
| J033224.06-274911.4 | - | - | - |
| J033218.81-274910.0 | - | - | - |
| J033248.13-274844.9 | - | - | - |
| J033237.54-274838.9 | - | - | - |
| J033240.67-274730.9 | - | - | - |
| J033216.53-274727.2 | - | - | - |
| J033238.60-274631.4 | - | - | - |
| J033226.87-274528.2 | - | - | - |
| J033241.32-274436.1 | - | - | - |
| J033233.82-274410.0 | - | - | - |
| J033230.03-274347.3 | - | - | - |
| J033221.42-274231.2 | - | - | - |
| J033231.58-274121.6 | - | - | - |
| J033236.21-275206.0 | - | - | - |
| J033253.26-275133.9 | - | - | - |
| J033229.41-275109.8 | - | - | - |

Tabla 4.3. Resultados de las galaxias espirales distantes barradas.

En la Figura 4.3, correspondiente a las galaxias de disco locales, se observa una clara tendencia positiva en la relación entre $R_{bar}$ de isofotas elípticas y $R_{bar}$ de Fourier. Los puntos rojos representan galaxias espirales y los puntos azules representan galaxias lenticulares. La línea roja ajustada indica una buena correlación entre los dos métodos, evidenciada por el coeficiente de correlación de Pearson, que es $r = 0.74$. Este valor sugiere una correlación positiva fuerte, lo que implica que los dos métodos son coherentes en la estimación del tamaño de la barra en las galaxias locales.

| Galaxias lenticulares distantes barradas | | | |
|---|---|---|---|
| *Galaxias* | $R_{bar}$ *(pixel)* *ISOFOTAS* | $R_{bar}$ *(pixel)* *FOURIER* | $S_b$ |
| J033217.29-274807.5 | 15 | 10 | 0.27 |
| J033205.97-274601.4 | 14 | 15 | 0.77 |
| J033212.31-274527.4 | 12 | 8 | 0.33 |
| J033230.07-275140.6 | 20 | 9 | 0.32 |
| J033221.54-275133.3 | 12 | - | 0.06 |
| J033233.11-275525.8 | - | - | - |
| J033246.37-274912.8 | - | - | - |
| J033215.81-274713.6 | - | - | - |
| J033217.77-274714.9 | - | - | - |
| J033238.79-274648.9 | - | - | - |
| J033219.24-274632.2 | - | - | - |
| J033214.44-274624.5 | - | - | - |
| J033233.02-274436.6 | - | - | - |
| J033217.12-274407.7 | - | - | - |
| J033225.47-274327.6 | - | - | - |
| J033239.17-274257.7 | - | - | - |
| J033212.47-274224.2 | - | - | - |
| J033227.36-274204.8 | - | - | - |
| J033242.39-274153.2 | - | - | - |
| J033228.88-274129.3 | - | - | - |
| J033235.02-275405.2 | - | - | - |
| J033238.27-275354.4 | - | - | - |
| J033241.64-275157.5 | - | - | - |
| J033229.94-275137.4 | - | - | - |
| J033232.73-275102.5 | - | - | - |
| J033230.09-275100.3 | - | - | - |

Tabla 4.4. Resultados de las galaxias lenticulares distantes barradas.

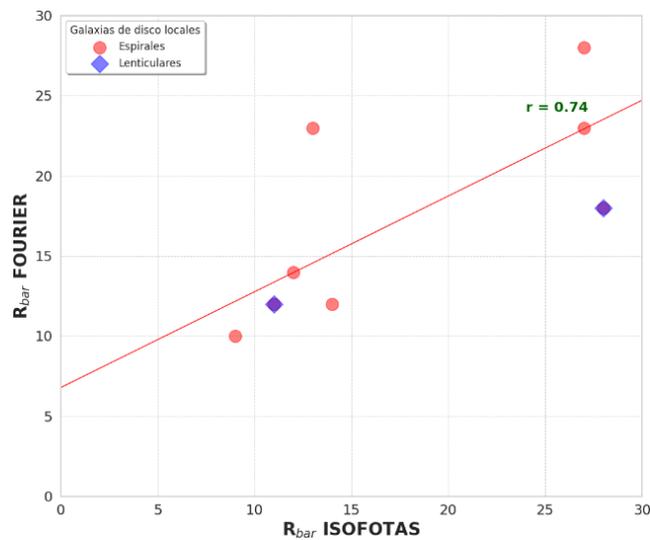

Figura 4.3. Muestra los gráficos de dispersión que comparan las medidas del radio de la barra ($R_{bar}$) obtenidas a través de los dos métodos en las galaxias de disco locales.

En la Figura 4.4, que muestra las galaxias de disco distantes, también se presenta una relación positiva entre las medidas. Sin embargo, la dispersión de los puntos es más pronunciada en comparación con el gráfico de galaxias locales. Aquí, también se utilizan puntos rojos para las galaxias espirales y puntos azules para las lenticulares, con una línea roja ajustada que indica la tendencia. El coeficiente de correlación de Pearson es $r = 0.60$, indicando una correlación positiva moderada. Este valor es relativamente más bajo que el de las galaxias locales, lo que sugiere que la relación entre los métodos es menos consistente en las galaxias distantes, posiblemente debido a factores como la menor resolución de las imágenes y el mayor ruido en los datos.

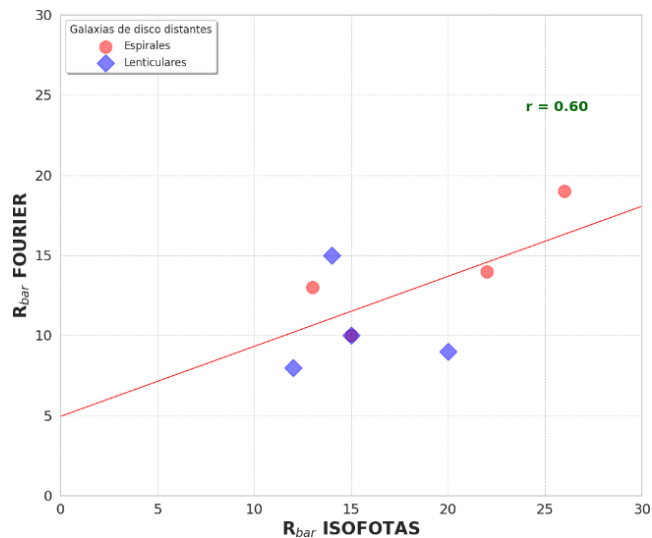

Figura 4.4. Muestra los gráficos de dispersión que comparan las medidas del radio de la barra ($R_{bar}$) obtenidas a través de los dos métodos en las galaxias de disco distantes.

En la investigación sobre la fracción de barras de [93], se evidenció que los métodos de clasificación afectan las fracciones de barras observadas en las galaxias, destacando especialmente la comparación entre el método de isofotas elípticas y el de Fourier. Encontraron que las correlaciones entre los tipos de barras y las propiedades de las galaxias son sensibles a los métodos empleados. En particular, el método de Fourier detecta barras fuertes predominantemente en galaxias de tipo temprano, mientras que el análisis mediante isofotas elípticas tiende a identificar barras en galaxias de tipo tardío, que a menudo presentan bulbos más pequeños. Este hallazgo se alinea con los coeficientes de correlación de Pearson encontrados, que mostraron una relación positiva entre $R_{bar}$ isofotas elípticas y $R_{bar}$ Fourier. Dado que el enfoque de isofotas elípticas permite obtener mediciones más precisas y robustas en galaxias de disco, evita las confusiones que pueden surgir al interpretar bulbos prominentes en galaxias de tipo temprano. Por lo tanto, a partir de este punto, todos los análisis presentados en esta tesis se centrarán en el método de isofotas elípticas.

### 4.1.2. Resultado de Fracción de Barras en Galaxias Locales y Distantes

A continuación, se amplía esta discusión al analizar la fracción de galaxias de disco con barras según su morfología. La Figura 4.5 muestra la fracción de galaxias espirales y lenticulares barradas en $z = 0.027$ y $z = 0.7$. Se observa que el $33\% \pm 12\%$ de las galaxias espirales locales presentan barras, mientras que en las galaxias espirales distantes este valor es de $23\% \pm 12\%$. En el caso de las galaxias lenticulares, el $25\% \pm 10\%$ de las locales y el $19\% \pm 10\%$ de las distantes presentan barras. Estos resultados fueron obtenidos mediante el método de isofotas elípticas, permitiendo una comparación directa entre galaxias locales y distantes

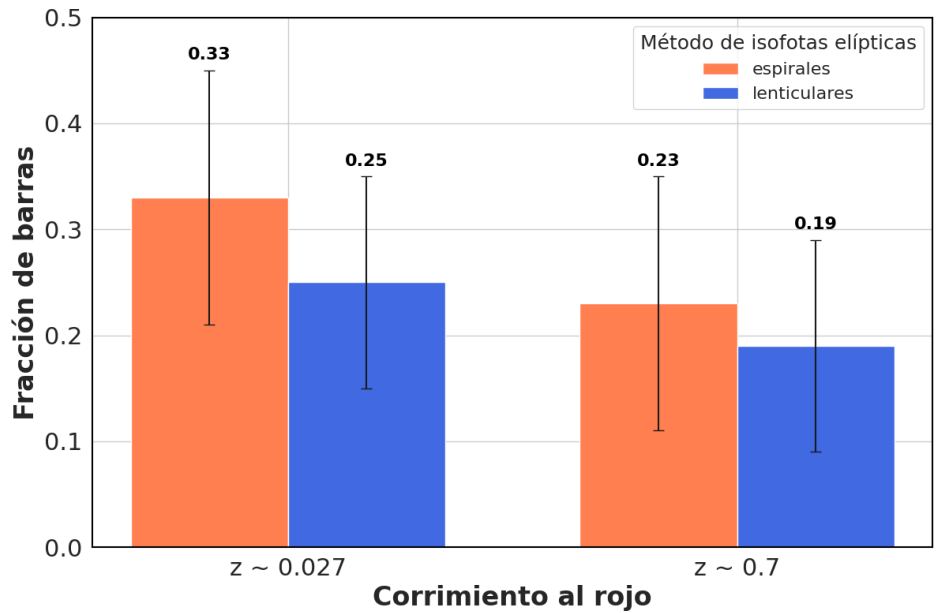

Figura 4.5. Fracción de galaxias barradas de disco (espirales y lenticulares) en $z = 0.027$ y $z = 0.7$.

En la figura 4.6 se observa la fracción, pero obtenido con el método de Fourier. Se observa que el 26% ± 11% de las galaxias espirales locales presentan barras, mientras que las galaxias espirales distantes son de 18% ± 11%; y para las galaxias lenticulares locales 25% ± 9% tienen barras y las distantes 15% ± 9%.

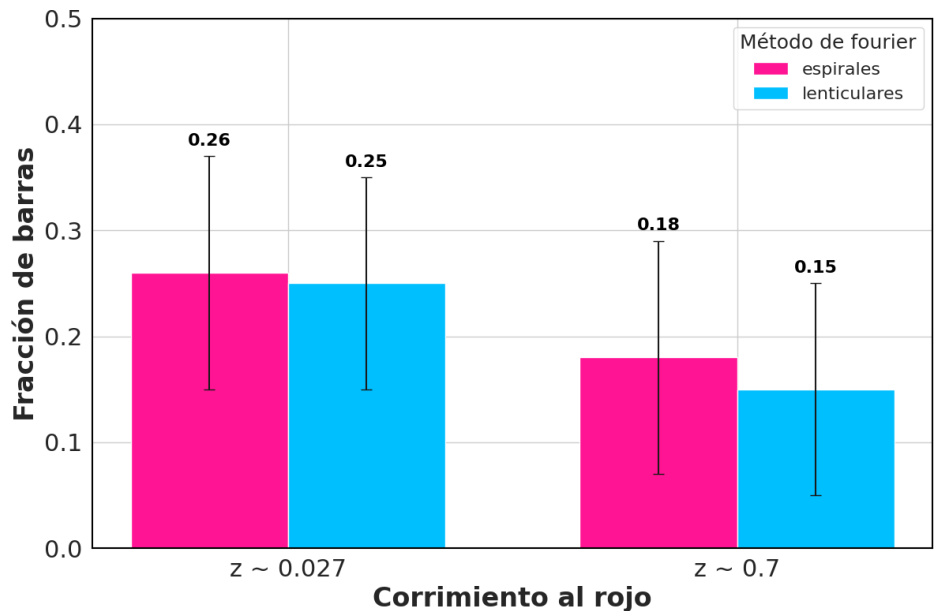

Figura 4.6. Fracción de galaxias barradas de disco (espirales y lenticulares) en $z = 0.027$ y $z = 0.7$.

En el caso de las galaxias espirales locales, la fracción de barras detectadas mediante Fourier (26%) es ligeramente inferior al valor obtenido mediante isofotas elípticas (33%), lo que refleja la diferencia en la sensibilidad de ambos métodos. Para las galaxias lenticulares locales, la fracción de barras detectadas mediante Fourier (25%) es igual al valor obtenido mediante isofotas elípticas (25%). Esto revela la eficiencia del método de Fourier para detectar barras en galaxias con bulbos prominentes. Para las espirales distantes, tanto el análisis de isofotas elípticas como la descomposición de Fourier coinciden en mostrar una reducción significativa en comparación con sus contrapartes locales, con fracciones de (23%) y (18%), respectivamente. Por otro lado, las galaxias lenticulares distantes presentan una disminución en su fracción de barras, pasando de (19%) mediante isofotas elípticas al (15%) utilizando Fourier. Estos resultados sugieren que, aunque las espirales y lenticulares barradas son menos comunes en galaxias distantes, las lenticulares presentan una leve disminución en su fracción de barras. El análisis comparativo en las Figuras 4.5 y 4.6 de la fracción de galaxias barradas en distintos corrimientos al rojo, revela las barras son menos comunes en las galaxias más distantes, lo cual es consistente con estudios previos que indican una menor frecuencia de barras en el pasado cósmico [47].

## 4.2 Fracción de Barras en Función de la Masa Estelar

- **Galaxias locales ($z = 0.027$)**

La Figura 4.7 muestra la dependencia de $f_{bar}$ con la masa estelar en nuestra muestra de galaxias de disco [134]. Las galaxias locales, representada en azul, se observa un incremento en la fracción de barras a medida que la masa estelar aumenta, con un claro ascenso a partir de $log(M_*/M_\odot) > 10.50$. Este comportamiento es consistente con el análisis de [135], quienes estudiaron galaxias en entornos mixtos utilizando datos en la banda $r$ del SDSS-DR7.

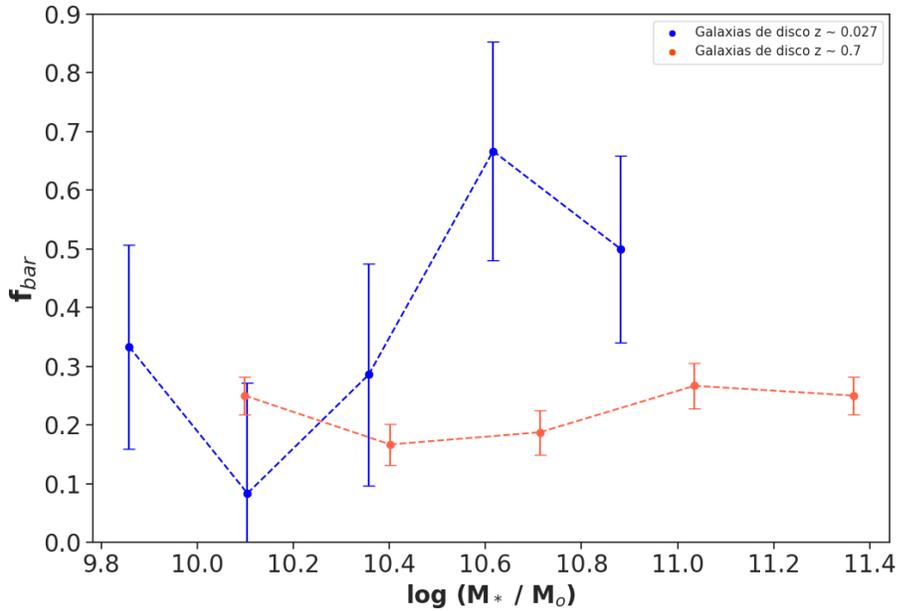

Figura 4.7. Fracción de barradas en función de la masa estelar en las galaxias de disco a $z = 0.027$ y $z = 0.7$.

Este resultado coincide con la tendencia reportada en [136], quienes investigaron

la fracción de barras utilizando imágenes en 3.6 μm del *Spitzer Survey of Stellar Structure*

*in Galaxies* (S[4]G) para galaxias en $z = 0.027$. Su estudio abarcó una amplia gama de masas estelares ($M_* \sim (10^8 - 10^{11})\,M_\odot$) y tipos morfológicos de Hubble ($-3 \leq T \leq 10$),

revelando una clara dependencia entre $f_{bar}$ y la masa estelar total. Sin embargo, no

observaron una fuerte correlación entre la fracción de barras y la clasificación visual de

galaxias con masas estelares $log(M_*/M_\odot) > 9$. En contraste, los métodos basados en la

descomposición de Fourier [137] y el ajuste de elipses para detectar barras en la muestra

S$^4$G mostraron que $f_{bar}$ aumenta con la masa estelar hasta $log(M_*/M_\odot)\sim9.5-10$, alcanzando un nivel constante de aproximadamente 50% en el rango $log(M_*/M_\odot)\sim10-11$, con un pico característico de 60% en $log(M_*/M_\odot)\sim10.5$. De manera similar, en nuestra muestra de galaxias de campo, observamos un comportamiento comparable, con un aumento significativo en la fracción de barras a $log(M_*/M_\odot)\sim10.6$.

- **Galaxias distantes** ($z = 0.7$)

En contraste con lo observado en la fracción de barras en galaxias $z = 0.7$ (Figura 4.7, línea roja) es notablemente menor en casi todo el rango de masa estelar, especialmente en el intervalo $log(M_*/M_\odot)\sim10.4-10.8$. Este comportamiento sugiere que las galaxias de disco en las primeras épocas del universo tuvieron una menor prevalencia de barras estelares. Las fusiones menores de galaxias desempeñan un papel crucial en la evolución dinámica de estas estructuras, ya que afectan la estabilidad y la morfología de las barras en galaxias discoides. Estudios recientes en alto corrimiento al rojo, mediante simulaciones [138], han encontrado que estas fusiones con satélites de menor masa, típicamente en una proporción de $1:10$ respecto a la galaxia principal, pueden inducir cambios significativos en la morfología de la galaxia. Durante los pasos pericéntricos, se genera una fase transitoria en la que la barra se amplifica, seguida de una estabilización en etapas posteriores a la fusión. Este proceso se debe a la acumulación de masa en el centro de la galaxia resultante, lo que incrementa la concentración de masa central y altera el momento angular específico, contribuyendo así a la estabilización de la barra estelar en el remanente de la fusión. Otro estudio de galaxias en alto corrimiento al rojo (entre $0.6$ y $2.6$), realizó un análisis de sus dinámicas internas, enfocándose en galaxias con formación estelar masiva. Estas galaxias tienen masas estelares significativas, en el rango de $log(M_*/M_\odot) \sim 10.6 - 11.1$, lo que las convierte en sistemas clave para estudiar la evolución galáctica durante el pico de formación estelar hace aproximadamente 10 mil millones de años [139]. Las curvas de rotación, obtenidas a partir de observaciones de la línea $H\alpha$ revelan que muchas de estas galaxias presentan una disminución de la velocidad en los radios externos, indicando una menor influencia de la materia oscura en comparación con las galaxias cercanas. Esto sugiere que, en las regiones internas, estas galaxias están dominadas por su contenido bariónico. Este resultado puede relacionarse con los hallazgos sobre galaxias de alto corrimiento al rojo, sugiriendo que las dinámicas internas y los procesos de formación de barras no dependen fuertemente de la masa estelar en ciertas épocas. En las galaxias de alto corrimiento al rojo, la disminución de la velocidad de rotación podría implicar que la formación de barras está más ligada a la masa bariónica interna que a la masa total. Esto indica que la estabilidad y dinámica del disco galáctico, como la dispersión de velocidades o la densidad de gas, son más determinantes que la masa estelar total. Así, mi estudio sugiere que la fracción de barras, al igual que las curvas de rotación descendentes, puede estar influenciada por factores como la turbulencia interna, que no varían significativamente con la masa estelar.

## 4.3 Fracción de Barras en Función del Índice de Color ($u - r$)

- **Galaxias locales ($z = 0.027$)**

La Figura 4.8 muestra la dependencia de la fracción de galaxias barradas ($f_{bar}$) con respecto al índice de color óptico ($u - r$). Para las galaxias locales (línea verde), se observa un aumento continuo en la fracción de galaxias barradas ($f_{bar}$) a medida que el índice de color $u - r$ incrementa, alcanzando un máximo alrededor de $(u - r){\sim}2.4$. Posteriormente, hay una ligera disminución en $f_{bar}$ en los extremos más rojos del rango de color. Este comportamiento es consistente con los hallazgos de estudios previos [33] [133], quienes observaron que las galaxias espirales con colores más rojos tienden a ser más propensas a albergar barras prominentes. Esto es particularmente evidente en galaxias que han experimentado una evolución pasiva, es decir, aquellas galaxias que ya no forman estrellas de manera significativa y cuya población estelar ha envejecido.

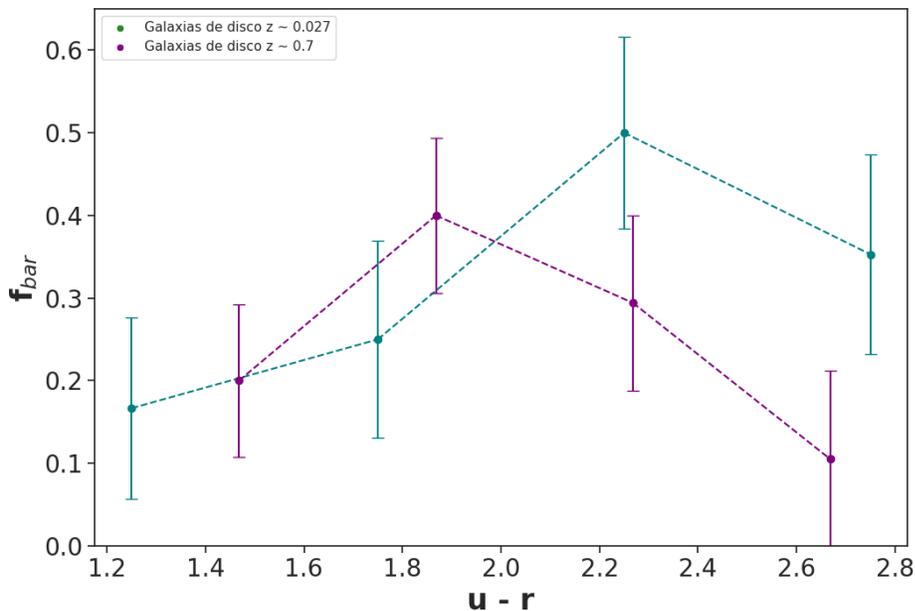

Figura 4.8. Fracción de barradas en función del color u- r en las galaxias de disco a $z = 0.027$ y $z = 0.7$.

- **Galaxias distantes ($z = 0.7$)**

En las galaxias distantes (Figura 4.8 línea morada), la fracción de galaxias barradas es notablemente menor a lo largo de todo el rango de colores en comparación con la muestra local. Aunque se observa un ligero incremento en la fracción de barras alrededor de $(u - r){\sim}1.9$, en galaxias más azules dentro de este grupo de estudio, es importante notar que estas galaxias no están dominadas por la formación estelar activa. En lugar de eso, este comportamiento podría estar vinculado a una evolución interna moderada o a la estabilización de los discos galácticos a través de procesos más sutiles, como la redistribución del gas y las estrellas. Las galaxias más azules, en este caso, podrían estar en una fase transitoria en la que sus discos han comenzado a estabilizarse.

El hecho de que estas galaxias no estén formando estrellas activamente sugiere que el incremento en la fracción de barras podría ser el resultado de la estabilización gradual del disco sin la intervención de procesos violentos, como fusiones o acreciones importantes, que podrían haber sido más comunes en etapas previas del universo. Esta estabilización permitiría la formación de barras, aunque estas podrían ser menos prominentes o más transitorias comparadas con aquellas observadas en galaxias más evolucionadas y de colores más rojos.

Por otro lado, los hallazgos en [140] contradicen esta tendencia al estudiar galaxias en cúmulos, donde reportan una correlación negativa entre la fracción de barras y el color $(B - V)$. Su muestra, basada en imágenes de la banda $V$ de cúmulos seleccionados del catálogo OmegaWINGS [141], sugiere que el entorno de cúmulo desempeña un papel crucial en las transformaciones morfológicas y estelares, disminuyendo la prevalencia de barras en galaxias más rojas. Este contraste refuerza la idea de que, en galaxias de campo como las que componen este estudio, la fracción de barras no parece verse afectada significativamente por el entorno, lo que sugiere que los procesos dinámicos internos son más determinantes que las interacciones y efectos ambientales en la evolución morfológica de estas galaxias.

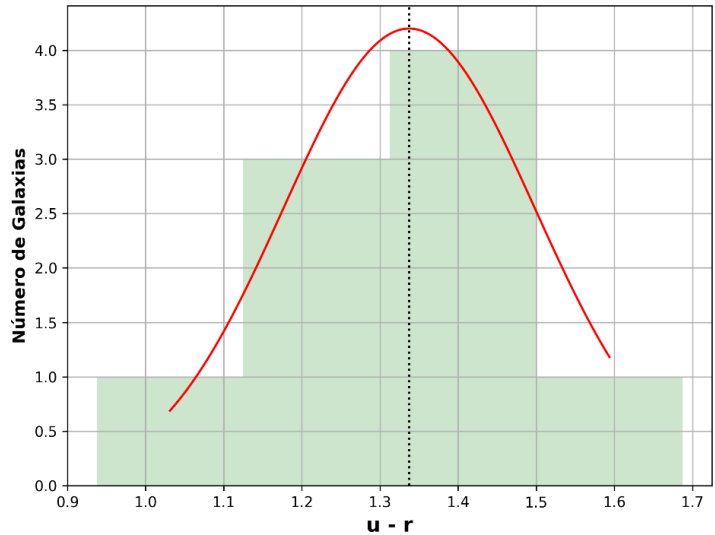

Figura 4.9. Distribución gaussiana de color (u-r) de galaxias espirales con formación estelar

La Figura 4.9 muestra la distribución gaussiana del color $(u-r)$ para galaxias espirales con formación estelar, con un pico en $(u-r)\sim1.34$. El color $(u-r)$ es un indicador de la actividad de formación estelar, donde galaxias más azules (con valores menores de $u-r$) presentan mayor actividad estelar. La ausencia de galaxias barradas indica que las galaxias con mayor formación estelar no presentan las condiciones dinámicas para el desarrollo de barras. Esto puede explicarse por la inestabilidad en los discos galácticos y la abundancia de gas frío, factores que inhiben la formación de barras. Al comparar con estudios locales, donde las barras son más comunes en galaxias rojas con menor formación estelar, se intuye que los procesos internos de estabilización del disco juegan un papel más relevante en galaxias distantes, mientras que en las locales las interacciones ambientales podrían tener mayor influencia.

En galaxias más evolucionadas y de colores más rojos, la prevalencia de barras es mayor en comparación con las galaxias más azules, que se encuentran en etapas más tempranas de su evolución. En estas galaxias jóvenes, la acreción de gas y fusiones menores tienden a desestabilizar los discos, dificultando la formación de barras. Esto refuerza la hipótesis de que las barras se desarrollan preferentemente en galaxias con discos estables y una evolución estelar más pasiva, típicamente cuando el gas ha sido agotado o expulsado, lo que permite la estabilización de la estructura interna. Este comportamiento es particularmente relevante en galaxias que se encuentran en la transición entre la secuencia azul y la roja, donde se ha observado una deficiencia de HI [142], lo que sugiere un enfriamiento en la formación estelar, posiblemente causado por procesos como el gas stripping. Aunque la transición de la secuencia azul a la roja ocurre relativamente rápido, el cese completo de la formación estelar y la estabilización del disco requieren más tiempo.

En nuestra muestra, las galaxias en esta región de transición muestran una fracción significativamente menor de barras en comparación con las galaxias más evolucionadas. Esto sugiere que los procesos de acreción de gas y fusiones menores continúan influyendo en estas galaxias, retrasando la estabilización de sus discos y, por tanto, la formación de barras.

### 4.3.1 Diferencias en la Formación Estelar en Galaxias con Barras Lentas y Rápidas

En estudios recientes [143], las barras galácticas se clasifican en lentas o rápidas según su cinemática, lo que influye directamente en la acumulación de gas y la formación estelar a lo largo de la barra. Las barras lentas tienen un radio de corrotación mayor que su longitud, mientras que las barras rápidas terminan cerca de dicho radio. Esto implica que las barras rápidas rotan a una velocidad similar a la de las estrellas del disco, mientras que las barras lentas rotan más lentamente (ver Figura 4.10).

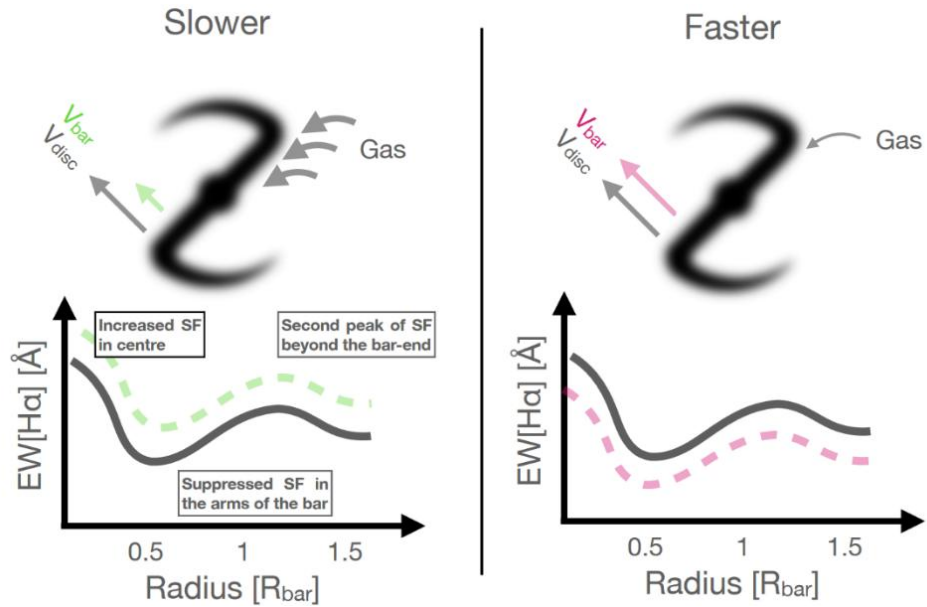

Figura 4.10. Este modelo ilustrativo muestra una posible interpretación física del estudio de barras lentas y rápidas. En el panel izquierdo, se muestra cómo el perfil radial de una galaxia con una barra fuerte cambia si esta es lenta. Esto provoca un aumento en los valores observados de EW[Hα] en las barras lentas. En contraste, el panel derecho muestra que, en las barras rápidas atrapa menos gas, resultando en menores valores de EW[Hα] en comparación con las barras lentas. Imagen tomada de [144].

Se ha observado que las barras lentas presentan una mayor diferencia de velocidad entre sus extremos y el gas en el disco, lo que facilita la concentración de gas a lo largo de la barra, resultando en valores más altos de EW[Hα]. En contraposición, las barras rápidas no logran concentrar tanto gas, lo que conlleva a valores de EW[Hα] más bajos. Esto sugiere que las barras lentas son más eficientes en barrer gas y crear vacíos [144] [145]. En simulaciones realizadas por [146] y [147], también se muestran estas regiones sin gas en galaxias con barras lentas. A pesar de estas diferencias en la concentración de gas, tanto las barras lentas como las rápidas muestran tasas globales de formación estelar (SFR) similares.

Aunque las barras lentas concentran más gas, otros factores, como el aumento en la dispersión de velocidad y el cizallamiento del gas en los brazos de la barra, actúan como fuerzas compensatorias, limitando el incremento esperado en la formación estelar. Esto es consistente con resultados de estudios que sugieren que la turbulencia inducida por la barra y el incremento en la dispersión de velocidad contribuyen al cese de la formación estelar en galaxias de disco [148]. En nuestra muestra, compuesta por galaxias de campo con formación estelar, no se observó la presencia de barras. Sin embargo, en nuestra muestra de galaxias de campo sin formación estelar, sí se encontraron galaxias con barras. En nuestra muestra representativas de galaxias con barra y sin formación estelar han experimentado este cese de formación estelar precisamente por la presencia de la barra, lo que muestra una importante de evolución fomentada por la barra.

## 4.4. Fracción de Barras en función del Cociente Bulbo-Disco ($B/T$)

- **Galaxias locales ($z = 0.027$)**

La Figura 4.11 muestra que, para las galaxias espirales cercanas (color amarillo), la fracción de galaxias barradas ($f_{bar}$) presenta una clara dependencia con el cociente bulbo-disco ($B/T$). La mayoría de estas galaxias tienen $B/T \leq 0.2$, coincidiendo con estudios anteriores que indican que galaxias con bajas concentraciones de masa en el bulbo tienen una alta probabilidad de albergar barras [149, 150]. Alrededor del 65% de las galaxias espirales con bulbos de baja concentración de masa ($n \leq 2$, donde $n$ es el índice de Sersic) contienen barras, mientras que el 68% de las galaxias con $B/T \leq 0.2$ tienden a ser barradas [151], lo que refuerza la relación observada entre estructuras de bulbo débiles y la presencia de barras. La fracción de barras tiende a aumentar en galaxias con $B/T = 0.7$, que corresponden a lenticulares barradas. Esto sugiere una evolución temporal en la formación de barras, donde un entorno dinámico más joven en galaxias distantes podría inhibir tanto la formación como el mantenimiento de barras con el paso del tiempo.

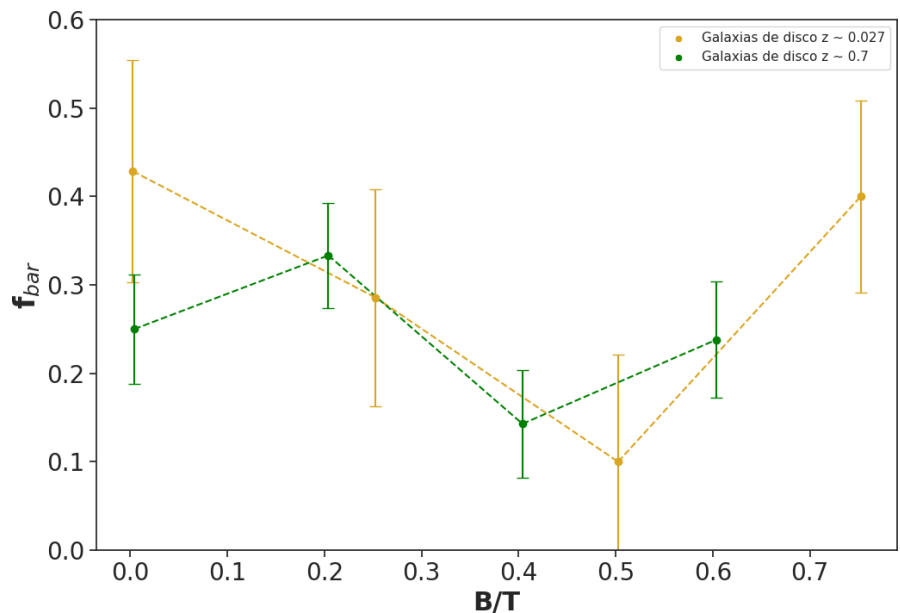

Figura 4.11. Fracción de barradas en función de B/T en las galaxias de disco a $z \sim 0.027$ y $z \sim 0.7$.

- **Galaxias distantes ($z = 0.7$)**

Las galaxias más distantes (color verde) y cercanas (color amarillo) muestran una tendencia similar en la fracción de barras en función de $B/T$, con una mayor fracción de barras en galaxias con $B/T$ bajos (discos prominentes). En galaxias con alto $B/T$, la fracción de barras disminuye notablemente en ambos casos, lo que sugiere que las galaxias con bulbos prominentes tienden a tener menos barras. Aunque en el universo temprano la formación de barras pudo haber sido afectada por un entorno más dinámico, este efecto parece ser más notable en galaxias con discos prominentes, donde la fracción de barras es ligeramente menor en comparación con el universo cercano.

Las simulaciones numéricas respaldan la hipótesis de que la presencia de un bulbo prominente o una alta concentración de masa central puede inhibir la formación de barras [152]. Las galaxias que no contienen bulbos prominentes o tienen concentraciones de gas bajas tienden a desarrollar barras más fuertes. En contraste, aquellas ricas en gas o con bulbos significativos presentan una menor tendencia a la formación de barras [153]. Las simulaciones numéricas también respaldan la idea de que la formación o mantenimiento de barras se inhibe en galaxias con un $B/T$ elevado o concentraciones de masa central altas [152].

En ambas muestras, la dependencia de $f_{bar}$ en función de $B/T$ es el factor más claro, mostrando que la estructura interna de la galaxia tiene un impacto significativo en la formación de barras, independientemente de la época cósmica.

## 4.5.    Comparación de Nuestros Resultados con otros Estudios

En esta sección, se explorará como nuestros resultados se comparan con las fracciones de barras observadas en otros estudios a diferentes corrimientos al rojo. En la Figura 4.12, se comparan los resultados de nuestra muestra representativa de galaxias con otros estudios previos. Como se mencionó en la sección 4.1, el $33\% \pm 12\%$ de las galaxias espirales locales ($z \sim 0.027$) presentan barras, en contraste con el $23\% \pm 12\%$ de las galaxias espirales distantes ($z \sim 0.7$). En el caso de las galaxias lenticulares, el 25% ± 10% de las locales ($z \sim 0.027$) y el 19% ± 10% de las distantes ($z \sim 0.7$) presentan barras.

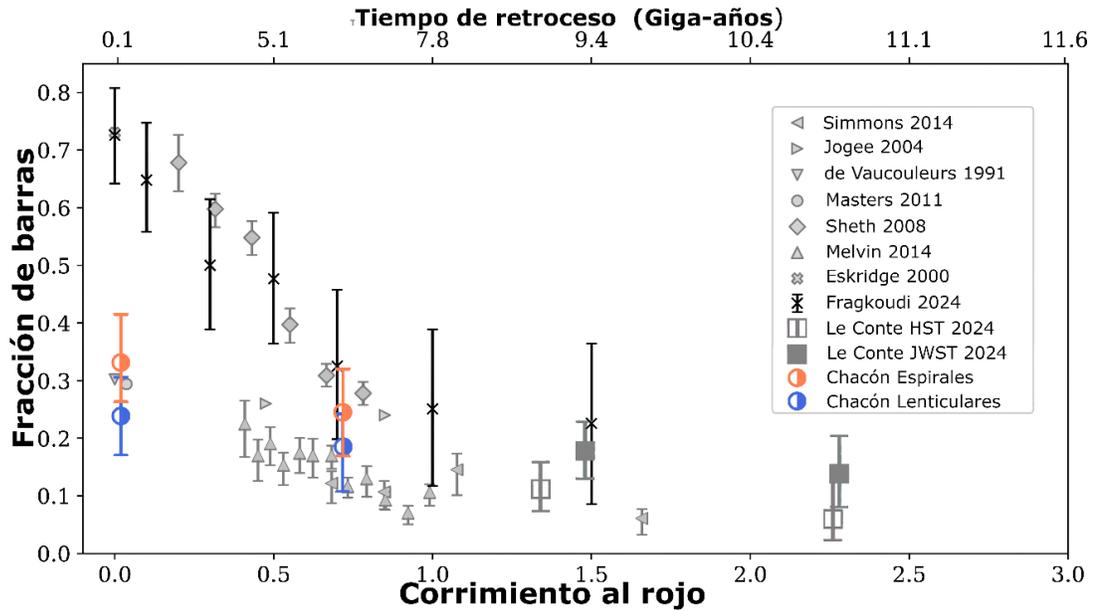

Figura 4.12. En la imagen se muestra la evolución de la fracción de barras estelares en galaxias de disco en relación con el corrimiento al rojo. Se observa una tendencia general de disminución en la fracción de barras a medida que aumenta el corrimiento, lo que sugiere que estas estructuras son menos comunes en épocas tempranas del universo. Esta observación refleja la evolución dinámica de las galaxias a lo largo del tiempo cósmico, abarcado por diferentes estudios.

A lo largo de los años, diversos estudios han investigado la evolución de las galaxias barradas en función del tiempo cósmico, destacando tanto las similitudes como las diferencias en relación con nuestra muestra. A continuación, comparamos nuestros hallazgos con los principales estudios realizados en esta área:

- Simmons, B. [38]: Este estudio utilizó imágenes del Cosmic Assembly Near-Infrared Deep Extragalactic Legacy Survey, en el filtro F160W se utiliza en la banda infrarroja cercana. Identificaron 123 galaxias barradas en una muestra de 876 galaxias de disco, en ambientes mixtos, clasificadas visualmente en *Galaxy Zoo*. Al centrarse en galaxias más brillantes que $L^*$ (galaxias relativamente brillantes y generalmente masivas), se encontró que la fracción de galaxias con barras en $0.5 \leq z \leq 2$ ($f_{bar} = 10.7\%$), no presenta una evolución significativa. Este hallazgo sugiere que las barras han permanecido estables en las galaxias de disco durante los últimos 11 mil millones de años. En nuestra muestra local, se observa una clara tendencia al aumento de la fracción de barras con la masa estelar, no en la muestra distante. Este comportamiento concuerda con el análisis de este estudio, quienes no observaron una evolución significativa de la fracción de barras en galaxias con masas estelares elevadas hasta -$z = 2$. Aunque, no enfocan tanto en la dependencia con la masa, ambas muestras subrayan que las galaxias más masivas tienden a formar y mantener barras más estables, lo que sugiere que la masa estelar es un factor crítico en la evolución de barras a lo largo del tiempo cósmico.

- Jogee, S. [91]: En este trabajo se estudiaron 627 galaxias espirales en ambientes mixtos, realizaron un corte de magnitud desde $M_V \leq -19.3$ a $M_V \leq -20.6$ (en la banda V) en el rango de $0.7 \leq z \leq 1.1$. Utilizaron los filtros F606W y F850LP y utilizaron una combinación de modelos matemáticos (Sérsic), índices de concentración, y análisis de color para clasificarlas como espirales. Las galaxias barradas fueron clasificadas gracias a las imágenes del Hubble Space Telescope Advanced Camera for Surveys (ACS), excluyendo galaxias muy inclinadas. Los resultados muestran que la fracción de barras fuertes se mantiene en torno al $f_{bar} \sim 27\%$, indicando que los discos ya eran inestables gravitacionalmente en esa época. Ellos, también detectaron que las galaxias con mayor masa estelar y bulbos prominentes muestran una mayor prevalencia de barras, lo que está alineado con nuestro resultado de que las galaxias masivas en nuestra muestra local presentan una fracción de barras más alta. Sin embargo, la disminución observada en las galaxias distantes de tu estudio contrasta con sus resultados, sugiere que, en corrimientos al rojo similares, la presencia de barras es más variable en función de otros factores como el tipo morfológico y la dinámica interna.

- Masters K. [33]: Este análisis de Galaxy Zoo 2 incluyó 13,665 galaxias de disco en ambiente mixto, en el rango $0.01 \leq z \leq 0.06$ y magnitud $M_r \leq -19.38$. La masa de las galaxias en la muestra estaba en el rango de $10^9 < M_* < 10^{11} M_\odot$. La fracción media de galaxias con barra fue de $f_{bar} = 29.4\%$ con un total de 4,020 galaxias barradas. La fracción de galaxias con barras aumenta en aquellas más rojas, menos luminosas y con bulbos prominentes, alcanzando el 50% en los sistemas más dominados por bulbos. Esto refuerza la idea de que las barras son clave en la evolución interna de las galaxias, coincidiendo con resultados de estudios tanto locales como a mayor corrimiento al rojo. En nuestros resultados, la tendencia de un aumento en la fracción de barras con la masa estelar coincide con el resultado de este estudio, donde se observa que la fracción de barras es mayor en galaxias más masivas, rojas y con bulbos prominentes. Este comportamiento es particularmente relevante para las galaxias locales de nuestra muestra, que exhiben un patrón similar, lo que refuerza la idea de que las propiedades internas, como la masa estelar y el bulbo, juegan un papel clave en la formación y estabilización de barras.

- Sheth, K. [37]: Este estudio utilizó una muestra de 2157 galaxias espirales brillantes en ambientes mixtos de COSMOS en el filtro F814W. Se excluyeron galaxias elípticas y lenticulares, seleccionando solo aquellas con un tipo fotométrico $T_{phot} > 2$, que correspondían a galaxias espirales. El resultado de la fracción de galaxias con barras disminuye significativamente con el corrimiento al rojo, en el rango $0.2 \leq z \leq 0.84$, de $f_{bar} = 65\%$ en el universo local hasta solo $f_{bar} = 20\%$ en el lejano. Viendo nuestros resultados con este estudio, también indican una disminución en la prevalencia de barras con el aumento del corrimiento al rojo. Sin embargo, muestran los resultados de la fracción de barras que muestra en el universo local son muchos más elevados que los nuestros. La evolución de la fracción de barras es más pronunciada en galaxias menos masivas y más azules, mientras que las más masivas mantienen barras a lo largo del tiempo. Esto sugiere que las galaxias masivas alcanzaron estabilidad dinámica más temprano, en paralelo con la regulación de su formación estelar.

- Melvin, T [47]: Este estudio sobre la evolución de las barras en galaxias de disco, en el Filtro F814W (I-band) del HST, a lo largo del tiempo cósmico, específicamente para galaxias con masas estelares superiores a $log(M_*/M_\odot) > 10.0$ y en un rango de corrimiento al rojo de $0.4 \leq z \leq 1.0$. Basada en imágenes del proyecto COSMOS y clasificaciones morfológicas visuales de Galaxy Zoo. Se analizó una muestra de 2380 galaxias de disco, de las cuales 317 son barradas.

Observando una disminución en $f_{bar} = 22\%$ a $z = 0.4$ al $f_{bar} = 11\%$ en $z = 1$, especialmente en galaxias masivas.

En comparación con este estudio, mis resultados también muestran una tendencia a la baja en la fracción de barras con el aumento del corrimiento al rojo. Sin embargo, ellos se centran en galaxias de disco masivas, nuestra muestra incluye la clasificación de espirales y lenticulares.

- Eskridge, P. [154]: En este estudio, se analizó muestra estadísticamente significativa de 186 galaxias espirales de (OSU Bright Spiral Galaxy Survey) revela que la fracción de galaxias fuertemente barradas es mucho mayor en el infrarrojo (banda H) $f_{bar} = 56\%$ presenta barras fuertes, en contraste con el $f_{bar} = 34\%$ detectado en óptico (banda B). Aproximadamente el $f_{bar} = 72\%$ de las galaxias tienen alguna forma de barra en el infrarrojo, en comparación con $f_{bar} = 64\%$ en el óptico. Además, el estudio muestra que entre 40% y 50% de las galaxias clasificadas como sin barra en estudios ópticos presentan barras detectables en el infrarrojo, lo que sugiere que la presencia de polvo y estrellas jóvenes afecta las clasificaciones ópticas. En comparación de este estudio, no incluimos las diferencias entre longitudes de onda, sino que se centra en la evolución de las barras con el corrimiento al rojo. Ellos, resaltan cómo el polvo y las estrellas jóvenes pueden ocultar barras en el óptico, nuestro trabajo utiliza un enfoque de clasificación robusto que integra diferentes parámetros (como la relación bulbo-total y los perfiles de galaxias), asegurando una detección coherente de barras a través de diferentes muestras.

- Fragkoudi, F. [155]: El estudio investiga la formación y evolución de 40 galaxias espirales barradas en un contexto cosmológico utilizando simulaciones de la suite Auriga, centradas en galaxias con masas alrededor de la Vía Láctea. Las simulaciones indican que la fracción de galaxias con barras es aproximadamente de $f_{bar} = 70\%$ en $z = 0$, y disminuye en aproximadamente $f_{bar} = 20\%$ en $z = 3$. Respalda los otros estudios que las barras en galaxias locales tienden a ser más fuertes, y su tiempo de formación se correlaciona con la dominancia de bariones de la galaxia. Los resultados de la fracción de barras de esta simulación son más altos que nuestros resultados de la fracción de barras en el Universo local. Además, nuestro estudio se basa en observaciones y una muestra representativa de galaxias espirales y lenticulares, aplicando un sistema robusto de clasificación. Ambos estudios coinciden en que las barras son estructuras que evolucionan a lo largo del tiempo, pero nuestro enfoque observacional ofrece una visión complementaria, enfocándose en cómo la fracción de barras varía en distintos tipos morfológicos. Sin embargo, es importante señalar que estas simulaciones cosmológicas podrían estar exagerando la evolución de las barras como función del corrimiento al rojo debido a problemas de resolución a altos valores de $z$.

- Le Conte, Z. [156]: En este estudio utilizan clasificaciones visuales de tipo Hubble para las galaxias que tienen clasificaciones visuales en el catálogo de CANDELS. Investigan la fracción de barras estelares en galaxias de disco a corrimientos al rojo altos de ($1 \leq z \leq 3$) utilizando imágenes del HST y del JWST. Se comparan los resultados de ambos telescopios para una muestra optimizada de galaxias, eliminando aquellas con morfología peculiar o inclinación elevada. La muestra final incluye 368 galaxias en el filtro JWST F444W y 126 galaxias en el filtro HST F160W. Los resultados muestran que, en el filtro JWST F444W, la fracción es de $f_{bar} = 17.8\%$ para $1 \leq z \leq 2$ y $f_{bar} = 13.8\%$ para $2 \leq z \leq 3$. En comparación, en el filtro HST F160W, la fracción de barras es menor: $f_{bar} = 11.2\%$ y $f_{bar} = 6.0\%$ en los mismos intervalos. Concluyen que la fracción de barras en imágenes NIR de JWST es aproximadamente el doble que en las imágenes ópticas de HST, lo que sugiere que la detectabilidad de las barras estelares depende de la longitud de onda. En comparación con este estudio, nuestros resultados se centran en una muestra más local y en corrimientos al rojo menores. Mientras que ellos destacan cómo la detectabilidad de las barras varía según la longitud de onda, mi estudio mantiene un enfoque más amplio y sistemático en la evolución de las barras en galaxias de disco, observando cómo su prevalencia disminuye a medida que aumenta el corrimiento al rojo. Ambos estudios refuerzan la idea de que las barras son estructuras comunes en galaxias de disco, pero sugieren que la longitud de onda utilizada puede jugar un papel crítico en su detección, lo que destaca la importancia de un análisis multibanda en la investigación de la evolución galáctica.

A continuación, se presentan las conclusiones sobre la detección y caracterización de barras estelares en galaxias de disco, utilizando los métodos de isofotas elípticas y descomposición de Fourier, destacando las fortalezas y limitaciones de cada técnica según la distancia, tipo de galaxia, masa, color y la relación $B/T$.

# Capítulo 5

## Conclusiones

### 5.1 Comparación de Métodos para la Detección de Barras en Galaxias de Disco

En galaxias locales, los métodos de detección de barras mediante isofotas elípticas y descomposición de Fourier presentan una alta correlación ($r = 0.74$), lo que refleja una gran consistencia en la estimación del tamaño de las barras. Esta coherencia parece estar relacionada con la mayor resolución y calidad de los datos en galaxias cercanas, lo que sugiere una mayor fiabilidad de ambos métodos en este contexto. En galaxias distantes, la correlación sigue siendo positiva ($r = 0.60$), pero más moderada, lo que indica una menor consistencia, posiblemente influida por las limitaciones de resolución y el ruido en los datos. Esto sugiere que las estimaciones del tamaño de las barras en objetos lejanos son menos precisas. Además, estudios previos destacan que los métodos de clasificación influyen significativamente en las fracciones de barras observadas.

El método de Fourier tiende a detectar barras fuertes en galaxias de tipo temprano, mientras que el análisis de isofotas elípticas es más eficiente en galaxias de tipo tardío, con bulbos más pequeños. Este comportamiento coincide con los coeficientes de correlación obtenidos. Dado que el método de isofotas elípticas ofrece mayor precisión y consistencia, se adoptará como principal en los análisis posteriores, garantizando una interpretación más robusta de las estructuras de barra, especialmente en galaxias con bulbos menos prominentes.

## 5.2 Diferencias en la Fracción de Barras según el Desplazamiento al Rojo

El análisis revela una tendencia de disminución en la fracción de galaxias barradas con el aumento del corrimiento al rojo, tanto en galaxias espirales como lenticulares. Sin embargo, aunque parece haber una diferencia en la magnitud de esta tendencia entre ambos tipos morfológicos, la cantidad de galaxias lenticulares observadas sugiere que se requiere mayor cautela al interpretar este resultado, especialmente en la comparación entre galaxias locales y distantes.

### 5.2.1 Galaxias locales ($z = 0.027$)

- **Galaxias espirales locales:** El 33% $\pm$ 12% de las galaxias espirales locales presentan barras, según el método de isofotas elípticas, mientras que la descomposición de Fourier indica un 26% $\pm$ 11%. Estos resultados sugieren que una proporción significativa de espirales en el universo local ha desarrollado barras, probablemente debido a la estabilización dinámica de sus discos con el tiempo.

- **Galaxias lenticulares locales:** En el caso de las lenticulares locales, el 25% $\pm$ 10% exhibe barras según las isofotas elípticas, con una cifra similar del 25% $\pm$ 9% a partir de la descomposición de Fourier. Esto sugiere que, con sus bulbos prominentes, las lenticulares también han experimentado una evolución que favorece la formación de barras.

### 5.2.2 Galaxias distantes ($z = 0.7$)

- **Galaxias espirales distantes:** En el universo distante, la fracción de espirales barradas disminuye considerablemente. Solo el $23\% \pm 12\%$ de estas galaxias muestran barras mediante isofotas elípticas, y un $18\% \pm 11\%$ según la descomposición de Fourier. Esto sugiere que, en épocas más tempranas, los discos eran menos estables, posiblemente debido a fusiones y acreción de gas frío que incrementaban la turbulencia, afectando la formación de barras.

- **Galaxias lenticulares distantes:** La disminución es aún más marcada en lenticulares distantes, con solo el $19\% \pm 10\%$ mostrando barras mediante isofotas elípticas, y un $15\% \pm 9\%$ según Fourier. Esto sugiere que las lenticulares, con mayores concentraciones de masa central, son más susceptibles a procesos dinámicos que inhiben la formación de barras, posiblemente influenciados por interacciones ambientales y fusiones en las primeras épocas del universo.

## 5.3 Tendencia de la Fracción de Barras con respecto a la Masa Estelar

El análisis de la fracción de barras en función de la masa estelar muestra diferencias significativas entre galaxias locales y distantes, lo que evidencia una estrecha relación entre la evolución galáctica y la masa estelar.

- **Galaxias locales ($z = 0.027$)**

  La fracción de barras aumenta con la masa estelar, mostrando un ascenso notable a partir de $log(M_*/M_\odot) > 10.5$, lo que sugiere que las galaxias más masivas y evolucionadas son más propensas a desarrollar barras prominentes.

A partir $log(M_*/M_\odot) \sim 9.5 - 10$, la fracción de barras se incrementa y alcanza un nivel constante del 50% en galaxias con $log(M_*/M_\odot) \sim 10. - 11$, llegando a un máximo del 60% en $log(M_*/M_\odot) \sim 10.5$. Este patrón en nuestra muestra refleja que las galaxias locales más masivas y estables son más propensas a desarrollar barras, probablemente debido a la mayor estabilidad de los discos galácticos en estas galaxias.

- **Galaxias distantes ($z = 0.7$):**

En contraste, la fracción de barras en galaxias distantes es significativamente menor, con una caída pronunciada en el rango de $log(M_*/M_\odot) \sim 10.4 - 10.8$. Este descenso sugiere que las condiciones dinámicas en el universo temprano, como una mayor tasa de fusiones y la acreción de gas frío, desestabilizaban los discos galácticos, impidiendo la formación de barras o reduciendo su prevalencia.

## 5.4 Tendencia de la Fracción de Barras con respecto al Color ($u - r$)

- **Galaxias locales ($z = 0.027$):**

La fracción de barras aumenta de forma continua con el incremento del índice de color ($u - r$), alcanzando un máximo en torno a ($u - r$) $\sim 2.4$. Este comportamiento sugiere que las galaxias más rojas, asociadas a una evolución estelar pasiva, son más propensas a desarrollar barras prominentes. Estas galaxias, con baja o nula formación estelar, presentan discos más estables que facilitan la aparición de barras. Después de ($u - r$) $\sim 2.4$, se observa una leve disminución en la fracción de barras, lo que sugiere que las galaxias extremadamente rojas han agotado las condiciones ideales para la formación de estas estructuras.

- **Galaxias distantes ($z = 0.7$):**

  En general, la fracción de barras es menor en comparación con las galaxias locales. Aunque, alrededor de $(u - r) \sim 1.9$, algunas galaxias más azules presentan un ligero aumento en la fracción de barras. Esto podría indicar que estas galaxias están en una fase de estabilización de sus discos, permitiendo la formación de barras menos prominentes o más transitorias. La inestabilidad de los discos, influenciada por la abundancia de gas frío y fusiones menores, afecta de manera más marcada a las galaxias distantes, lo que reduce la fracción de barras en etapas tempranas.

- **Galaxias con formación estelar activa**

  Las galaxias de campo con formación estelar activa, caracterizadas por un índice de color azul $(u - r) \sim 1.34$, no muestran barras en nuestra muestra. La dinámica de estos discos jóvenes e inestables, afectada por la redistribución de gas y fusiones, dificulta la formación de barras en estas galaxias. Esto refuerza la idea de que las condiciones necesarias para su formación están ausentes en etapas evolutivas tempranas debido a la inestabilidad del disco.

- **Relación entre barras y formación estelar**

  En nuestra muestra, las galaxias con índices de color bajos, característicos de formación estelar activa, no muestran barras. Por el contrario, las galaxias sin formación estelar exhiben una mayor fracción de barras lentas, favoreciendo su estabilidad en galaxias evolucionadas con escasa o nula formación estelar. Las barras rápidas, que rotan a velocidades similares a las de las estrellas del disco, no logran concentrar tanto gas, lo que explica su menor presencia en galaxias activas.

## 5.5 Tendencia entre la Fracción de Barras y la Evolución Estructural Galáctica ($B/T$)

- **Galaxias locales ($z = 0.027$)**

Las barras se forman más comúnmente en galaxias con bajos cocientes bulbo-disco ($B/T \leq 0.2$), lo que sugiere que un bulbo menos masivo favorece la formación de barras prominentes. Esto se debe a que un disco más estable puede soportar una barra sin la interferencia gravitacional significativa del bulbo. Aproximadamente el 68 % de las galaxias con bajo $B/T$ muestran barras, resaltando la fuerte relación entre la estructura morfológica y la presencia de barras.

- **Galaxias distantes ($z = 0.7$)**

Aunque algunas galaxias distantes con bajo $B/T$ también muestran barras, la fracción total de galaxias barradas es menor en comparación con las galaxias locales. Esto sugiere que, en las galaxias de campo distantes, las condiciones dinámicas eran menos favorables para la formación de barras debido a la mayor inestabilidad de los discos y un entorno cósmico más activo, lo que dificultaba tanto la formación como el mantenimiento de barras.

- **Evolución temporal de la fracción de barras**

Con el envejecimiento de las galaxias y la estabilización de sus discos, la formación de barras se vuelve más común. En épocas recientes, esta fracción aumenta, mostrando que las barras se asocian a galaxias maduras con discos más estables. A medida que el universo evoluciona, las condiciones para la formación de barras mejoran.

- **Impacto del gas y concentración central**

Las galaxias con altos cocientes B/T o con grandes cantidades de gas frío tienden a tener menos barras, ya que la presencia de gas o bulbos prominentes desestabiliza los discos galácticos, impidiendo la formación o el mantenimiento de barras. Las simulaciones numéricas apoyan esta hipótesis, mostrando que bulbos prominentes y la presencia de gas denso suprimen la formación de barras. En contraste, las galaxias con discos menos perturbados por el gas o el bulbo son más propensas a desarrollar barras fuertes y duraderas.

# Capítulo 6

## Perspectivas para Trabajos Futuros

### 6.1 Galaxias en Cúmulos y el Impacto en la Evolución de Barras

Una interesante dirección para futuros trabajos es investigar la formación y evolución de barras en galaxias que habitan cúmulos. Dado que las galaxias en cúmulos experimentan interacciones ambientales significativas, como la presión de arrastre (ram-pressure stripping) y las interacciones de marea con otras galaxias y con el cúmulo en sí, es probable que el entorno juegue un papel crucial en la dinámica interna y, por ende, en la formación de barras [157]. Estudios previos han señalado una tendencia negativa entre la fracción de barras y el color en cúmulos, lo que sugiere que las barras podrían ser menos comunes en galaxias más evolucionadas o rojas que han experimentado procesos como el "quenching" estelar [33]. Sin embargo, queda por determinar con mayor claridad cómo estos procesos afectan la estabilidad de los discos galácticos y la presencia de barras en diferentes etapas evolutivas. Otro tema relevante sería el estudio de galaxias medusas (Figura 6.1), sistemas galácticos afectados por la presión de arrastre en los cúmulos, lo que provoca que sus discos pierdan gas y formen características distintivas como filamentos de gas y estrellas [158]. Estas interacciones ambientales no solo influyen en la formación estelar, sino también en la evolución morfológica de la galaxia, incluyendo la formación y estabilidad de las barras. Comprender la relación entre la presión de arrastre y la posible inhibición o desarrollo de barras en estas galaxias podría proporcionar nuevas perspectivas sobre la interacción entre las fuerzas del entorno y la dinámica interna.

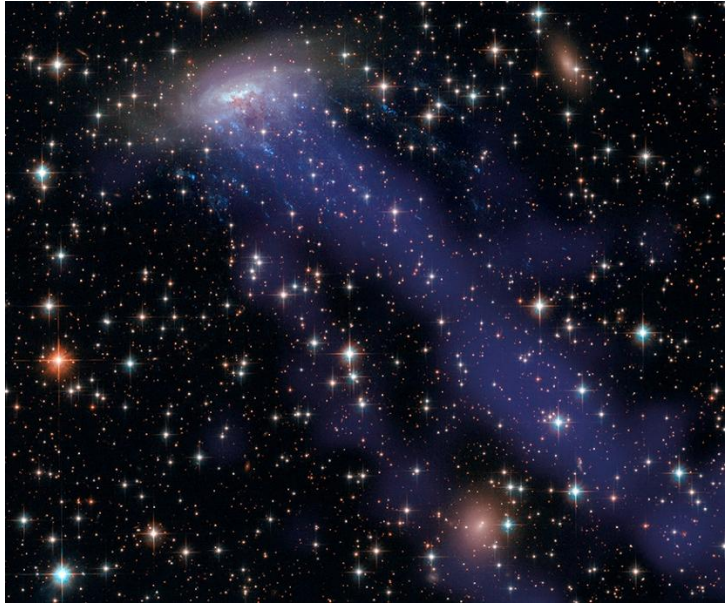

Figura 6.1. Las galaxias medusa tienen un cuerpo estelar luminoso principal con colas de gas que se extienden en una dirección, como ESO 137-001. Imagen tomada de [159]

Un enfoque interesante para investigaciones venideras podría incluir el análisis de galaxias de cúmulos a diversos corrimientos al rojo, utilizando imágenes profundas de telescopios espaciales y simulaciones numéricas. Esto permitiría explorar cómo la evolución de las barras se ve afectada por el entorno del cúmulo en comparación con galaxias de campo, arrojando luz sobre las diferencias fundamentales en la formación de barras y proporcionando un contexto más amplio sobre la influencia del entorno en la morfología galáctica [160]. Por otro lado, el análisis detallado de galaxias en cúmulos podría centrarse en cómo las interacciones ambientales afectan la formación de barras en función de su distancia al centro del cúmulo, donde las galaxias experimentan diferentes grados de influencia del potencial gravitacional del cúmulo. Este enfoque puede incluir la relación entre la fracción de galaxias barradas y el radio virial del cúmulo, una medida clave para reflejar la intensidad de las fuerzas ambientales [161].

## 6.2 Legacy Survey y el Telescopio Espacial James Webb (JWST)

El Legacy Survey [162] y el James Webb Space Telescope (JWST) [163] proporcionan una oportunidad única para estudiar la fracción de galaxias barradas desde el universo local hasta el lejano. El Legacy Survey, con su amplia cobertura y alta resolución, permite un análisis detallado de las barras en galaxias cercanas y en diversos entornos. Por otro lado, el JWST, con su capacidad para observar el universo temprano en el infrarrojo, permitirá detectar barras en galaxias distantes, aportando datos clave sobre su formación y evolución en las primeras etapas del universo. Juntos, estos observatorios ofrecerán una visión completa de la evolución de las barras galácticas a lo largo del tiempo cósmico. A conuación una descripción general de ambos proyectos astronómicos.

### 6.2.1. Legacy Survey para el Universo Local

El Legacy Survey es un ambicioso proyecto astronómico que combina observaciones en el rango óptico e infrarrojo cercano, utilizando la Dark Energy Camera (DECam) y el Beijing-Arizona Sky Survey (BASS), con el objetivo de estudiar el universo local con mayor profundidad y detalle [164]. A diferencia del Sloan Digital Sky Survey (SDSS), este proyecto ofrece una cobertura más extensa y detallada de las estructuras galácticas, lo permitirá realizar investigaciones sobre la fracción de galaxias barradas, la estructura interna de las galaxias y sus propiedades dinámicas. Además, las observaciones en el infrarrojo cercano permiten penetrar el polvo galáctico, lo que facilita una visión más clara de la estabilidad de las barras en galaxias espirales y lenticulares. Entre los principales avances que permite el Legacy Survey se destacan los siguientes:

- Estudio más completo de la fracción de barras: Gracias a su mayor sensibilidad y resolución en múltiples bandas, es posible revisar la fracción de galaxias barradas en el universo local con un conjunto de datos más amplio y diverso, que incluye una mayor variedad de tipos de galaxias y entornos galácticos. El análisis de las barras en diferentes entornos, como cúmulos de galaxias y regiones menos densas, proporcionará una comprensión más profunda del impacto del entorno en la evolución de las barras.

- Exploración de la estructura galáctica en el infrarrojo: Las observaciones en el infrarrojo permiten penetrar el polvo galáctico, facilitando una mejor observación de las estructuras internas de las galaxias, lo que resulta crucial para un análisis detallado de la presencia y estabilidad de las barras en galaxias espirales y lenticulares.

- Agregar sobre el estudio de las propiedades dinámicas: Al combinar imágenes profundas con datos espectroscópicos, el Legacy Survey podría revelar detalles sobre las propiedades dinámicas, como la cinemática galáctica, que influyen en la formación y evolución de las barras.

### 6.2.2.  James Webb Space Telescope (JWST) para el Universo Lejano

Las Observaciones de Lanzamiento Temprano (EROs) del Telescopio Espacial James Webb (JWST), también conocidas como las "Primeras Imágenes y Espectros de Webb" [165], marcaron el inicio de las operaciones científicas del telescopio. Publicadas el 12 de julio de 2022, estas primeras imágenes demostraron las capacidades del JWST al capturar una variedad de objetos celestiales. Estas observaciones resaltaron el potencial del JWST para realizar descubrimientos científicos revolucionarios. El JWST ofrece una visión sin precedentes del universo temprano, abriendo nuevas oportunidades para estudiar la fracción de galaxias barradas a altos desplazamientos al rojo ($z > 1$):

- Detección de barras en galaxias más tempranas: A diferencia de GOODS, el JWST tiene una resolución espacial mucho mayor en el infrarrojo medio y cercano, lo que permitirá detectar barras en galaxias jóvenes y distantes, donde las observaciones actuales carecen de detalles suficientes para identificar estructuras internas. Esto ayudará a determinar cuándo y cómo aparecen las primeras barras en el universo.

- Estudio de la evolución temprana de las barras: Con el JWST, se podrá investigar si las barras juegan un papel crucial en la estabilización de las galaxias durante sus etapas de formación, y explorar si las barras en galaxias de alto desplazamiento al rojo se forman a través de mecanismos diferentes a los observados en el universo local.

- Relación entre formación estelar y barras: Gracias a las detalladas observaciones en el infrarrojo del JWST, se podría correlacionar la presencia de barras con episodios de formación estelar y el crecimiento de los bulbos galácticos, refinando los modelos sobre cómo las barras influyen en la evolución galáctica en diferentes épocas cósmicas.

## 6.3 IlustrisTNG

En esta sección se muestra un estudio preliminar en la detección y análisis de barras estelares en galaxias de disco, utilizando simulaciones cosmológicas avanzadas, como las del proyecto IllustrisTNG. Aquí se presentan resultados iniciales que sentarán las bases para investigaciones futuras, donde se abordará con mayor profundidad la detección de barras en galaxias de campo y la evolución de la fracción de galaxias barradas a lo largo de diversos corrimientos al rojo. Se discutirá cómo los resultados obtenidos de la simulación TNG50 pueden ayudar a entender la formación de barras galácticas, estableciendo un marco sólido para estudios posteriores.

El proyecto IllustrisTNG [166][167][168] es un referente en simulaciones cosmológicas que modelan la evolución del Universo desde sus inicios hasta la actualidad, utilizando el código AREPO [169], basado en una malla dinámica de Voronoi para resolver las ecuaciones de magnetohidrodinámica y gravedad. Este enfoque permite simular con precisión tanto interacciones galácticas pequeñas como grandes cúmulos, facilitando el estudio de la formación galáctica a diversas escalas.

IllustrisTNG, sucesor de Illustris [170], introduce mejoras como la inclusión de campos magnéticos y una optimización del modelo de formación estelar, representando mejor procesos clave como la retroalimentación de agujeros negros y vientos galácticos [171]. Las simulaciones TNG50, TNG100 y TNG300 cubren volúmenes de 50, 100 y 300 Mpc cúbicos (Figura 6.2), permitiendo estudiar la evolución de materia oscura, gas, estrellas y agujeros negros desde $z = 127$ hasta el presente [167].

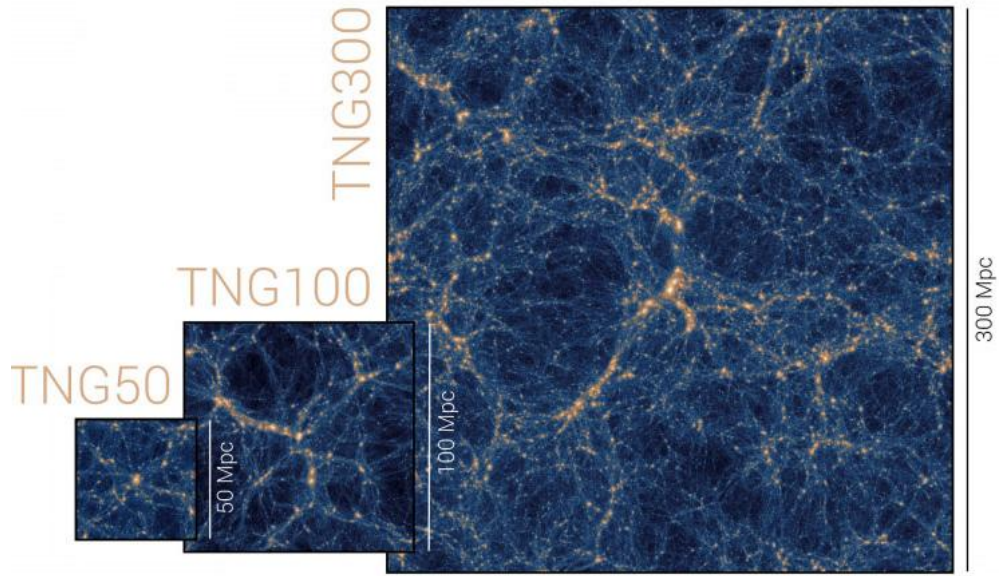

Figura 6.2. Representación de los tres volúmenes de simulación que conforman IllustrisTNG: TNG50, TNG100 y TNG300. Como se aprecia, el nombre de cada una denota la longitud del lado de la caja en megapársecs (Mpc). Imagen tomada de [172].

En esta sección, hemos seleccionado TNG50 debido a su alta resolución, lo que permite un análisis detallado de galaxias individuales, crucial para la detección y estudio de barras galácticas, fundamentales para comprender la dinámica y evolución de estas estructuras.

### 6.3.1 Uso de TNG50 para la Detección de Barras Estelares

TNG50, la simulación más detallada, abarca un volumen de 50 Mpc y simula la evolución de $2 \times 2160^3$ partículas de materia oscura y gas. Las condiciones iniciales se establecieron en $z = 127$ usando la aproximación de Zeldovich y el código N-GENIC [168], y los parámetros cosmológicos se alinean con los resultados de la colaboración Planck [173]. Una de las principales ventajas de TNG50 es su capacidad para generar imágenes de alta resolución de estructuras galácticas, lo que permite el análisis detallado de barras estelares, una característica común en galaxias espirales. Estas barras son cruciales para estudiar la dinámica galáctica y los procesos de formación estelar. La alta resolución de TNG50 facilita el uso de métodos morfológicos para caracterizar estas barras en términos de longitud, fortaleza e influencia en la distribución de estrellas y gas.

Utilizamos datos públicos de la simulación TNG50-1 y los catálogos adicionales provistos por [174], que emplean el código MORDOR para descomponer las componentes cinemáticas estelares en disco fino, disco grueso, bulbo y pseudobulbos. Para asegurar una descomposición robusta, seleccionamos galaxias con más de $1e^4$ partículas estelares, lo que da lugar a galaxias con una masa estelar de $log(M_*/M_\odot) > 8.5$, siendo la mayoría $log(M_*/M_\odot) > 9.5$.

Los catálogos entregados por MORDOR también incluyen la fracción de estrellas en el disco fino, disco grueso y bulbo, junto con la amplitud del modo de Fourier $m = 2$, que permite identificar tanto barras débiles como fuertes. Aunque en este trabajo contabilizamos ambas barras indistintamente, en estudios futuros podríamos separarlas según la amplitud del modo de Fourier. Para la clasificación morfológica, seguimos la metodología presentada en [175], dónde separan las galaxias espirales, lenticulares y elípticas en base a la fracción de estrellas pertenecientes al disco fino, de forma que:

Galaxias espirales: Disco Fino / Masa total > 0.5

Galaxias lenticulares: Disco Fino / Masa total > 0.3

Elípticas: Disco Fino / Masa total < 0.3

Con estas consideraciones, encontramos un notable acuerdo con las observaciones a $z = 0$. La Figura 6.3 muestra que la fracción de barras en nuestra muestra observacional de galaxias repesentativas, determinada a través de análisis de Fourier, fue del 26% para galaxias espirales y del 18% para lenticulares. En comparación, las simulaciones de IllustrisTNG mostraron una fracción del 27% para espirales y del 24% para lenticulares. Estos resultados sugieren que nuestras simulaciones están alineadas con las tendencias observadas en el universo real, lo que refuerza la validez de nuestros métodos y análisis.

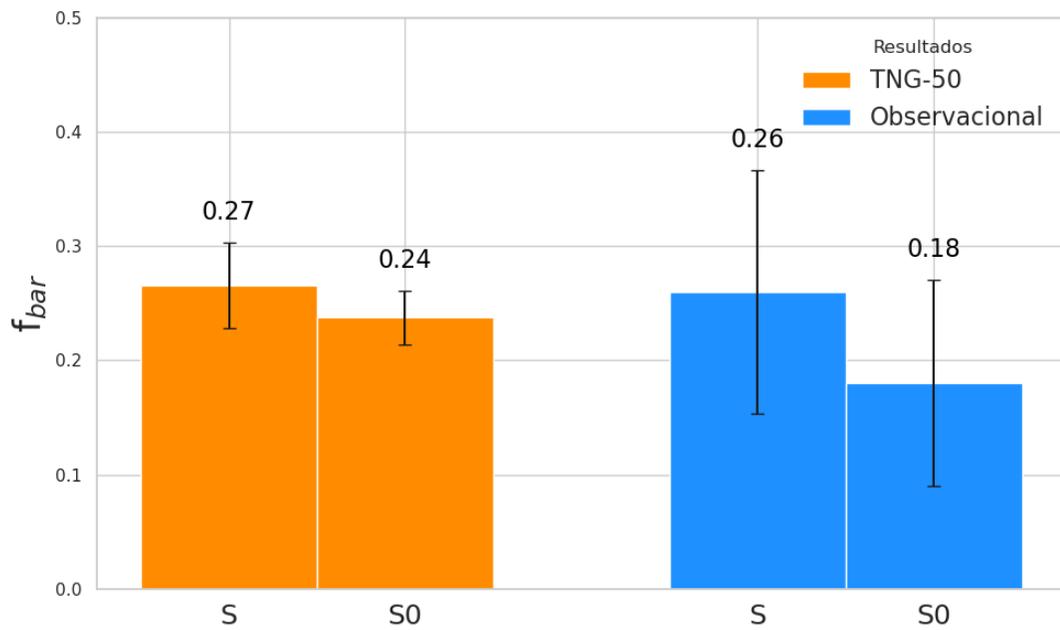

Figura 6.3. Muestra la fracción de barras determinadas en el Universo local $z = 0$: por las simulaciones de IllustrisTNG y por nuestra muestra representativa de galaxias locales.

# Apéndices

# A. Una Visión General de la Formación y Evolución de las Galaxias

Las galaxias son objetos formados por millones de estrellas, gas, polvo y materia oscura, ligados gravitacionalmente. Se pueden dividir en dos tipos principales, a saber, galaxias de disco aplanadas y soportadas rotacionalmente, y galaxias elípticas que están dominadas por movimientos aleatorios (en la siguiente sección se describen las clasificaciones de galaxias).

## A.1 Fluctuaciones iniciales de Densidad hasta las Primeras Galaxias

Según la teoría más ampliamente aceptada en la actualidad, el Big Bang, el universo se originó a partir de una gran explosión hace aproximadamente 13.8 mil millones de años [173]. La base para esta idea fue establecida por el descubrimiento de que nuestro Universo está expansión. En 1922, Alexander Friedmann desarrolló modelos de un Universo estático y varios en expansión basados en la teoría de la relatividad general de Albert Einstein [176][177]. Independientemente, Georges Lemaître derivó una de las soluciones obtenidas por Friedmann y fue el primero en demostrar que las velocidades de recesión de las galaxias están correlacionadas linealmente con sus distancias; evidencia observacional de la expansión del Universo [178].

Según la teoría del Big Bang 9 (Figura A.1), el Universo comenzó a expandirse y

enfriarse a partir de un estado inicial caliente y denso. Después de aproximadamente $10^{-36}$ segundos, la expansión se aceleró de manera exponencial impulsada por la energía del vacío en un campo cuántico, un período llamado inflación que duró hasta t ≈ $10^{-32}$ segundos [179][180]. Las fluctuaciones de densidad cuántica en ese campo, permitidas por el principio de incertidumbre de Heisenberg [181], pudieron crecer durante la inflación hasta convertirse en perturbaciones macroscópicas. Después de eso, la expansión y enfriamiento más lento pero continuo del Universo condujo a la formación de bariones como protones y neutrones, y eventualmente a isótopos ligeros, como el helio-3, deuterio y litio, durante la nucleosíntesis primordial [182]. Después de aproximadamente 360,000 años, la temperatura era lo suficientemente baja (aproximadamente 3,000 K) como para que los protones capturaran electrones para formar hidrógeno neutro [183]. Esta época se suele llamar recombinación.

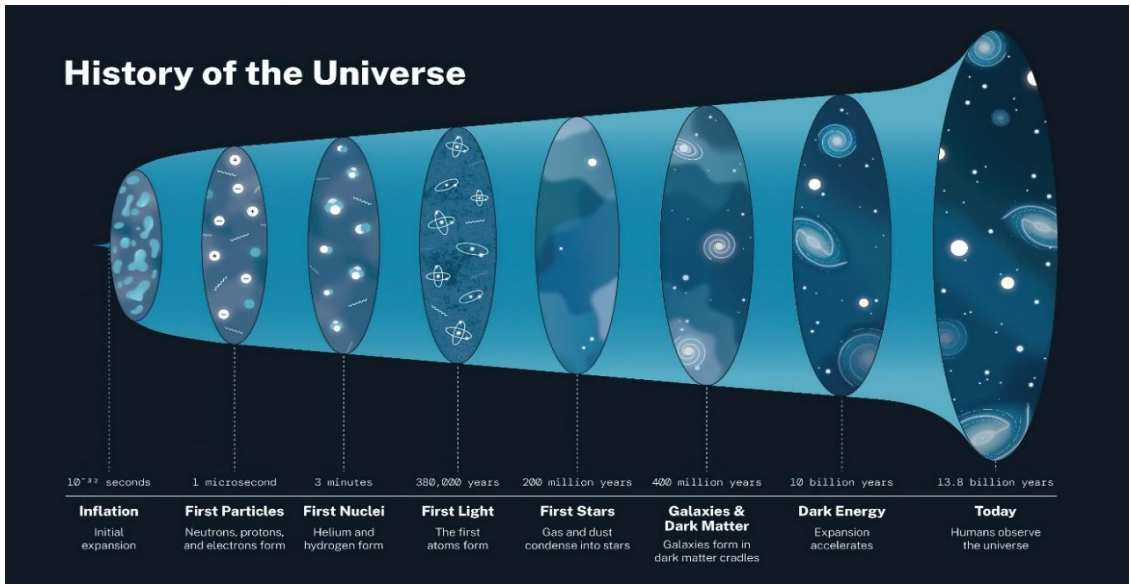

Figura A.1. La infografía traza la historia del universo. NASA[1]

---

[1] https://science.nasa.gov/universe/overview/

Una consecuencia fundamental de la recombinación es que la dispersión de Thomson por electrones se redujo considerablemente y el Universo se volvió transparente a los fotones. El desacoplamiento de los fotones y la materia liberó un campo de radiación que viajó a través del Universo y se enfrió hasta el día de hoy a aproximadamente 2.7 K. Fue predicho por Alpher y Herman en 1948 [184] y descubierto por Penzias y Wilson en 1965 [185] como el Fondo Cósmico de Microondas (CMB por sus siglas en inglés). Un avance en las mediciones del CMB se logró muchas décadas después mediante tres misiones satelitales consecutivas: el Cosmic Background Explorer (COBE) [186], el Wilkinson Microwave Anisotropy Probe (WMAP) [187]. Los mapas del CMB (Figura A.2) entregados por estas misiones muestran un grado muy alto de isotropía en escalas grandes, lo que confirma el principio cosmológico de un Universo homogéneo e isotrópico.

Estas fluctuaciones de temperatura en el CMB son cruciales para nuestro entendimiento de la formación de estructuras a gran escala en el Universo, incluyendo las galaxias. Las pequeñas variaciones en la densidad del CMB actuaron como las semillas a partir de las cuales se formaron las primeras estructuras cósmicas, tales como galaxias y cúmulos de galaxias. Comprender estas fluctuaciones nos proporciona una visión detallada de las condiciones iniciales del Universo y de cómo las galaxias se formaron y evolucionaron a lo largo del tiempo cósmico.

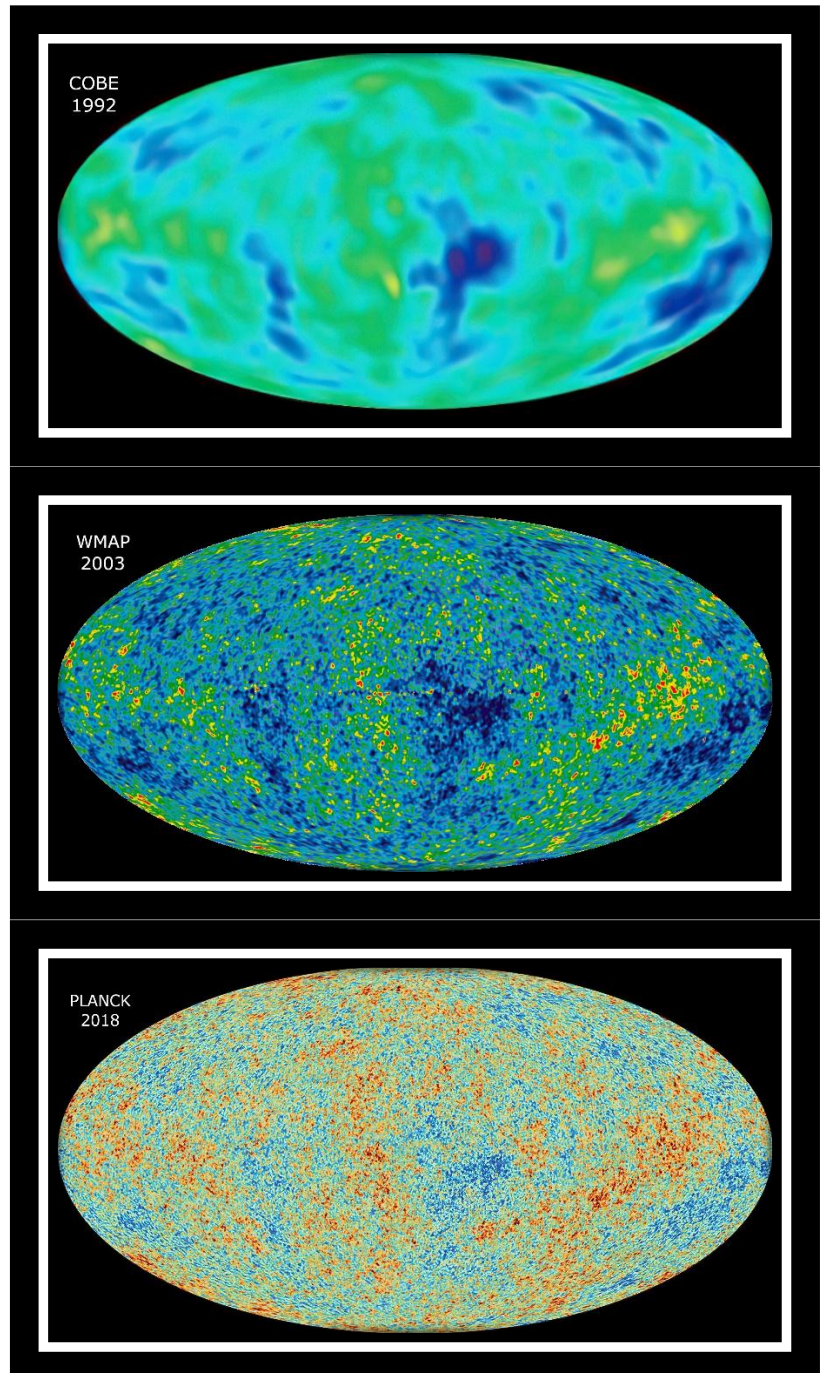

Figura A.2. Comparación de la mejora en la resolución de la imagen de la radiación de fondo de microondas cósmicas (CMB) obtenida por los satélites COBE[2], WMAP[3] y Planck[4].

---

## A.2 Formación y Evolución Galáctica

En la cosmología moderna, un desafío central es comprender el proceso de formación y evolución de las galaxias. Cualquier teoría o simulación de la evolución galáctica debería tener como objetivo final igualar las observaciones de galaxias en todos los períodos cósmicos. En el siguiente párrafo, resumiré algunos resultados observacionales clave que debemos intentar conectar y explicar con esta teoría.

Las primeras limitaciones observacionales sobre la distribución de la estructura después de la recombinación, aproximadamente al corrimiento al rojo $z \approx 1100$, se basan en el Fondo Cósmico de Microondas (CMB). Las observaciones directas de galaxias comienzan mucho más tarde en el tiempo y abarcan un amplio rango de $11.1 < z < 0$, es decir, aproximadamente 400 millones de años después del Big Bang, durante la época de la reionización hasta la actualidad (ver [188] para la galaxia actualmente más distante GN-z11 con corrimiento al rojo confirmado espectroscópicamente).

Actualmente, existen dos modelos propuestos para la formación de galaxias:

El Modelo Clásico: también conocido como el modelo de colapso "monolítico" o "Secular", se observa de manera esquemática en la Figura A.3 en el lado izquierdo. Este modelo propone que las galaxias se desarrollan a partir del colapso rápido de grandes nubes de gas. Este colapso inicial de gas es crucial para la formación de estructuras galácticas. Para comprender cómo y por qué estas nubes se colapsan, debemos considerar el papel de las fluctuaciones de densidad en el universo primitivo. Las fluctuaciones en la densidad del Fondo Cósmico de Microondas (CMB) proporcionan las semillas de inhomogeneidades que eventualmente llevan al colapso de grandes nubes de gas. Estas pequeñas variaciones en la densidad primordial, detectadas en el CMB, crearon regiones de mayor densidad que se convirtieron en el lugar de formación de las primeras estructuras galácticas. A medida que estas regiones densas colapsaban bajo su propia gravedad, se formaban las primeras galaxias. Este proceso de formación galáctica, donde grandes estructuras colapsan primero y luego se organizan en componentes más pequeños, refleja el enfoque de "arriba hacia abajo" del Modelo Clásico. En este modelo, el entorno circundante tiene poco impacto en el desarrollo de las galaxias, ya que el énfasis está en la evolución interna de las grandes nubes de gas y su colapso bajo su propia gravedad. El modelo propuesto por Eggen, Lynden-Bell y Sandage en 1962 [189] representó un paso significativo en el intento de explicar la formación de galaxias. Este enfoque ha sido instrumental en la comprensión de la formación tanto de galaxias elípticas como de los bulbos presentes en las galaxias espirales.

El Modelo Jerárquico: Representado en el lado derecho de la Figura A.3, este modelo propone que las galaxias se forman y evolucionan gradualmente a través de una serie de fusiones y agregaciones de estructuras más pequeñas, siguiendo un enfoque de "abajo hacia arriba". Según esta teoría, la formación de las galaxias no ocurre de manera instantánea a partir del colapso de grandes nubes de gas, sino que resulta de la acumulación progresiva de materia en halos de materia oscura que posteriormente se fusionan y crecen.

En el Modelo Jerárquico, las galaxias se forman primero en halos de materia oscura más pequeños que colapsan debido a la influencia gravitatoria. La masa bariónica, que se encuentra en estos halos, se acumula gradualmente y da lugar a la formación de estrellas y otras estructuras galácticas. A medida que estos halos de materia oscura se fusionan, forman galaxias más grandes y complejas. Este proceso de crecimiento y fusión sucesiva permite que las galaxias evoluciones a lo largo del tiempo cósmico y explica la variedad observada en las galaxias de diferentes épocas del universo. El modelo propuesto por White y Rees en 1978 [190] fue un avance significativo en nuestra comprensión de la formación y evolución galáctica. Este enfoque jerárquico también ofrece una explicación para la presencia de galaxias en el universo temprano, así como para las características estructurales observadas en galaxias con discos. Además, el Modelo Jerárquico ayuda a interpretar cómo las interacciones y fusiones de galaxias han contribuido a la formación de estructuras más grandes y complejas en el universo, como cúmulos de galaxias.

Observacionalmente, las galaxias en corrimientos al rojo altos exhiben características como compacidad, fragmentación y una alta tasa de formación estelar, con dinámicas turbulentas [191]. En contraste, en el universo cercano, las galaxias muestran una distribución bimodal en diversos parámetros, como color [192], densidad de masa estelar [193], tasa de formación estelar y concentración [194]. Aunque existe una gran cantidad de investigaciones sobre las galaxias, la formación de discos sigue siendo un desafío importante para nuestra comprensión actual de la evolución galáctica.

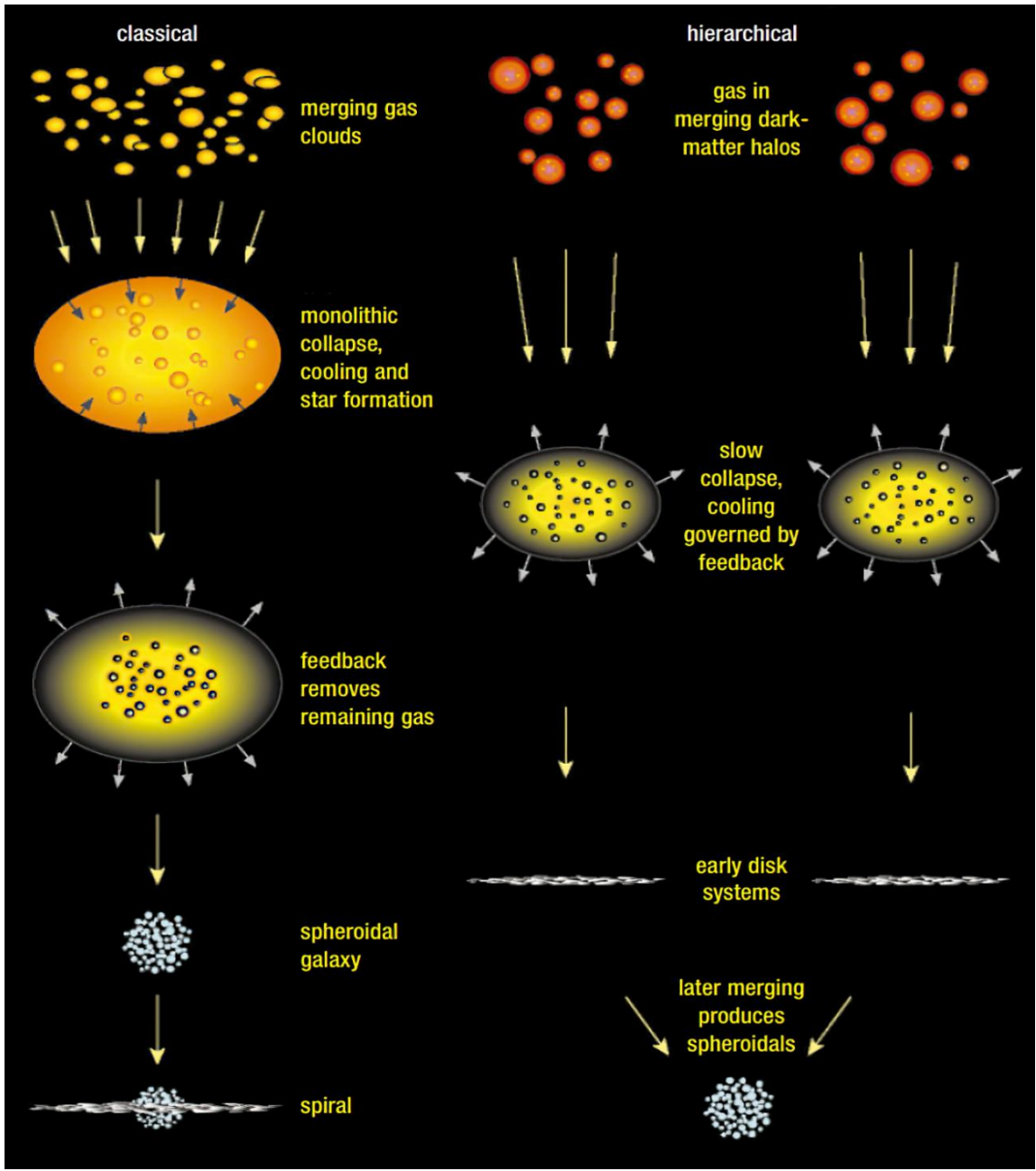

Figura A.3. La figura ilustra dos modelos principales: el Modelo Clásico, que describe la formación de galaxias a partir del colapso de grandes nubes de gas (enfoque de "arriba hacia abajo"), y el Modelo Jerárquico, que explica la evolución galáctica a través de fusiones de halos de materia oscura más pequeños (enfoque de "abajo hacia arriba"). Créditos de Imagen [195].

# B.  Bootstrap

El bootstrap o bootstrapping es una técnica estadística empleada para estimar una distribución de muestreo a través del remuestreo de los datos originales. Este método es especialmente útil cuando es complicado o incluso imposible obtener una expresión exacta para la distribución de muestreo. En la inferencia estadística tradicional, se suelen hacer suposiciones sobre la distribución de la población, usando dichas suposiciones para calcular la distribución de muestreo de una estadística. Sin embargo, en muchas situaciones reales, la distribución poblacional es desconocida o demasiado compleja para modelarse con precisión, lo que hace que el bootstrapping sea una herramienta valiosa.

El término "bootstrapping" fue introducido por [196], y se ha convertido en una herramienta estadística poderosa para cuantificar la incertidumbre asociada con un estimador. Este método consiste en tomar múltiples muestras con reemplazo de los datos observados, generando muchas remuestras. Cada una de estas remuestras tiene el mismo tamaño que el conjunto de datos original, lo que implica que algunos datos se repiten varias veces en las remuestras, mientras que otros pueden no ser seleccionados.

Mediante la repetición de este proceso, se obtiene una distribución de remuestras, conocida como distribución bootstrap. Esta distribución sirve como aproximación de la distribución de muestreo, permitiendo derivar métricas como el error estándar y los intervalos de confianza, sin necesidad de asumir una distribución específica para la población subyacente.

El procedimiento puede resumirse en los siguientes pasos:

1. Seleccionar la muestra original.

2. Definir el número de submuestras bootstrap a generar (usualmente N = 1,000).

3. Extraer N submuestras con reemplazo.

4. Calcular la estadística de interés (como la desviación estándar) para cada submuestra.

5. Promediar las estadísticas obtenidas de las 1,000 submuestras.

6. El promedio calculado se utiliza como el error aplicado a la muestra original.

En este estudio, utilicé el bootstrapping para estimar los errores en las fracciones de barras observadas tanto en el universo local como en el universo lejano. Además, apliqué este método para analizar la fracción de barras en función de propiedades como la masa estelar, el color u-r y la relación B/T (bulge-to-total). Las barras de error representan un intervalo de confianza del 90%, obtenido a partir de 1,000 realizaciones aleatorias de la muestra original.

# C.   Resultados destacados del

# Método de Isofotas Elípticas

# C.1. Galaxias Espirales a $z = 0.027$

**Galaxia: J143411.25+003656.6**
**Filtro: r_band_SDSS**

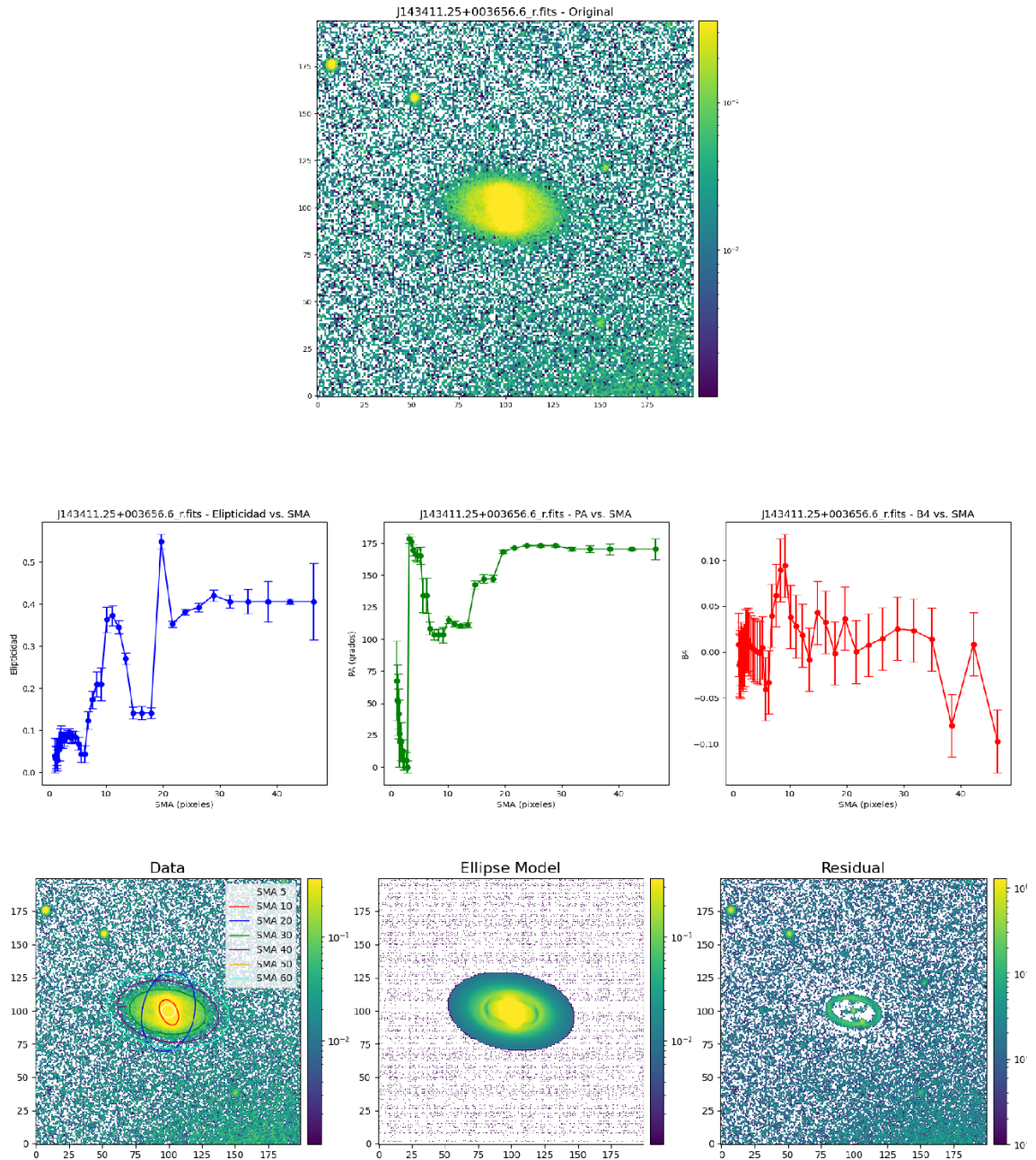

Galaxia barrada: cumple con los criterios del método de isofotas elípticas, confirmando la presencia de una barra estelar.

**Galaxia: J120127.92-004306.1**
**Filtro: r_band_SDSS**

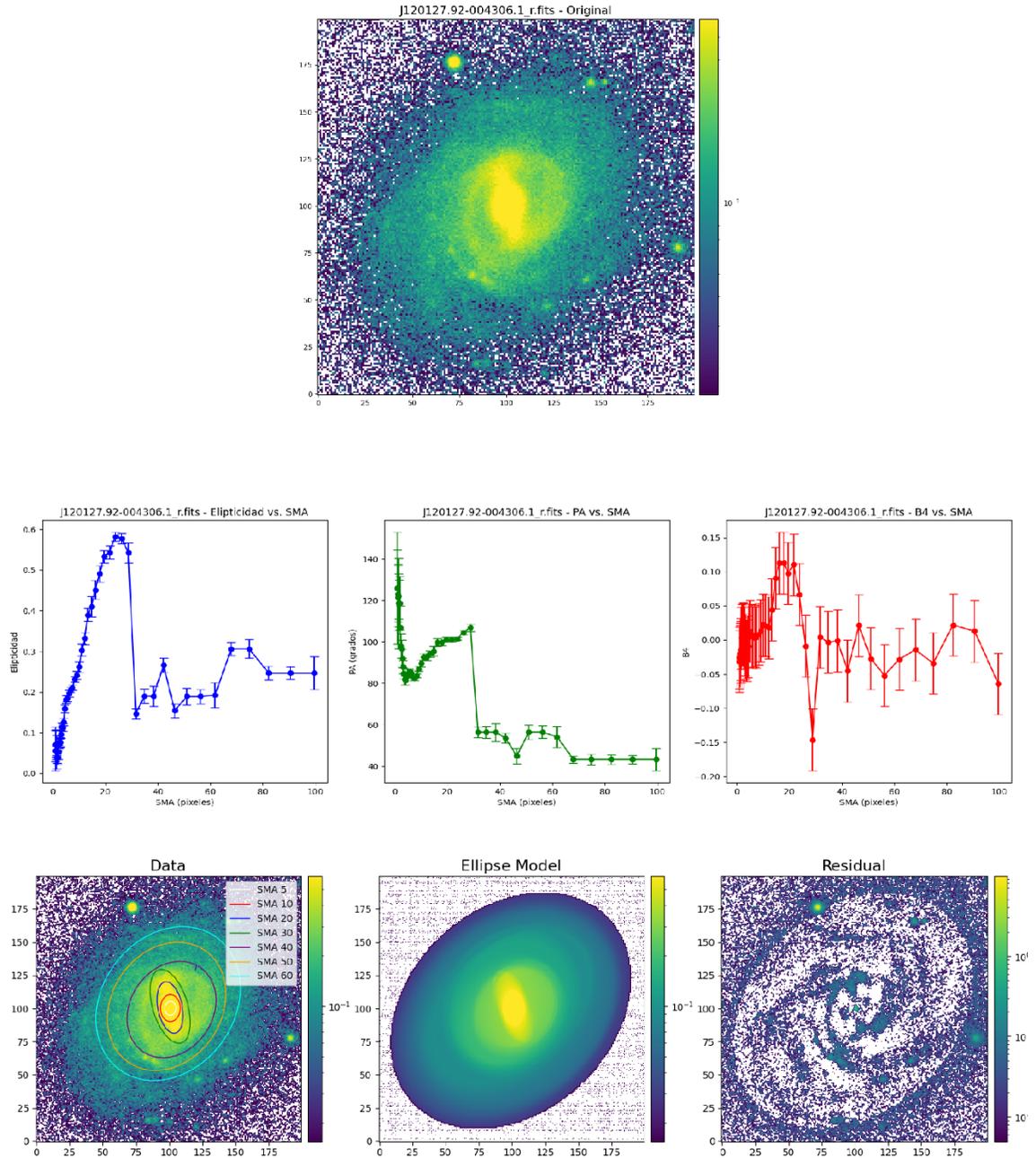

Galaxia barrada: cumple con los criterios del método de isofotas elípticas, confirmando la presencia de una barra estelar.

**Galaxia: J135102.22-000915.1**
**Filtro: r_band_SDSS**

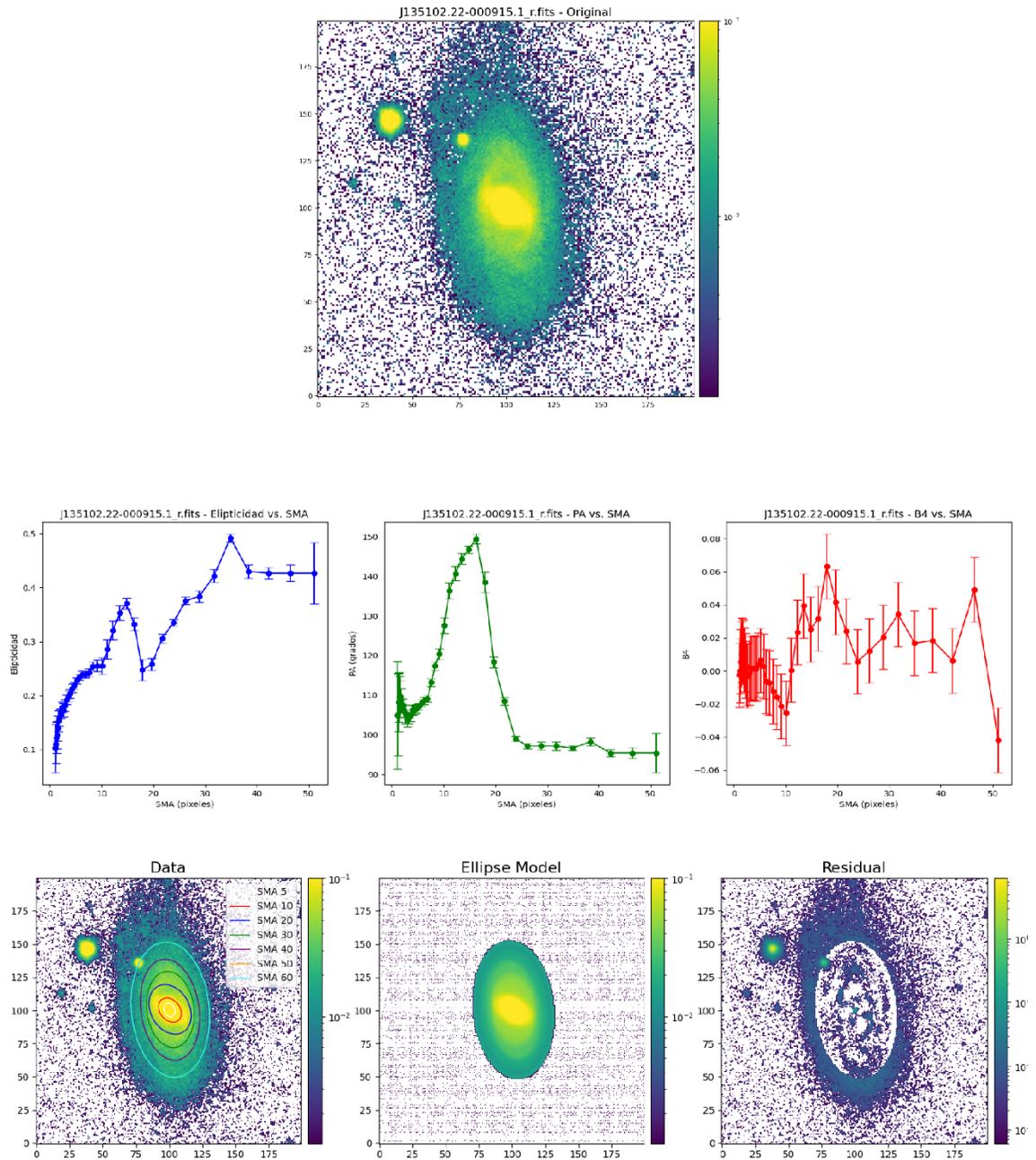

Galaxia barrada: cumple con los criterios del método de isofotas elípticas, confirmando la presencia de una barra estelar.

**Galaxia: J140320.74-003259.7**
**Filtro: r_band_SDSS**

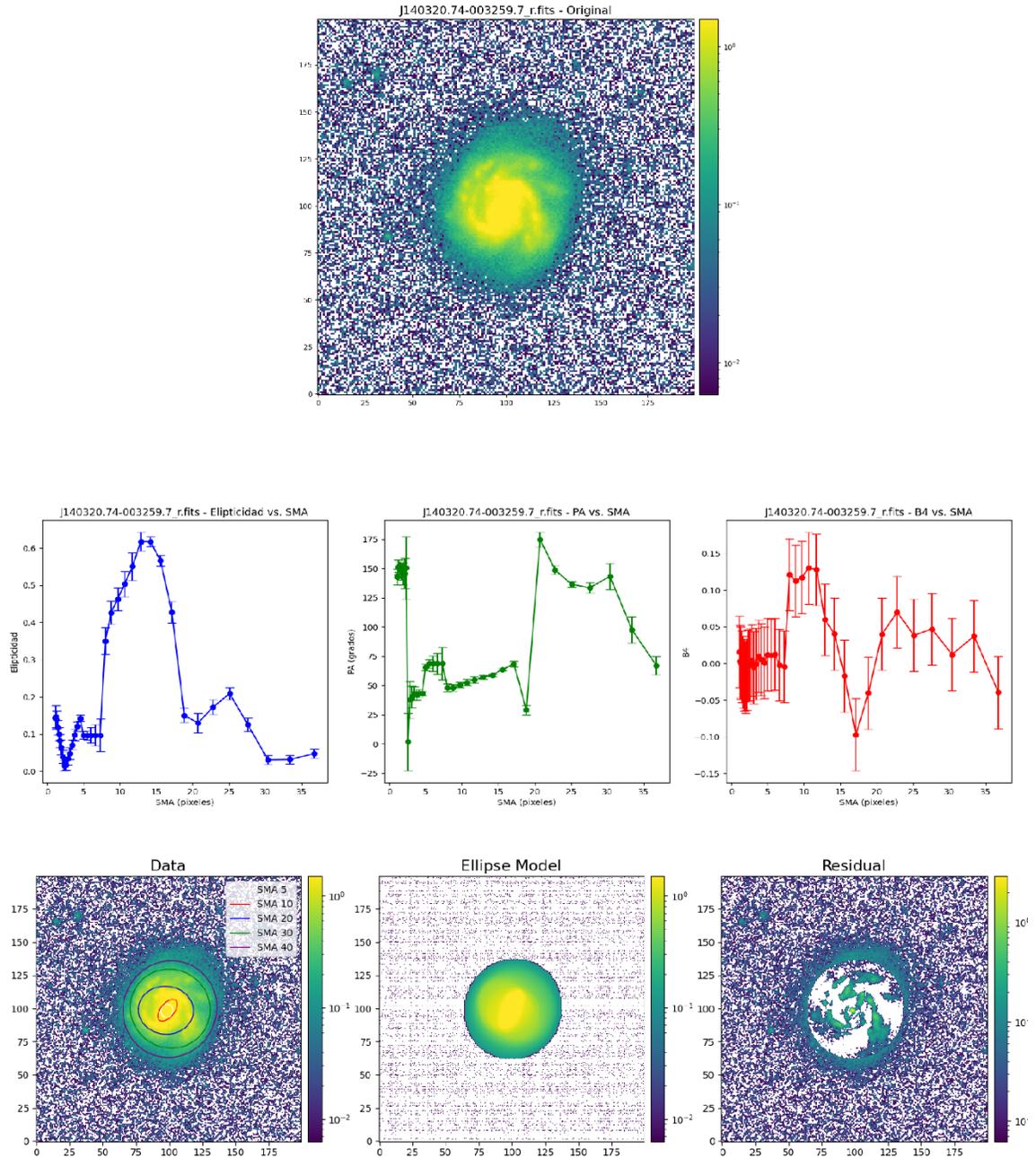

Galaxia barrada: cumple con los criterios del método de isofotas elípticas, confirmando la presencia de una barra estelar.

**Galaxia: J113833.27-011104.1**
**Filtro: r_band_SDSS**

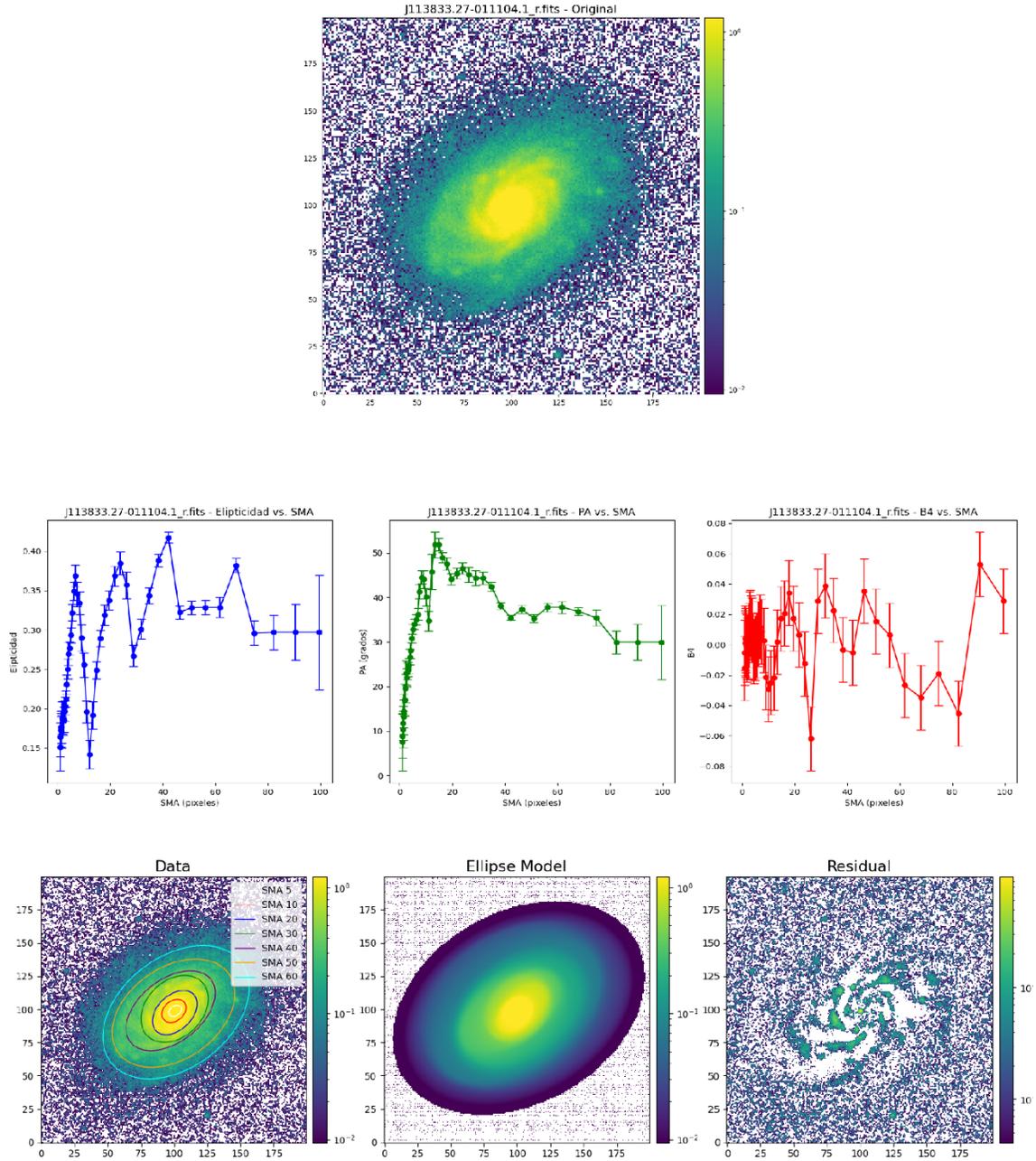

Galaxia no barrada: no cumple con los criterios del método de isofotas elípticas, descartando la presencia de una barra estelar.

**Galaxia: J141814.91+005327.9**
**Filtro: r_band_SDSS**

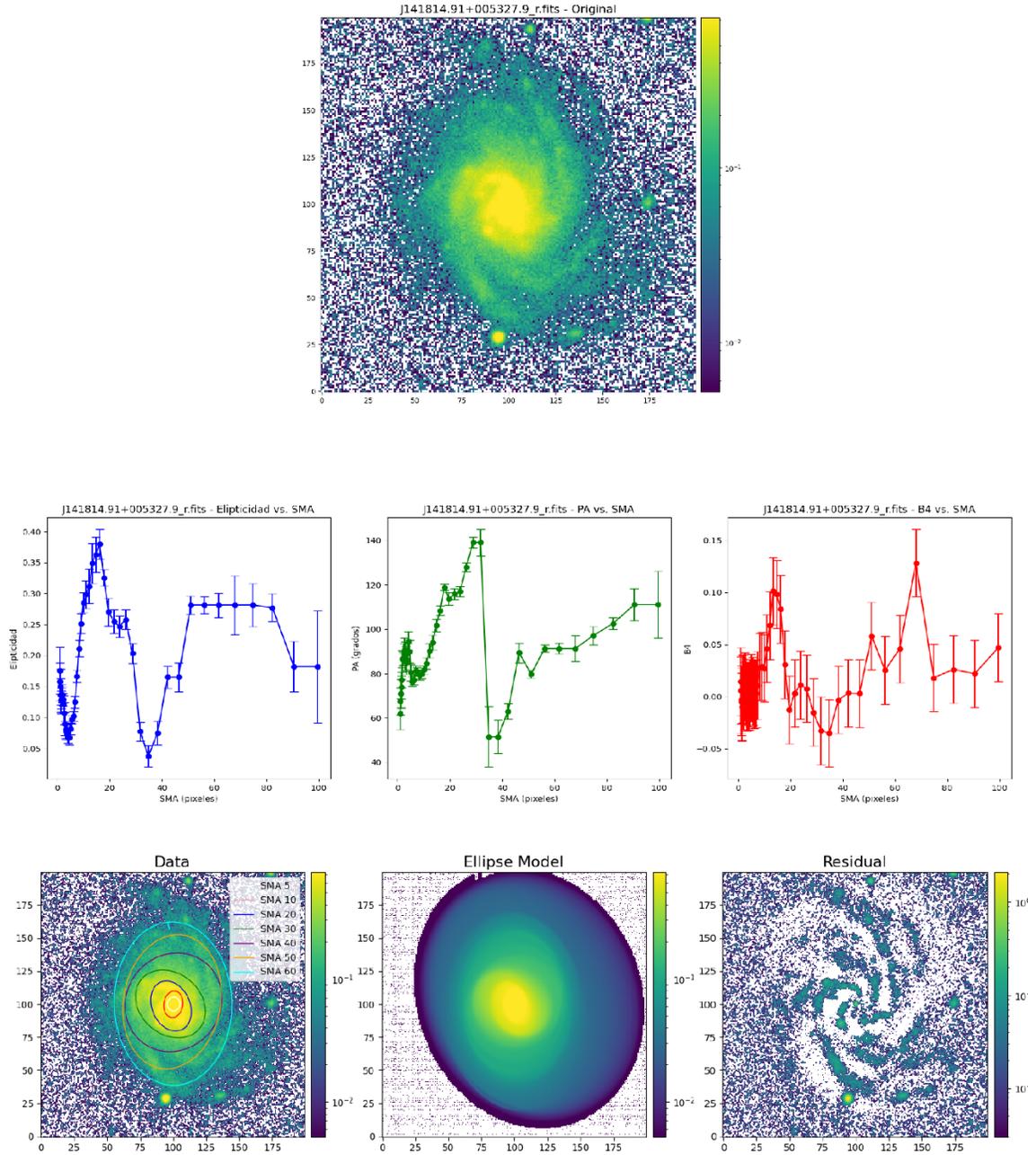

Galaxia no barrada: no cumple con los criterios del método de isofotas elípticas, descartando la presencia de una barra estelar.

**Galaxia: J111849.55+003709.3**
**Filtro: r_band_SDSS**

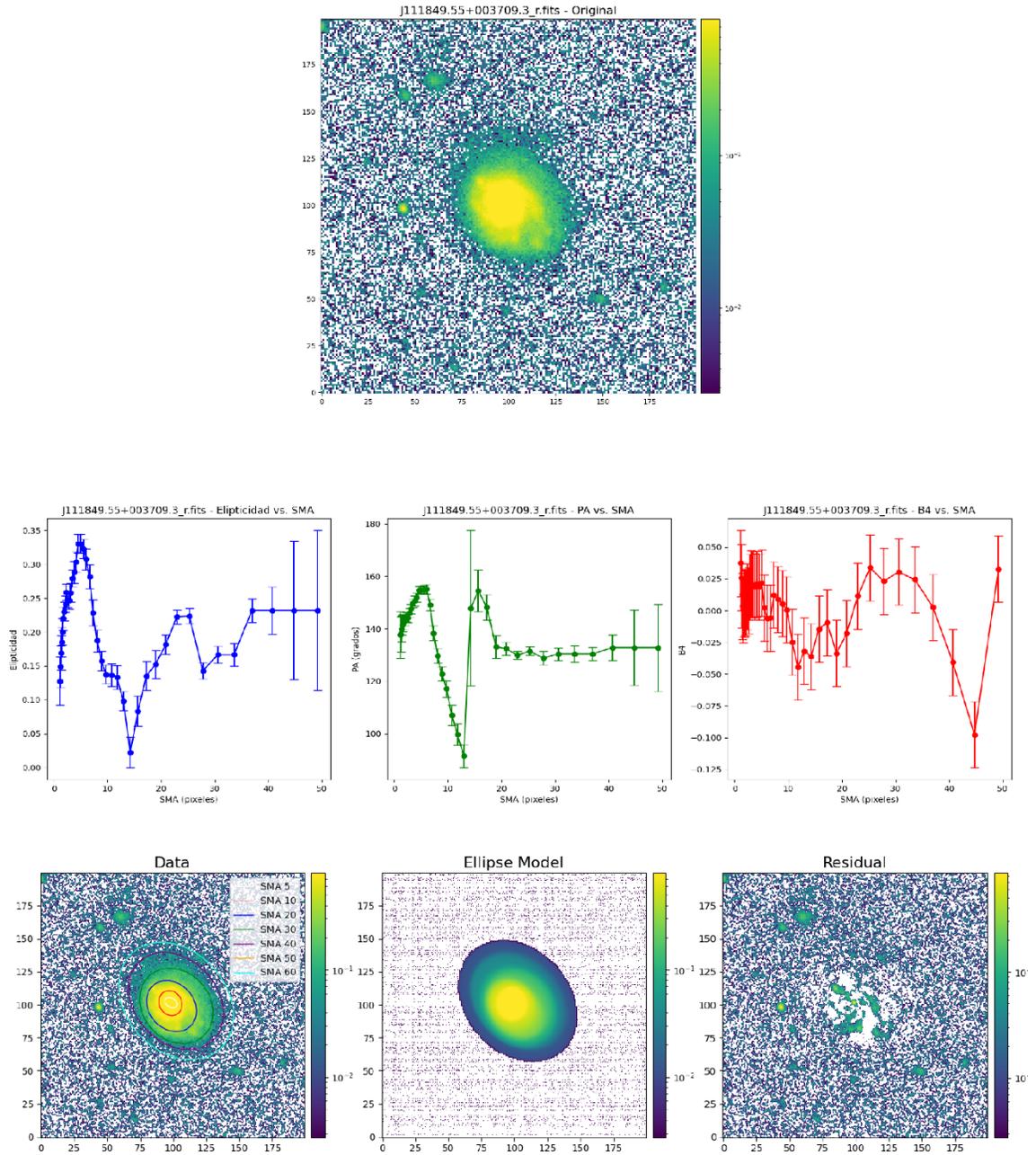

Galaxia no barrada: no cumple con los criterios del método de isofotas elípticas, descartando la presencia de una barra estelar.

**Galaxia: J144613.34+005157.7**
**Filtro: r_band_SDSS**

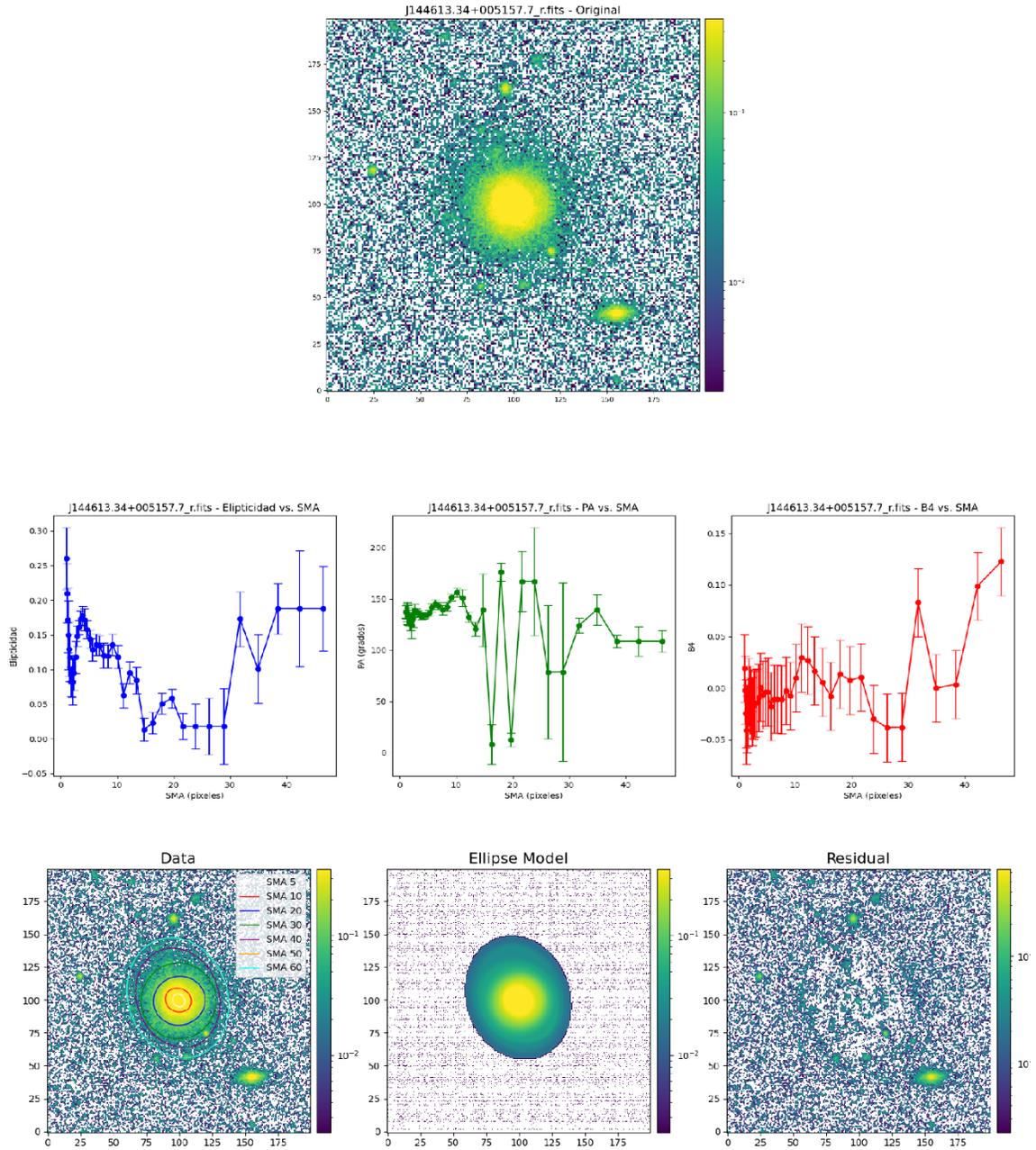

Galaxia no barrada: no cumple con los criterios del método de isofotas elípticas, descartando la presencia de una barra estelar.

## C.2. Galaxias Lenticulares a $z = 0.027$

**Galaxia: J143124.59+011403.7**
**Filtro: r_band_SDSS**

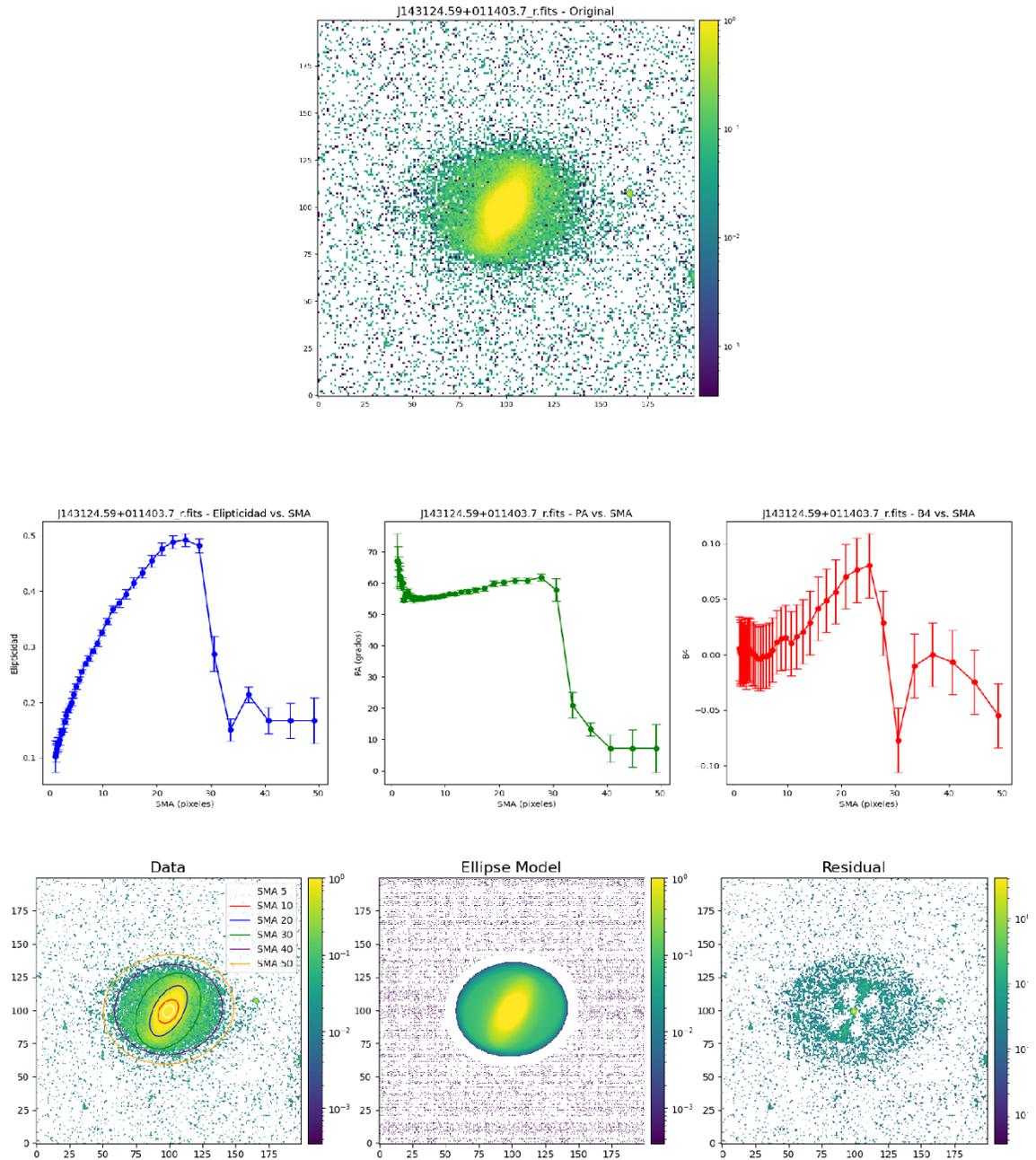

Galaxia barrada: cumple con los criterios del método de isofotas elípticas, confirmando la presencia de una barra estelar.

**Galaxia: J143540.07+001217.7**
**Filtro: r_band_SDSS**

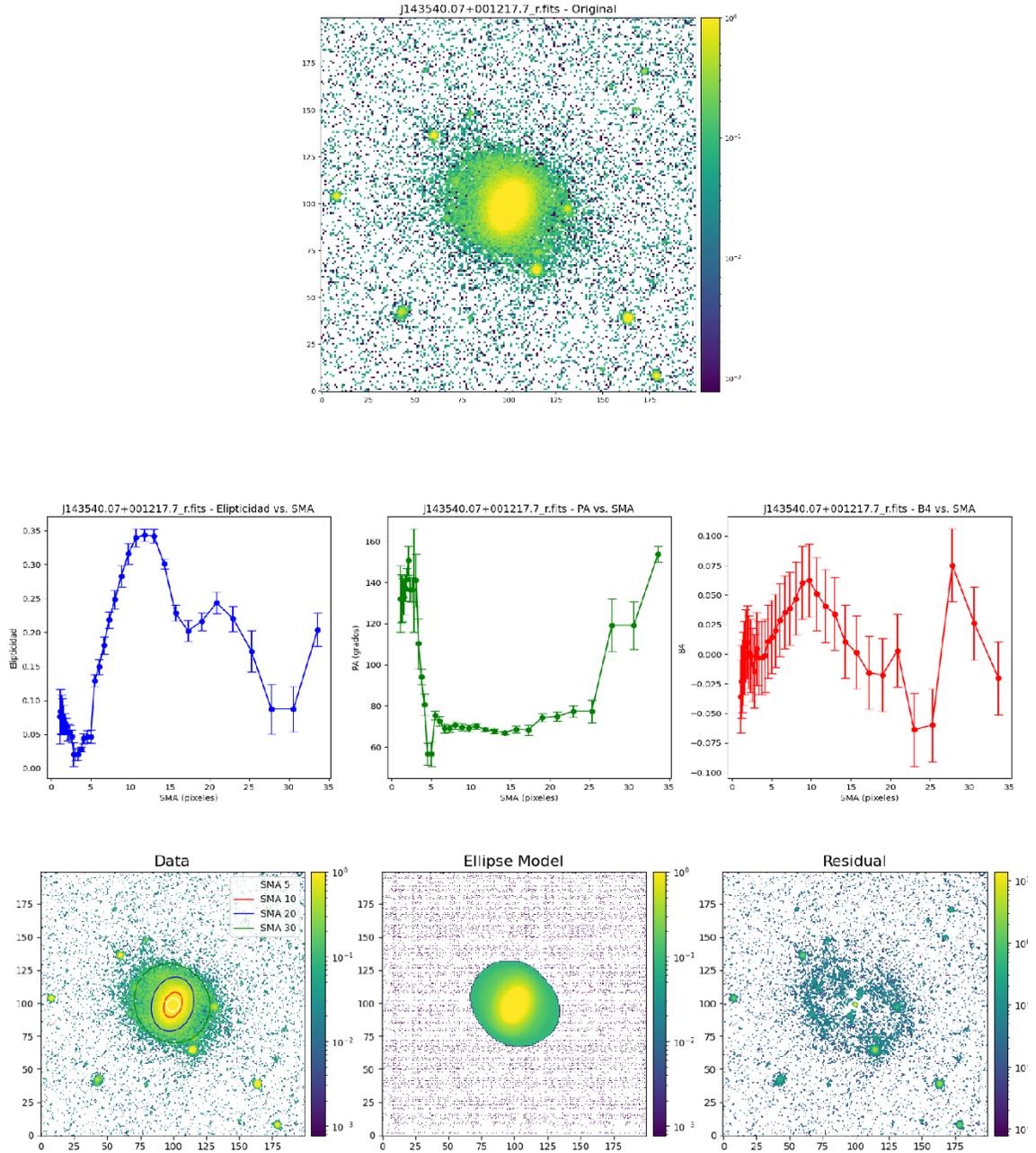

Galaxia barrada: cumple con los criterios del método de isofotas elípticas, confirmando la presencia de una barra estelar.

**Galaxia: J113523.27+000525.9**
**Filtro: r_band_SDSS**

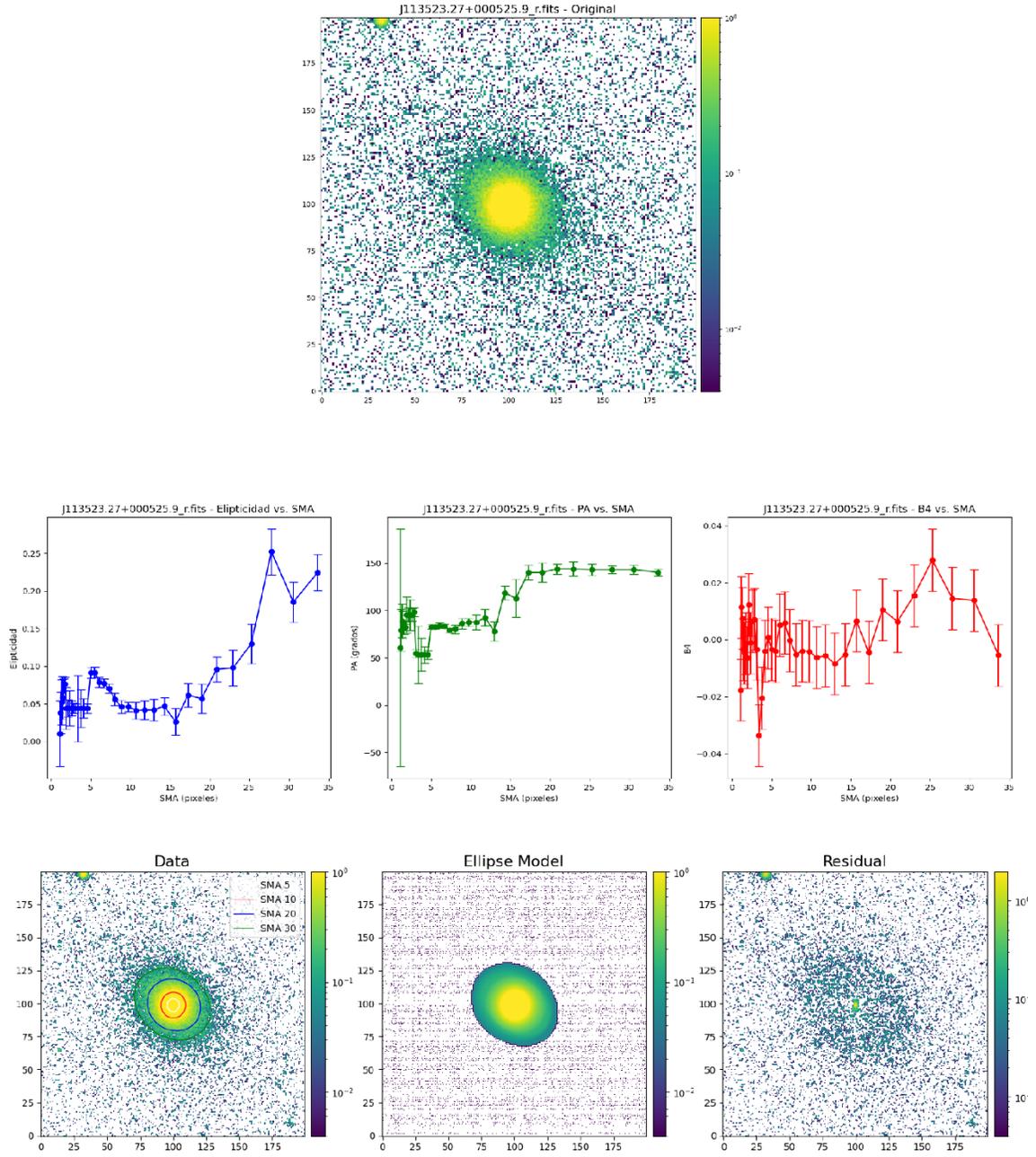

Galaxia no barrada: no cumple con los criterios del método de isofotas elípticas, descartando la presencia de una barra estelar.

**Galaxia: J103534.47-002116.2**
**Filtro: r_band_SDSS**

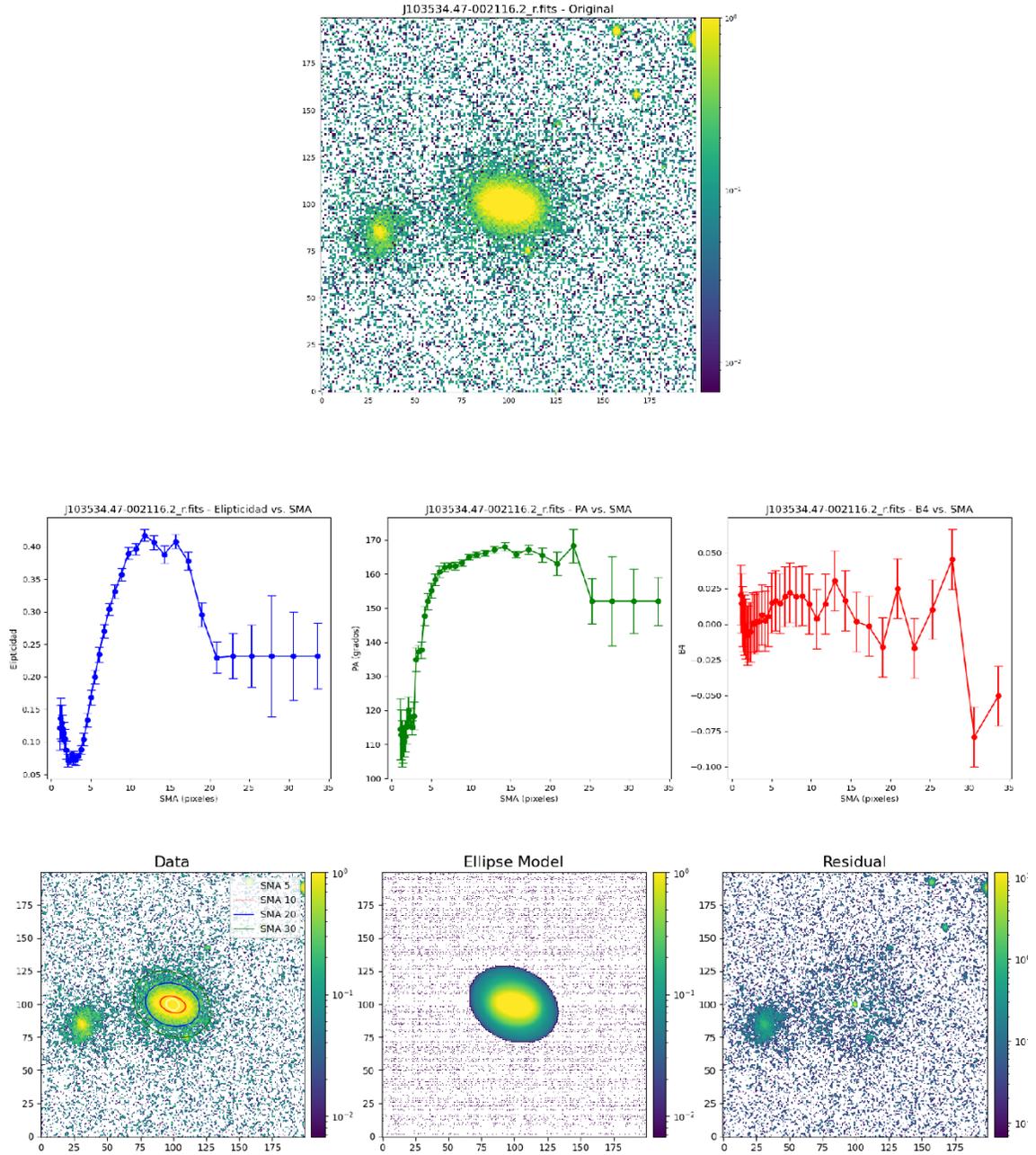

Galaxia no barrada: no cumple con los criterios del método de isofotas elípticas, descartando la presencia de una barra estelar.

**Galaxia: J113420.50+001856.4**
**Filtro: r_band_SDSS**

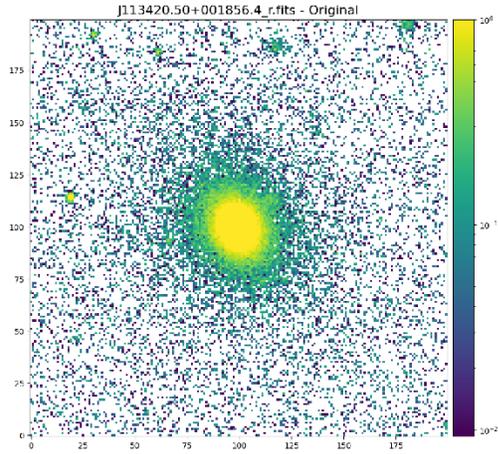

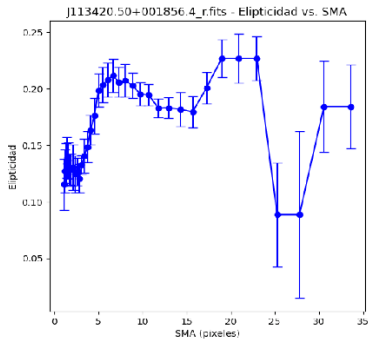
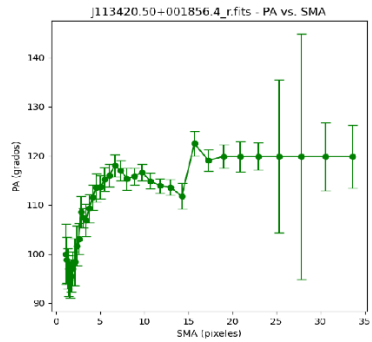
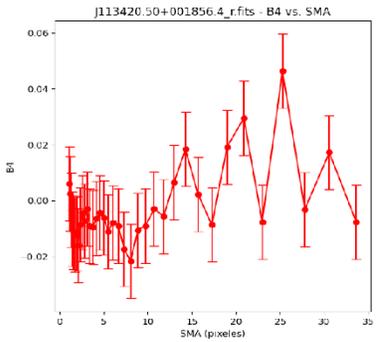

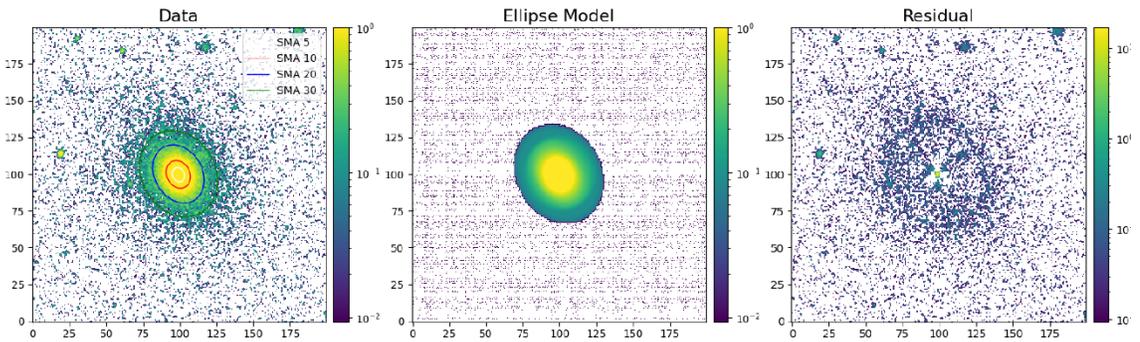

Galaxia no barrada: no cumple con los criterios del método de isofotas elípticas, descartando la presencia de una barra estelar.

# C.3. Galaxias Espirales a $z = 0.7$

**Galaxia: J033221.99-274655.9**
**Filtro: z_band_GOODS**

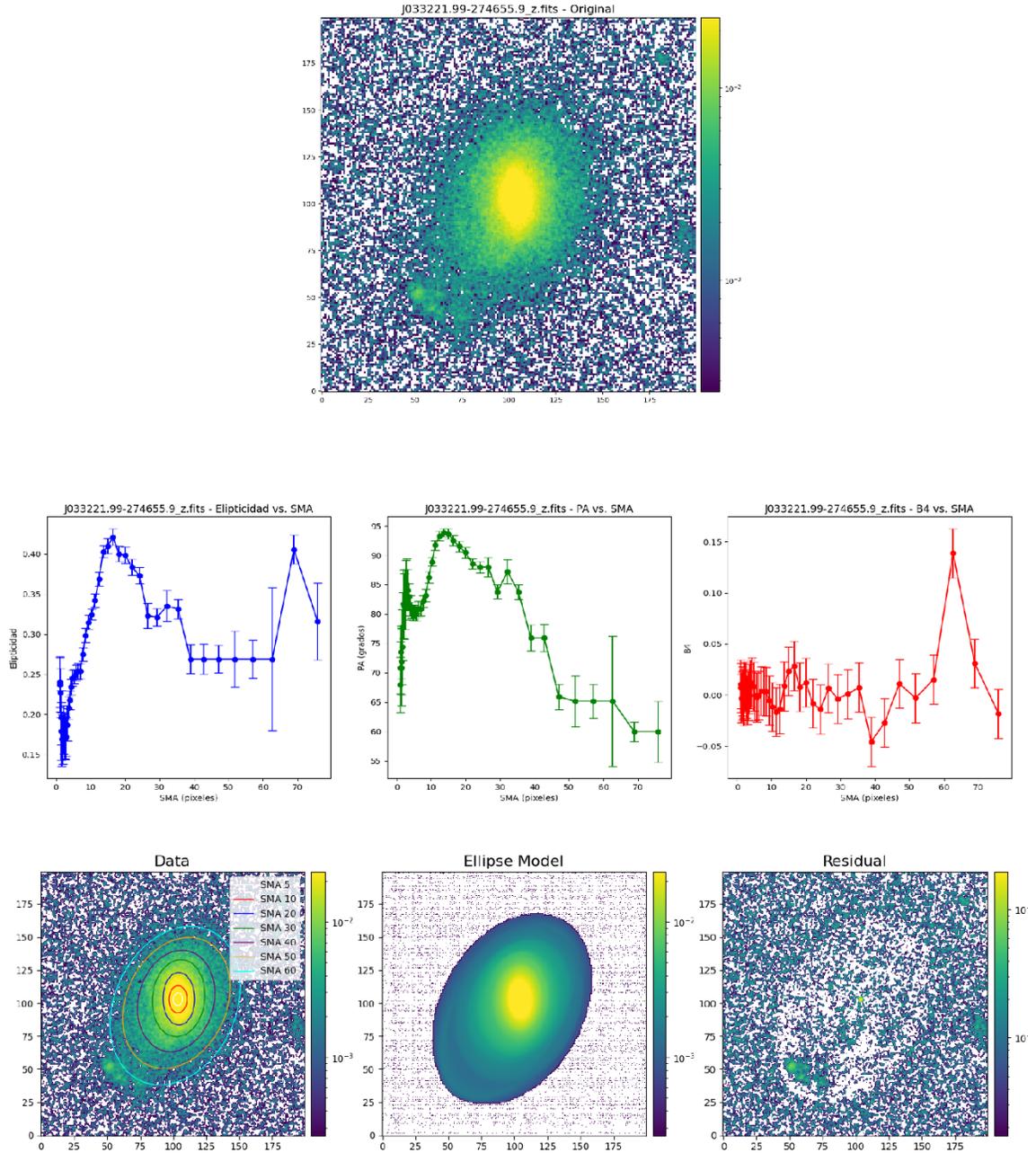

Galaxia barrada: cumple con los criterios del método de isofotas elípticas, confirmando
la presencia de una barra estelar.

**Galaxia: J033223.40-274316.6**
**Filtro: z_band_GOODS**

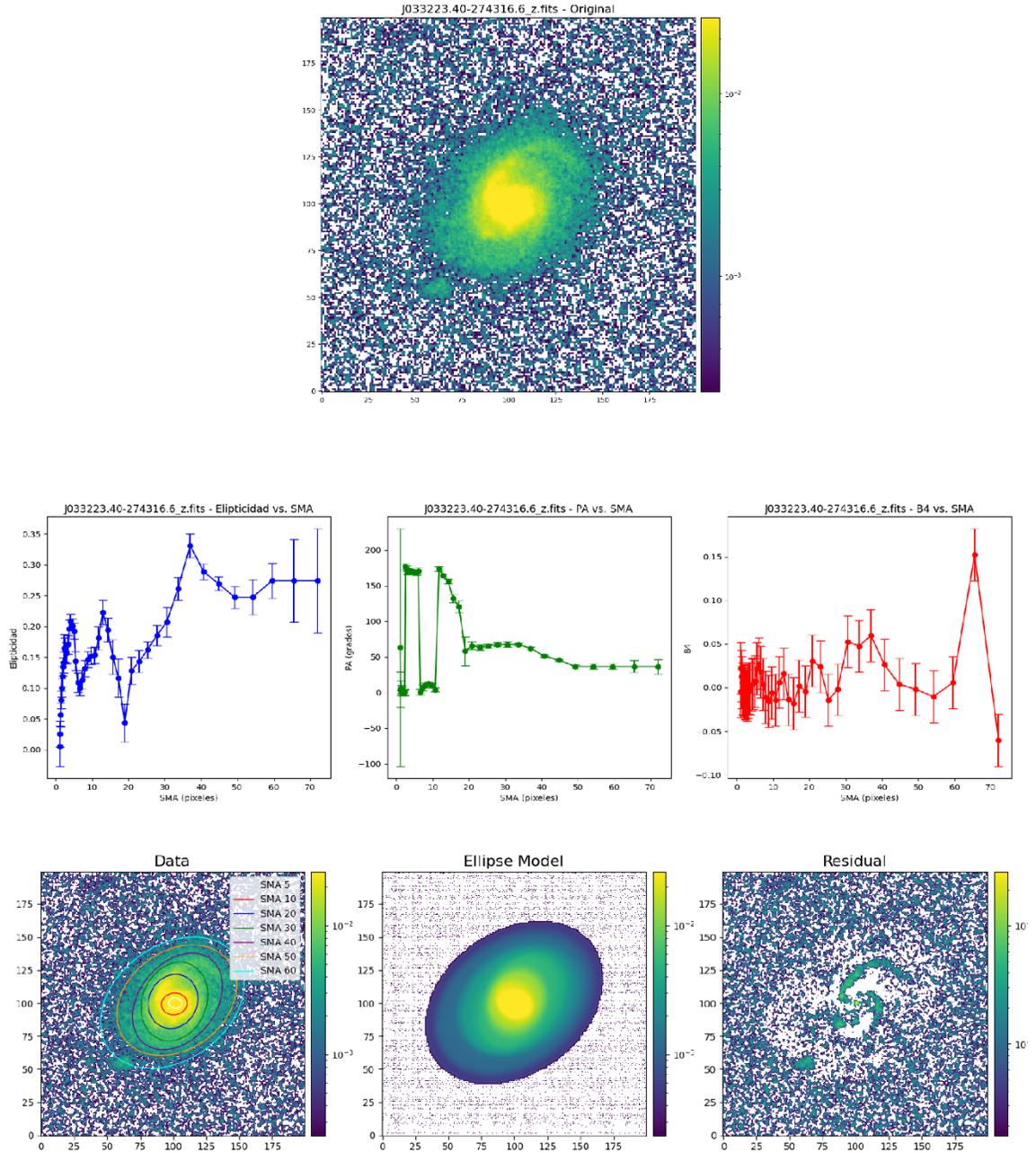

Galaxia barrada: cumple con los criterios del método de isofotas elípticas, confirmando la presencia de una barra estelar.

**Galaxia: J033233.08-275123.9**
**Filtro: z_band_GOODS**

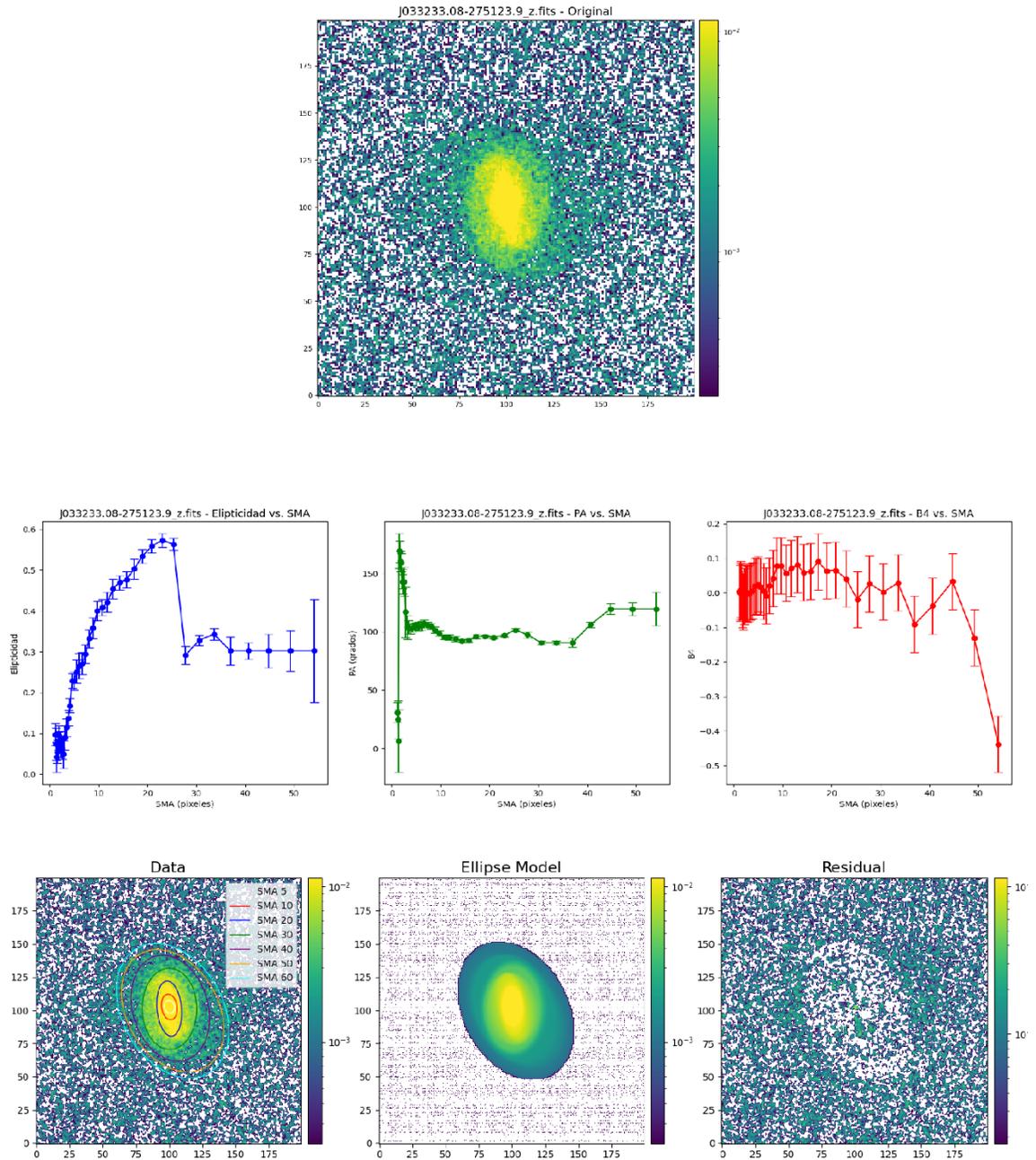

Galaxia barrada: cumple con los criterios del método de isofotas elípticas, confirmando la presencia de una barra estelar.

**Galaxia: J033224.06-274911.4**
**Filtro: z_band_GOODS**

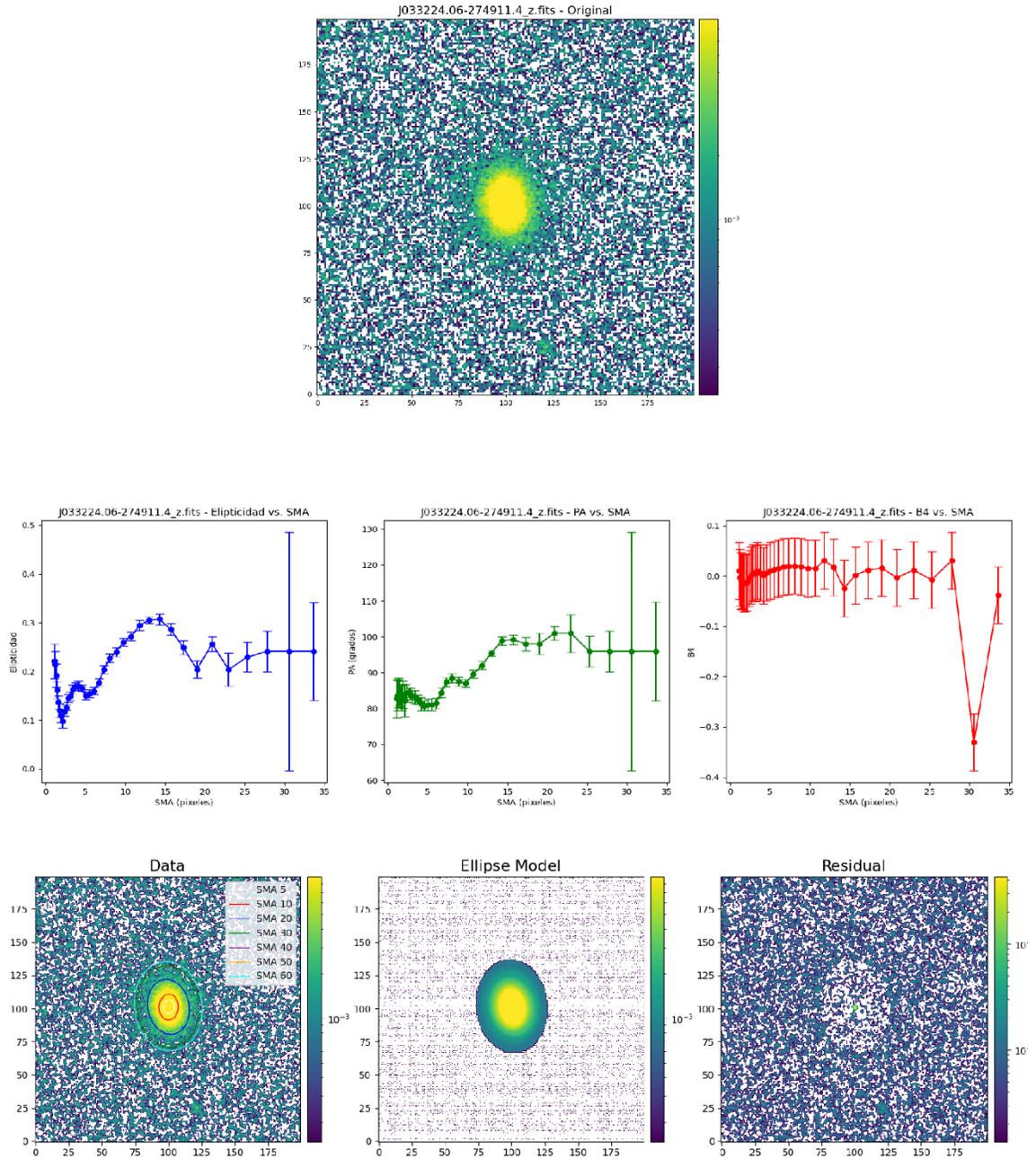

Galaxia no barrada: no cumple con los criterios del método de isofotas elípticas,

descartando la presencia de una barra estelar.

**Galaxia: J033230.03-274347.3**
**Filtro: z_band_GOODS**

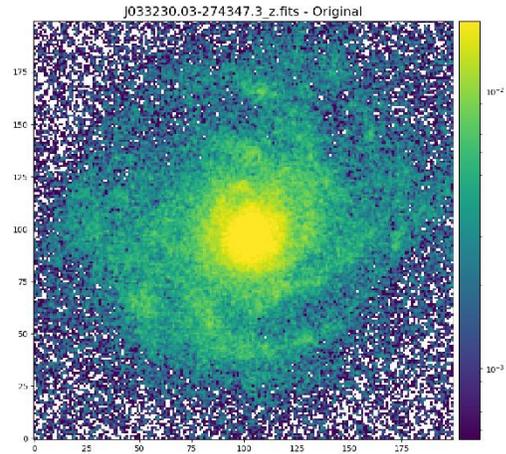

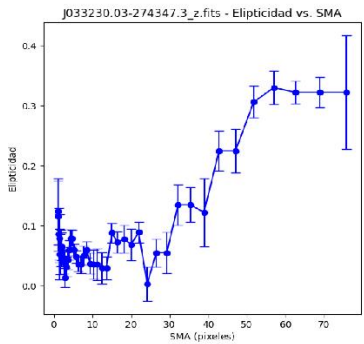
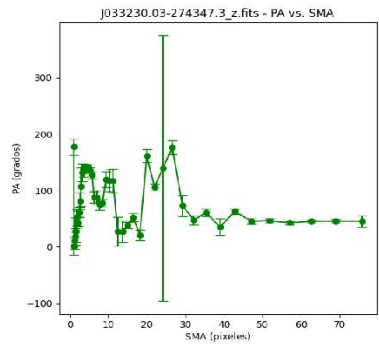
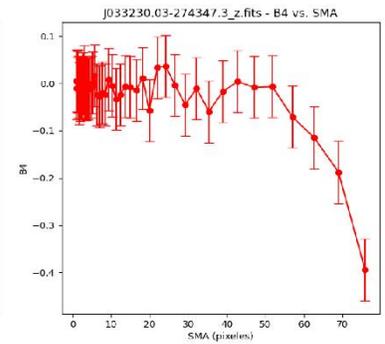

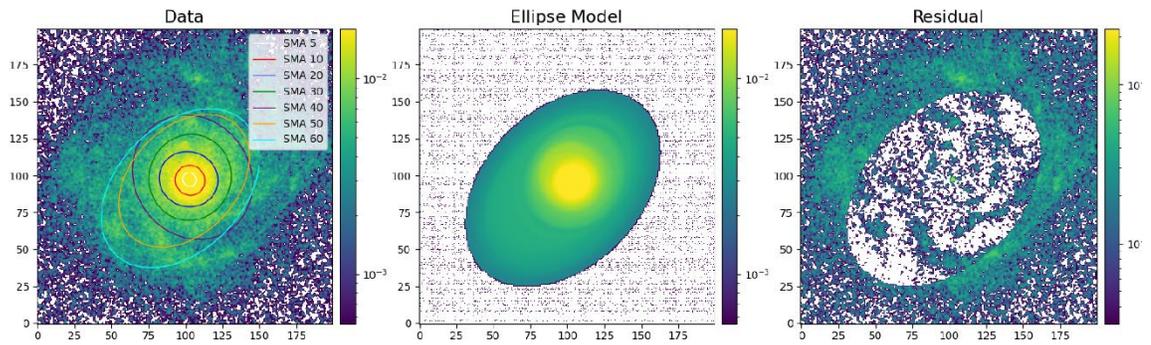

Galaxia no barrada: no cumple con los criterios del método de isofotas elípticas, descartando la presencia de una barra estelar.

**Galaxia: J033219.68-275023.6**
**Filtro: z_band_GOODS**

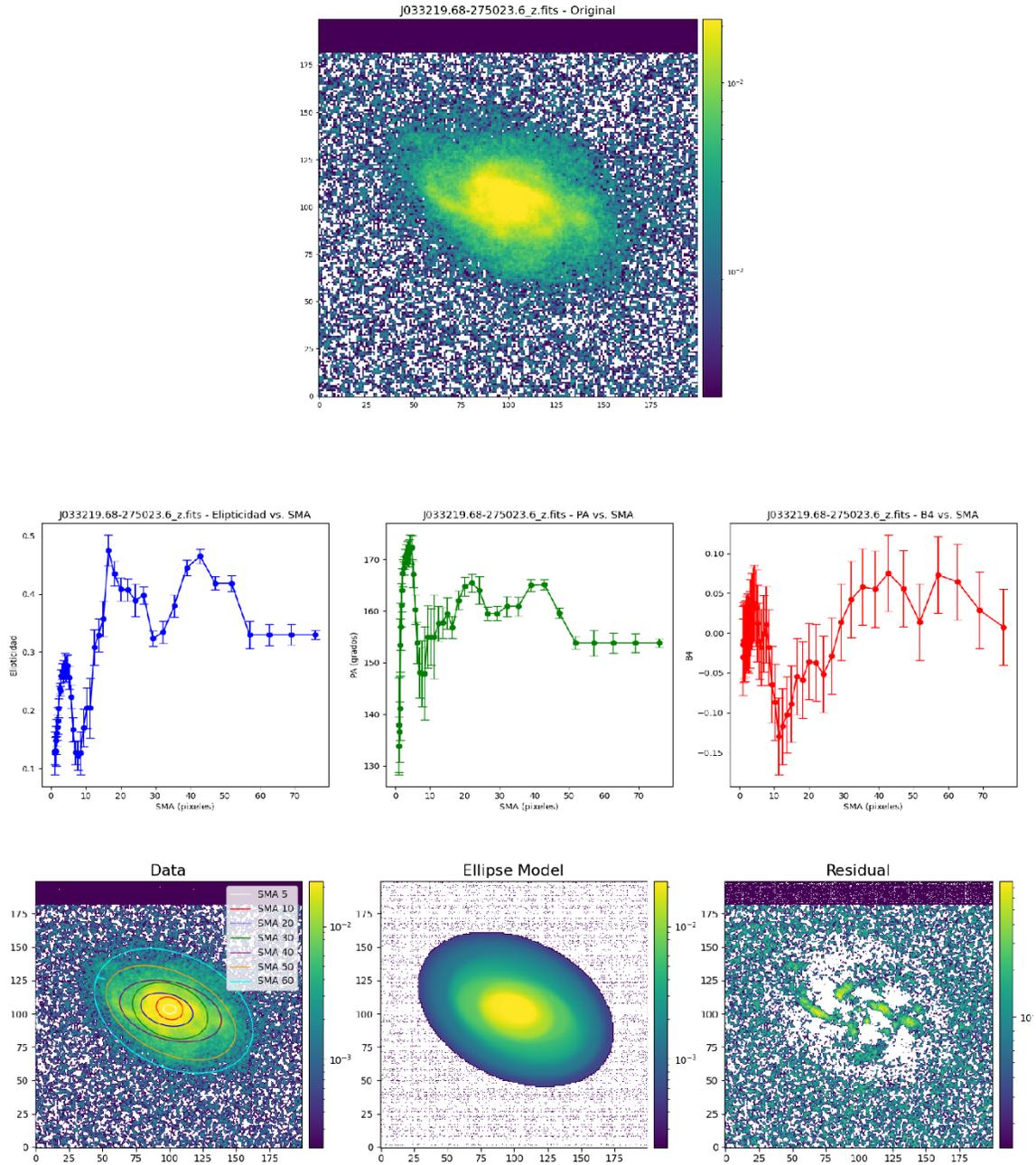

Galaxia no barrada: no cumple con los criterios del método de isofotas elípticas, descartando la presencia de una barra estelar.

# C.4. Galaxias Lenticulares a $z = 0.7$

**Galaxia: J033205.97-274601.4**
**Filtro: z_band_GOODS**

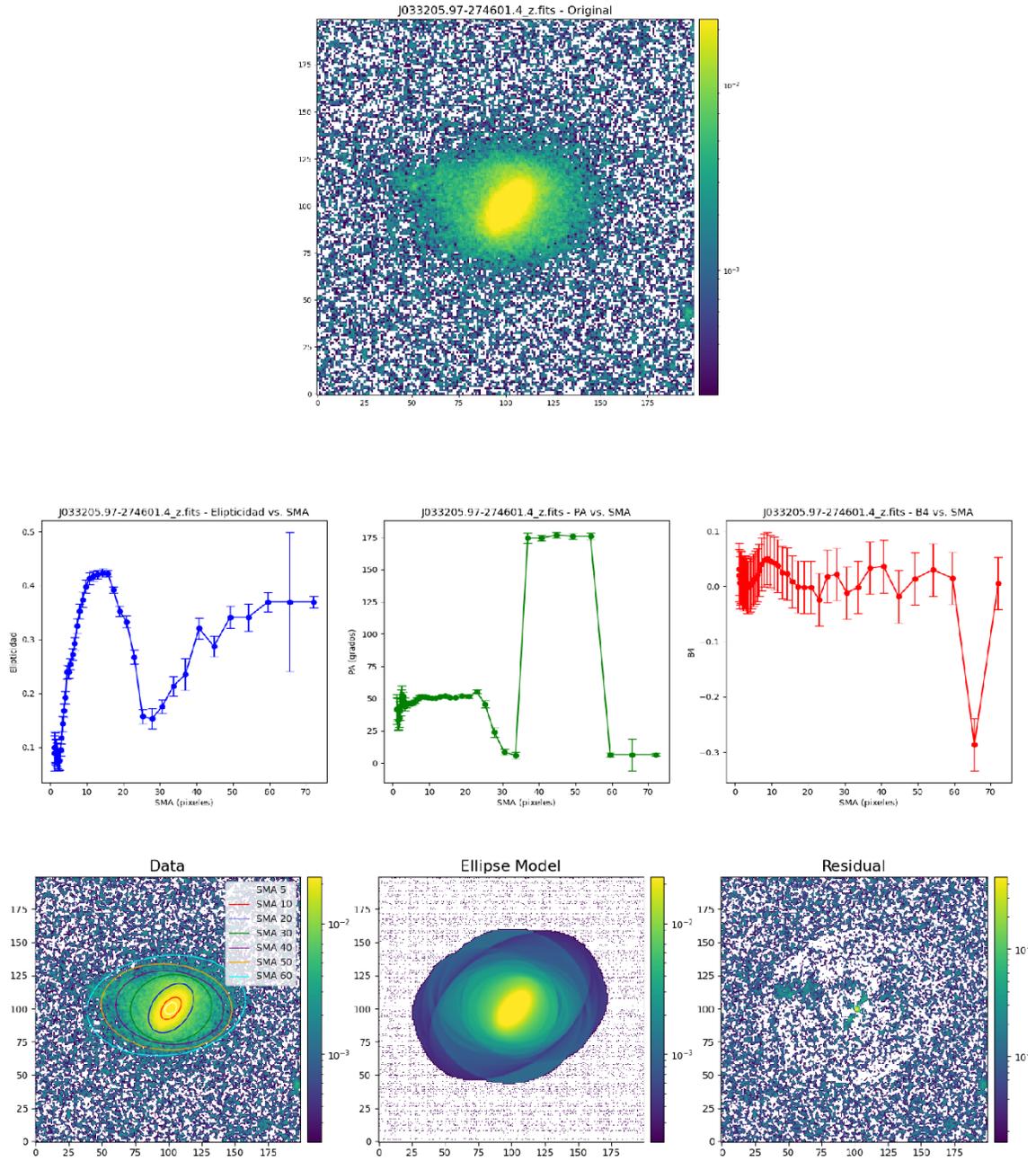

Galaxia barrada: cumple con los criterios del método de isofotas elípticas, confirmando la presencia de una barra estelar.

**Galaxia: J033212.31-274527.4**
**Filtro: z_band_GOODS**

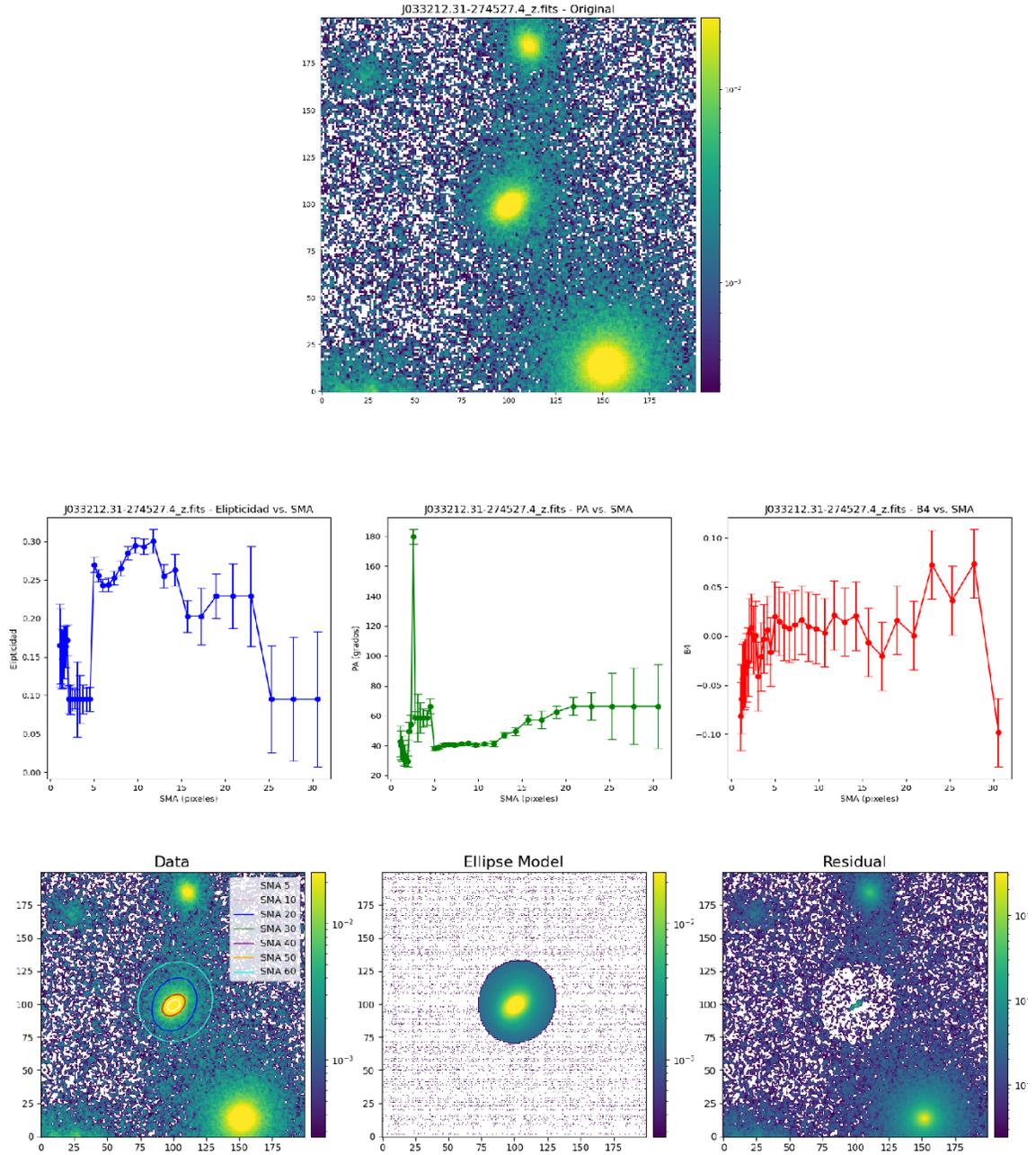

Galaxia barrada: cumple con los criterios del método de isofotas elípticas, confirmando la presencia de una barra estelar.

**Galaxia: J033230.07-275140.6**
**Filtro: z_band_GOODS**

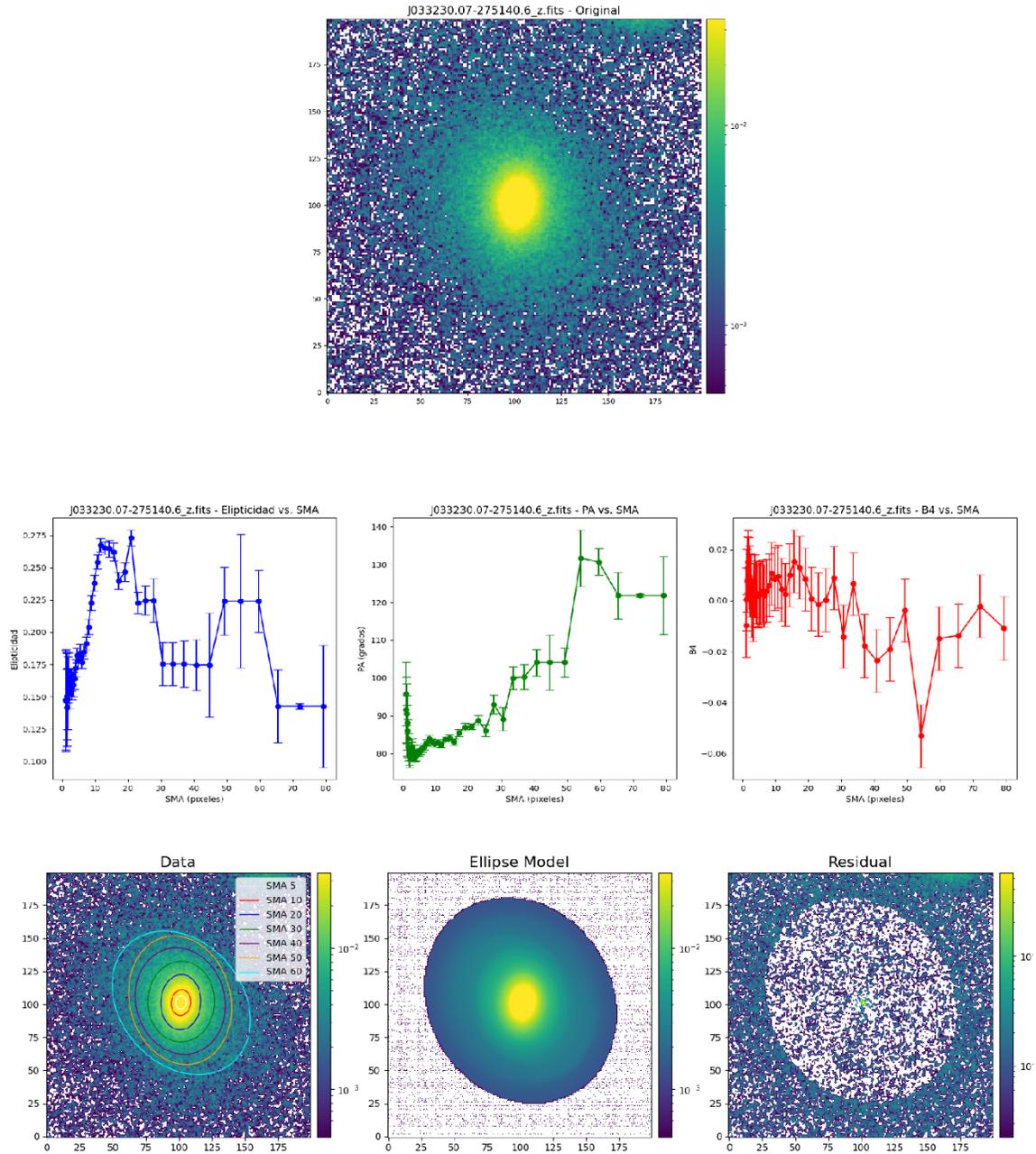

Galaxia barrada: cumple con los criterios del método de isofotas elípticas, confirmando la presencia de una barra estelar.

**Galaxia: J033235.02-275405.2**
**Filtro: z_band_GOODS**

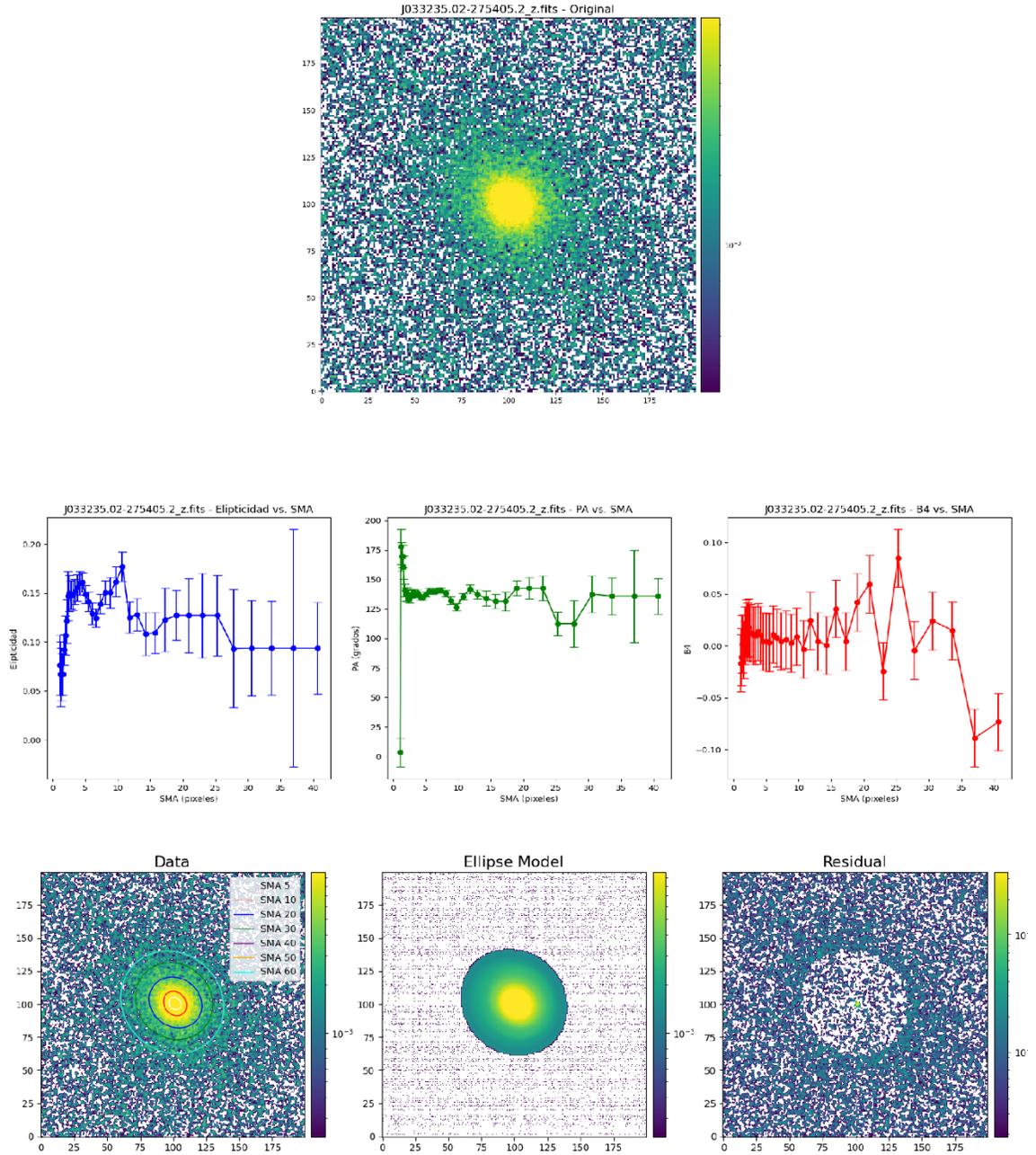

Galaxia no barrada: no cumple con los criterios del método de isofotas elípticas, descartando la presencia de una barra estelar.

**Galaxia: J033238.27-275354.4**
**Filtro: z_band_GOODS**

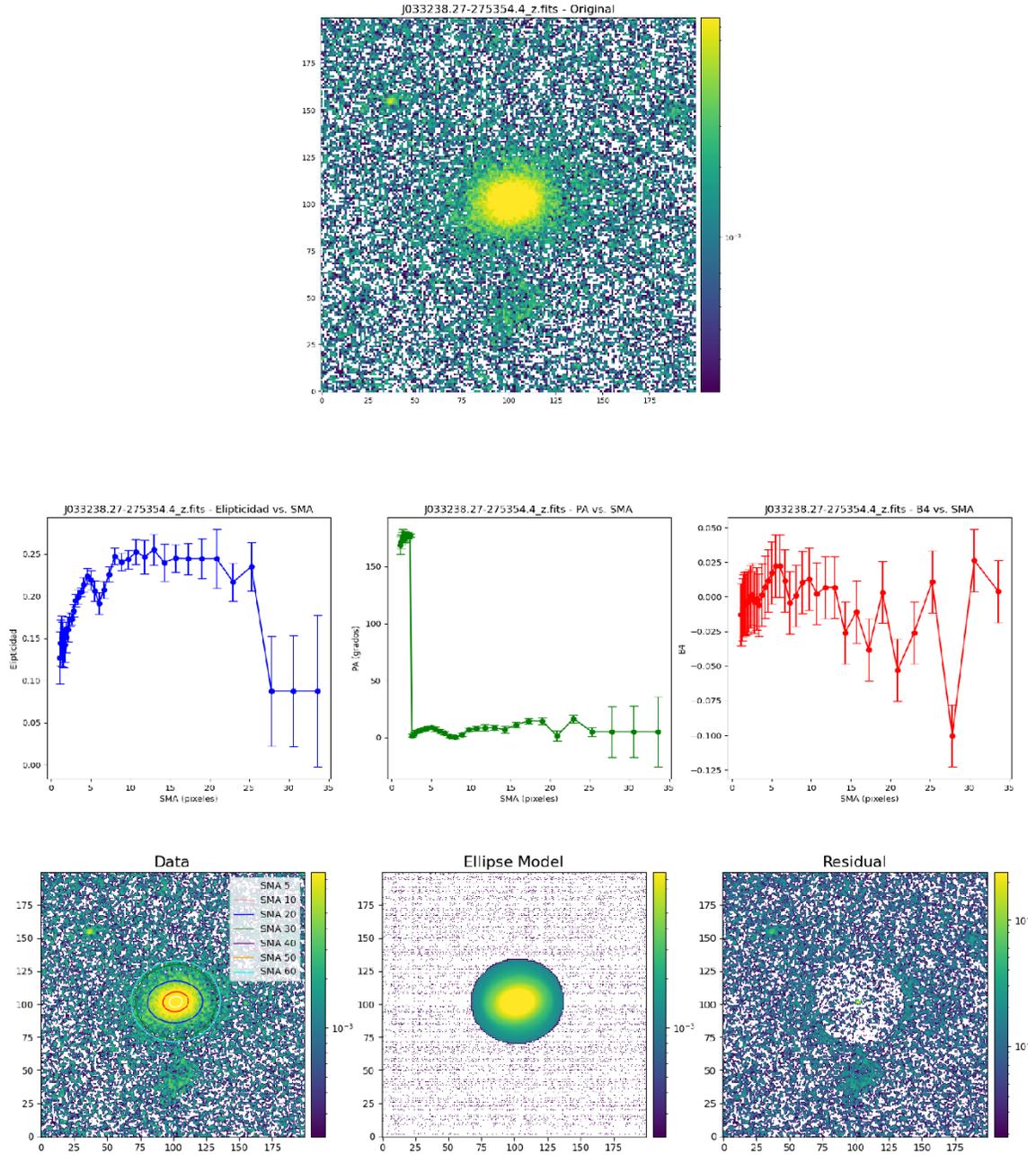

Galaxia no barrada: no cumple con los criterios del método de isofotas elípticas, descartando la presencia de una barra estelar.

**Galaxia: J033242.39-274153.2**
**Filtro: z_band_GOODS**

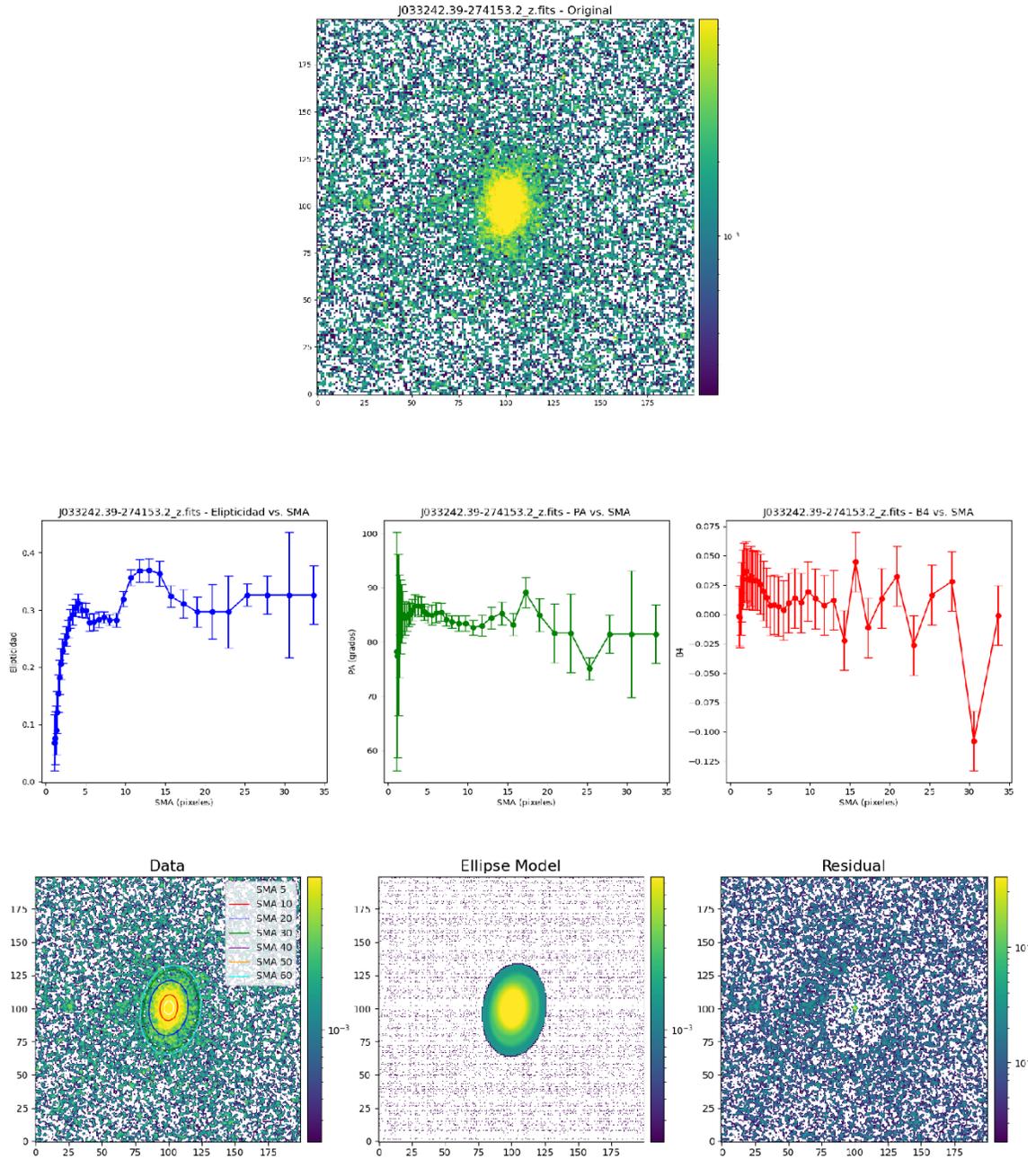

Galaxia no barrada: no cumple con los criterios del método de isofotas elípticas, descartando la presencia de una barra estelar.

# D. Resultados del Método de

# Fourier

# D.1. Galaxias Espirales a $z = 0.027$

**Galaxia: J143411.25+003656.6**

**Filtro: r_band_SDSS**

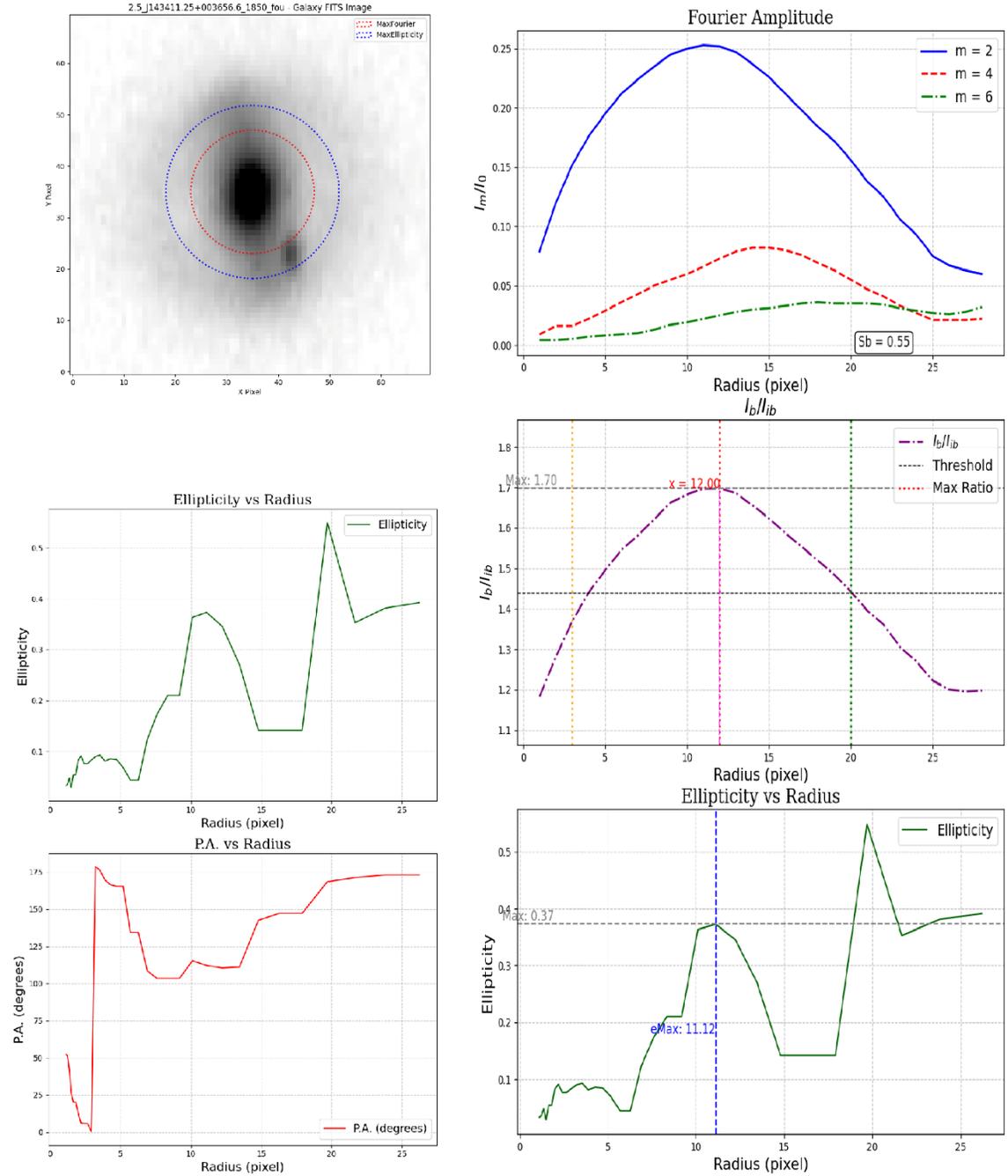

Galaxia barrada: cumple con los criterios del análisis de Fourier, revelando una estructura barrada significativa.

**Galaxia: J120127.92-004306.1**

**Filtro: r_band_SDSS**

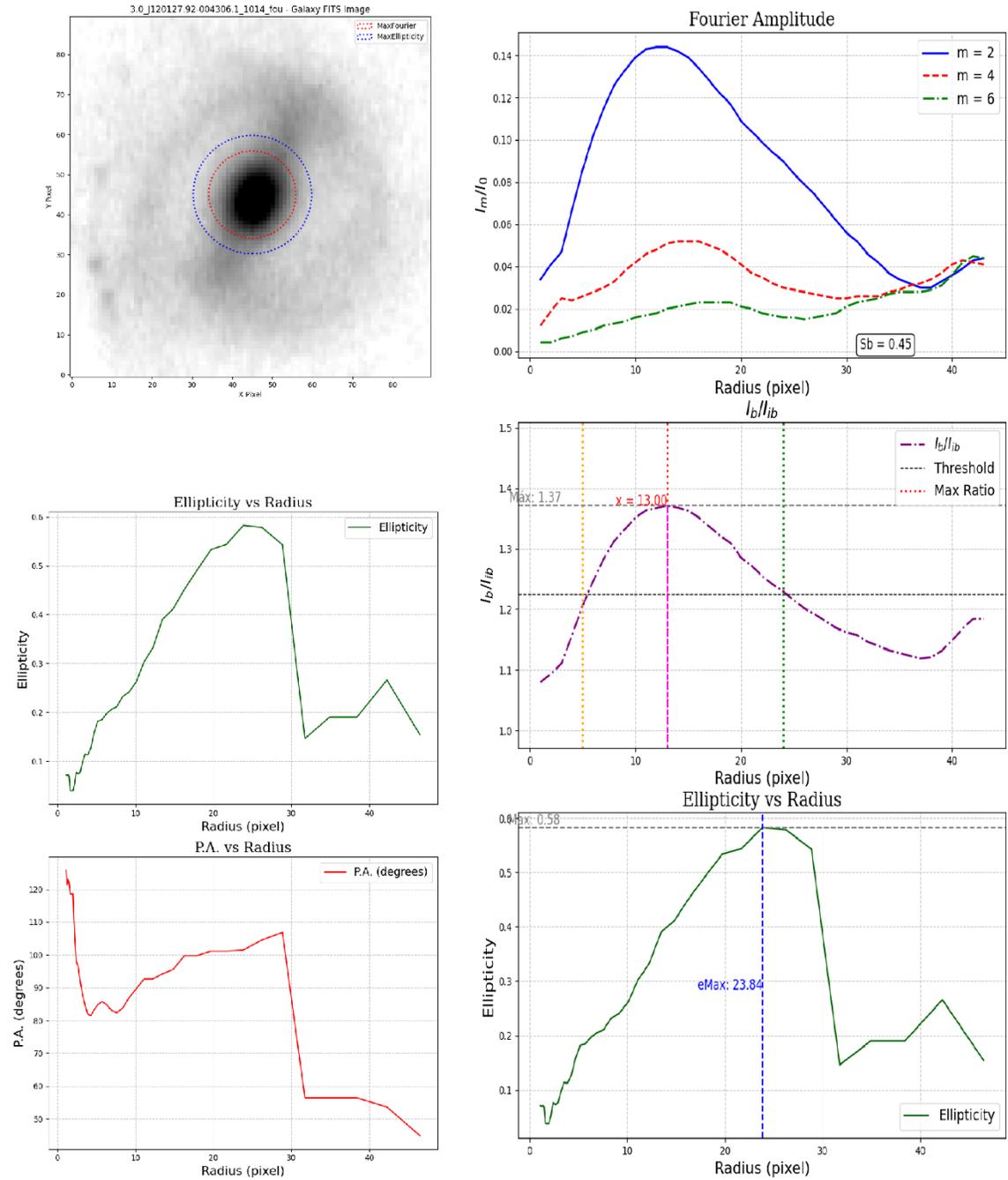

Galaxia barrada: cumple con los criterios del análisis de Fourier, revelando una estructura barrada significativa.

**Galaxia: J135102.22-000915.1**

**Filtro: r_band_SDSS**

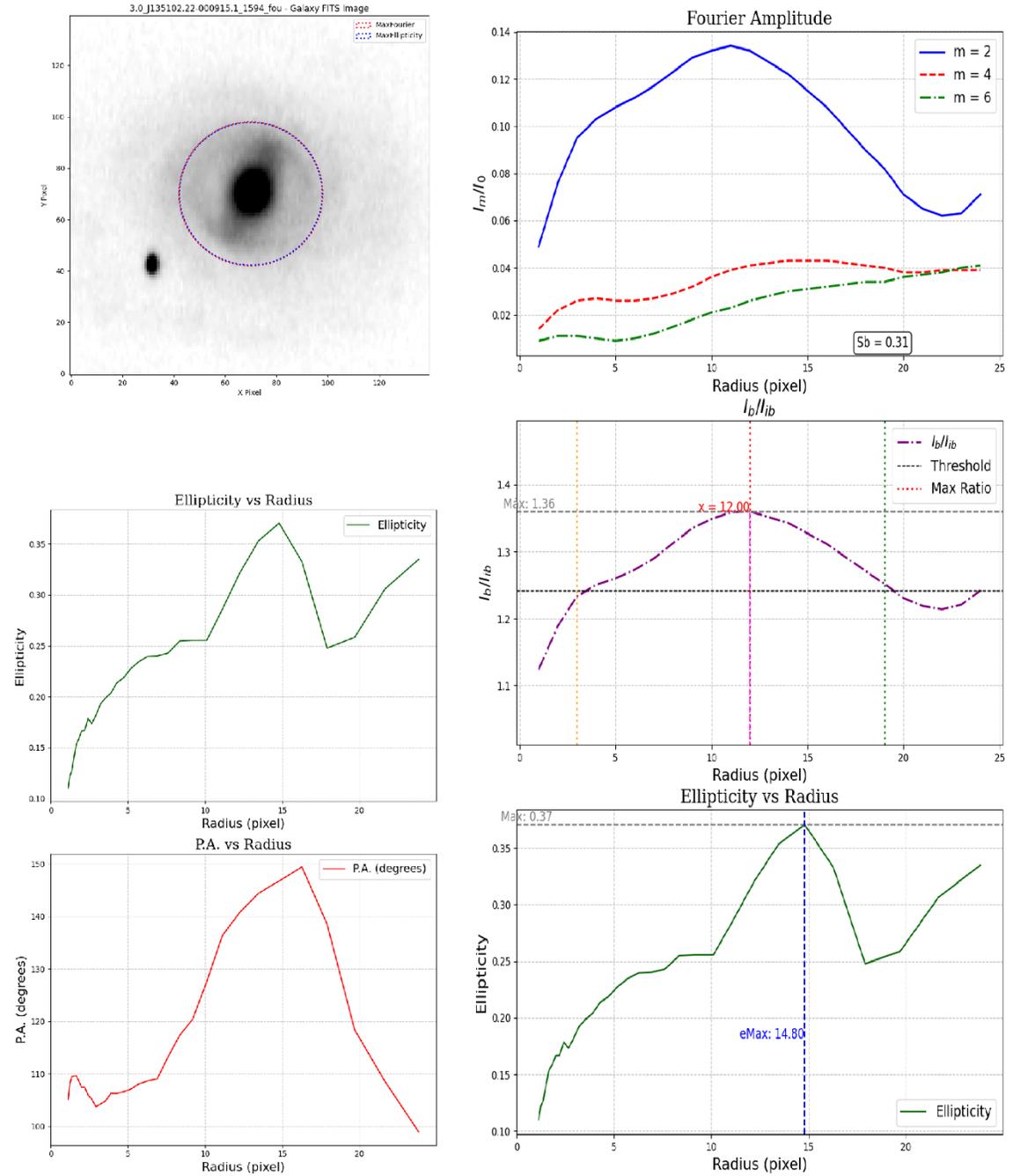

Galaxia barrada: cumple con los criterios del análisis de Fourier, revelando una estructura barrada significativa.

**Galaxia: J140320.74-003259.7**

**Filtro: r_band_SDSS**

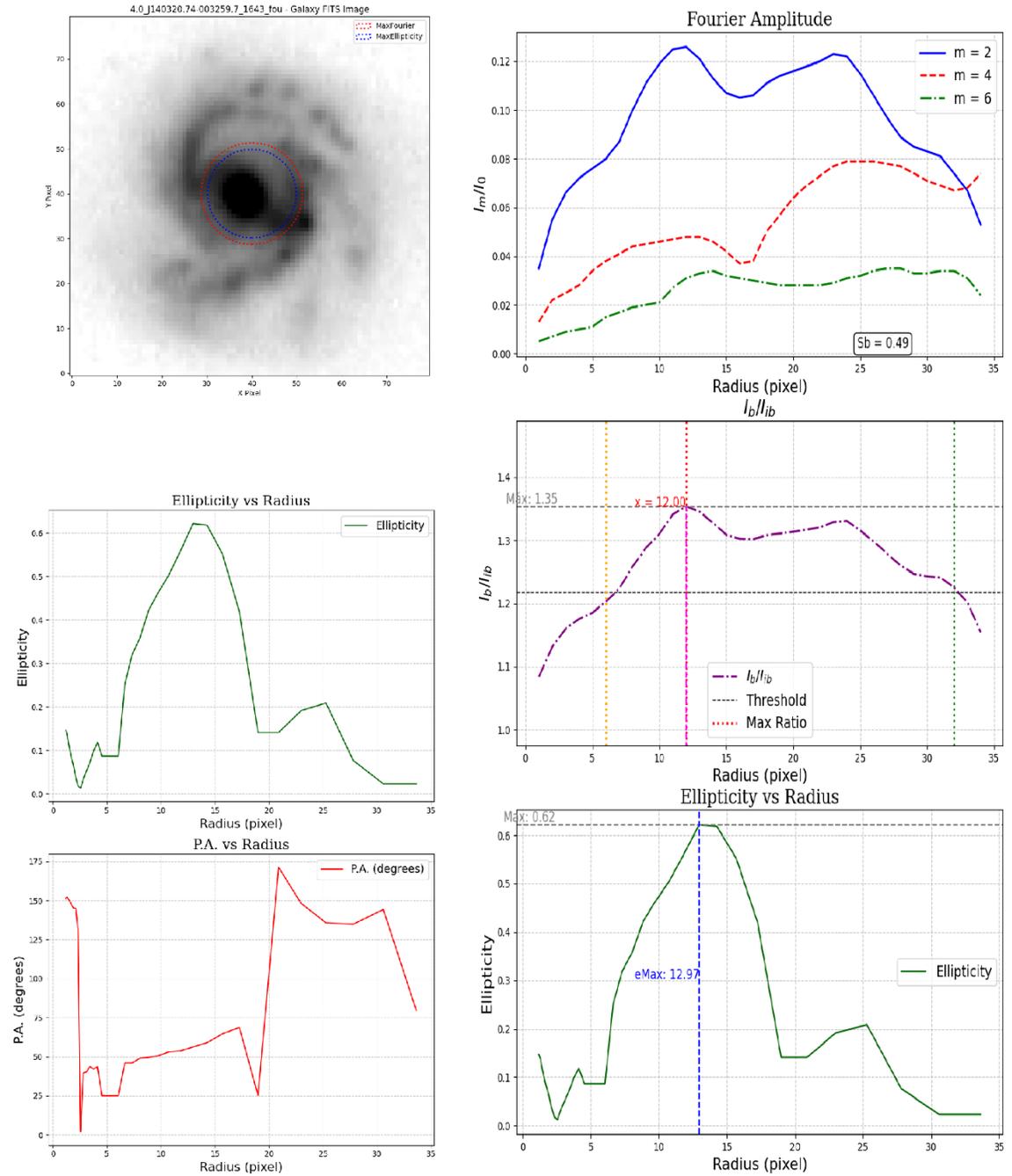

Galaxia barrada: cumple con los criterios del análisis de Fourier, revelando una estructura barrada significativa.

**Galaxia: J113833.27-011104.1**

**Filtro: r_band_SDSS**

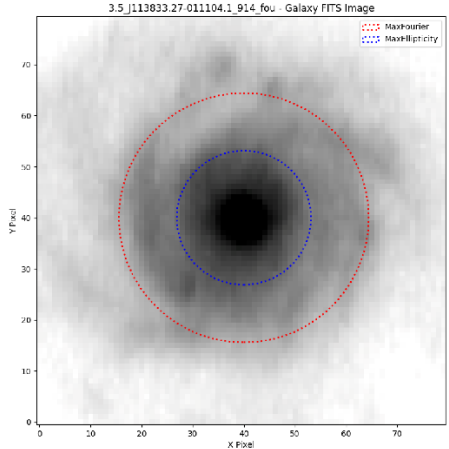

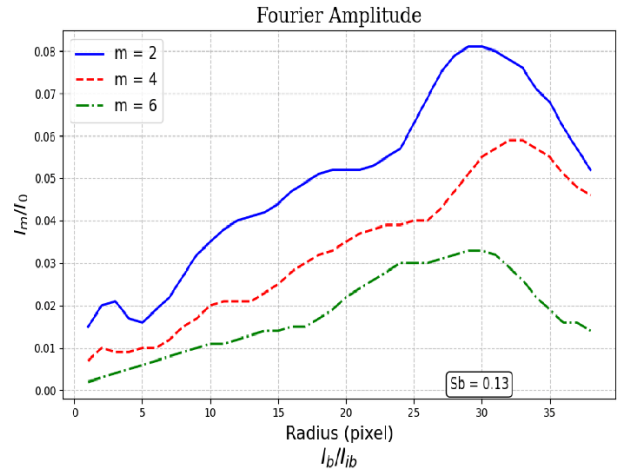

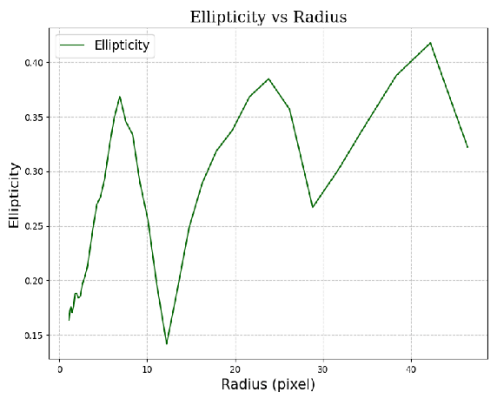

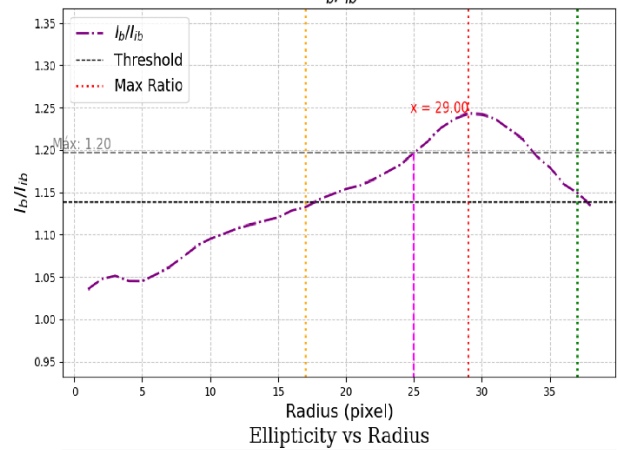

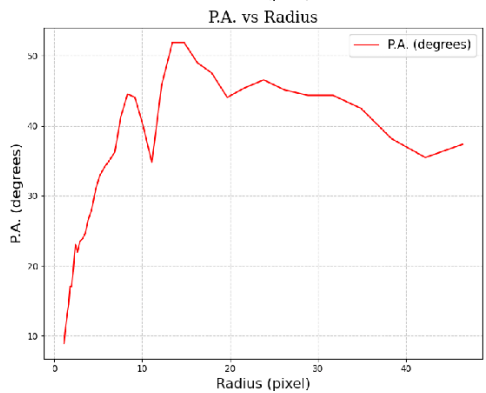

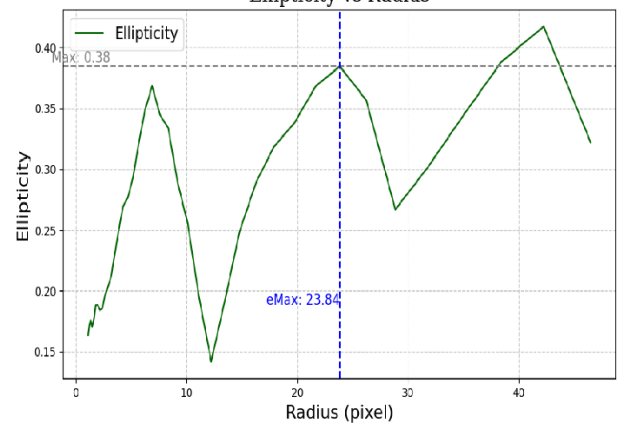

Galaxia no barrada: no cumple con los criterios del análisis de Fourier, descartando la presencia de una barra estelar.

**Galaxia: J141814.91+005327.9**

**Filtro: r_band_SDSS**

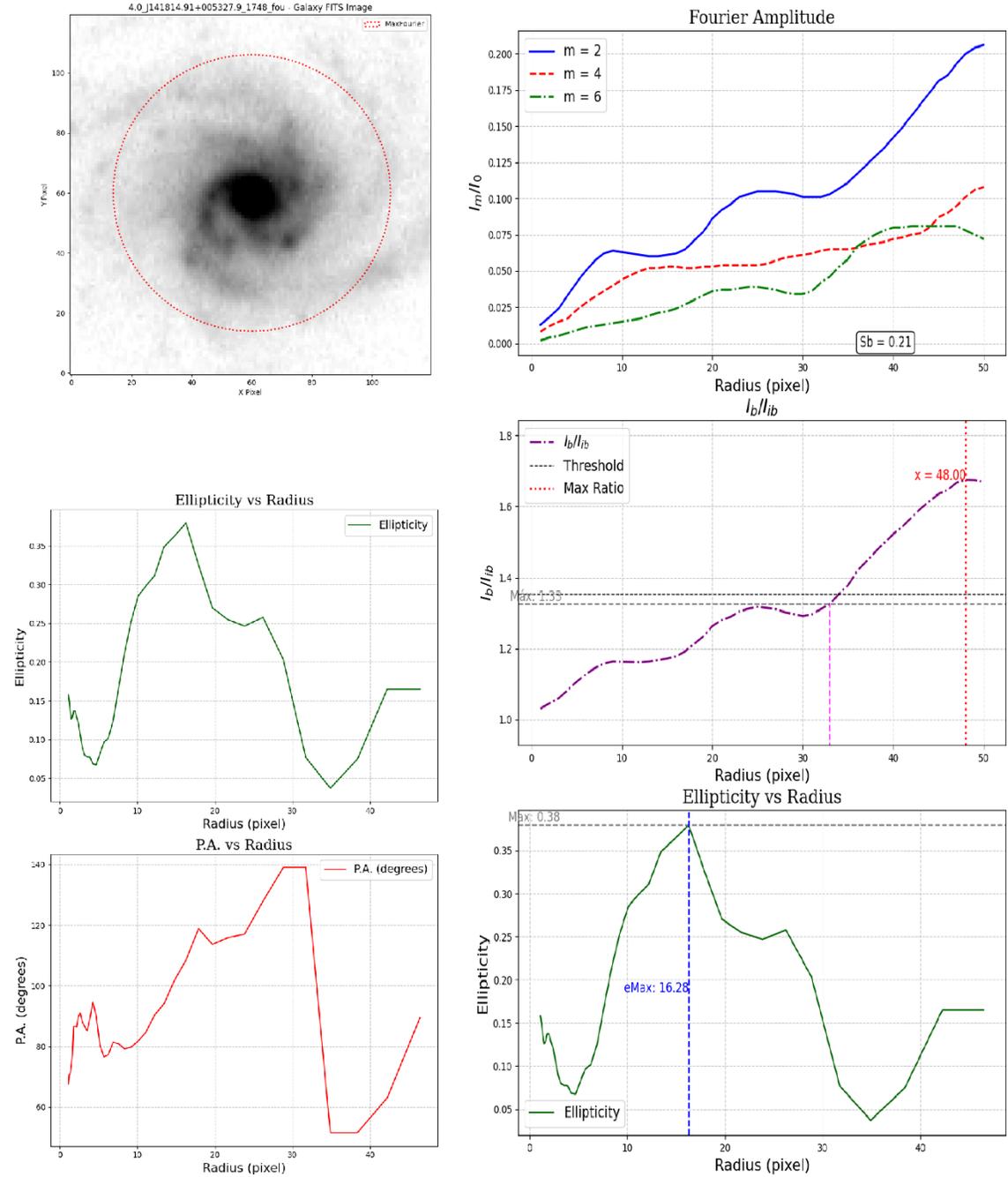

Galaxia no barrada: no cumple con los criterios del análisis de Fourier, descartando la presencia de una barra estelar.

**Galaxia: J111849.55+003709.3**

**Filtro: r_band_SDSS**

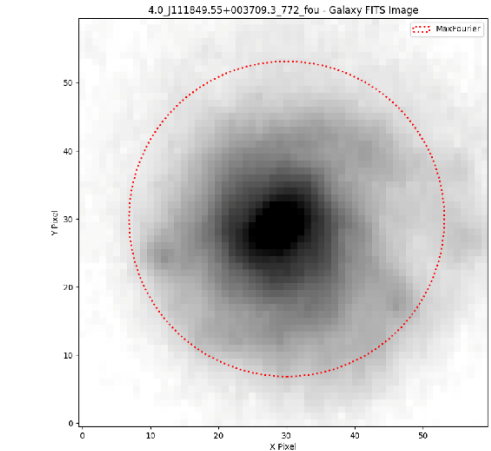

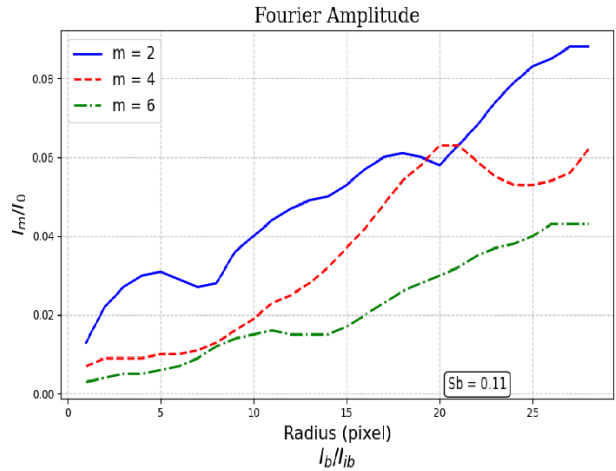

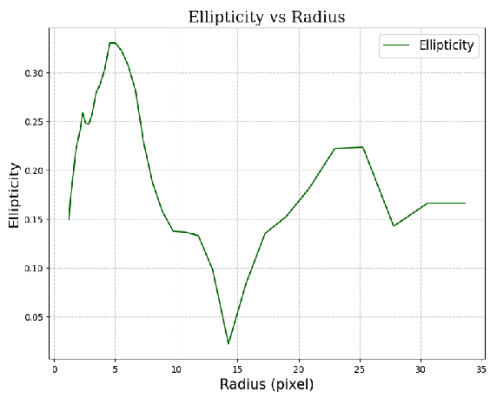

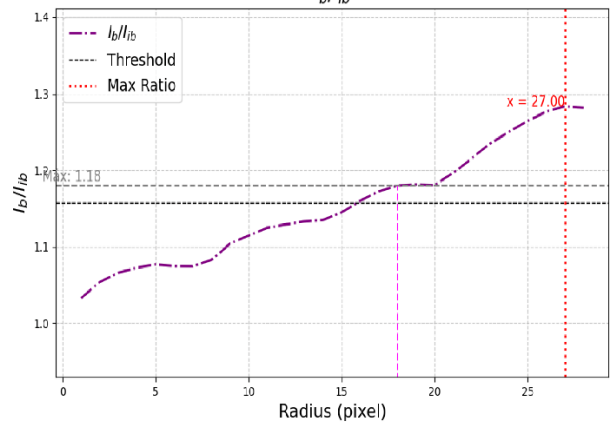

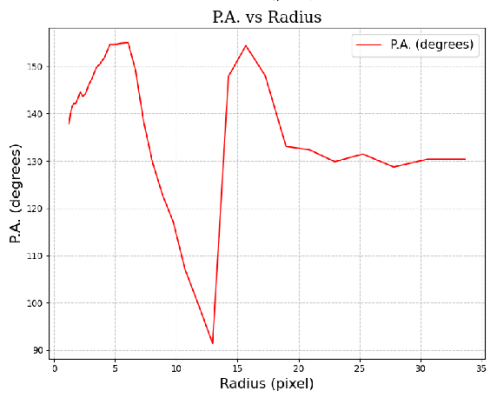

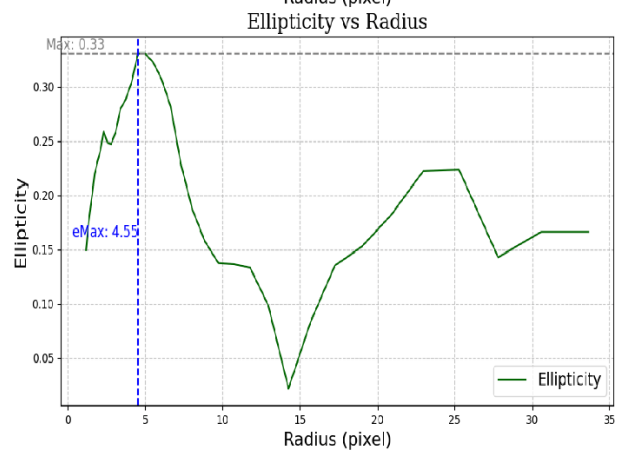

Galaxia no barrada: no cumple con los criterios del análisis de Fourier, descartando la presencia de una barra estelar.

**Galaxia: J144613.34+005157.7**

**Filtro: r_band_SDSS**

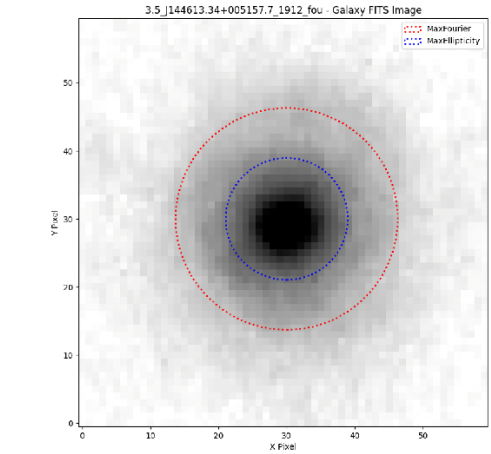

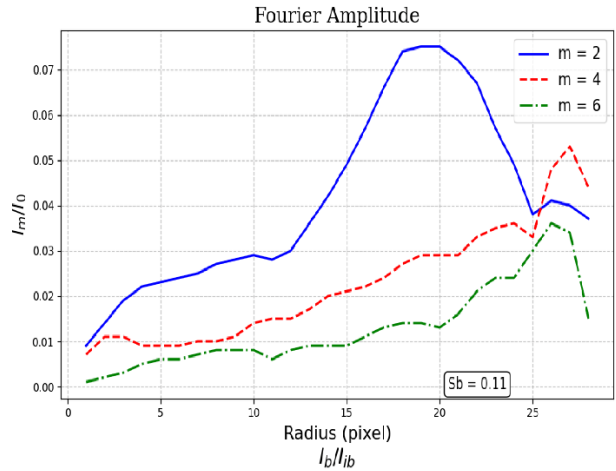

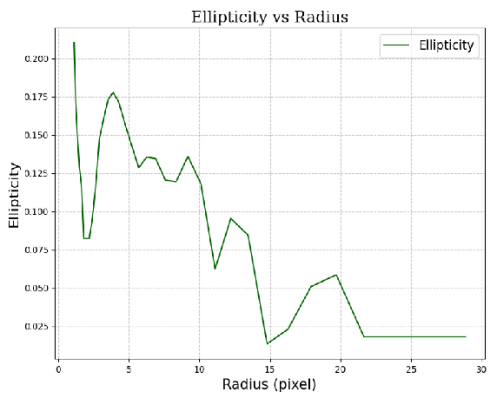

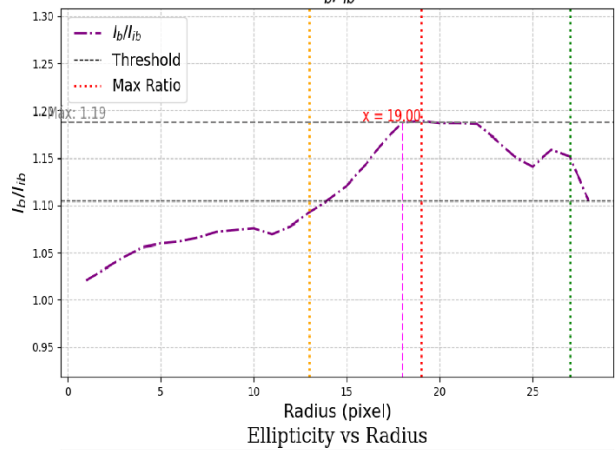

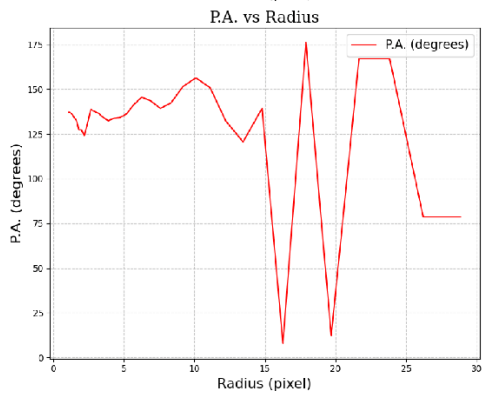

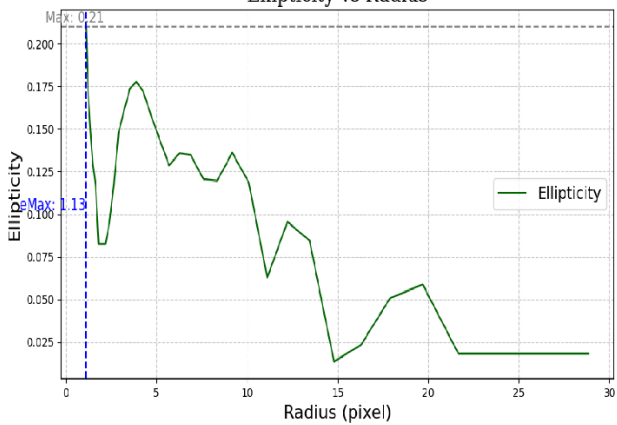

Galaxia no barrada: no cumple con los criterios del análisis de Fourier, descartando la presencia de una barra estelar.

# D.2. Galaxias Lenticulares a $z = 0.027$

**Galaxia: J143124.59+011403.7**

**Filtro: r_band_SDSS**

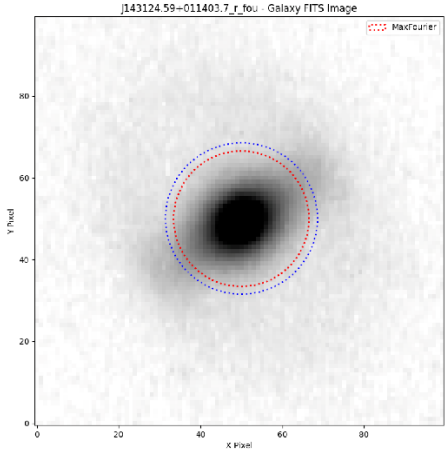

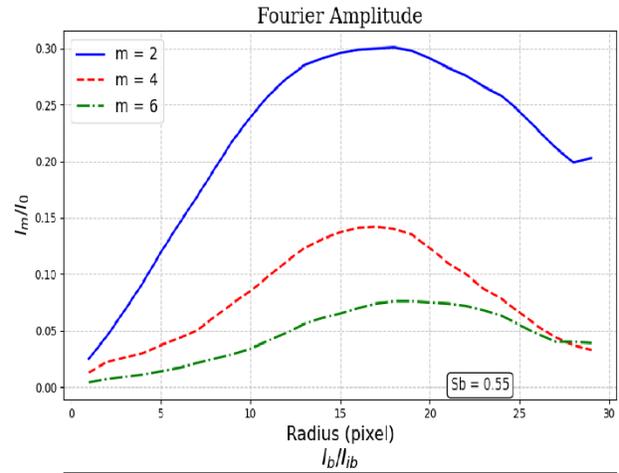

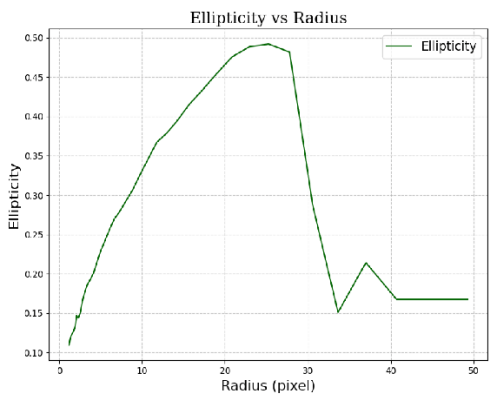

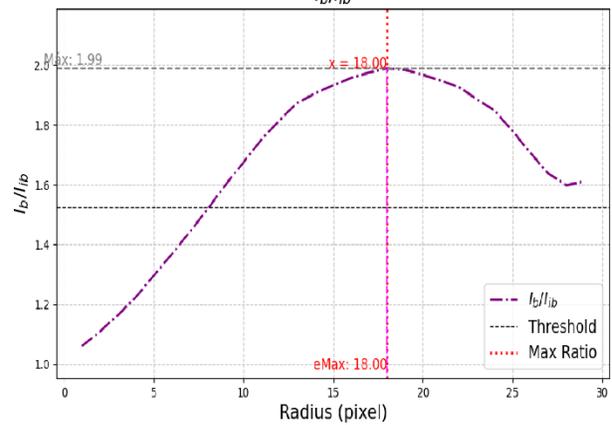

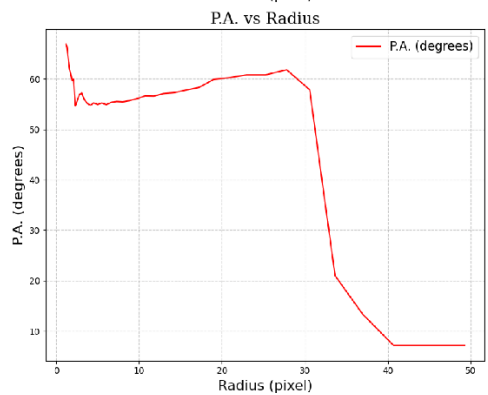

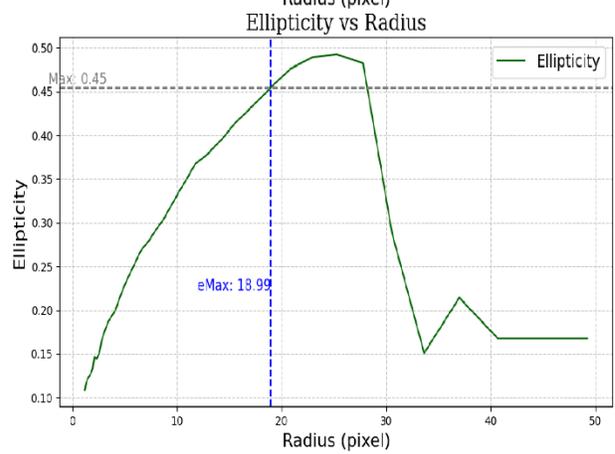

Galaxia barrada: cumple con los criterios del análisis de Fourier, revelando una estructura barrada significativa.

**Galaxia: J143540.07+001217.7**

**Filtro: r_band_SDSS**

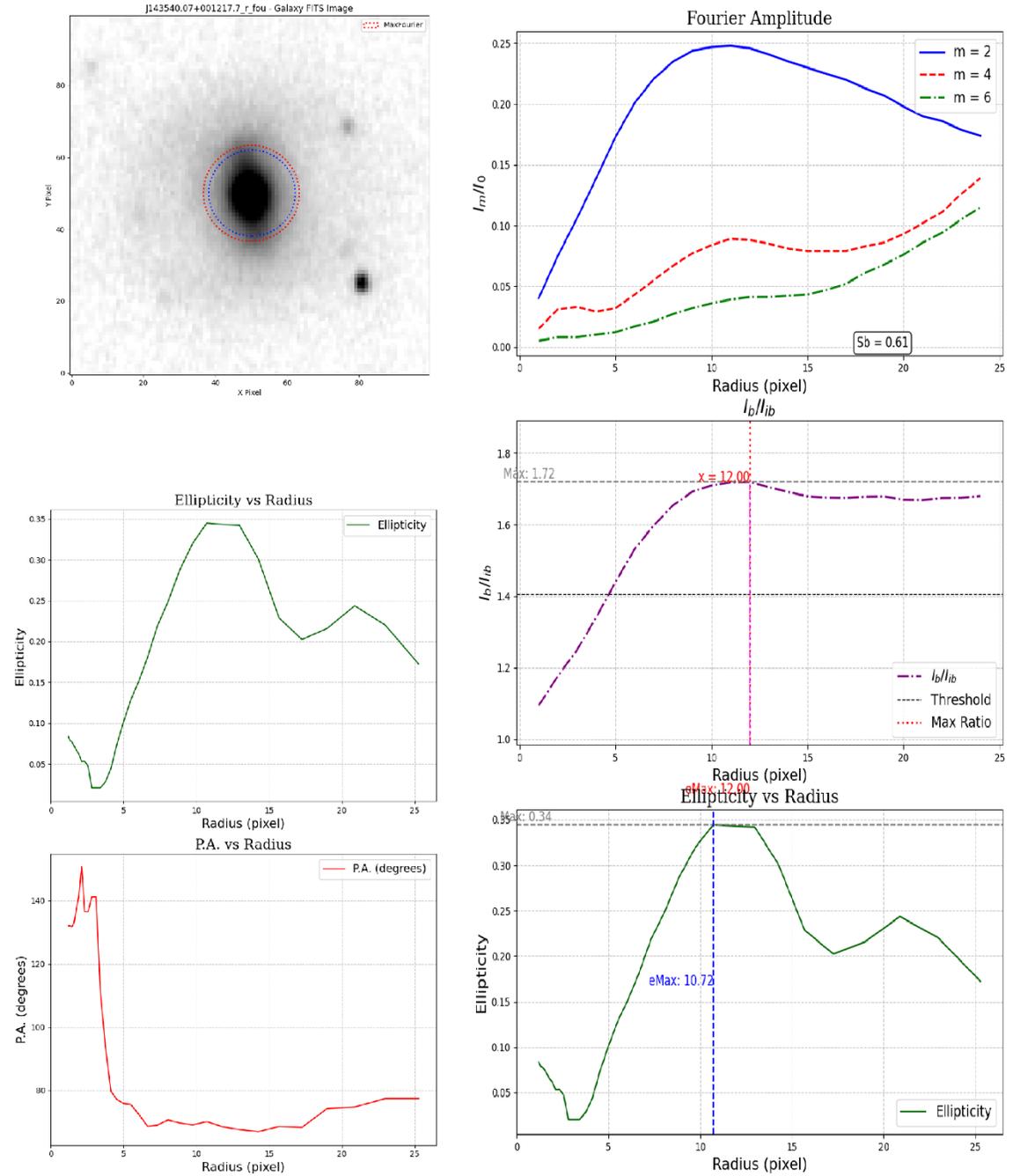

Galaxia barrada: cumple con los criterios del análisis de Fourier, revelando una estructura barrada significativa.

**Galaxia: J113523.27+000525.9**

**Filtro: r_band_SDSS**

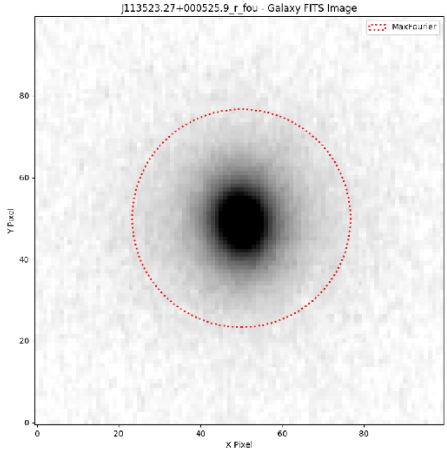

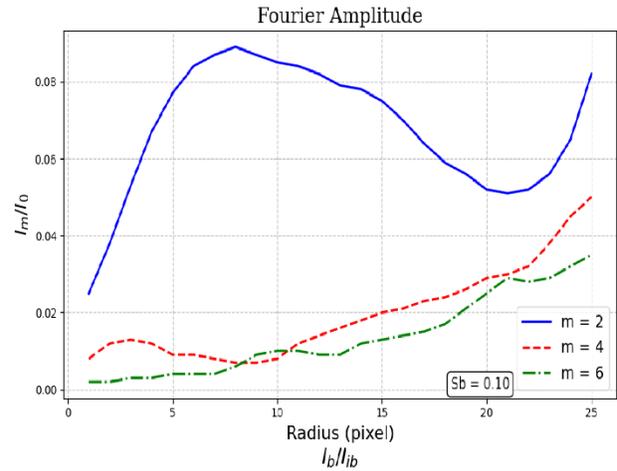

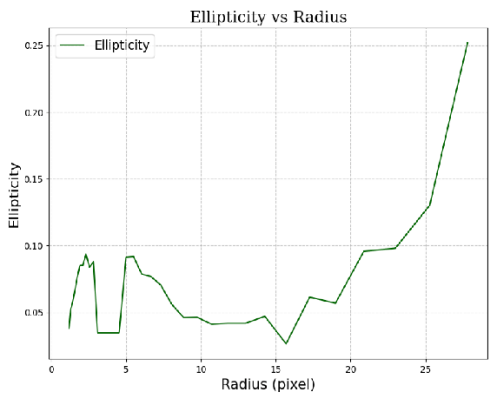

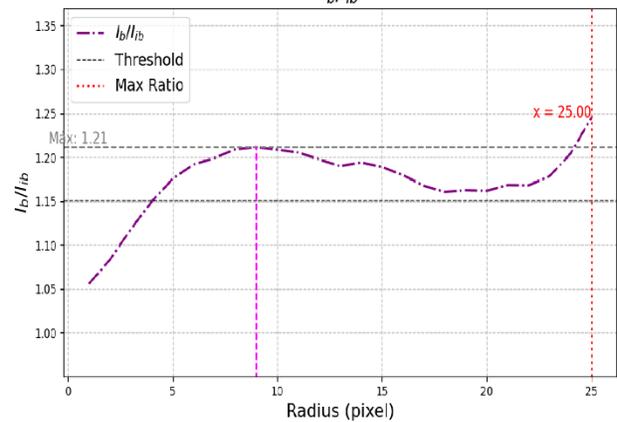

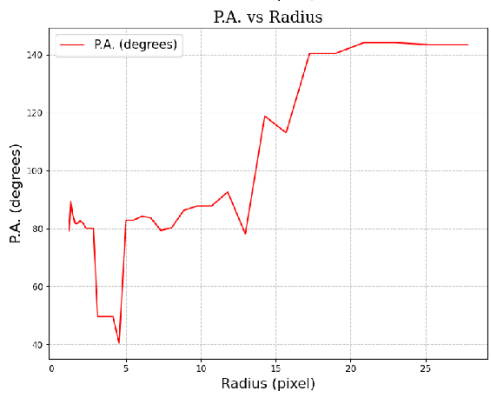

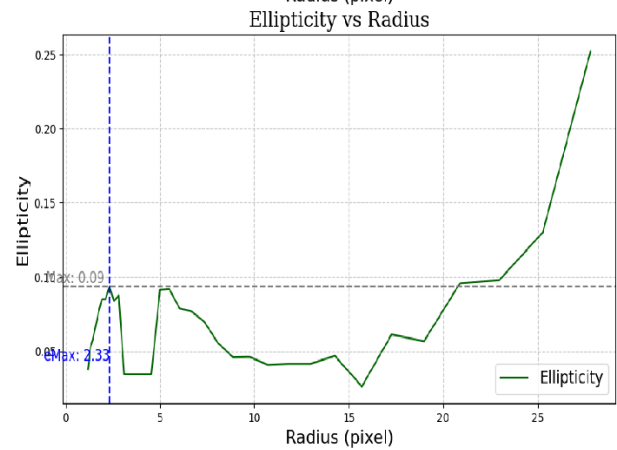

Galaxia no barrada: no cumple con los criterios del análisis de Fourier, descartando la presencia de una barra estelar.

**Galaxia: J103534.47-002116.2**

**Filtro: r_band_SDSS**

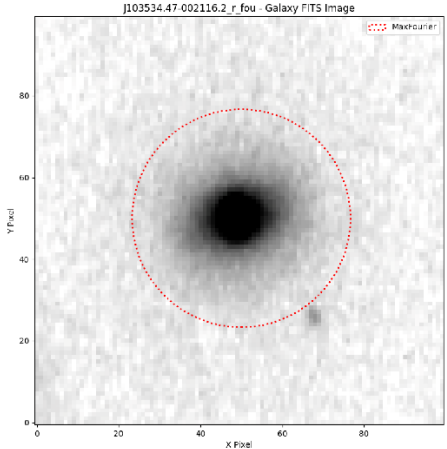

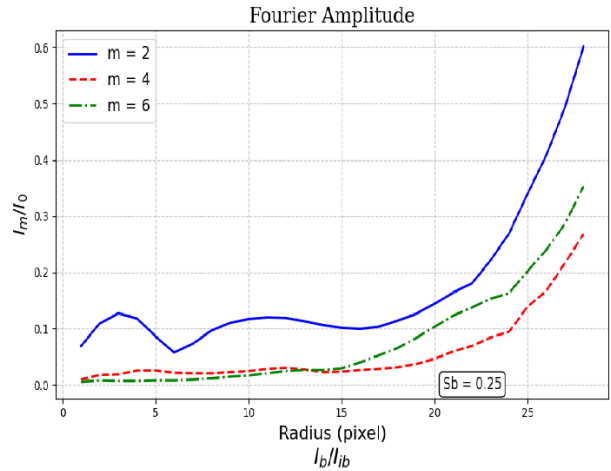

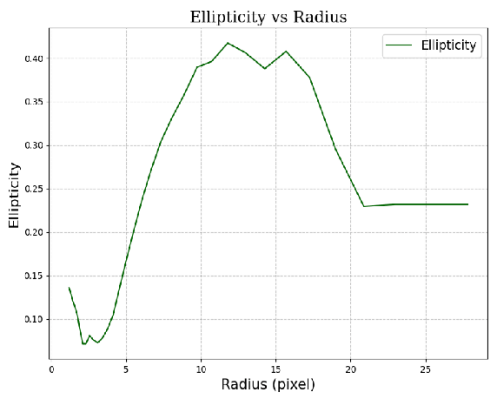

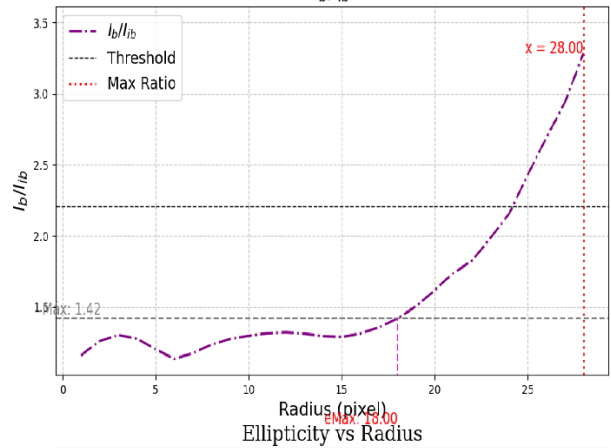

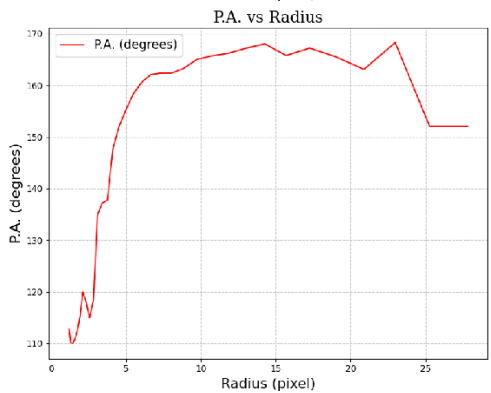

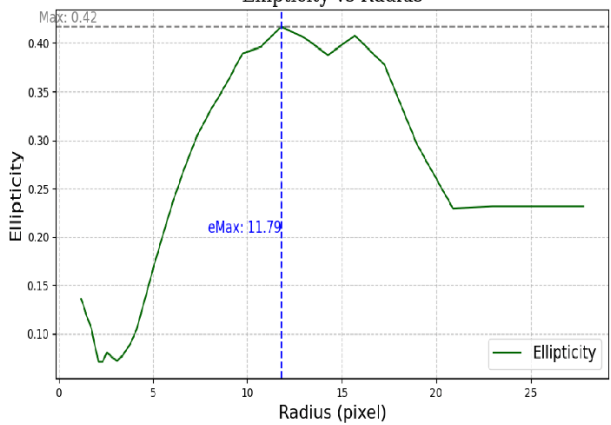

Galaxia no barrada: no cumple con los criterios del análisis de Fourier, descartando la presencia de una barra estelar.

**Galaxia: J113420.50+001856.4**

**Filtro: r_band_SDSS**

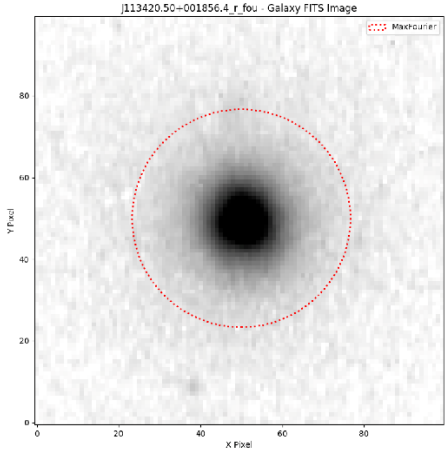

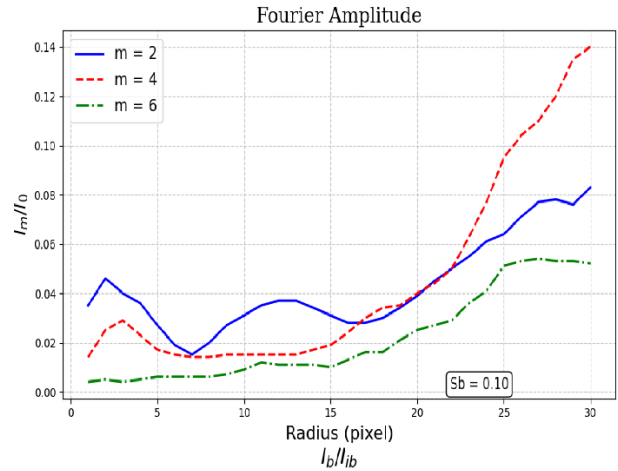

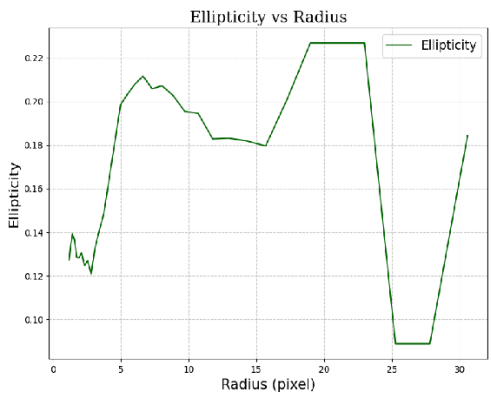

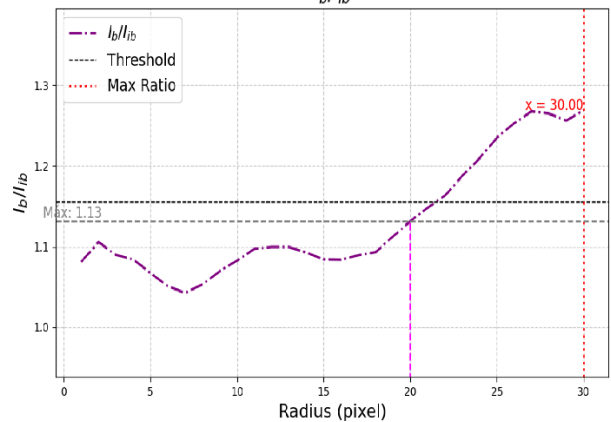

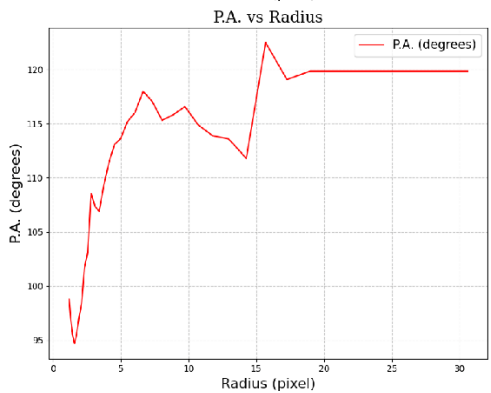

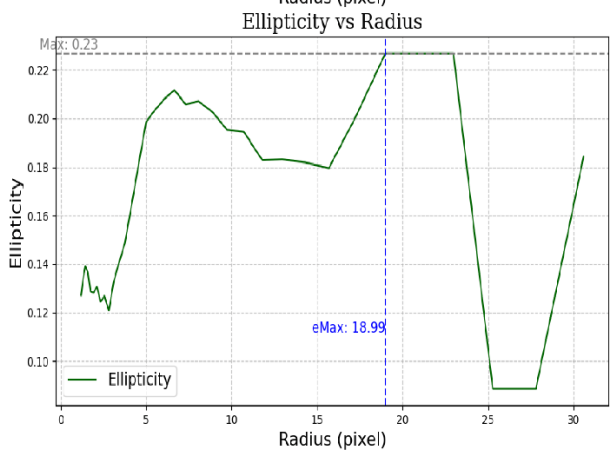

Galaxia no barrada: no cumple con los criterios del análisis de Fourier, descartando la presencia de una barra estelar.

# D.3. Galaxias Espirales a $z = 0.7$

**Galaxia: J033221.99-274655.9**

**Filtro: z_band_GOODS**

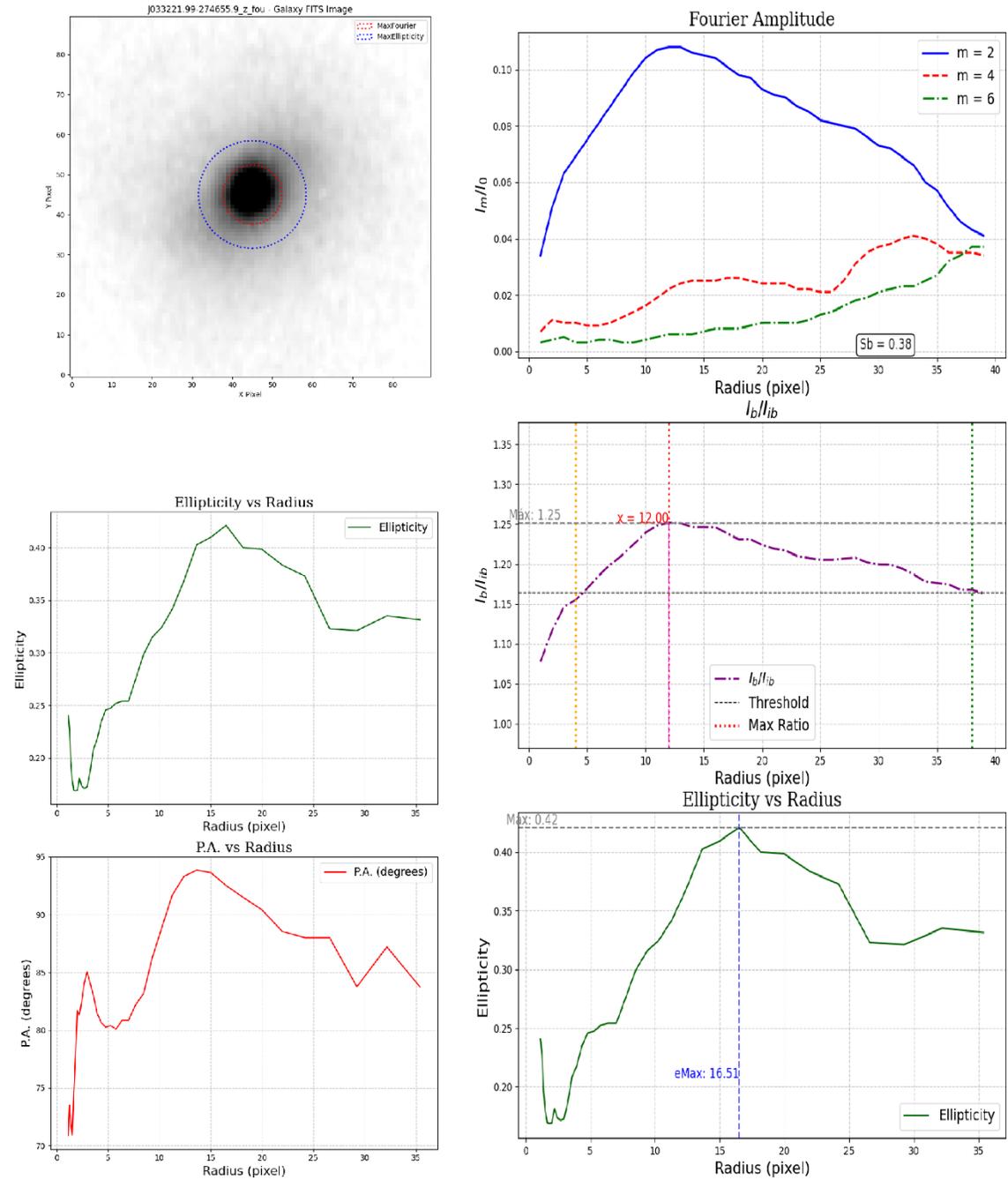

Galaxia barrada: cumple con los criterios del análisis de Fourier, revelando una estructura barrada significativa.

**Galaxia: J033223.40-274316.6**

**Filtro: z_band_GOODS**

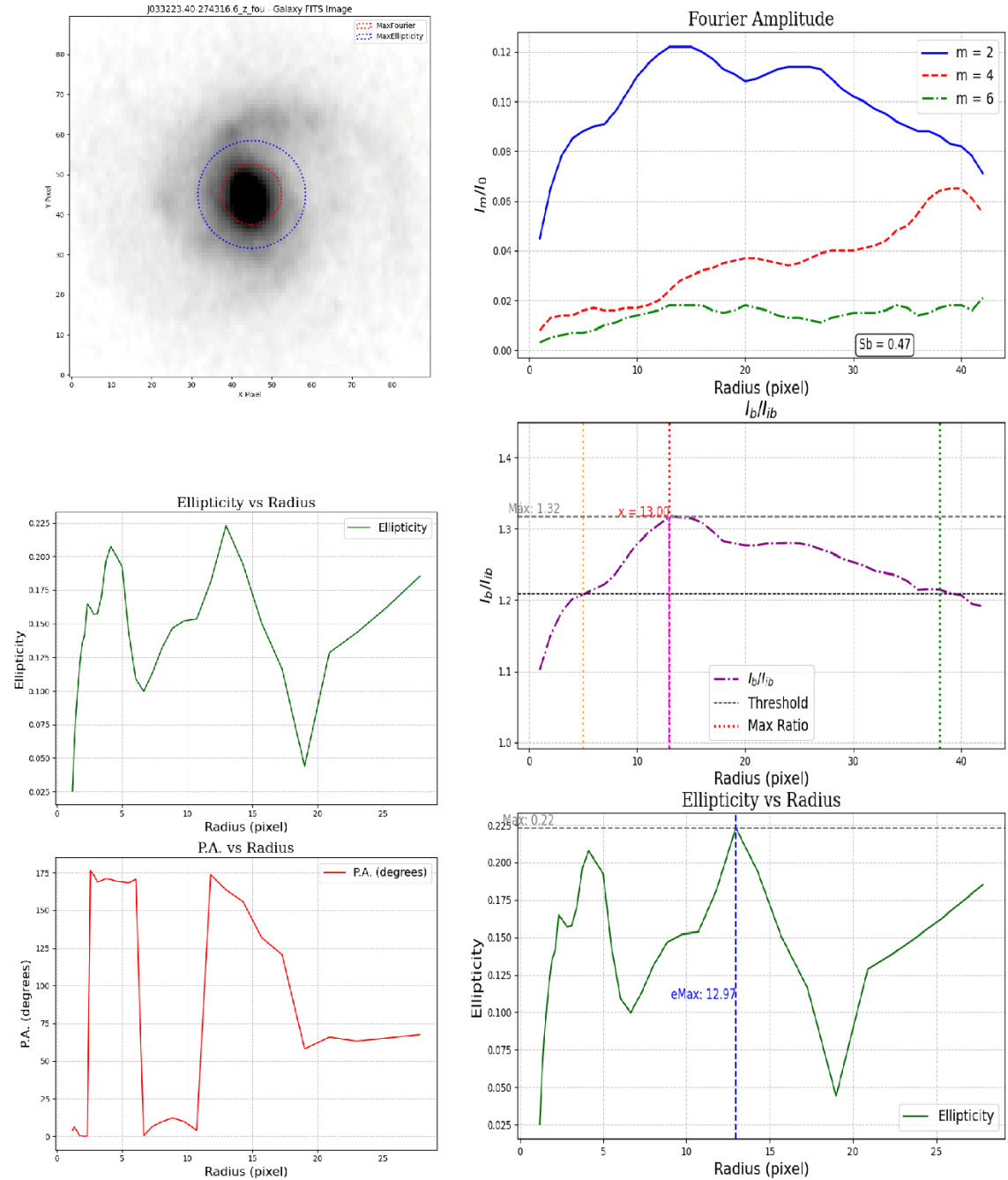

Galaxia barrada: cumple con los criterios del análisis de Fourier, revelando una estructura barrada significativa.

**Galaxia: J033233.08-275123.9**

**Filtro: z_band_GOODS**

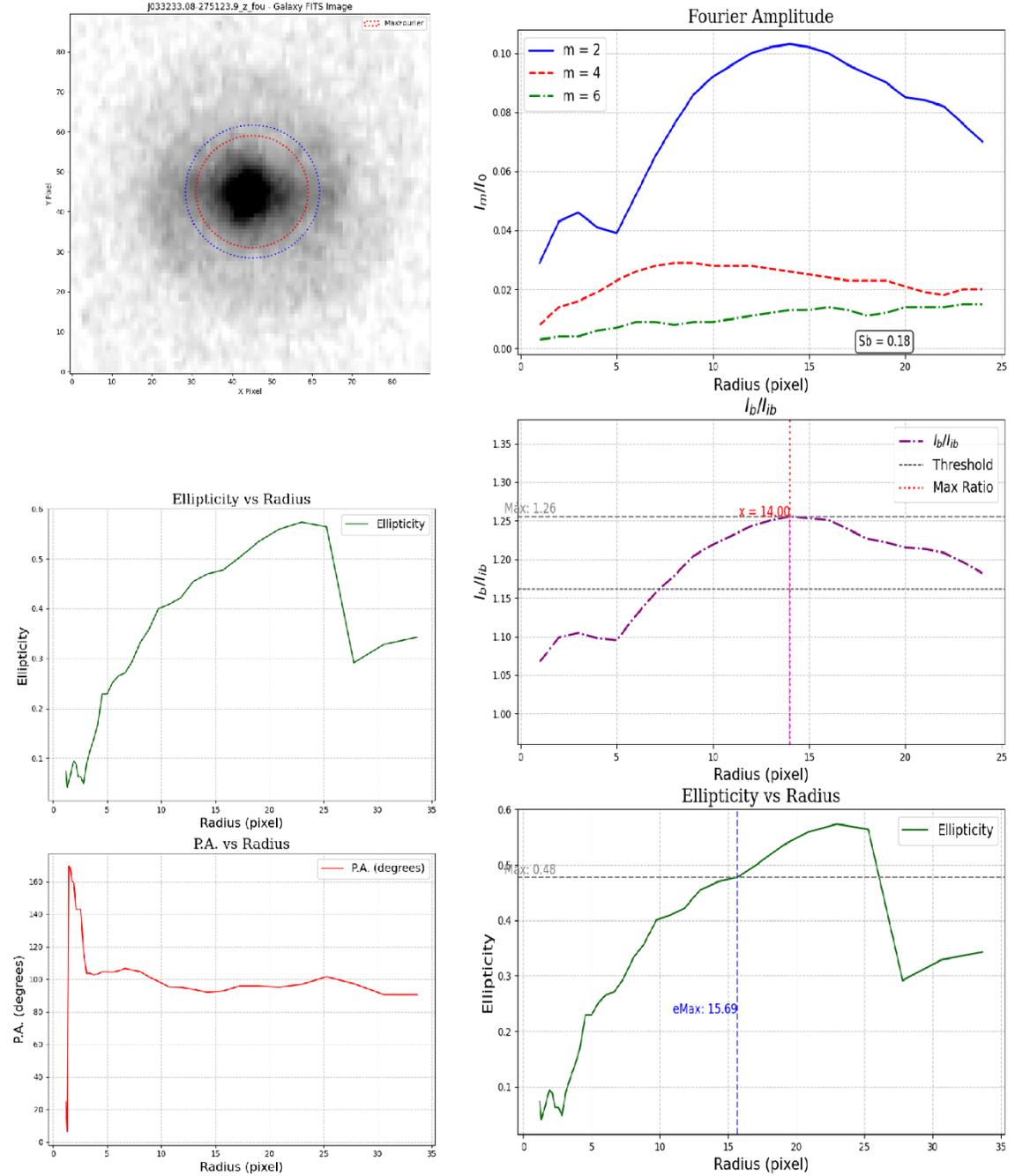

Galaxia barrada: cumple con los criterios del análisis de Fourier, revelando una estructura barrada significativa.

**Galaxia: J033224.06-274911.4**

**Filtro: z_band_GOODS**

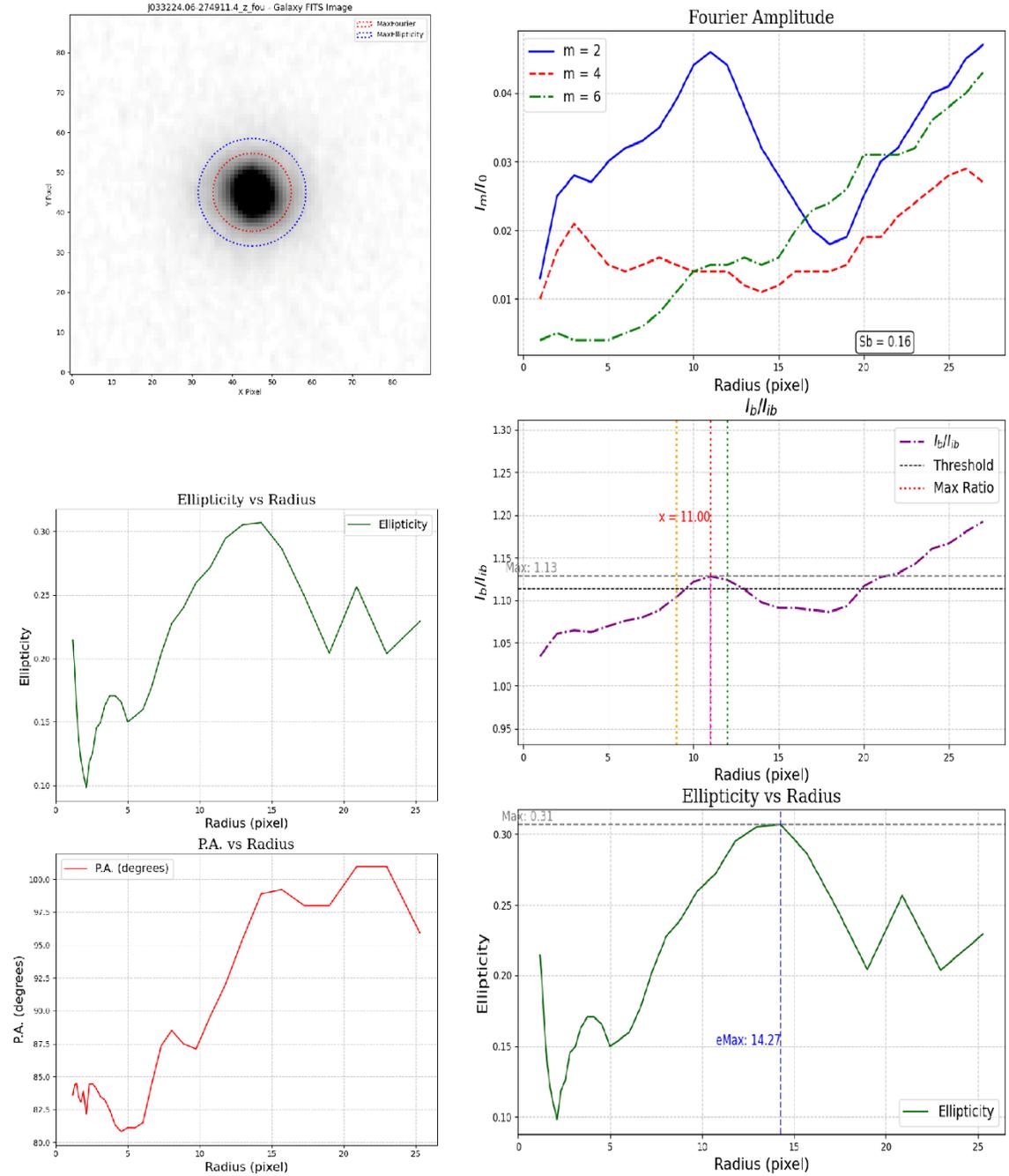

Galaxia no barrada: no cumple con los criterios del análisis de Fourier, descartando la presencia de una barra estelar.

**Galaxia: J033230.03-274347.3**

**Filtro: z_band_GOODS**

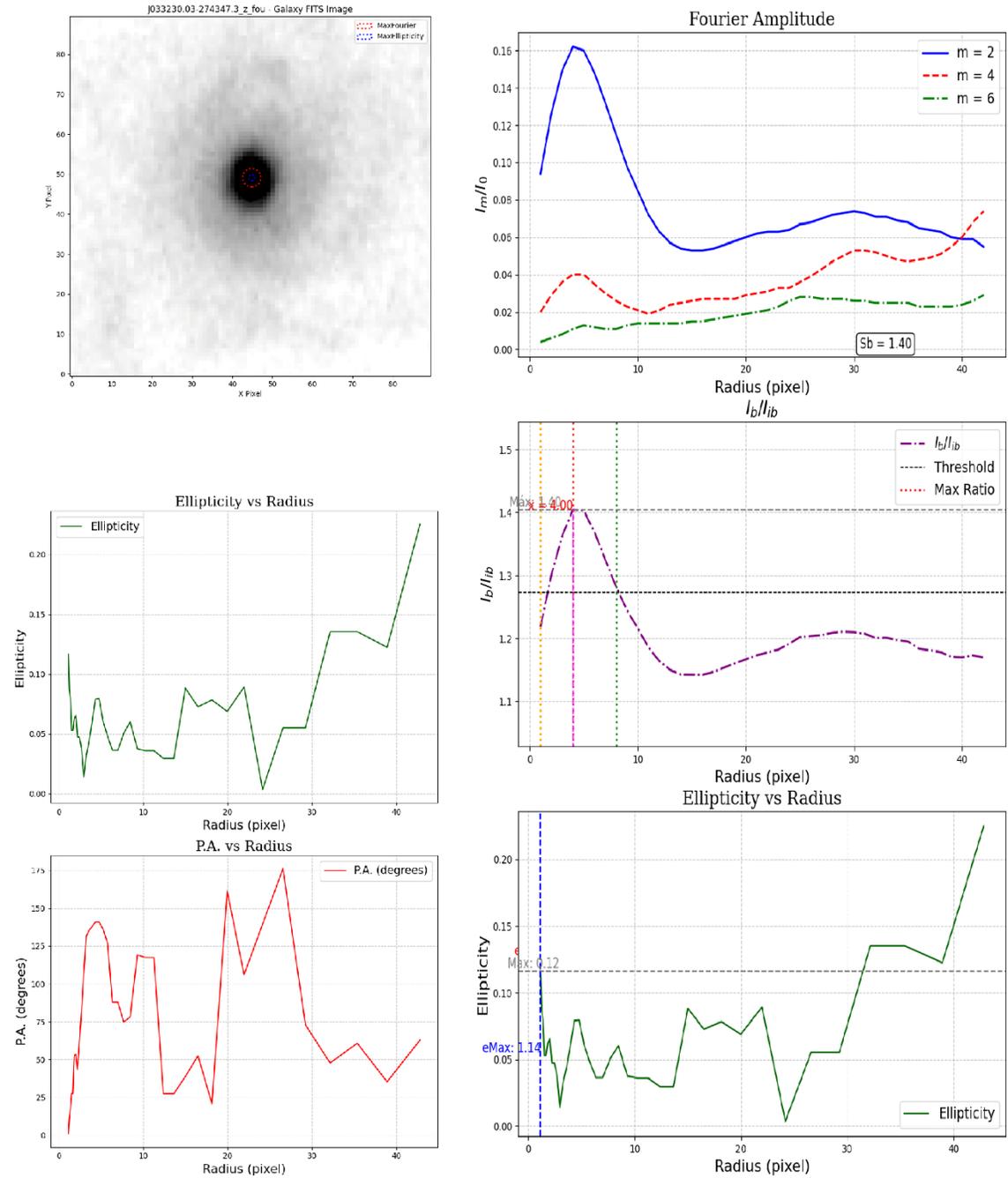

Galaxia no barrada: no cumple con los criterios del análisis de Fourier, descartando la presencia de una barra estelar.

**Galaxia: J033219.68-275023.6**

**Filtro: z_band_GOODS**

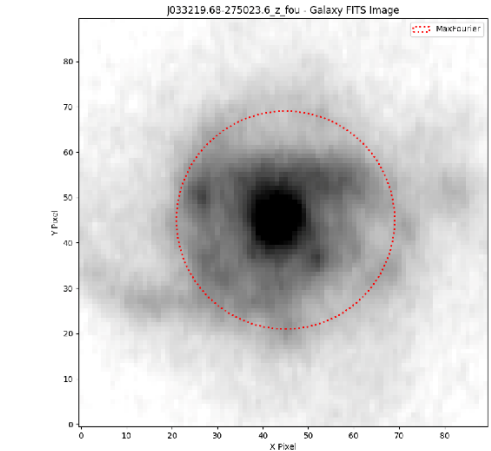

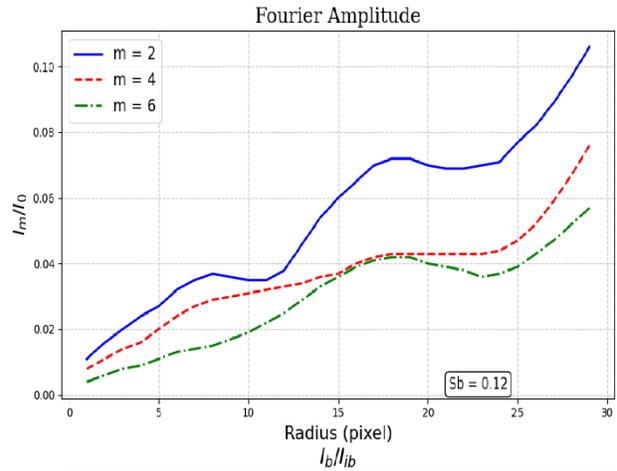

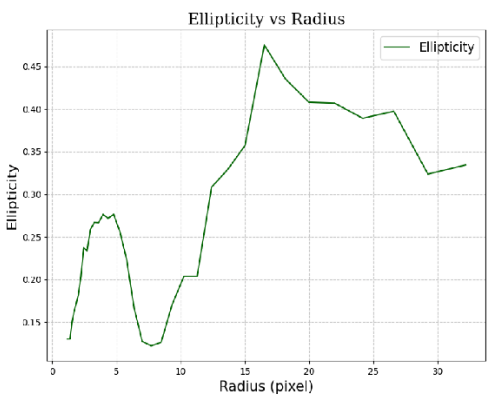

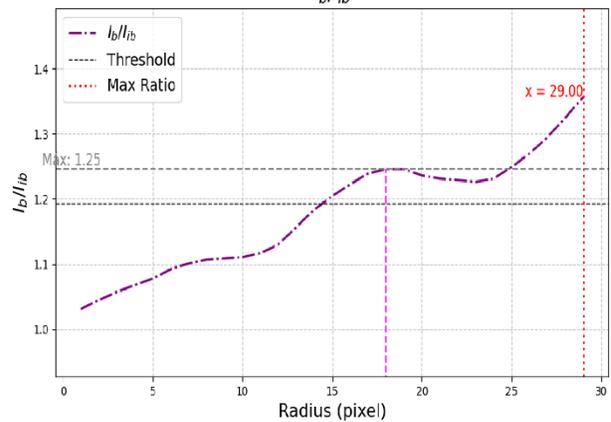

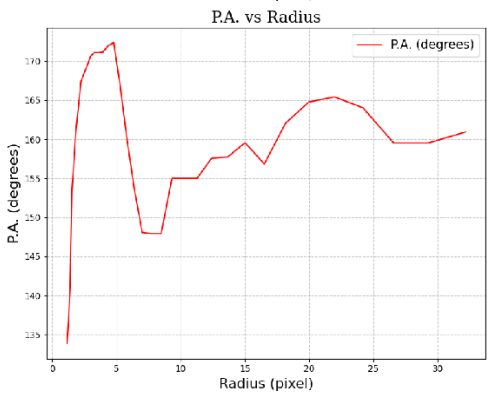

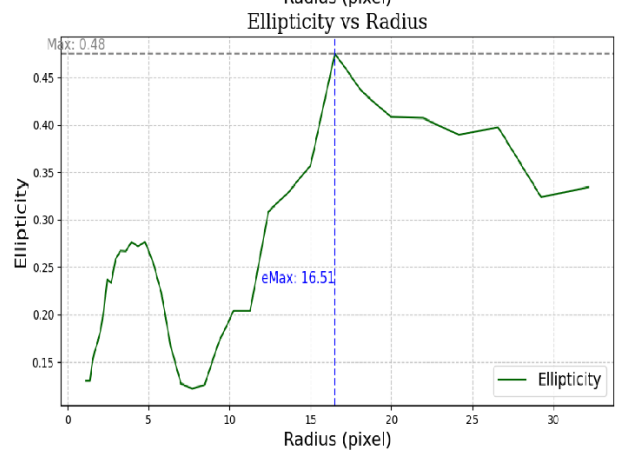

Galaxia no barrada: no cumple con los criterios del análisis de Fourier, descartando la presencia de una barra estelar.

# D.4. Galaxias Lenticulares a $z = 0.7$

**Galaxia: J033205.97-274601.4**

**Filtro: z_band_GOODS**

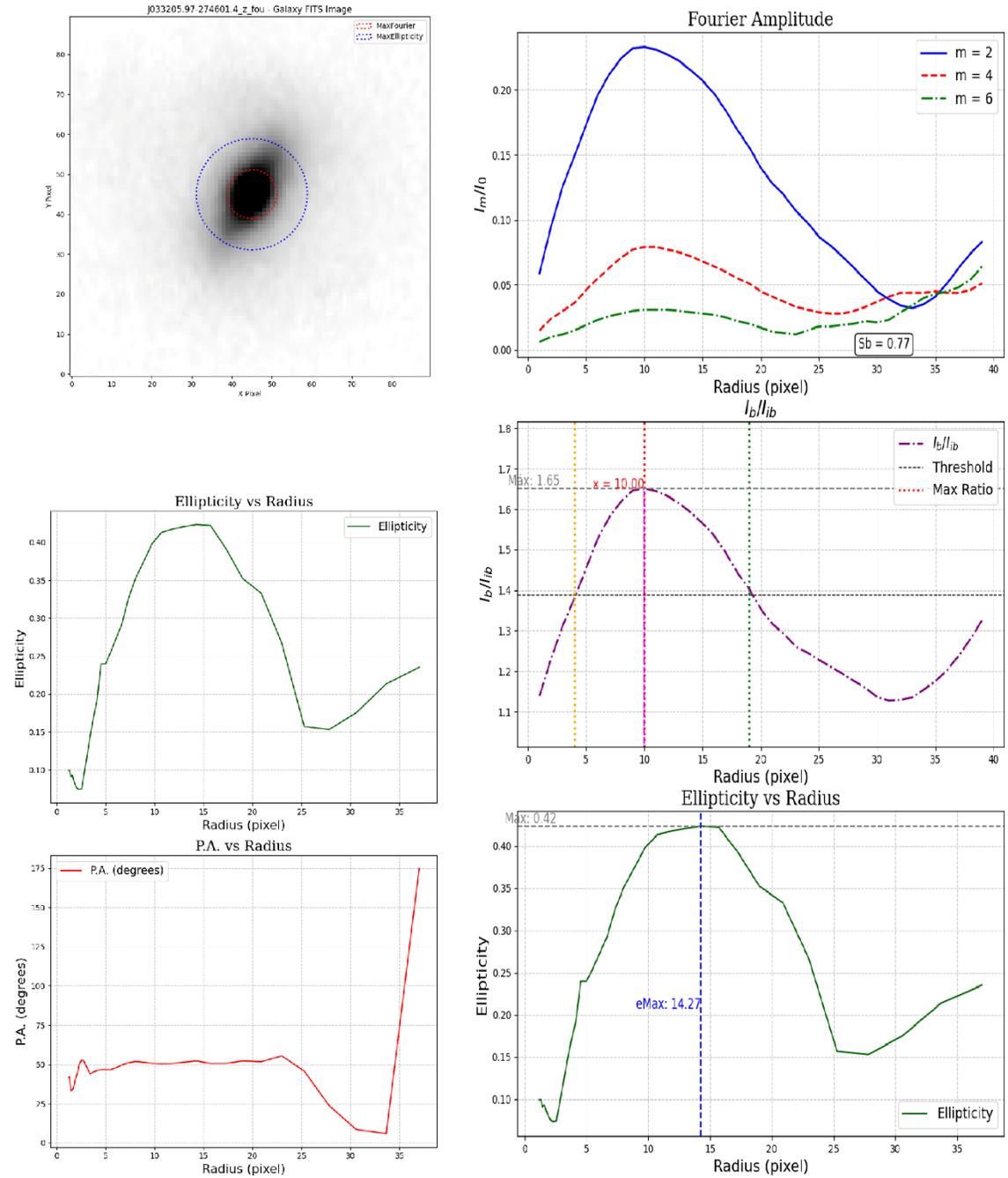

Galaxia barrada: cumple con los criterios del análisis de Fourier, revelando una estructura barrada significativa.

**Galaxia: J033212.31-274527.4**

**Filtro: z_band_GOODS**

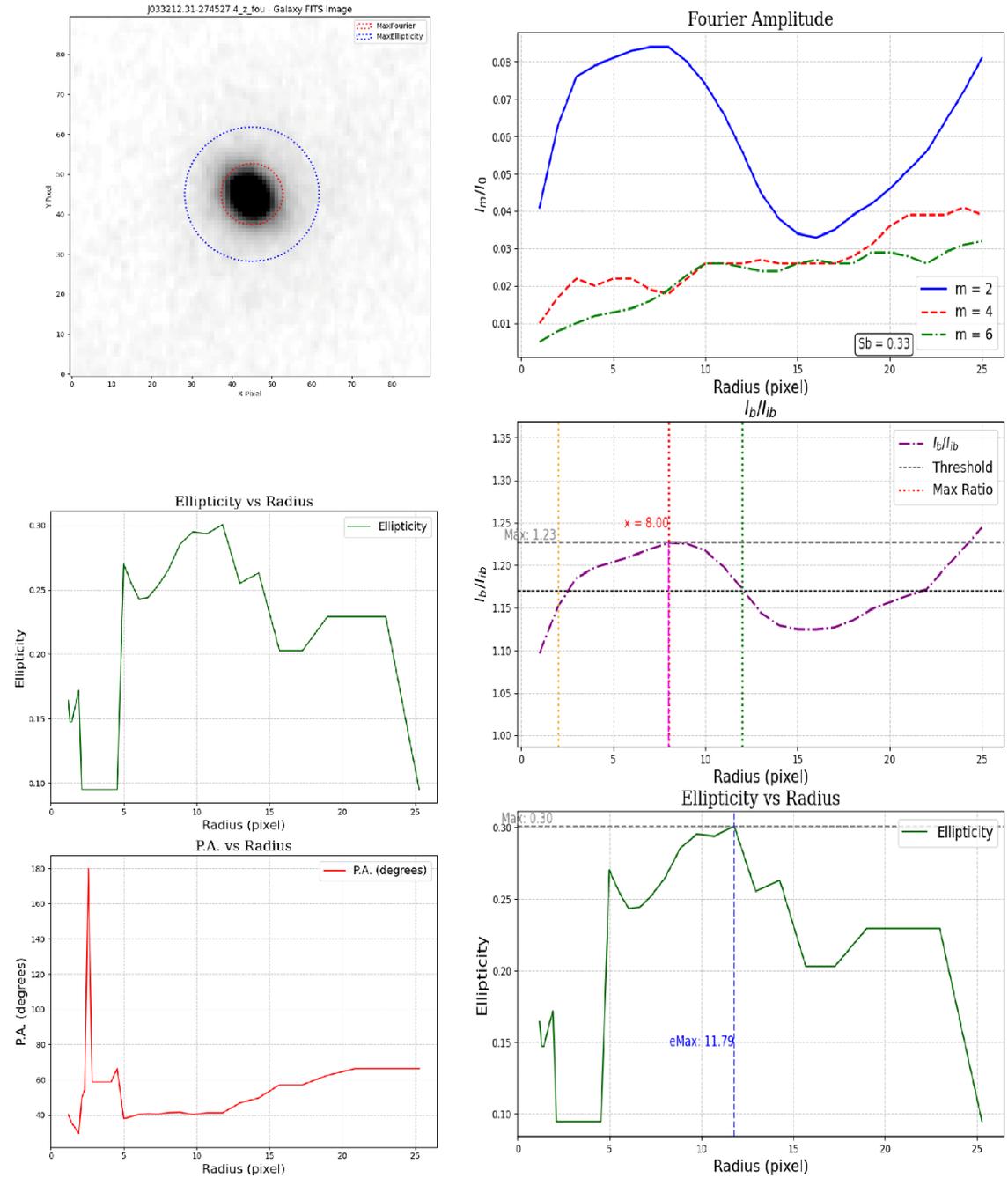

Galaxia barrada: cumple con los criterios del análisis de Fourier, revelando una estructura barrada significativa.

**Galaxia: J033230.07-275140.6**

**Filtro: z_band_GOODS**

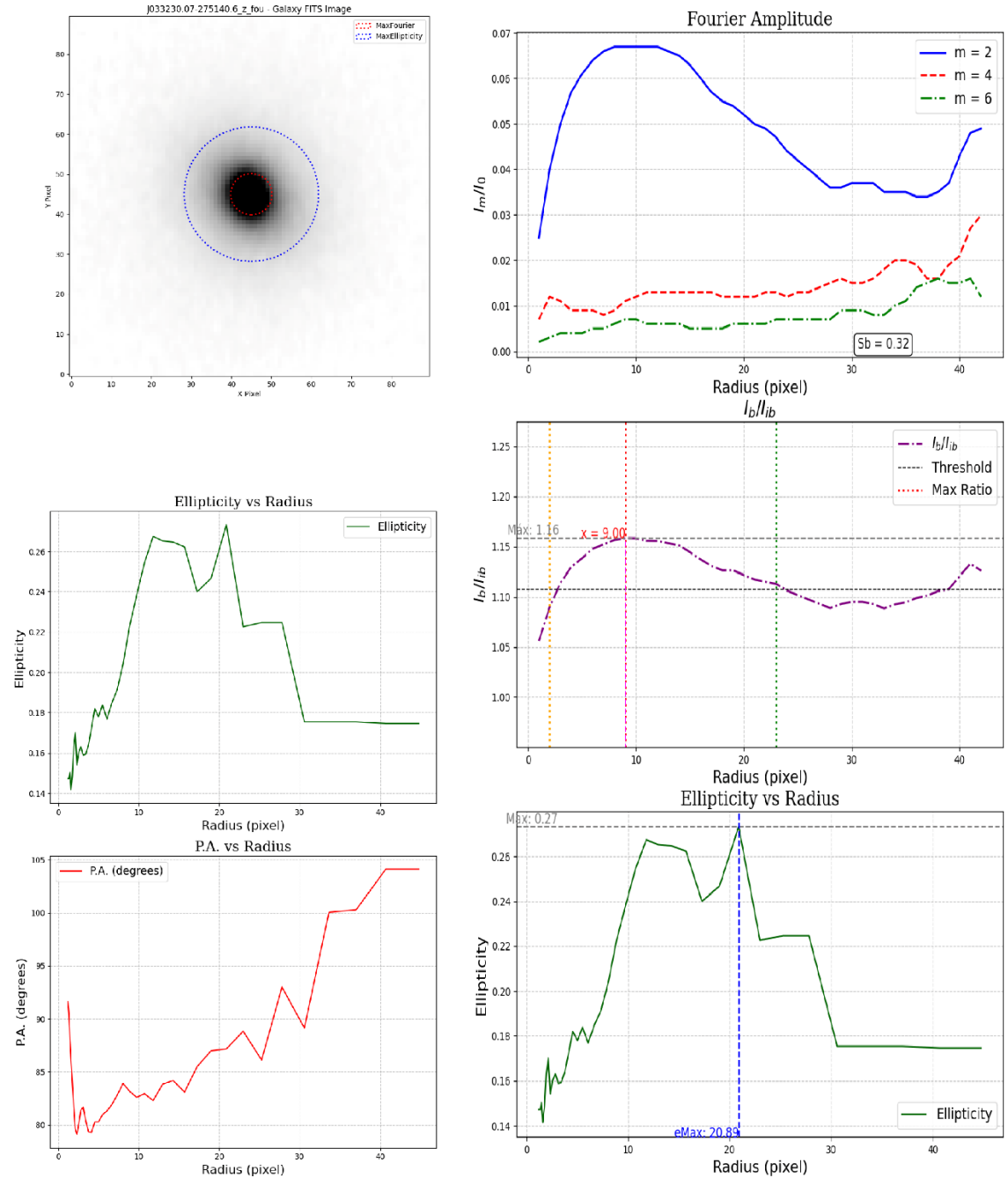

Galaxia barrada: cumple con los criterios del análisis de Fourier, revelando una estructura barrada significativa.

**Galaxia: J033235.02-275405.2**

**Filtro: z_band_GOODS**

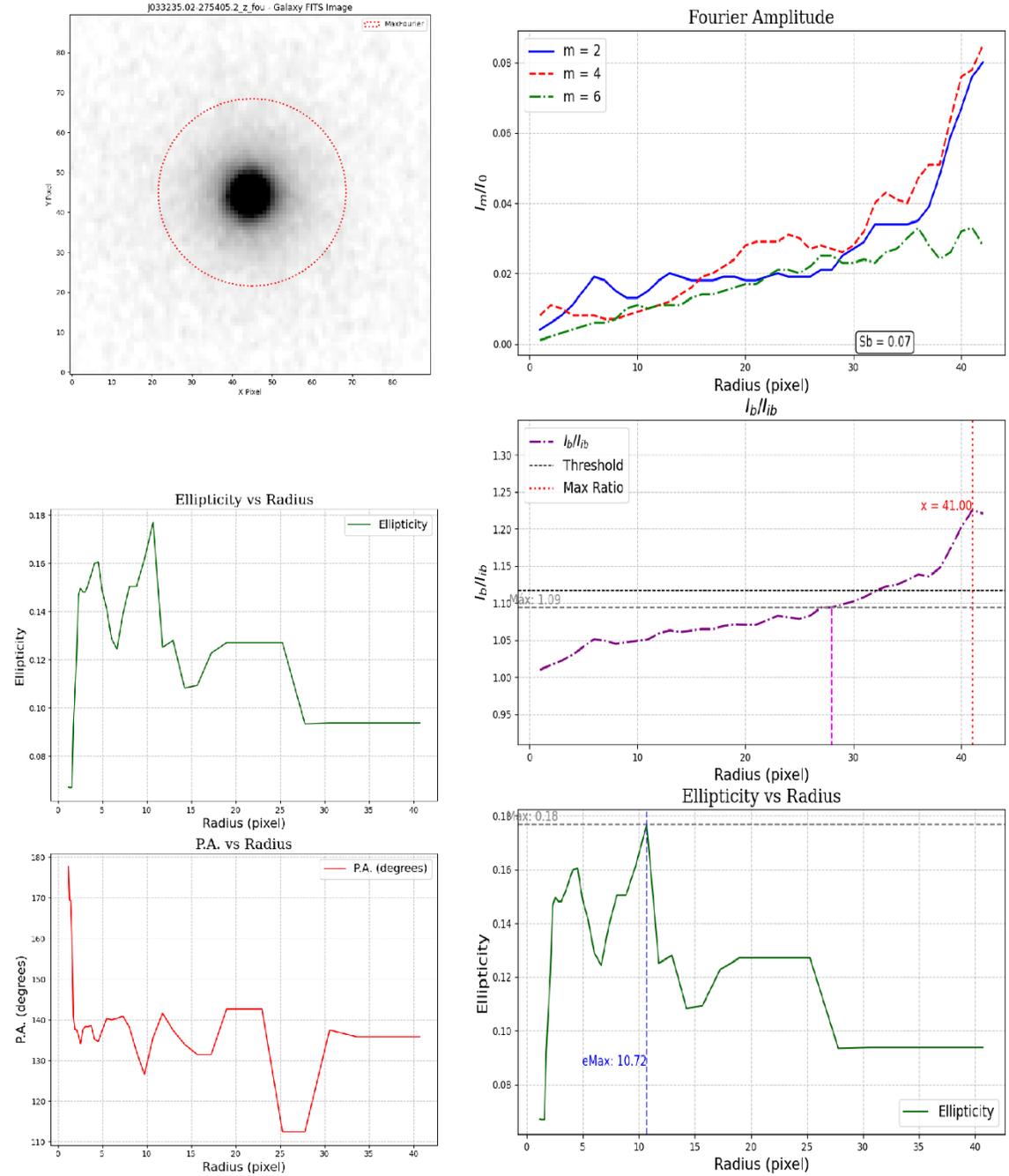

Galaxia no barrada: no cumple con los criterios del análisis de Fourier, descartando la presencia de una barra estelar.

**Galaxia: J033238.27-275354.4**

**Filtro: z_band_GOODS**

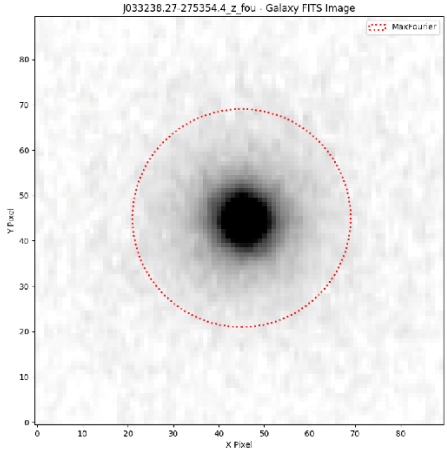

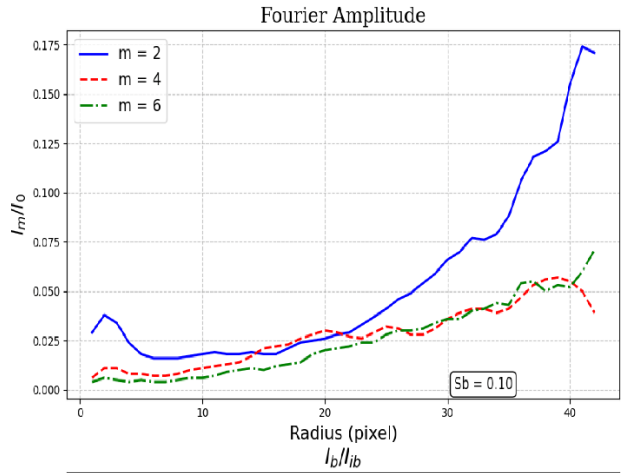

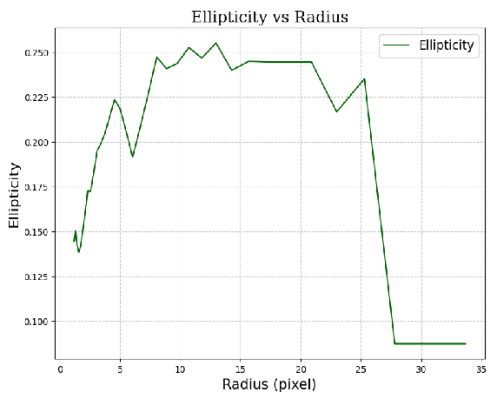

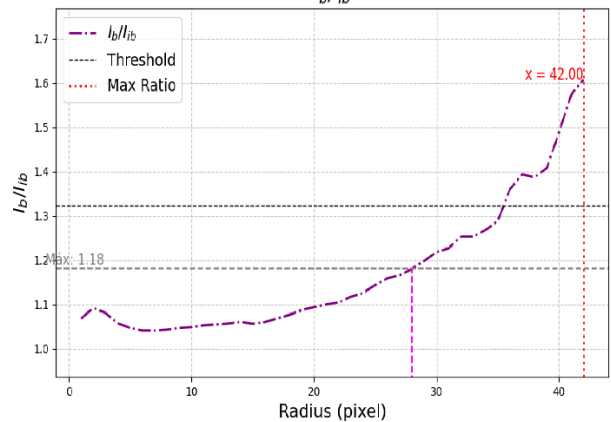

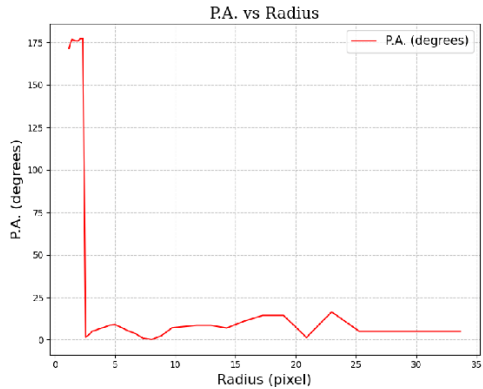

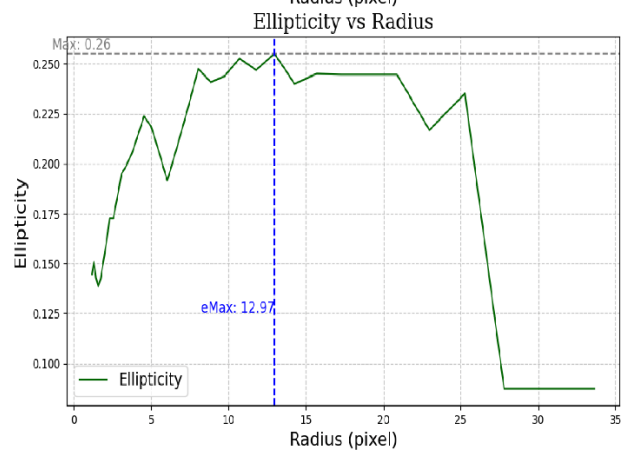

Galaxia no barrada: no cumple con los criterios del análisis de Fourier, descartando la presencia de una barra estelar.

**Galaxia: J033242.39-274153.2**

**Filtro: z_band_GOODS**

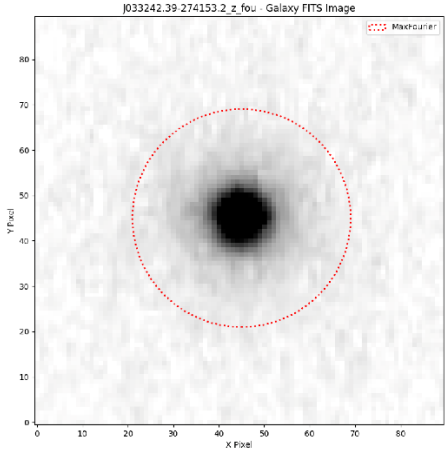

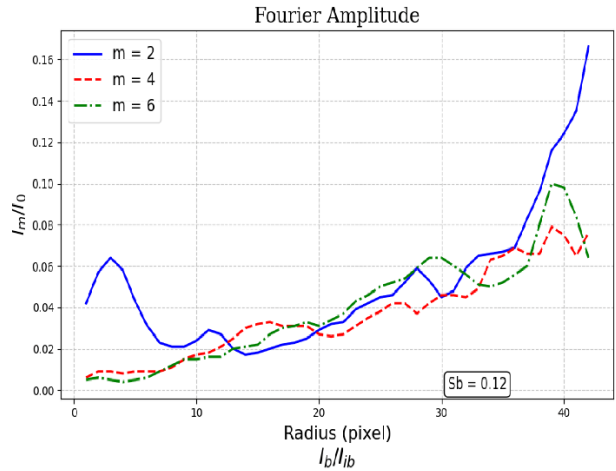

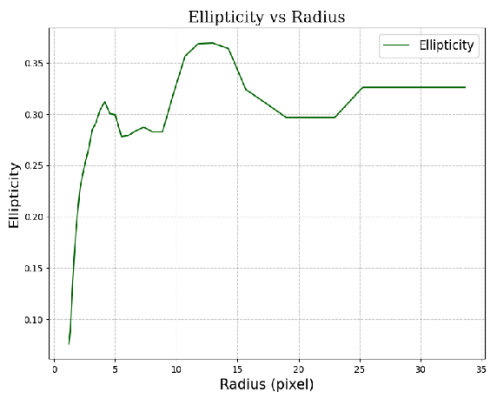

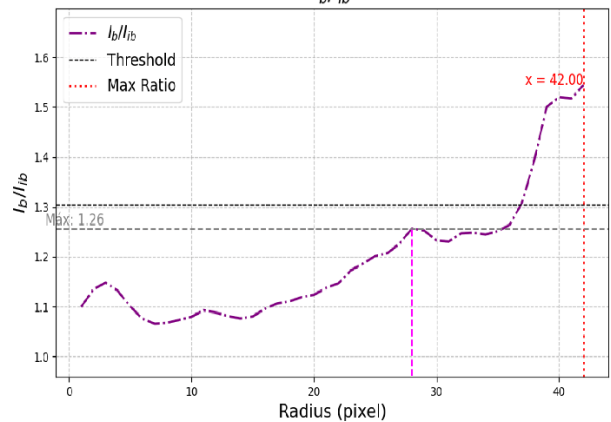

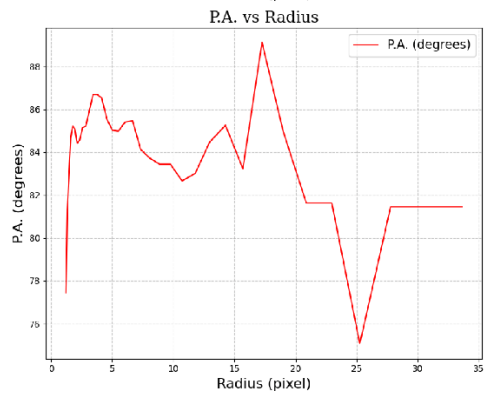

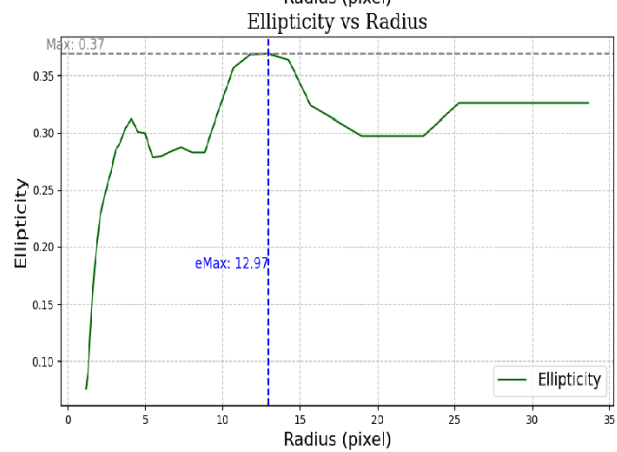

Galaxia no barrada: no cumple con los criterios del análisis de Fourier, descartando la presencia de una barra estelar.

# E. Pasantía de Investigación

Esta pasantía de investigación fue financiada por la Secretaria Nacional de Ciencias y Tecnología e Innovación (SENACYT) a través del proyecto "Formación de Capacidades Investigativas en la Línea de Astronomía Extragaláctica". Como parte de este programa, realicé estancias en diversas instituciones de renombre internacional en México, Chile y Argentina, donde adquirí habilidades avanzadas en el análisis de datos astronómicos y participé en proyectos científicos que contribuyeron significativamente al desarrollo de mi tesis doctoral. Durante estas pasantías, tuve acceso a recursos de clase mundial y colaboré con investigadores expertos, lo que enriqueció mi formación académica y profesional, fortaleciendo las bases de mis investigaciones en Astronomía Extragaláctica.

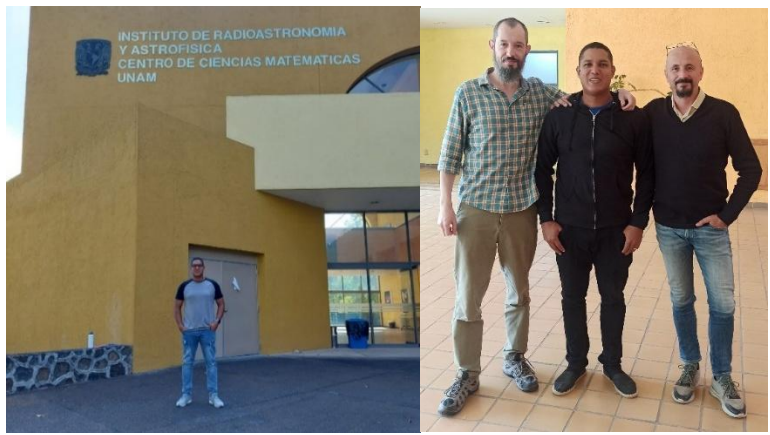

Figura E.1. Con parte del grupo de Extragaláctica de IRyA.

La primera parte de mi pasantía de investigación se desarrolló en México, donde tuve la oportunidad de colaborar con instituciones académicas y científicas de gran relevancia. Mi estancia principal fue en el Instituto de Radioastronomía y Astrofísica (IRyA) de la Universidad Nacional Autónoma de México (UNAM), bajo la supervisión del Dr. Bernardo Cervantes Sodi. Con el apoyo del Programa de Apoyo a Proyectos de Investigación e Innovación Tecnológica (PAPIIT) IN108323 (Galaxias de bajo brillo superficial, medio ambiente, evolución y el caso de galaxias barradas), de la Dirección General de Asuntos del Personal Académico (DGAPA) de la Universidad Nacional Autónoma de México (UNAM). Participé en reuniones semanales con su equipo, lo que me permitió integrarme en proyectos de investigación enfocados en astronomía extragaláctica. Además, asistí a varios cursos y capacitaciones que ampliaron significativamente mis conocimientos en el análisis de datos astronómicos.

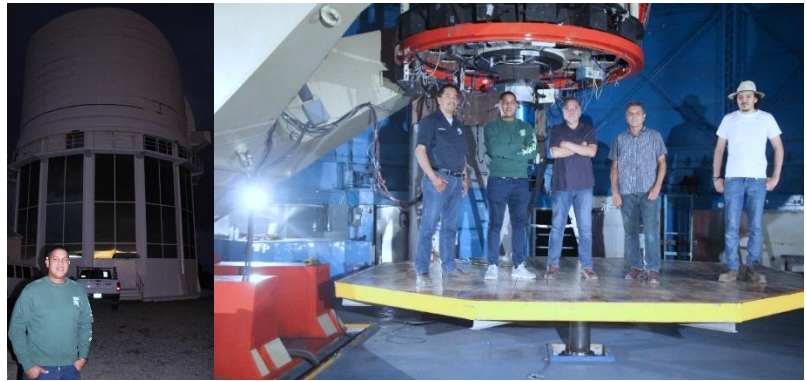

Figura E.2. Izquierda junto Edificio del Observatorio Astronómico Guillermo Haro (OAGH); y derecha el Telescopio de 2.1 m junto al personal que opera las sesiones de observación.

Un aspecto clave de mi experiencia en México fue la visita a observatorios astronómicos. En el Instituto de Geofísica de la UNAM (IGUM), visité el Observatorio de Centelleo Interplanetario MEXART, donde conocí los avances en la investigación del clima espacial y la propagación de tormentas solares. Posteriormente, viajé a Hermosillo, Sonora, donde fui recibido por el Dr. Lorenzo Olguín y el Dr. Julio Saucedo en la Universidad de Sonora. Durante esta visita, recorrí el Observatorio Astronómico de la Universidad de Sonora y observé de primera mano su trabajo en la investigación de objetos astronómicos nocturnos. Además, visité el Observatorio Astronómico Guillermo

Haro en Cananea, donde tuve la oportunidad de participar en sesiones de observación nocturna utilizando el telescopio de 2.1 metros de diámetro, lo que me permitió profundizar en las técnicas de observación profesional y en el procesamiento de datos astronómicos.

A lo largo de mi pasantía, también participé en diversos cursos y capacitaciones que fueron clave para mi formación académica. Uno de los cursos más destacados fue Programación en Python para astrónomos, impartido por el MCs Daniel Díaz. Este curso me permitió profundizar en herramientas esenciales como NumPy, Matplotlib y Astropy, que son fundamentales para el análisis y la visualización de datos astronómicos. La combinación de estas herramientas me proporcionó una base sólida para realizar cálculos numéricos complejos y crear representaciones gráficas precisas de los resultados obtenidos en las investigaciones. Otra área en la que recibí capacitación fue en radioastronomía, específicamente en multifrecuencia en la banda de radio, dirigida por la Dra. Aina Palau. Durante estas sesiones, aprendí los principios básicos de la radioastronomía y su importancia en la detección de señales provenientes de fenómenos astronómicos, como la formación de estrellas y la dinámica de sistemas protoplanetarios. Posteriormente, recibí capacitación en la banda de infrarrojo, a cargo del Dr. Jacopo Fritz, donde estudiamos la distribución de energía espectral (SED) de galaxias. Este entrenamiento me brindó una comprensión más profunda sobre cómo utilizar la banda infrarroja para investigar objetos celestes, detectar regiones de formación estelar, y estudiar la evolución de galaxias a través de datos obtenidos en esas longitudes de onda. Además, el Dr. Gustavo Bruzual, experto en poblaciones estelares, me brindó un curso sobre poblaciones estelares, en el que estudiamos la síntesis de sus propiedades y la evolución espectral de las galaxias. Esta capacitación me permitió entender cómo las propiedades físicas de las galaxias, como su masa estelar y el contenido de metales, pueden derivarse a partir de modelos espectrales.

Esta experiencia en México fue profundamente enriquecedora tanto en términos académicos como profesionales. Me permitió adquirir conocimientos avanzados en astronomía, establecer colaboraciones con investigadores de renombre, y participar en proyectos científicos de gran relevancia.

Esta pasantía fue una experiencia enriquecedora tanto a nivel académico como profesional. Me permitió adquirir conocimientos avanzados en astronomía, establecer colaboraciones con investigadores de renombre, y participar en proyectos científicos de gran relevancia.

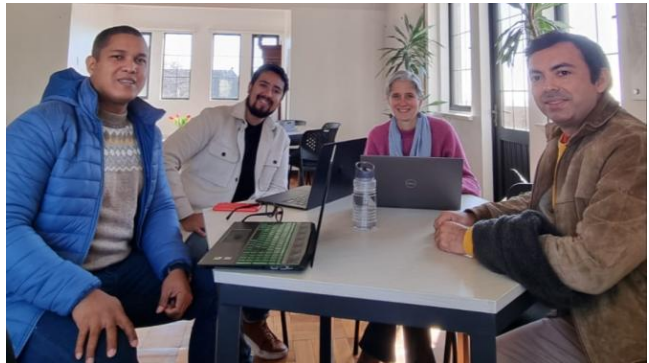

Figura E.3. Reunión grupal de grupo de Extragaláctica de DINACE-UTP con parte del grupo de USM.

Después de concluir mi pasantía en México, continué con la segunda parte en Chile y Argentina, donde llevé a cabo diversas actividades académicas y científicas en colaboración con reconocidas instituciones. Mi estancia principal fue en el Departamento de Física de la Universidad Técnica Federico Santa María (USM) en Valparaíso, donde trabajé bajo la supervisión del Dr. Hugo Alarcón y la Dra. Yara Jaffé. Me enfoqué en el estudio de la formación de barras en galaxias de campo y la evolución galáctica. Participé en reuniones grupales y coloquios con investigadores doctorales y posdoctorales, lo que me permitió adquirir habilidades avanzadas en técnicas de análisis de datos astronómicos. Estando en Chile, tuve la oportunidad de visitar dos de los observatorios astronómicos más importantes del mundo: el Very Large Telescope (VLT) en el Observatorio Paranal y el Atacama Large Millimeter/submillimeter Array (ALMA), ambos ubicados en el desierto de Atacama, Chile. Estas visitas me brindaron una comprensión profunda del funcionamiento de instalaciones astronómicas de vanguardia. En el VLT, conocí los detalles técnicos sobre el uso de telescopios ópticos de gran tamaño para la observación de galaxias distantes. En ALMA, observé cómo el conjunto de antenas de alta precisión permite estudiar el cosmos en longitudes de onda milimétricas y submilimétricas, lo cual es crucial para investigaciones sobre la formación estelar y la evolución de galaxias.

Además, recibí una capacitación intensiva sobre el uso de Illustris en Jupyter Notebook, por el Dr. Diego Pallero, una de las simulaciones más avanzadas en cosmología y astrofísica, que modela la formación y evolución de galaxias a diferentes escalas. Aprendí a manejar esta herramienta para generar datos y visualizaciones, lo que me permitió comparar los resultados de las simulaciones con observaciones astronómicas reales. Este conocimiento será fundamental para nuestras investigaciones futuras en Astronomía Extragaláctica.

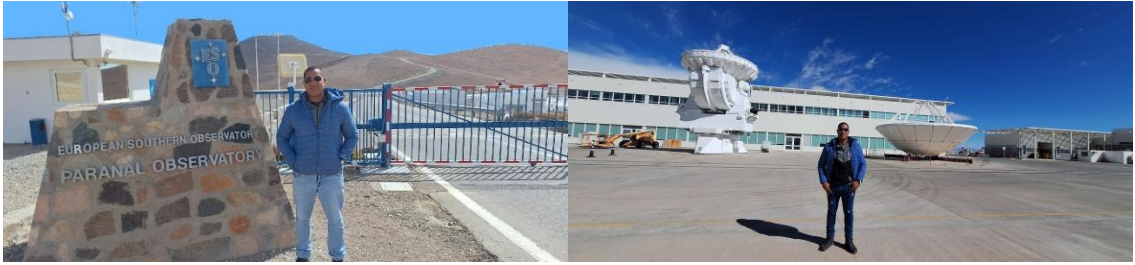

Figura E.4.Izquierda:Visita al telecopio Very Large Telescope Project (VLT); y derecha: el Atacama Large Millimeter/submillimeter Array (ALMA)

Posteriormente, me trasladé a Argentina, donde visité la Universidad Nacional de Cuyo en Mendoza y colaboré con la Dra. María Belén Planes en simulaciones de impactos de partículas en cometas y polvo planetario. También visité la Universidad de Mendoza, donde tuve la oportunidad de discutir temas relacionados con heliofísica y el estudio del clima espacial, así como la física de materiales aplicada a la investigación planetaria.

Esta pasantía en México, Chile y Argentina fue profundamente enriquecedora tanto en términos académicos como profesionales, y ha sido fundamental para el desarrollo de mi tesis doctoral. La oportunidad de adquirir conocimientos avanzados en astronomía extragaláctica, combinada con el acceso a recursos de clase mundial y la colaboración con investigadores de renombre, ha fortalecido mi formación académica y mi capacidad para abordar desafíos científicos complejos.

En México, mi participación en proyectos científicos de relevancia internacional y mi acceso a observatorios e instalaciones de primer nivel me proporcionaron una visión amplia de las técnicas y metodologías aplicadas en la investigación astrofísica. Estas experiencias me permitieron adquirir habilidades clave en el análisis de datos astronómicos, lo que será esencial para avanzar en mi investigación doctoral. El trabajo colaborativo con expertos en el Instituto de Radioastronomía y Astrofísica (IRyA) de la UNAM me brindó una plataforma sólida para profundizar en temas como la evolución galáctica y la formación de barras, elementos clave en mi tesis.

Asimismo, la pasantía en Chile y Argentina consolidó mi formación, permitiéndome aplicar lo aprendido en escenarios internacionales de vanguardia. Las visitas a observatorios como el Very Large Telescope (VLT) y el Atacama Large Millimeter/submillimeter Array (ALMA), así como mi participación en simulaciones astrofísicas avanzadas como Illustris, ampliaron mi entendimiento sobre el comportamiento y la evolución de las galaxias. Esta capacitación, sumada a mi interacción con expertos en heliofísica y física de materiales en Argentina, complementó de manera crucial mi investigación doctoral.

Las habilidades adquiridas y los vínculos internacionales establecidos durante estas pasantías no solo son un aporte valioso para el desarrollo de mi tesis, sino que también abren la puerta a futuras colaboraciones que fortalecerán la capacidad de Panamá en el campo de la astronomía extragaláctica. Todo este aprendizaje será esencial para aplicar enfoques innovadores en mi trabajo de investigación y posicionar nuestras investigaciones dentro de la comunidad científica internacional.

# F. Publicaciones

Los primeros resultados de esta investigación fueron publicados en la revista *Research in Astronomy and Astrophysics* bajo el título: "Bar Presence in Local Galaxies: Dependence on Morphology in Field Galaxies". La publicación fue aceptada en octubre de 2024, destacando la relevancia del estudio y aportando valiosas observaciones sobre la evolución de las galaxias barradas en el universo local.

**IOP**science    Q    Journals ▾   Books   Publishing Support   Login ▾

Research in Astronomy and Astrophysics

ACCEPTED MANUSCRIPT

## Bar Presence in Local Galaxies: Dependence on Morphology in Field Galaxies

Manuel Alejandro Chacón[1], Rodney Delgado Serrano[2] and Bernardo Cervantes Sodi[3]

📄 Accepted Manuscript PDF

▾ Article and author information

**Article metrics**
4 Total downloads

**Permissions**
Get permission to re-use this article

**Share this article**
✉ f 𝕏 ᗰ

Abstract

## Abstract

We analyzed the fractions of barred galaxies in the local universe using a volume-limited sample of galaxies from the Sloan Digital Sky Survey Data Release 3. We examined 116 field galaxies with redshifts between 0.0207 and 0.030, using r and z-band images. Overall, the bar fraction was 26% in the r-band and 19% in the z-band. For distinct morphological groups, barred spiral galaxies had fractions of 33% in the r-band and 22% in the z-band, while barred lenticular galaxies had 25% in the r-band and 12% in the z-band. We observed that the bar fraction in spiral galaxies increases for stellar masses log(M∗/M☉) > 10.5 and for galaxies with red colors (u − r) > 2.0. Additionally, most barred galaxies have a bulge-to-total ratio B/T ≤ 0.2. Our results indicate that the bar fraction is more dependent on internal morphology than on the galaxy environment.

Export citation and abstract   BibTeX   RIS

**You may also like**

Además de los primeros resultados publicados en revistas, los avances de esta investigación también fueron presentados en la conferencia 2019 International Engineering, Sciences and Technology Conference (IESTEC), bajo el título: "Study about the Bars Detection in Spiral Galaxies Using the Isophotal Method".

Como parte de la exploración de nuevas líneas de investigación en astronomía, participé como coautor en el diseño y construcción de una antena para la detección de meteoros, presentando los resultados en la 2022 International Engineering, Sciences and Technology Conference (IESTEC), bajo el título: "Design and Construction of Antenna for Meteor Detection"

# Referencias Bibliográficas